%% file: thesis.tex
\documentclass{book}
\usepackage[dvips]{graphicx,color}

\def\baselinestretch{1.5} 
\newcommand{\delr}{{\stackrel{\rightarrow}{\nabla}}}
\newcommand{\dell}{{\stackrel{ \leftarrow}{\nabla}}}
\newcommand{\bfbet}{{\mbox{\boldmath $\beta$}}}
\newcommand{\bftau}{{\mbox{\boldmath $\tau $}}}
\newcommand{\mytitle}[1]{\vspace{0.5in}\renewcommand{\baselinestretch}{2}{\huge\bf\begin{center}{#1}\end{center}}}
\newcommand{\authors}[1]{\vspace{0.2in}\renewcommand{\baselinestretch}{1.3}{\large\begin{center}{#1}\end{center}}}
\newcommand{\abstract}[1]{\vspace{0.5in}\renewcommand{\baselinestretch}{1.3}\noindent\ignorespaces {#1} \par\vfill\pagebreak}

\begin{document}
\pagenumbering{roman}
\begin{titlepage} 
\setcounter{page}{1}
\thispagestyle{plain}
\mytitle{
\phantom{a}
\vspace{-1.25truein}
Covariant Lagrangian Methods of Relativistic Plasma Theory
}
\vspace{-0.250truein}
\authors{
By\\
\smallskip
BRUCE MICHAEL BOGHOSIAN\\
B.S. (Massachusetts Institute of Technology) 1978\\
M.S. (Massachusetts Institute of Technology) 1978\\
\medskip
DISSERTATION\\
\smallskip
Submitted in partial satisfaction of the requirements for the degree of\\
DOCTOR OF PHILOSOPHY\\
in\\
\smallskip
Engineering -- Applied Science\\
\smallskip
in the\\
GRADUATE DIVISION\\
of the\\
UNIVERSITY OF CALIFORNIA\\
DAVIS\\
\smallskip
\begin{flushleft}Approved:\end{flushleft}
\rule{3.9truein}{.012truein}\\
\rule{3.9truein}{.012truein}\\
\rule{3.9truein}{.012truein}\\
\smallskip
Committee in Charge\\
\medskip
1987
}
\end{titlepage}
\nopagebreak


\chapter*{\phantom{Dedication}}
\setcounter{page}{2}
\thispagestyle{plain}
\begin{center}
Copyright by\\
Bruce Michael Boghosian\\
1987
\end{center}

\newpage

\setcounter{page}{3}
\thispagestyle{plain}
\mytitle{
\vspace{-1.0in} 
Covariant Lagrangian Methods of Relativistic Plasma Theory
}
\authors{
Bruce M. Boghosian\\ 
University of California at Davis\\
Department of Applied Science\\
Livermore, California 94550\\ 
\bigskip 
May, 1987\\
\bigskip
\bigskip
ABSTRACT
}
\abstract{
The relativistic electromagnetic projection operators discovered by
Fradkin are used to obtain a covariant decomposition of the motion of a
relativistic charged particle into parallel motion and perpendicular
gyration.  The Lagrangian Lie transform method of Littlejohn is used to
achieve a transformation to guiding-center coordinates in which the
rapid oscillatory motion is removed.  The natural guiding-center Poisson
bracket structure and Hamiltonian are derived.  The guiding-center
equations of motion are presented to one order higher than the usual
drifts, and the correction to the gyromomentum is given.  Correspondence
with the usual noncovariant results, as given by Northrop, is
demonstrated.

It is possible to add one or more eikonal wave perturbations to the
Lagrangian action for a single particle before performing the
guiding-center transformation.  It is shown that such perturbations can
be written in manifestly gauge-invariant form in guiding-center
coordinates; this observation allows us to develop a manifestly
gauge-invariant oscillation-center theory to arbitrarily high order.  In
this way, again using Lagrangian Lie transforms, we obtain the
ponderomotive Hamiltonian.

By summing the guiding-center Lagrangian action over the full
distribution of guiding centers and adding the Maxwell action, we obtain
the total action of a guiding-center plasma.  Upon variation of this
total action, we find a self-consistent set of covariant relativistic
kinetic and field equations; from these we can identify the
guiding-center current density and the guiding-center magnetization.
Upon application of Noether's theorem, the total action yields covariant
conservation laws for the momentum-energy and the angular momentum of a
relativistic guiding-center plasma; from these we can identify the
guiding-center stress-energy tensor and the guiding-center spin angular
momentum tensor.

By summing the Lagrangian action for a guiding/oscillation center over
the full distribution and adding the Maxwell action, variation yields
self-consistent relativistic kinetic and field equations for the plasma
in the wave field, including the dispersion relation for the wave; from
these we can identify the wave magnetization and susceptibility, thereby
demonstrating the K-$\chi$ theorem.  Noether's theorem then yields
conservation laws for the guiding-center plasma in the presence of a
wave field, including the wave contribution to the stress-energy and
spin angular momentum tensors.
}


\chapter*{\phantom{Dedication}}
\thispagestyle{plain}
\begin{center}
{\it To my family.}
\end{center}
\nopagebreak

\include{acknowledgements}

\tableofcontents
\include{introduction}
\include{mathprelim}
\include{relgc}
\include{reloc}
\include{relgcp}
\include{questions}

\include{biblio}
\appendix
\include{glossary}
\include{vecspaces}
\include{gyrofreq}
\include{specfuns}
\include{besssums}

\end{document}

%% file: acknowledgements.tex
\chapter*{Acknowledgements}
\setcounter{page}{6}
This thesis is the culmination of a six-year PhD program
at the University of California at Davis.  During this period,
I worked full-time as a plasma physicist in the Magnetic Fusion Energy
Division of the Lawrence Livermore National Laboratory.  Much of my research
was carried out in collaboration with the Plasma Theory group at the
Lawrence Berkeley Laboratory.  All of this gives rise to a long list of
people to thank.

We'll start with the personal one$\ldots$

First, and foremost, I would like to thank my wife, Laura, for her love,
patience and support during the course of my research.  Full-time employment
plus thesis research is a stressful combination, and I couldn't have done it
without her.

Then comes the thesis committee$\ldots$

More than anyone else, my thesis committee chairman, William A. Newcomb, taught
me how to think.  I consider myself privileged to have had the benefit of
Bill's insights on a variety of topics (from physics, to mathematics,
to philosophy, to jogging).  The precision of his thinking and the clarity
of his explanations will remain a role model for me throughout my career.

Most all of the research in this thesis was carried out with Allan Kaufman
and his Plasma Theory group at the Lawrence Berkeley Laboratory.  Allan has
been the principal pioneer of the action principle formulation of plasma physics,
which is the main underlying theme of this thesis.  He has the distinction of
being not only the founder of a new school of thought within plasma physics
that emphasizes a differential geometric approach to the classical problems of
the field, but also of being the founder of a new school of plasma physicists
among whom I am proud to be included.  Allan's
Socratic style of teaching was indispensible in helping me think my way through
the maze of concepts encountered in the course of my research.  He also
taught me the importance of a broadened perspective in physics research.  As a
result, I became familiar with many areas of physics and mathematics to which I
would not have otherwise been exposed, and I was able to apply much of that
knowledge in this thesis.

Gary Smith has been a source of learning and guidance for me from the very
beginning of my program of study.  When one first enters a field as 
complex as plasma physics, it is indispensible to
have someone who is willing to act as a tutor, answering the myriad
questions (some stupid, some not-so-stupid) that arise during the learning
process.  Gary has never failed me in this regard.  His uncanny intuition about
plasma waves and stability theory has shaped the way I think about the whole
field.  He always found time to
discuss things with me, and was always genuinely concerned about my welfare.
Wherever our paths may take us, I know that he will always be my friend.

Next, all the folks at Livermore$\ldots$

I would like to thank my supervisors at the
Lawrence Livermore National Laboratory,
Gus Carlson from 1978 to 1982 and L. Donald Pearlstein from 1982 to the
present, for generously allowing me the time I needed to complete this thesis.

The entire Mirror Theory and Computations group at Livermore has been a
great source of help, but the following names stand out in particular:

Steven Auerbach for innumerable discussions about everything under the sun
(plus some things inside the sun) including lots of stuff in this thesis.
It is a pleasure to have a friend who appreciates the love of learning and
the joy of figuring things out as much as Steve does.

Bruce Cohen for his concern and guidance during several critical stages of 
this thesis, and especially for his many excellent suggestions for the
application of this work.

Brendan McNamara for teaching me much about the application of Lie transforms
to mode-coupling analyses in weak turbulence theory in general, and in the
B\'{e}nard convection problem in particular.

Dana Richards and Jan Wikkerink for their continued violations of the
second law of thermodynamics, without which everything would have fallen
into utter disorganization and chaos.

Next, all the folks in Berkeley$\ldots$

Robert Littlejohn for millions of wonderful discussions on guiding-center
theory and differential geometry.  Robert's influence permeates this thesis
from begining to end.  In many ways, this thesis is the relativistic
generalization of the revolution in plasma physics that was initiated by
him.

Jerrold Marsden for numerous enlightening discussions about reduction,
gauge groups, nonlinear stability theory, etc.  His course on the theory
and applications of symplectic and Poisson reduction has molded the way I think
about those topics.

Richard Montgomery for lots of helpful discussions about symplectic geometry,
the method of averaging, reduction, and the energy-Casimir method.

Stephen Omohundro for teaching me the importance of geometrical intuition.
Many was the blackboard discussion at which Steve would label something
``obvious'' that became clear to me only after much calculation.
It did not take many such discussions to convince me of the importance of
differential geometrical intuition.  With Steve's help, I devoured books on
the subject, and my entire outlook in mathematical physics has not been
the same since.  The prevalence of such geometrical techniques in this thesis
attests to their influence on my thinking.

Philippe Similon for guidance and helpful suggestions on just about
everything in this thesis, especially the Poisson brackets and conservation
laws for the guiding-center plasma.  Philippe has the rare combination of 
abilities required to understand and use formalism, while never losing sight 
of the fact that its purpose is to solve real problems in the real world.

Jonathan Wurtele for many discussions about nonlinear dynamics and
differential geometric techniques in physics.

Huanchun Ye for many helpful discussions, especially on the problem of
resonances.  This problem was largely ignored in my thesis, but
Huanchun's thesis should fill in the gap and make the theory more complete.
His observations on the relationship between torsion and spin seem to 
underlie the theory leading to the conservation law for angular momentum
in the guiding-center plasma.  Once again, it is quite possible that there is
much left to be discovered in this area, and it may be in Huanchun's thesis.

Next, the folks at the University of California at Davis Department of
Applied Science$\ldots$

My oral exam committee which consisted of G. Donald Chakerian, Abraham
Goldberg, Myron (Mike) A. Hoffman, John Killeen and Frederick Wooten.

Abe Goldberg for teaching me quantum mechanics, and for lots of good advice, 
discussions and vanpool conversation on everything from
Bell's theorem to quantum field theory.

John Killeen for teaching me much advanced mathematics, and for several
enlightening discussions on computational plasma physics.

Fred Wooten for much guidance and advice about how to survive a PhD
program.

William Hoover for teaching me statistical mechanics and for several good
discussions about nonlinear dynamics.

Robert White for serving as a sounding board for zillions of ideas, some of
which appeared in this thesis and some of which did not.  As the only other
``plasma type'' in my class at D.A.S. with an interest in nonlinear dynamics,
Robert and I consulted with each other a great deal.

Donna Clifford for making sure I didn't run afoul of the bureaucracy.  I'm
not even sure I'd be a registered student now were it not for her (not to
mention the innumerable late fees that only her reminders saved me from having
to pay).

Last, but certainly not least, some folks who don't fit neatly into any of
the above categories$\ldots$

Bedros Afeyan for numerous interesting discussions while he was at Livermore.

John Cary for his insights on ponderomotive theory and the K-$\chi$ theorem.

John David Crawford for helping me better understand Van Kampen modes, and for
several other enlightening discussions.

Charles Karney and Bruce Nemnich for helping me with \LaTeX.

This work was supported by the US DOE under contract numbers W-7405-ENG-48
and DE-AC03-76SF00098.

%% file: introduction.tex
\chapter{Introduction}
\addtolength{\textheight}{0.5truein}
\pagenumbering{arabic}
\pagestyle{myheadings}
\markboth{Introduction}{Introduction}

There was a time when a thorough working knowledge of geometry was
considered an indispensible ingredient in the education of a
natural philosopher.  From Euclid's first systematization of the subject
more than two thousand years ago to well after the end of the Renaissance,
the study of the {\it Elements} was considered a critically important
part of mathematical instruction.
Indeed, when reading Newton's {\it Principia} or {\it Opticks,} one is struck
by the prevalence of geometrical arguments and descriptions.

Alas, the introduction of coordinate systems by Descartes and the
concommitant analyticization of geometry changed all this.  Using
coordinates, geometrical problems could be reduced to algebraic problems.
The perceived need for good geometrical intuition gradually disappeared.
By the time Whittaker's {\it Treatise on the Analytical Dynamics of
Particles and Rigid Bodies} was first published in 1904, this attitude
had taken hold to the extent that Whittaker apparently felt no need to
include illustrations in his nearly five-hundred-page-long 
(and otherwise excellent) document.
At present, one can obtain an undergraduate degree in physics or even applied
mathematics with little more geometry background than is found in a
secondary school textbook.

That this trend is disastrous has been appreciated only for the past
couple of decades.  This appreciation has been due, in large part, to modern
developments in the general theory of relativity.  The entire lesson of
relativity theory is that physical laws ought not to depend upon the
coordinate system chosen to describe them; that is,
the meaning of physical laws transcends their coordinate description.
Conversely, coordinate descriptions can have a way of masking
fundamental physical reality.  Thus, a 
coordinate-free description of physical laws can have the
beneficial effect of allowing one more easily to glimpse the underlying
fundamental physical reality.  Such coordinate-free mathematical language
is available, thanks in large part to the works of Cartan and Lie.
Modern differential geometry, including the exterior calculus and the
theory of Lie groups, is capable of providing a coordinate-free
description of physical law.  Please note that what is being argued here
is that such a coordinate-free description is far more than just an
alternative mathematical notation; the contention is that it yields an improved
understanding of the {\it physics} involved.  A physicist who takes the time
to learn how, say, electromagnetic theory can be described in terms of
differential forms will have, as a result, an improved understanding of the
electromagnetic field.

There is an additional benefit to the geometrical point of view.  Just as
Descartes found that algebra can be used as a tool for obtaining
geometrical results, likewise geometry can be used as a tool for obtaining
analytical results that would be far more difficult to obtain any other way.
Several examples of this phenomenon will be pointed out in the course of this
thesis.

Since the 1960's it has been known that classical mechanics is describable
in terms of symplectic geometry.  This observation paved the way for powerful
generalizations of some of the traditional methodologies of mechanics.  For
example, whereas Hamiltonian mechanics had been originally formulated in
terms of canonically conjugate pairs of coordinates, it was found that
noncanonical coordinates could be used instead, oftentimes to great advantage.
Powerful new types of perturbation theory, based on Lie transforms, were
introduced; this made higher-order perturbative treatments less laborious 
and more systematic.

Nowhere was the impact of this revolution more profound and beneficial than in
the field of plasma physics.  Because the motion of charged particles in 
complicated electromagnetic geometries and in wave fields requires a 
perturbative treatment, it is not surprising that
Lie transform perturbation theory was shown to be
a natural tool for systematizing, simplifying and {\it better understanding}
many of the calculations of plasma physics.  Furthermore, it was shown that
the most natural treatment of the guiding-center problem (i.e. the ubiquitous 
problem of computing the drifts of
a charged particle gyrating in a slowly-varying electromagnetic field)
involved the use of noncanonical coordinates and noncanonical coordinate
transformations.  All of this will become more clear as we proceed.

During the late nineteen seventies, Dewar~\cite{zbw} introduced the idea of
canonical oscillation-center transformations.  Johnston and Kaufman~\cite{zbj}
and Johnston~\cite{zbi} used canonical perturbation
theory to perform oscillation-center and
mode coupling analyses for the Vlasov plasma.  In Cary's PhD thesis~\cite{zbk},
Lie transforms were shown to be a useful tool for ponderomotive theory, and
the K-$\chi$ theorem~\cite{zbl} relating the ponderomotive Hamiltonian with
the linear susceptibility was formulated.  

The extension of these techniques 
to magnetized plasma was made possible, or at least greatly facilitated, by
Littlejohn's work on the guiding-center problem in his PhD thesis~\cite{zbm}.
Littlejohn made the key observation that the transformation from 
single-particle to guiding-center coordinates was best done using noncanonical
methods.  This noncanonical transformation was done in his thesis by using
the Darboux theorem constructively, and it was followed by a canonical Lie 
transformation that averaged over the rapid gyromotion.  
Subsequently, Littlejohn~\cite{zaq} discovered that the entire transformation 
could be done by a single Lie transform with a vector generator.  This is
the approach followed in this thesis.

Ponderomotive theory for a magnetized relativistic plasma was then done by
Grebogi and Littlejohn~\cite{zbn}, who used canonical Lie transforms.  They
pointed out that the oscillation-center transformation for a magnetized plasma
might best be handled by noncanonical Lie methods, but they did not do it this
way.  Their result was subsequently simplified by Cary and 
Newberger~\cite{zbo}.

Meanwhile, Dubin, Krommes, Oberman and Lee~\cite{zbp} showed how to use
Littlejohn's methods to derive self-consistent gyrokinetic equations for an
electrostatic plasma, including the Poisson equation whose source term was
written in terms of the
guiding-center distribution function.  Kaufman and Boghosian~\cite{zbq} 
showed that this calculation could be done by summing the guiding-center 
action over the entire distribution and coupling it to the Maxwell action;
variation with respect to the coordinate fields (considered to be functions
of their initial conditions) then yields the gyrokinetic equation, and
variation with respect to the vector potential then yields the self-consistent
field equaton.  Finally, Similon~\cite{zbr} showed that conservation laws
for the guiding-center plasma could be obtained by application of Noether's
theorem to this system action.

The above-mentioned work by Grebogi and Littlejohn was done for a relativistic
plasma, but was not manifestly covariant in that it was done in 
``$1+3$'' notation.  A manifestly covariant treatment
is made possible with the help of certain projection operators which
were introduced by Fradkin~\cite{zae} who obtained the
drifts for a relativistic guiding center (but did not use Lie methods),
and by Dumais~\cite{zbs}.

The general plan of this thesis is as follows:  

Chapter~\ref{yaa} will cover the
mathematical preliminaries necessary to understand the differential
geometric arguments used in this thesis.  It should be emphasized that
this constitutes no more than a sketchy introduction, and is no substitute
for a good text on the subject; nevertheless it is probably sufficient to 
enable a persistent person with an undergraduate background in physics to
read and understand this entire text.  Chapter~\ref{yaa} also describes the
application of these techniques to Hamiltonian and Lagrangian mechanics;
specifically, Lie transform perturbation theory is introduced here and
many simple examples of its use are presented.

Chapter~\ref{yac} will treat the guiding-center problem for a relativistic
charged particle.  We shall begin by examining the geometry of the
electromagnetic field in four-dimensional spacetime,
and we shall find that there is a 
covariant way to isolate the rapidly-gyrating component of the particle's four 
velocity.  Lie transform perturbation theory is then applied to the particle's 
phase-space Lagrangian in order to remove this rapidly-gyrating component and 
thus obtain the residual parallel and drift motion.  The perturbative 
calculation is carried out to one order higher than the usual drifts, the 
natural guiding-center Poisson bracket structure and Hamiltonian are presented,
and the correction to the gyromomentum is given.  Finally, it is shown how to 
cast these results in a {\it manifestly} gyrogauge invariant format.

In Chapter~\ref{yap} we shall study the effects of
eikonal wave perturbations on a
guiding center, once again using Lie transform perturbation theory.  The
result is a complete ponderomotive description of the relativistic guiding
center in an eikonal wave field, and we show how to cast this in
{\it manifestly} gauge-invariant form.  To achieve manifest gauge-invariance,
we shall find it necessary to abandon
the usual approach of expanding the eikonal wave
perturbation in a series of Bessel functions of $k_\perp\rho.$  Instead, we
shall first perform a Lagrangian gauge transformation, and then we shall
expand in a series of special functions that are related to indefinite
integrals of Bessel functions.  The required Lagrangian gauge transformation
is not obvious, and it would never have been discovered without the use of
differential geometric techniques.  Finally, the ponderomotive Hamiltonian is
derived using Lie transforms.

In Chapter~\ref{yaq} we shall sum the resulting
guiding-center Lagrangian over the entire distribution of particles present in
a plasma, and couple with the Maxwell field to obtain the total
Lagrangian for a Vlasov plasma of relativistic guiding centers.  By varying
this it is possible to derive a self-consistent
gyrokinetic description of such a plasma, including the magnetic 
moment tensor, in manifestly-covariant format.  Application of Noether's
theorem then yields conservation laws for the guiding-center plasma, and these
are also cast in manifestly covariant form.  Finally, using the results of 
Chapter~\ref{yap}, the conservation laws are derived for a guiding-center
plasma in the presence of a wave field.

In Chapter~\ref{ybf} we discuss some of the unanswered questions raised
by this study.  These could be topics for future research.

Appendix~\ref{yuk} is a glossary of the mathematical symbols and notation
used in this thesis.

Appendix~\ref{yav} is a review of some of the more primitive mathematical
concepts used in this thesis, such as {\it vector spaces}, {\it dual
spaces}, {\it algebras}, and {\it modules}.

Appendix~\ref{ybd} applies vector Lie transforms to the nonrelativistic
guiding-center problem in two dimensions, and derives the shift in
gyrofrequency due to spatial gradients in the magnetic and
(perpendicular) electric fields.  This is useful both as a demonstration
of the vector Lie transform technique, and as a comparison to the
techniques and results of Chapter~\ref{yac}.

Appendix~\ref{yaw} derives and discusses the properties of a pair of special
functions that were introduced in Chapter~\ref{yap}.

Appendix~\ref{ybe} is a short tutorial on how to derive Bessel function sum
rules, including (but not limited to) those that were useful in
Chapter~\ref{yap}.

%% file: mathprelim.tex
\chapter{Mathematical Preliminaries}
\pagestyle{myheadings}
\markboth{Mathematical Preliminaries}{Mathematical Preliminaries}
\label{yaa}

\section{Discussion}

This chapter divides naturally into three sections.  The first covers
the basic results of differential geometry that are necessary to understand
the rest of this thesis.  This includes the calculus of tensors and the
exterior algebra.  To reiterate, the exposition here is not intended to
replace a good introductory book on the subject (see, for example, the
excellent introductory texts by Schutz~\cite{zaa}, Edelen~\cite{zab},
Singer and Thorpe~\cite{zac}, or Burke~\cite{zad}), but it does present
enough material to make the thesis self-contained, and to establish
notational conventions.  The theory of Lie groups has been omitted from
this section because it is not absolutely essential to the understanding 
of what follows, but the reader with background in this area will be at a
definite advantage.

Next, these tools are used to reformulate Hamiltonian and Lagrangian mechanics.
The generalization to noncanonical coordinates is discussed, including those
with singular Poisson structures.  Noether's theorem is formulated, and
numerous worked examples are given.  Mechanical systems with constraints are
examined from this new point of view.

Finally, Lie transform perturbation theory is presented, and its use for
noncanonical coordinates is discussed.  Because we shall use Lie transforms
in a more general context than that in which they are usually presented, I
recommend that this section be read even by those already familiar with the
subject.

\section{Differential Geometric Concepts}
\subsection{Manifolds, Vectors, and Covectors}

In this subsection, we shall discuss the ideas that are necessary to
reformulate tensor calculus in a fashion that more directly
illustrates the geometrical foundations of the subject.
Appendix~\ref{yav} goes one level deeper, and gives set-theoretical
definitions for many of the primitive terms that we shall use here
(such as {\it vector space} and {\it algebra}).

A {\it manifold} is a space that is locally Euclidean and in which there is a
notion of differentiation.  This can be made more precise as follows:  
There must be a differentiable one-to-one map, or {\it diffeomorphism}, from
the neighborhood of any point of a manifold to the points of $\Re^n,$ for some 
$n.$  Such a map is called a {\it chart}, and the collection of all such maps
for a given manifold is called an {\it atlas}.  There is an additional 
requirement that two maps in the same atlas that overlap must do so smoothly;
this means, among other things, that all charts in the same atlas must map to
$\Re^n$ with the same $n.$  The number $n$ is thus characteristic of the 
entire manifold, and is called the {\it dimension} of the manifold.

A chart is realized by (local) coordinates on the manifold.  Since an
$n$-dimensional manifold, $M,$ must map smoothly onto $\Re^n,$ it must be 
possible to label the points of $M,$ at least locally, by $n$ numbers, say
$z^1,\ldots,z^n.$  Then the map is given by expressing these
numbers as functions of the coordinates, $x^1,\ldots,x^n,$ on $\Re^n.$  
Specifically, we write $z^\alpha (x^1,\ldots,x^n),$ for 
$\alpha=1,\ldots,n.$

It is generally not possible to cover an entire manifold with one chart.
For example the surface of a sphere is a manifold called $S^2,$ and, as is
well known, coordinate charts on $S^2$ must break down somewhere.
The chart
\begin{equation}
x=\frac{4\theta}{3\pi}\cos\phi
\end{equation}
\begin{equation}
y=\frac{4\theta}{3\pi}\sin\phi,
\end{equation}
where $\theta$ and $\phi$ are the usual spherical coordinates (colatitude and
azimuthal angles, respectively), maps the region $0\leq\theta < 3\pi/4$ onto
the open unit disk in $\Re^2.$  The chart
\begin{equation}
x=\frac{4(\pi-\theta)}{3\pi}\cos\phi
\end{equation}
\begin{equation}
y=\frac{4(\pi-\theta)}{3\pi}\sin\phi
\end{equation}
then maps the region $\pi/4 < \theta\leq\pi$ onto the open unit disk in
$\Re^2.$  These two charts are thus sufficient to cover all of $S^2,$ and
therefore constitute an atlas.  Any atlas for $S^2$ must contain at least
two charts.  In general, the number of charts needed to cover a manifold depends
on its global topological properties.

A mapping from an $m$-dimensional manifold onto an $n$-dimensional manifold
is called an {\it injection} if $m<n,$ a {\it projection} if $m>n,$ and a
{\it bijection} if $m=n.$  Consider a map from $\Re$ to an 
$n$-dimensional manifold, $M.$  That is, $\Re \mapsto M.$  Note
that this is an injection if $n>1,$ and a bijection if $n=1.$  This map
defines a {\it path} through the manifold, $M.$  The points in $M$ that are on
the path are those in the range of the map.  The realization of this mapping
is given by expressing each of the coordinates on $M$ as functions of the
coordinate, $x,$ on $\Re.$  That is, we write $z^\alpha (x)$ 
for $\alpha=1,\ldots,n.$  As $x$ varies along $\Re,$ 
the coordinates $z^\alpha$ trace out the path in $M.$
Note that although we keep writing down the coordinate realizations of these
things, the notion of a map from one manifold to another has an intrinsic
geometrical meaning as an association of members of one set of points with
members of another set of points, consistent with local topological properties
of nearness, etc.

Now let $P$ be a point on the above-mentioned path through the manifold, $M.$
Denote its coordinates by $z^1_P,\ldots,z^n_P.$  Since it lies along the path,
there must exist a coordinate, $x_0,$ of a point in $\Re,$ such that
$z^\alpha_P=z^\alpha (x_0)$ for $\alpha=1,\ldots,n.$  
Now consider the derivatives of the functions, $z^\alpha(x),$ with respect 
to the path parameter, $x.$  Denote these by $dz^\alpha /dx.$  Evaluate these 
at the point $P.$  This gives the $n$ numbers,
\begin{equation}
V^\alpha\equiv \frac{dz^\alpha}{dx} (x_0),
\end{equation}
associated with the point, $P.$

It is clear that there are many different curves passing through point $P$ 
that will yield the same set of $n$ numbers.  Indeed, any curve whose 
coordinates near $P$ are given by
\begin{equation}
z^\alpha=z^\alpha_P+V^\alpha\delta x+{\cal O}(\delta x^2)
\end{equation}
where $\delta x\equiv x-x_0,$ will do so.  The identification of these $n$
numbers thus gives us a way to partition the set of all curves passing through
point $P$ into (an infinity of) equivalence classes; two curves are said to
be equivalent if they yield the same set of $n$ numbers.  That is, two curves
are equivalent if they both have the form given in the above equation (with
the same $V^\alpha$'s).

Consider the set of equivalence classes of curves thus obtained.
We can define addition and scalar multiplication among the elements of this set
in the following very natural way:  The equivalence class of curves with the
$n$ numbers $V^\alpha$ adds to the equivalence class of curves with the $n$
numbers $U^\alpha$ to yield the equivalence class of curves with the $n$
numbers $V^\alpha+U^\alpha.$  The scalar $a$ multiplies the equivalence class
of curves with the $n$ numbers $V^\alpha$ to yield the equivalence class of
curves with the $n$ numbers $aV^\alpha.$  With these operations, we have
converted the space of all equivalence classes of curves through the point $P$
into a vector space.  This vector space will be called the {\it tangent space}
at point $P$ of the manifold.  Its elements have been introduced as equivalence
classes of curves, but it will become clear momentarily that these may be 
identified with the usual notion of vectors as arrows with a certain 
magnitude and direction and with certain transformation properties.  Note,
however, that the base of the arrow is not free to move around, but rather is
``pinned down'' at the point $P.$  There is a different tangent space at each
point of a manifold, and vectors in one tangent space may not be added to
vectors in another different tangent space.  Note that the dimension of a
tangent space is equal to the dimension of the manifold (in the above case,
the dimension is $n$).

It is evident that the above-described $n$ numbers $V^\alpha$ associated with 
an equivalence class of curves depend on our choice of coordinates for $M.$  
If our coordinates on $M$ had been ${z'}^\alpha,$ then the $n$ numbers 
would have been
\begin{equation}
{V'}^\alpha=\frac{d{z'}^\alpha}{dx} (x_0)
           =\frac{\partial {z'}^\alpha}{\partial z^\beta}
            \frac{dz^\beta}{dx} (x_0)
           =\frac{\partial {z'}^\alpha}{\partial z^\beta}V^\beta,
\label{mck}
\end{equation}
where we have adopted the convention of summation over repeated indices.
Readers familiar with traditional presentations of tensor calculus will
recognize this as the transformation law for components of contravariant
vectors.  

Recall that even though the components of a vector may vary from
one coordinate system to another, the vector itself, as an abstract
mathematical object, is an invariant geometrical concept.  
That is, given two sets of basis vectors, $\hat{\bf e}_\alpha$ 
and ${{\hat{\bf e}}'}_\alpha,$ we can write the
components of a vector ${\bf V}$ as $V^\alpha$ in the first system and as
${V'}^\alpha$ in the second.  Though these will, in general, be different,
the abstract vector 
${\bf V}=V^\alpha{\hat{\bf e}}_\alpha={V'}^\alpha{{\hat{\bf e}}'}_\alpha$
retains its form under the change of basis.

So how can we introduce bases in our tangent spaces that will reflect this 
idea?  Despite the fact that the above-described $n$ numbers are 
coordinate-dependent, if we form a first-order linear differential operator 
by using them as coefficients
\begin{equation}
\hat{{\bf V}} \equiv V^\alpha \frac{\partial}{\partial    z^\alpha}
                = {V'}^\alpha \frac{\partial}{\partial {z'}^\alpha},
\end{equation}
we see that this operator retains its form under a coordinate transformation.
This much is clear from the above equation.  By analogy with the argument in
the preceeding paragraph, we can thus {\it identify} the
operator $\hat{{\bf V}}$ with the vector ${\bf V},$ and the $n$ operators
$\partial/\partial z^\alpha$ with basis vectors that span the tangent space.
Thus the idea of vectors as arrows, as equivalence classes of curves, and as
first order linear differential operators are all valid descriptions of the 
same concept!

A word is in order concerning the basis vectors that we have used above.  Note
that they were induced by the coordinate system that we used.  The choice of
a coordinate system $z^\alpha$ on the manifold $M$ gives rise to a natural 
basis $\partial/\partial z^\alpha$ in each tangent space at each point of the
manifold (or, more precisely, at each point of $M$ where the chart
$z^\alpha$ is operative).  A change in
coordinate system thus gives rise to a change of basis; this is in accordance
with the usual transformation properties of contravariant vectors.  A basis
that is thus induced by a coordinate system is called a {\it coordinate basis}.
In the ``arrow'' picture, the basis vectors lie along the local coordinate
axes.  In the ``equivalence class of curves'' picture, they are curves that
are locally coincident with the coordinate axes.  In the ``operator'' picture,
they are directional derivatives along the coordinate directions.

One might well ask if {\it all} possible bases are coordinate bases.  The
answer is ``no.''  If we start from a coordinate basis and make a change
of basis by taking various linearly independent combinations of basis vectors
in each tangent space, where the combinations may vary from point to point in
the manifold, we may arrive at a new basis that is not the coordinate basis for
any coordinate system on $M.$  Thus, starting from the coordinate basis,
$\partial/\partial z^\alpha,$ we may define the new basis
\begin{equation}
{\hat{\bf e}}_\beta=\Lambda^{\phantom{\beta}\alpha}_\beta
                    \frac{\partial}{\partial z^\alpha},
\end{equation}
where $(\Lambda^{\phantom{\beta}\alpha}_\beta)$ is any nonsingular matrix.
This new basis is perfectly good for resolving vectors into coordinates.  For
example, the vector ${\bf V}$ may be written
\begin{equation}
{\bf V}=V^\alpha\frac{\partial}{\partial z^\alpha}
       =(V^\alpha\Lambda^\beta_{\phantom{\beta}\alpha}){\hat{\bf e}}_\beta
\end{equation}
where the matrix $(\Lambda^\beta_{\phantom{\beta}\alpha})$ is the inverse of
the matrix $(\Lambda^{\phantom{\beta}\alpha}_\gamma).$  So the components of
${\bf V}$ in the new basis are $V^\alpha\Lambda^\beta_{\phantom{\beta}\alpha}.$
The only different thing about this new basis is that there
may not be any system of coordinates $Z^\alpha$ such that
${\hat{\bf e}}_\alpha=\partial/\partial Z^\alpha.$  In this case,
such a basis is called a {\it noncoordinate
basis}.  This idea will become more clear and examples will be given in
Subsection~\ref{yay}.

Meanwhile, since we have now attached vector spaces to every point of a
manifold, we can go on to construct their dual spaces.  The dual space
to the tangent space of vectors at point $P$ is called the {\it cotangent
space} at point $P.$  Its elements are called {\it covectors} or
{\it covariant vectors} or {\it one forms}.  Once again, the cotangent space
has the same dimension as the manifold.  

Once we have a set of basis vectors in the tangent space, say 
${\hat{\bf e}}_\alpha,$ there is induced a preferred set of basis covectors 
in the cotangent space, call them ${\tilde{\omega}}^\alpha,$ such that
$\langle {\tilde{\omega}}^\alpha,{\hat{\bf e}}_\beta \rangle 
= \delta^\alpha_\beta.$  Thus we can represent a covector
at point $P$ by $n$ numbers, say $a_\alpha,$ where, as usual, $\alpha$ can
range from $1$ to $n.$  The abstract covector is then 
${\bf a}=a_\alpha{\tilde{\omega}}^\alpha.$  The covector $a$ pairs with the
vector $V$ to yield
\begin{equation}
\langle {\bf a},{\bf V} \rangle=\langle a_\alpha{\tilde{\omega}}^\alpha,
  V^\beta {\hat{\bf e}}_\beta \rangle
  =a_\alpha V^\beta\langle {\tilde{\omega}}^\alpha,
  {\hat{\bf e}}_\beta \rangle
  =a_\alpha V^\beta \delta^\alpha_\beta
  =a_\alpha V^\alpha.
\end{equation}

Note that even though there is a naturally induced covector basis corresponding
to a given vector basis, there is no natural correspondence between individual
vectors and individual covectors.  That is, there is no natural map from the
tangent space to the cotangent space.  Later on, we shall see that if we
endow our manifold with a {\it metric}, such a map is established.  The 
addition of a metric thus gives the manifold much more structure than it would
otherwise have.  At this point in our discussion, we are not assuming the
existence of a metric on our manifold.  As we shall see, even without a metric,
a manifold has lots of interesting structure to study.  The general
philosophy of this discussion is to start simply and slowly add structure;
thus a discussion of metrics is deferred to the end of this section.

To make our discussion of covectors more concrete, let us suppose that we
have a coordinate system $z^\alpha$ on our manifold, $M.$  This induces the
coordinate basis vectors $\partial/\partial z^\alpha$ on each tangent space
of $M.$  If we transform coordinates to another system $Z^\alpha,$ the 
components of the vector ${\bf V}$ transform according to Eq.~(\ref{mck}).
Now say the covector ${\bf a}$ has components $a_\alpha$ in the first 
coordinate system.  The components of the covector must transform in such a 
way as to leave the scalar $\langle {\bf a},{\bf V}\rangle$ invariant.  Thus
\begin{equation}
a_\beta V^\beta
   ={a'}_\alpha {V'}^\alpha
   ={a'}_\alpha \frac{\partial {z'}^\alpha}{\partial z^\beta}V^\beta
\end{equation}
so
\begin{equation}
{a'}_\alpha=\frac{\partial z^\beta}{\partial {z'}^\alpha}a_\beta.
\label{mcl}
\end{equation}
Once again, readers familiar with traditional presentations of tensor calculus 
will recognize this as the transformation law for components of covariant
vectors.  

Now, how can we introduce bases in our cotangent spaces that will reflect the 
above ideas?  Despite the fact that the $n$ numbers $a_\alpha$ are 
coordinate-dependent, if we form the differential that has them as coefficients
\begin{equation}
\tilde{{\bf a}} \equiv a_\alpha dz^\alpha
                  = {a'}_\alpha d{z'}^\alpha,
\end{equation}
we see that this retains its form under a coordinate transformation.
This much is clear from the above equation.  We can thus {\it identify} the
differential form $\tilde{{\bf a}}$ with the covector ${\bf a},$ and the $n$
differentials $dz^\alpha$ with basis covectors that span the cotangent space.

Thus, just as contravariant vectors could be identified with first order linear
differential operators, we see that covectors can be identified with
differential forms.  These descriptions are dual to each other, so
\begin{equation}
\langle dz^\alpha,\frac{\partial}{\partial z^\beta}\rangle=\delta^\alpha_\beta.
\end{equation}

Finally we note that the same distinction between coordinate and noncoordinate
bases that applied to our discussion of tangent space bases also applies to
cotangent space bases.  Up until now, we have restricted our attention to
coordinate cotangent bases, but we could define new basis one forms by taking
linear combinations of the $dz^\alpha$ where the combinations may vary from 
point to point in the manifold.  In this way, we may arrive at a new basis 
that is not the coordinate cotangent basis for any coordinate system on $M.$  
Thus, starting from the coordinate cotangent space basis,
$dz^\alpha,$ we may define the new cotangent space basis
\begin{equation}
{\tilde{\omega}}^\beta=\Lambda^\beta_{\phantom{\beta}\alpha}dz^\alpha
\end{equation}
where $(\Lambda^\beta_{\phantom{\beta}\alpha})$ is any nonsingular matrix.
This new basis is perfectly good for resolving covectors into coordinates.  For
example, the covector ${\bf a}$ may be written
\begin{equation}
{\bf a}=a_\alpha dz^\alpha
       =(a_\alpha\Lambda^{\phantom{\beta}\alpha}_\beta)
        {\tilde{\omega}}^\beta
\end{equation}
where the matrix $(\Lambda^{\phantom{\beta}\alpha}_\gamma)$ is the inverse of
the matrix $(\Lambda^\beta_{\phantom{\beta}\alpha}).$  So the components of
${\bf a}$ in the new basis are $a_\alpha\Lambda^{\phantom{\beta}\alpha}_\beta.$
The only different thing about this new basis is that there
may not be any system of coordinates $Z^\alpha$ such that
${\tilde{\omega}}^\alpha=dZ^\alpha.$  Once again, this idea will become more 
clear and examples will be given in Subsection~\ref{yay}.

\subsection{General Tensors and the Tensor Product}
Now that we have a tangent space and a cotangent space associated with each
and every point of our manifold, we can create still bigger spaces at each
point by taking the Cartesian product of some number of tangent spaces and
some number of cotangent spaces.  Suppose we define the space $\Pi^s_r$ to
be the Cartesian product of $s$ copies of the tangent space and $r$ copies of
the cotangent space at point $P$ of a manifold $M.$  Consider a multilinear map
$\Pi^s_r\mapsto\Re.$  That is, we are considering a map that takes $s$
vectors and $r$ covectors at point $P$ and returns a real number.  If the
$s$ vectors are denoted ${\bf V}_1,\ldots,{\bf V}_s,$ and the $r$ covectors
are denoted ${\bf a}^1,\ldots,{\bf a}^r,$ then the real number will be
denoted by ${\bf T}({\bf a}^1,\ldots,{\bf a}^r,{\bf V}_1,\ldots,{\bf V}_s).$
By a ``multilinear'' map, we mean that ${\bf T}$ is linear in all of its 
arguments.  Such a map is said to be a {\it tensor of type} $(r,s).$
Note that a vector is a tensor of type $(1,0),$ and a covector is a tensor of
type $(0,1)$; this is because a vector can take a covector and return a real
number (by the pairing), and vice versa.

There is an obvious way to define addition among tensors:  Given two tensors,
${\bf T}_1$ and ${\bf T}_2,$ we can define a new tensor, ${\bf T}_3,$ by
the prescription
\begin{eqnarray}
\lefteqn{
{\bf T}_3({\bf a}^1,\ldots,{\bf a}^r,{\bf V}_1,\ldots,{\bf V}_s)}\nonumber\\
&=&{\bf T}_1({\bf a}^1,\ldots,{\bf a}^r,{\bf V}_1,\ldots,{\bf V}_s)+
  {\bf T}_2({\bf a}^1,\ldots,{\bf a}^r,{\bf V}_1,\ldots,{\bf V}_s),
\label{mcq}
\end{eqnarray}
for all possible arguments.  In this case, we write 
${\bf T}_3={\bf T}_1+{\bf T}_2.$  This operation of addition makes the space
of all tensors of type $(r,s)$ a vector space.

Suppose we have two vectors, ${\bf U}_1$ and ${\bf U}_2,$ and a covector,
${\bf b}^1,$ at some point of a manifold.  Suppose we are given anew a 
pair of covectors, ${\bf a}^1$ and ${\bf a}^2$, and a vector, ${\bf V}_1$
(at the same point of the manifold).  
Consider the following recipe for obtaining a real number:  
Pair the two covectors with ${\bf U}_1$ and ${\bf U}_2,$
respectively, and pair the vector with ${\bf b}^1.$  This gives us three real
numbers.  Multiply them together to get a single real number.  In this way, the
presence of ${\bf U}_1, {\bf U}_2,$ and ${\bf b}^1$ provides us with a map from
$\Pi^1_2$ to $\Re.$  It is easily seen that this map is multilinear.  Thus, the
presence of ${\bf U}_1, {\bf U}_2,$ and ${\bf b}^1$ provides us with the
following tensor of type $(2,1)$:
\begin{equation}
{\bf T}({\bf a}^1,{\bf a}^2,{\bf V}_1)
  =\langle {\bf a}^1,{\bf U}_1 \rangle
   \langle {\bf a}^2,{\bf U}_2 \rangle
   \langle {\bf b}^1,{\bf V}_1 \rangle.
\end{equation}
A tensor formed in this way is said to be the {\it tensor product} of
${\bf U}_1, {\bf U}_2,$ and ${\bf b}^1.$  This is denoted
\begin{equation}
{\bf T}={\bf U}_1\otimes {\bf U}_2\otimes {\bf b}^1.
\end{equation}

More generally, given $r$ vectors, ${\bf U}_1,\ldots,{\bf U}_r,$ and
$s$ covectors, ${\bf b}^1,\ldots,{\bf b}^s,$ we can form a tensor of
type $(r,s)$ by taking the tensor product
\begin{equation}
{\bf T}={\bf U}_1\otimes\ldots\otimes {\bf U}_r\otimes
        {\bf b}^1\otimes\ldots\otimes {\bf b}^s.
\end{equation}
If we feed this tensor the $r$ covectors, ${\bf a}_1,\ldots,{\bf a}_r,$
and the $s$ vectors, ${\bf V}_1,\ldots,{\bf V}_s,$ then we get the
scalar
\begin{equation}
{\bf T}({\bf a}_1,\ldots,{\bf a}_r,{\bf V}_1,\ldots,{\bf V}_s)
   =\langle {\bf a}_1,{\bf U}_1 \rangle\ldots
    \langle {\bf a}_r,{\bf U}_r \rangle
    \langle {\bf b}_1,{\bf V}_1 \rangle\ldots
    \langle {\bf b}_s,{\bf V}_s \rangle.
\end{equation}

The space of all possible tensors (of any type)
at some point in an $n$-dimensional manifold
may be thought of as an infinite dimensional vector space, although it is
somewhat strange in that two of its elements can be added if and only if they
are tensors of the same type.  In any event, the tensor product makes this
space an algebra.

It is straightforward to see that the vector space of all tensors of
type $(r,s)$ is $n^{r+s}$-dimensional.  That is, a tensor of type
$(r,s)$ has $n^{r+s}$ independent components.  A moment's
thought convinces one that a basis for this space is given by the
$n^{r+s}$ basis tensors
\begin{equation}
{\hat{\bf e}}_{\alpha_1}\otimes\ldots\otimes 
  {\hat{\bf e}}_{\alpha_r}\otimes
{\tilde{\bf \omega}}^{\beta_1}\otimes\ldots\otimes 
  {\tilde{\bf \omega}}^{\beta_s},
\end{equation}
where the $\hat{\bf e}$'s and $\tilde{\bf \omega}$'s are the basis vectors
and basis covectors in the tangent and cotangent spaces, respectively, and 
where the $\alpha$ and $\beta$ indices all range from $1$ to $n.$
Thus, a general tensor may be written
\begin{equation}
{\bf T}=T^{\alpha_1\ldots\alpha_r}_{\beta_1\ldots\beta_s}
{\hat{\bf e}}_{\alpha_1}\otimes\ldots\otimes 
  {\hat{\bf e}}_{\alpha_r}\otimes
{\tilde{\bf \omega}}^{\beta_1}\otimes\ldots\otimes 
  {\tilde{\bf \omega}}^{\beta_s}.
\end{equation}

Finally, we consider the transformation properties of the components of
these general tensors.  We know how vector and covector components 
transform, and we know that a tensor of
type $(r,s)$ takes $r$ covectors and $s$ vectors and returns a scalar 
invariant.  Thus, by an argument identical to that which led to 
Eq.~(\ref{mcl}), we find that for a transformation from one coordinate
basis to another coordinate basis
\begin{equation}
{T'}^{\alpha_1\ldots\alpha_r}_{\beta_1\ldots\beta_s}
  =\frac{\partial {z'}^{\alpha_1}}{\partial z^{\mu_1}}\cdots
   \frac{\partial {z'}^{\alpha_r}}{\partial z^{\mu_r}}
   \frac{\partial z^{\nu_1}}{\partial {z'}^{\beta_1}}\cdots
   \frac{\partial z^{\nu_s}}{\partial {z'}^{\beta_s}}
   T^{\mu_1\ldots\mu_r}_{\nu_1\ldots\nu_s}.
\end{equation}
The usual distinction between coordinate and noncoordinate bases applies here
as well, so that for a transformation between general bases the above equation
generalizes to
\begin{equation}
{T'}^{\alpha_1\ldots\alpha_r}_{\beta_1\ldots\beta_s}
  =\Lambda^{\alpha_1}_{\phantom{\alpha_1}\mu_1}\cdots
   \Lambda^{\alpha_r}_{\phantom{\alpha_r}\mu_r}
   \Lambda^{\phantom{\beta_1}\nu_1}_{\beta_1}\cdots
   \Lambda^{\phantom{\beta_1}\nu_s}_{\beta_s}
   T^{\mu_1\ldots\mu_r}_{\nu_1\ldots\nu_s}.
\end{equation}

\subsection{The Lie Bracket}
\label{ybc}
Given a vector field, ${\bf V},$ the corresponding
first-order linear differential operator is:
\begin{equation}
\hat{{\bf V}} \equiv V^\alpha \frac{\partial}{\partial z^\alpha}.
\end{equation}
Notice that the $\alpha$th component of the vector can be recovered by
applying the operator to $z^\alpha$:
\begin{equation}
V^\alpha=\hat{{\bf V}}z^\alpha .
\end{equation}

As has been mentioned, it is possible to actually {\it identify} the vector 
with its corresponding operator.  Many mathematics texts actually do this,
and it is perfectly permissible since there is
an obvious one to one correspondence between vectors and first-order
linear differential operators by the above equations.  Indeed, there
are numerous advantages to such identification, but we shall continue
to use the circumflex to distinguish the operator in order to avoid 
any ambiguity.  

It is important to note that the operators corresponding
to two different vector fields do not, in general, commute.
Indeed, the commutator of two first-order linear differential operators
is another first-order linear differential operator.  At first this may
seem surprising because it is not obvious that this commutator is a {\it first}
order operator.  By writing it in terms of the components of ${\bf V}$
and ${\bf U},$ however, we see that the second order terms do indeed cancel:
\begin{eqnarray}
\left[ \hat{{\bf V}} , \hat{{\bf U}} \right]
  &=&V^\beta \frac{\partial}{\partial z^\beta}
     \left( U^\alpha \frac{\partial}{\partial z^\alpha} \right)
    -U^\beta \frac{\partial}{\partial z^\beta}
     \left( V^\alpha \frac{\partial}{\partial z^\alpha} \right) \nonumber\\
  &=&\left( V^\beta U^\alpha_{\phantom{\alpha},\beta}
          - U^\beta V^\alpha_{\phantom{\alpha},\beta} \right) 
     \frac{\partial}{\partial z^\alpha}.
\label{maa}
\end{eqnarray}

The vector whose operator is the commutator of the operators of two other
vectors, ${\bf V}$ and ${\bf U},$ is said to be the {\it Lie bracket} of 
those two vectors, and is denoted by $[{\bf V},{\bf U}].$  Note that the
Lie bracket operation makes the space of all vector fields into a Lie
algebra.

Using the Lie bracket, it is possible to give a simple test that will
determine whether or not any given set of basis vectors is a coordinate basis:
A set of $n$ linearly independent vectors constitutes a coordinate basis
if and only if the Lie bracket of any two elements of the set vanishes.
The ``only if'' part of this theorem is obvious, since coordinate basis vectors
are partial derivatives and these always commute with each other.  The
converse, however, is a special case of something called Frobenius' theorem, and is
somewhat harder to see.  To prove it algebraically, we must show that it is 
possible to actually construct a coordinate system (at least locally) given the
$n$ linearly independent commuting vectors.  We shall not follow this approach 
here (see Schutz~\cite{zaa} for details on how to prove it this way).  Instead,
we shall follow a more geometrical line of reasoning that will make the theorem
almost obvious.  To do this, however, we first need to learn about the Lie
derivative.


\subsection{Lie Derivatives}
\begin{figure}[p]
\center{
\vspace{2.46truein}
\mbox{\includegraphics[bbllx=0,bblly=0,bburx=259,bbury=115,width=5.82truein]{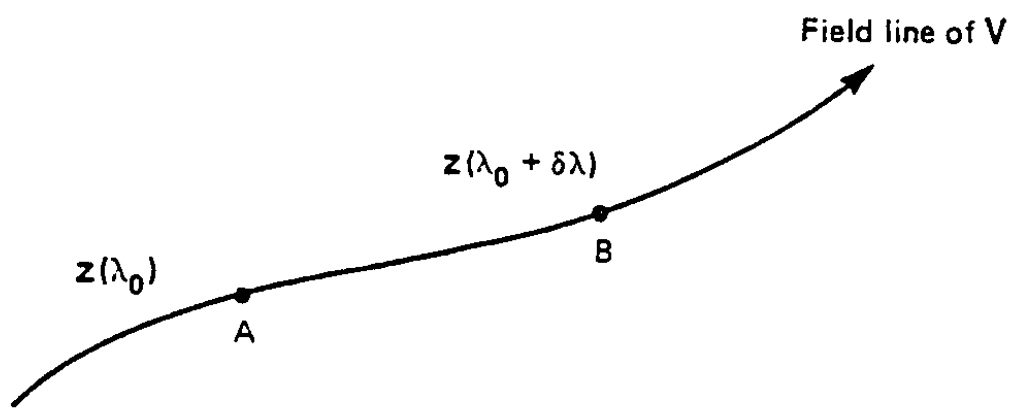}}
\vspace{2.46truein}
}
\caption{Lie Differentiation of a Scalar Field}
\label{mce}
\end{figure}

The {\it Lie derivative} of a scalar field, $f(z),$ with respect to the vector
field, ${\bf V},$ is a new scalar field denoted by ${\cal L}_V f,$ and 
is given by:
\begin{equation}
{\cal L}_V f=\hat{{\bf V}} f=V^\alpha \frac{\partial f}{\partial z^\alpha}.
\label{mbo}
\end{equation}
This is recognized as the directional derivative of $f$ along the vector
field, ${\bf V}.$  Along any given field line of ${\bf V},$ it is possible to
define a coordinate, $\lambda,$ such that:
\begin{equation}
V^\alpha=\frac{dz^\alpha}{d\lambda}
\end{equation}
so
\begin{equation}
\hat{{\bf V}}=\frac{d}{d \lambda},
\end{equation}
and so the Lie derivative of $f$ with respect to ${\bf V}$ is simply
$df/d\lambda.$  That is, we evaluate the scalar field at the points
$z(\lambda_0)$ and $z(\lambda_0+\delta \lambda)$ along the field line,
subtract the first value from the second, divide the result by
$\delta \lambda,$ and let $\delta \lambda$ go to zero to get the Lie 
derivative.  In Fig.~\ref{mce}, these two points of evaluation are
denoted by $A$ and $B.$

A scalar field, $f,$ whose Lie derivative with respect to ${\bf V}$ vanishes
is said to be a {\it Lie dragged} scalar field with respect to the vector
field ${\bf V}.$  Intuitively, this means that the scalar field is constant
along the field lines of ${\bf V}.$  Alternatively stated, it means that
the scalar field satisfies the first-order linear differential equation 
$\hat{{\bf V}}f=0,$ whose characteristics are the field lines of ${\bf V}.$
Thus, if the value of a Lie dragged scalar field is specified at any 
one point of a field line of ${\bf V},$ its value everywhere else on that
same field line is determined (it's the same value).  Using this concept, we 
can reword our definition of a Lie derivative:  Begin by evaluating the scalar 
field $f$ at point $A.$  Next, drag the scalar $f$ at point $B$ back to point
$A$ to get the scalar $f^*$ at $A$ (note $f^*(A)=f(B)$).  Now at 
the point $A$ we subtract $f$ from $f^*,$ divide the result by 
$\delta \lambda,$ and let $\delta \lambda$ go to zero to get ${\cal L}_V f.$
This may sound like a fancy way of saying the same thing, but it will aid in
our efforts to generalize the Lie derivative to act on other things 
besides scalars.

Consider the problem of trying to define an analogous derivative that
acts on contravariant vectors.  We could begin by evaluating a vector field, 
say ${\bf U},$ at the same two 
points, $z(\lambda_0)$ and $z(\lambda_0+\delta \lambda),$ along a field
line of ${\bf V}.$  Unfortunately, however, we cannot subtract them
because they live in two different vector spaces:  The first lives in
the space of all vectors at the point $z(\lambda_0),$ while the
second lives in the space of all vectors at $z(\lambda_0+\delta \lambda).$
We are dealing with spaces in which there may be no notion of parallel 
transport, so there is no natural way of comparing vectors located at
two different points.

\begin{figure}[p]
\center{
\vspace{1.47truein}
\mbox{\includegraphics[bbllx=0,bblly=0,bburx=260,bbury=204,width=5.82truein]{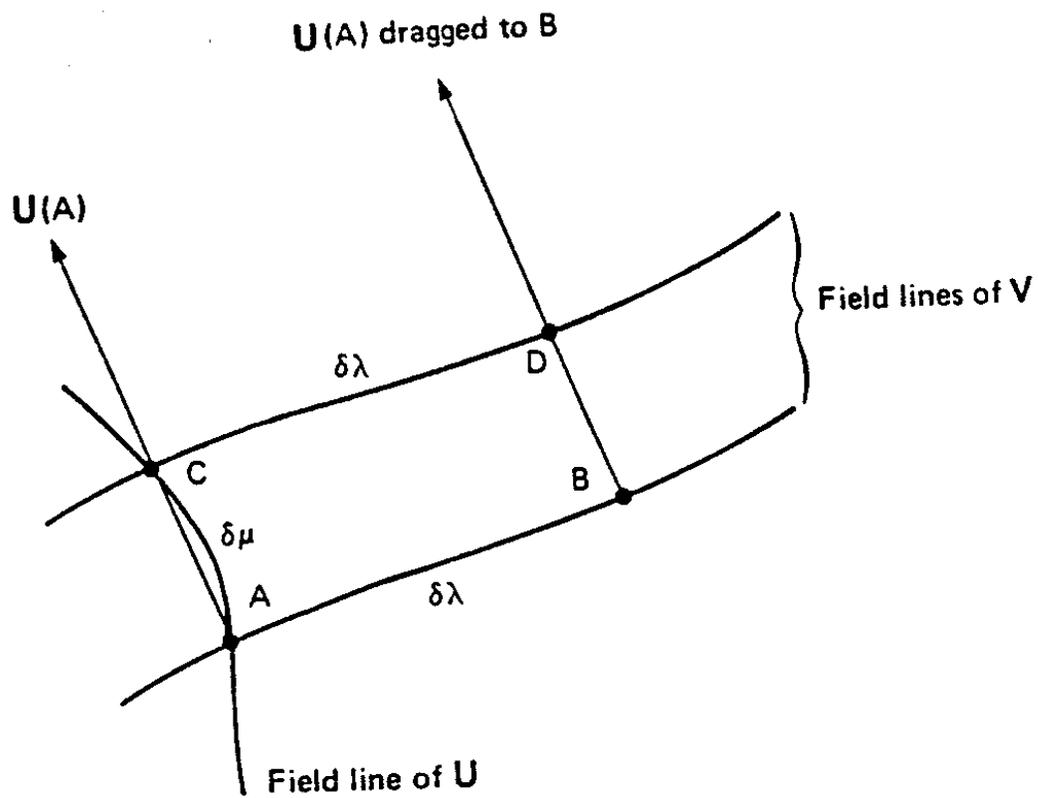}}
\vspace{1.47truein}
}
\caption{Lie Dragging a Vector Field Along Another Vector Field}
\label{mcf}
\end{figure}

So we must be a little more clever.  Refer to Fig.~\ref{mcf}.
Just as we have coordinatized a given field line of ${\bf V}$ by $\lambda,$
we shall use $\mu$ to coordinatize a given field line of ${\bf U}.$  The
operator $\hat{{\bf U}}$ is then $d/d\mu.$  It acts on scalars by evaluating
them at $z(\mu_0)$ and $z(\mu_0+\delta \mu),$ subtracting the first value
from the second, dividing by $\delta \mu,$ and letting $\delta \mu$ go to
zero.  In Fig.~\ref{mcf},
these two points of evaluation are denoted by $A$ and $C$; note that we have 
arranged things in this figure so that point $A$ is parametrized by both
$\lambda_0$ on the ${\bf V}$ field line, and $\mu_0$ on the ${\bf U}$
field line.  

Now we can imagine sliding the points $A$ and $C$ along the
${\bf V}$ field lines for an increment $\delta \lambda,$ to arrive at the
new points $B$ and $D,$ respectively.  These new points define a new
first-order linear differential operator based at the point $B.$  It acts
on scalars by evaluating them at the points $B$ and $D,$ subtracting the
first value from the second, dividing by $\delta \mu$ (it is clear that
points $B$ and $D$ coincide as $\delta \mu \rightarrow 0$), and letting
$\delta \mu$ go to zero.  This first-order linear differential operator at
$B$ corresponds to a vector at point $B,$ and so we see that we have 
found a natural way to drag the vector field ${\bf U}$ along the 
vector field ${\bf V}$.  If a vector field ${\bf U}$ is unchanged by 
dragging it along ${\bf V},$ then it is said to be a {\it Lie
dragged} vector field with respect to ${\bf V}.$

Armed with this insight, we are ready to define the Lie derivative of a
vector field, ${\bf U}$ with respect to another vector field, ${\bf V}.$
We begin by evaluating ${\bf U}$ at point $A.$  Next, we drag the vector
${\bf U}$ at point $B$ back to point $A$ to get the vector ${\bf U}^*$
at $A.$  Now we can subtract ${\bf U}$ from ${\bf U}^*,$ divide the result
by $\delta \lambda,$ and let $\delta \lambda$ go to zero to get 
${\cal L}_V {\bf U}.$  It should be clear from this description that the Lie 
derivative of a Lie dragged vector field vanishes, just as was the case for 
scalars.

\begin{figure}[p]
\center{
\vspace{1.34truein}
\mbox{\includegraphics[bbllx=0,bblly=0,bburx=255,bbury=211,width=5.82truein]{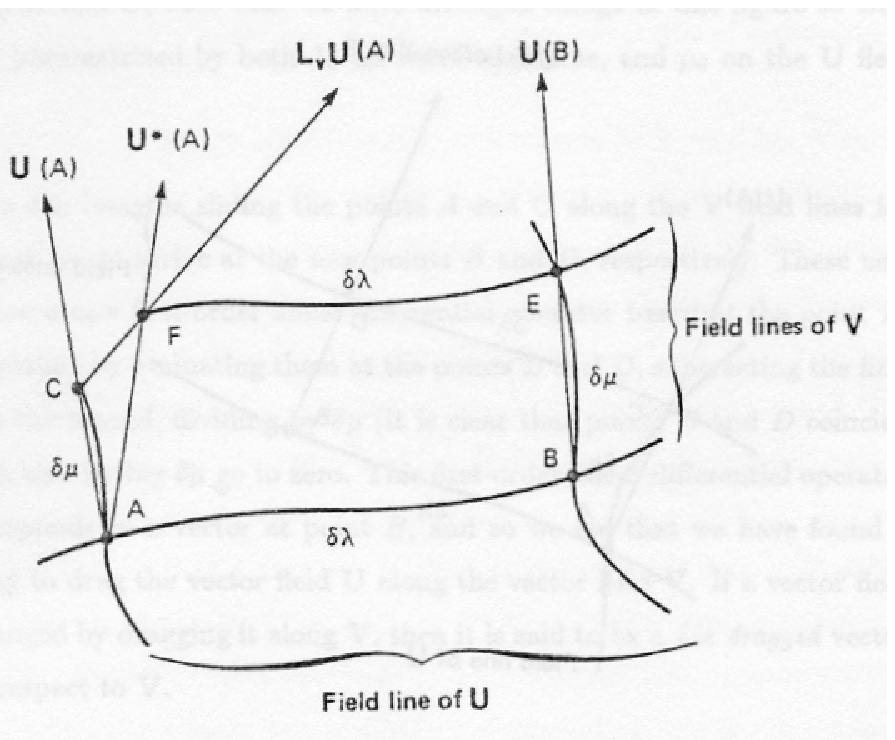}}
\vspace{1.34truein}
}
\caption{Lie Differentiation of a Vector Field}
\label{mcg}
\end{figure}

Now that we have the geometrical picture of what is happening, we need to
find an analytic expression for ${\cal L}_V {\bf U}.$  Refer to Fig.~\ref{mcg}.
It is clear that we may write:
\begin{eqnarray}
U^\alpha(A)   
&=&\lim_{\delta \mu \rightarrow 0}\frac{z^\alpha_C-z^\alpha_A}{\delta \mu}, \nonumber\\
U^\alpha(B)   
&=&\lim_{\delta \mu \rightarrow 0}\frac{z^\alpha_E-z^\alpha_B}{\delta \mu}, \nonumber\\
U^{*\alpha}(A)
&=&\lim_{\delta \mu \rightarrow 0}\frac{z^\alpha_F-z^\alpha_A}{\delta \mu}, \nonumber\\
\noalign{\hbox{and}}
({\cal L}_V {\bf U})^\alpha(A)
&=&\lim_{\delta \lambda \rightarrow 0}
\frac{U^{*\alpha}(A)-U^\alpha(A)}{\delta \lambda}\nonumber\\
&=&\lim_{\delta \lambda \rightarrow 0} \lim_{\delta \mu \rightarrow 0}
\frac{z^\alpha_F-z^\alpha_C}{\delta \mu \delta \lambda},
\label{mab}
\end{eqnarray}
where $z^\alpha_A$ through $z^\alpha_F$ are the coordinates at the points $A$
through $F,$ respectively.  To find these coordinates, we use Taylor expansion.
Thus, to express the coordinates of point $B$ in terms of quantities at
point $A,$ we write:
\begin{eqnarray}
z^\alpha_B
&=&z^\alpha(\lambda_0+\delta \lambda)\nonumber\\
&=&z^\alpha_A + {\left. \frac{dz^\alpha}{d\lambda} \right| }_A \delta \lambda
   +\frac{1}{2}{\left. \frac{d^2z^\alpha}{d\lambda^2} \right| }_A 
    \delta \lambda^2 + \cdots 
\label{mac}
\end{eqnarray}
Similarly, the coordinates of point $C$ are given by:
\begin{eqnarray}
z^\alpha_C&=&z^\alpha(\mu_0+\delta \mu)\nonumber\\
     &=&z^\alpha_A + {\left. \frac{dz^\alpha}{d\mu} \right| }_A \delta \mu
      +\frac{1}{2}{\left. \frac{d^2z^\alpha}{d\mu^2} \right| }_A 
      \delta \mu^2 + \cdots 
\label{mad}
\end{eqnarray}
Next, the coordinates of point $E$ can be expressed in terms of quantities at
point $B$:
\begin{equation}
z^\alpha_E=z^\alpha_B + {\left. \frac{dz^\alpha}{d\mu} \right| }_B \delta \mu
      +\frac{1}{2}{\left. \frac{d^2z^\alpha}{d\mu^2} \right| }_B 
      \delta \mu^2 + \cdots ,
\end{equation}
and these in turn may be expressed in terms of quantities at point $A$:
\begin{eqnarray}
z^\alpha_E
&=&z^\alpha_A + {\left. \frac{dz^\alpha}{d\lambda} \right| }_A \delta \lambda
      +\frac{1}{2}{\left. \frac{d^2z^\alpha}{d\lambda^2} \right| }_A 
      \delta \lambda^2 + \cdots \nonumber\\
     &&\qquad + {\left. \frac{dz^\alpha}{d\mu} \right| }_A \delta \mu
      +{\left. \frac{d^2z^\alpha}{d\lambda d\mu} \right| }_A
      \delta \lambda \delta \mu + \cdots \nonumber\\
     &&\qquad + \frac{1}{2}{\left. \frac{d^2z^\alpha}{d\mu^2} \right| }_A 
      \delta \mu^2 + \cdots
     \label{mae}
\end{eqnarray}
Finally, the coordinates of point $F$ can be expressed in terms of quantities
at point $E,$ which in turn can be expressed in terms of quantities at point
$B,$ which in turn can be expressed in terms of quantities at point $A$:
\begin{eqnarray}
z^\alpha_F
&=&z^\alpha_E - {\left. \frac{dz^\alpha}{d\lambda} \right| }_E \delta \lambda
      +\frac{1}{2}{\left. \frac{d^2z^\alpha}{d\lambda^2} \right| }_E
      \delta \lambda^2 + \cdots \nonumber\\
     &=&z^\alpha_B + {\left. \frac{dz^\alpha}{d\mu} \right| }_B \delta \mu
      +\frac{1}{2}{\left. \frac{d^2z^\alpha}{d\mu^2} \right| }_B 
      \delta \mu^2 + \cdots \nonumber\\
     &&\qquad - {\left. \frac{dz^\alpha}{d\lambda} \right| }_B \delta \lambda
      -{\left. \frac{d^2z^\alpha}{d\mu d\lambda} \right| }_B 
      \delta \mu \delta \lambda - \cdots \nonumber\\
     &&\qquad +\frac{1}{2}{\left. \frac{d^2z^\alpha}{d\lambda^2} \right| }_B
      \delta \lambda^2 + \cdots \nonumber\\
     &=&z^\alpha_A 
      + {\left. \frac{dz^\alpha}{d\lambda} \right| }_A \delta \lambda
      +\frac{1}{2}{\left. \frac{d^2z^\alpha}{d\lambda^2} \right| }_A 
      \delta \lambda^2 + \cdots \nonumber\\
     &&\qquad + {\left. \frac{dz^\alpha}{d\mu} \right| }_A \delta \mu
      +{\left. \frac{d^2z^\alpha}{d\lambda d\mu} \right| }_A 
      \delta \lambda \delta \mu + \cdots \nonumber\\
     &&\qquad + \frac{1}{2}{\left. \frac{d^2z^\alpha}{d\mu^2} \right| }_A
      \delta \mu^2 + \cdots \nonumber\\
     &&\qquad - {\left. \frac{dz^\alpha}{d\lambda} \right| }_A \delta \lambda
      -{\left. \frac{d^2z^\alpha}{d\lambda^2} \right| }_A 
      \delta \lambda^2 - \cdots \nonumber\\
     &&\qquad -{\left. \frac{d^2z^\alpha}{d\mu d\lambda} \right| }_A 
      \delta \mu \delta \lambda - \cdots \nonumber\\
     &&\qquad +\frac{1}{2}{\left. \frac{d^2z^\alpha}{d\lambda^2} \right| }_A
      \delta \lambda^2 + \cdots \nonumber\\
     &=&z^\alpha_A + {\left. \frac{dz^\alpha}{d\mu} \right| }_A \delta \mu
      + \frac{1}{2}{\left. \frac{d^2z^\alpha}{d\mu^2} \right| }_A
      \delta \mu^2 + \cdots \nonumber\\
     &&\qquad +{\left. \frac{d^2z^\alpha}{d\lambda d\mu} \right| }_A 
      \delta \lambda \delta \mu
      -{\left. \frac{d^2z^\alpha}{d\mu d\lambda} \right| }_A 
      \delta \mu \delta \lambda + \cdots
\label{maf}
\end{eqnarray}
Thus, using Eqs.~(\ref{mab}), we find:
\begin{eqnarray}
U^\alpha(A)   &=&{\left. \frac{dz^\alpha}{d\mu} \right|}_A, \nonumber\\
\noalign{\hbox{and}}
U^{*\alpha}(A)&=&{\left. \frac{dz^\alpha}{d\mu} \right|}_A
   +\left( {\left. \frac{d^2z^\alpha}{d\lambda d\mu} \right|}_A
   -       {\left. \frac{d^2z^\alpha}{d\mu d\lambda} \right|}_A \right)
   \delta \lambda + \cdots \nonumber\\
\noalign{\hbox{and so}}
({\cal L}_V {\bf U})^\alpha(A)
   &=&{\left. \frac{d^2z^\alpha}{d\lambda d\mu} \right|}_A
    -{\left. \frac{d^2z^\alpha}{d\mu d\lambda} \right|}_A \nonumber\\
   &=&{\left. \left( \hat{{\bf V}}\hat{{\bf U}}
                   -\hat{{\bf U}}\hat{{\bf V}}\right) z^\alpha \right|}_A \nonumber\\
   &=&{\left. \left[ \hat{{\bf V}} , \hat{{\bf U}} \right] z^\alpha \right|}_A.
\label{mag}
\end{eqnarray}
We have just demonstrated that the Lie derivative of ${\bf U}$ with respect to
${\bf V}$ is simply the Lie bracket of ${\bf V}$ and ${\bf U}$:
\begin{equation}
{\cal L}_V {\bf U} = \left[ {\bf V} , {\bf U} \right].
\end{equation}

In a coordinate basis, this result may be written
\begin{equation}
({\cal L}_V {\bf U})^\alpha=
  V^\beta U^\alpha_{\phantom{\alpha},\beta}
 -U^\beta V^\alpha_{\phantom{\alpha},\beta}.
\end{equation}
Note that this way of writing the result may be taken as valid for a
noncoordinate basis as well if we reinterpret the commas as meaning
``operation by the basis vector.''  That is, $f_{,\alpha}$ denotes the
result of applying to $f$ the operator corresponding to the basis vector
${\hat{\bf e}}_\alpha.$  For a coordinate basis, the operators corresponding
to basis vectors are simply partial derivatives with respect to the
coordinates, so this reduces to the usual meaning of the comma.  This
generalization of what the comma means will be useful in everything that
follows.

Now that we know how to take the Lie derivative of a contravariant vector 
field, we shall try to extend this process to covector fields.  
Recall that covectors contract with contravariant vectors to give 
scalars.  We define a {\it Lie dragged} covector field to be one which
when contracted with any Lie dragged contravariant vector field yields a Lie 
dragged scalar field.  To take the Lie derivative of a covector field ${\bf a}$
with respect to ${\bf V},$ we evaluate ${\bf a}$ at the points $A$ and
$B$ in Fig.~\ref{mce}, drag ${\bf a}(B)$ back to $A$ 
to get ${\bf a}^*(A),$ subtract ${\bf a}(A)$ from ${\bf a}^*(A),$ 
divide by $\delta \lambda ,$ and let $\delta \lambda$ go to zero.
The result is:
\begin{equation}
({\cal L}_V {\bf a})_\alpha=
  V^\beta a_{\alpha,\beta}
 +V^\beta_{\phantom{\beta},\alpha} a_\beta,
\end{equation}
Once again, this result is valid for noncoordinate bases if we generalize the
meaning of the commas.

Next we consider the Lie derivative of a general tensor.  We first define
a {\it Lie dragged} tensor of type $(r,s)$ as one which yields a Lie dragged
scalar field when fed $r$ Lie dragged covectors and $s$ Lie dragged vectors.
To take the Lie derivative of a tensor ${\bf T}$ of type $(r,s)$ with respect
to ${\bf V}$, we evaluate ${\bf T}$ at the points $A$ and $B$ in 
Fig.~\ref{mce}, drag ${\bf T}(B)$ back to $A$ 
to get ${\bf T}^*(A),$ subtract ${\bf T}(A)$ from ${\bf T}^*(A),$ 
divide by $\delta \lambda ,$ and let $\delta \lambda$ go to zero.
The result is:
\begin{eqnarray}
({\cal L}_V {\bf T})^{\alpha_1\ldots\alpha_r}_{\beta_1\ldots\beta_s}&=&
  V^\gamma 
  T^{\alpha_1\ldots\alpha_r}_{\beta_1\ldots\beta_s,\gamma}\nonumber\\
 &&\qquad -V^{\alpha_1}_{\phantom{\alpha_1},\gamma}
  T^{\gamma\alpha_2\ldots\alpha_r}_{\beta_1\ldots\beta_s}-\cdots\nonumber\\
 &&\qquad -V^{\alpha_r}_{\phantom{\alpha_r},\gamma}
  T^{\alpha_1\ldots\alpha_{r-1}\gamma}_{\beta_1\ldots\beta_s}\nonumber\\
 &&\qquad +V^{\gamma}_{\phantom{\gamma},\beta_1}
  T^{\alpha_1\ldots\alpha_r}_{\gamma\beta_2\ldots\beta_s}+\cdots\nonumber\\
 &&\qquad +V^{\gamma}_{\phantom{\gamma},\beta_s}
  T^{\alpha_1\ldots\alpha_r}_{\beta_1\ldots\beta_{s-1}\gamma}
 \label{mcn}
\end{eqnarray}
Note that the above geometrical picture for Lie derivatives of general tensors
is equivalent to the neat coordinate-free algebraic formula
\begin{eqnarray}
{\cal L}_V
  \left( {\bf T}({\bf a}_1,\ldots,{\bf a}_r,{\bf U}_1,\ldots,{\bf U}_s)\right)
&=&({\cal L}_V{\bf T})({\bf a}_1,\ldots,{\bf a}_r,{\bf U}_1,\ldots,{\bf U}_s)\nonumber\\
&&\qquad +{\bf T}
  ({\cal L}_V{\bf a}_1,\ldots,{\bf a}_r,{\bf U}_1,\ldots,{\bf U}_s)+\cdots\nonumber\\
&&\qquad +{\bf T}
  ({\bf a}_1,\ldots,{\cal L}_V{\bf a}_r,{\bf U}_1,\ldots,{\bf U}_s)\nonumber\\
&&\qquad +{\bf T}
  ({\bf a}_1,\ldots,{\bf a}_r,{\cal L}_V{\bf U}_1,\ldots,{\bf U}_s)+\cdots\nonumber\\
&&\qquad +{\bf T}
  ({\bf a}_1,\ldots,{\bf a}_r,{\bf U}_1,\ldots,{\cal L}_V{\bf U}_s).
\label{mco}
\end{eqnarray}

Finally, it is straightforward to show that Lie derivatives obey the Leibniz
rule over the tensor product.  That is
\begin{equation}
{\cal L}_V ({\bf T}_1\otimes {\bf T}_2)
  =({\cal L}_V {\bf T}_1)\otimes {\bf T}_2
  +{\bf T}_1\otimes ({\cal L}_V {\bf T}_2).
\end{equation}

Before leaving this subsection, it is important to emphasize that the same
geometrical notions that led us to the Lie derivative of a vector field
still apply for arbitrary tensors:  The Lie derivative is
the natural way to ``drag'' any tensorial object along
the field lines of a vector field.  Just as we dragged
the vector ${\bf U}$ along the field line of the vector ${\bf V}$
for a parameter interval $\delta\lambda$ to get the vector
\begin{equation}
{\bf U}^*={\bf U}+({\cal L}_V {\bf U})\delta\lambda
\end{equation}
(see Eq.~(\ref{mab})), so we can drag the tensor ${\bf T}$ in exactly the same
way to get
\begin{equation}
{\bf T}^*={\bf T}+({\cal L}_V {\bf T})\delta\lambda.
\end{equation}
This geometrical insight is crucial to the understanding of Lie transforms.

\subsection{Examples of Coordinate and Noncoordinate Bases}
\label{yay}
We are now in a position to understand the theorem presented at the end
of Section~\ref{ybc} from a geometrical
point of view.  Fig.~\ref{mcg} and Eq.~(\ref{mab}) make it clear that the
Lie bracket of two vector fields is related to the infinitesimal difference
in position resulting from the operation of moving along the first vector 
field for a certain parameter interval, then along the second vector field,
then backwards for the same parameter interval along the first, then
backwards along the second.  Clearly, if the vector fields involved are
basis elements of a coordinate basis, this operation will simply take one
around a square right back to the original position.  The sides of the
square are the contours of constant values of the two coordinates involved.
Conversely, if two members of a set of $n$ linearly independent vectors
have nonvanishing Lie bracket, then it is impossible to construct a
coordinate system that has those vectors as a basis because moving around
the above-described infinitesimal loop does not return one to the starting
point; the changing parameters do not ``hook together'' in the manner necessary
for them to be coordinates.

\begin{figure}[p]
\center{
\vspace{1.53truein}
\mbox{\includegraphics[bbllx=0,bblly=0,bburx=232,bbury=177,width=5.82truein]{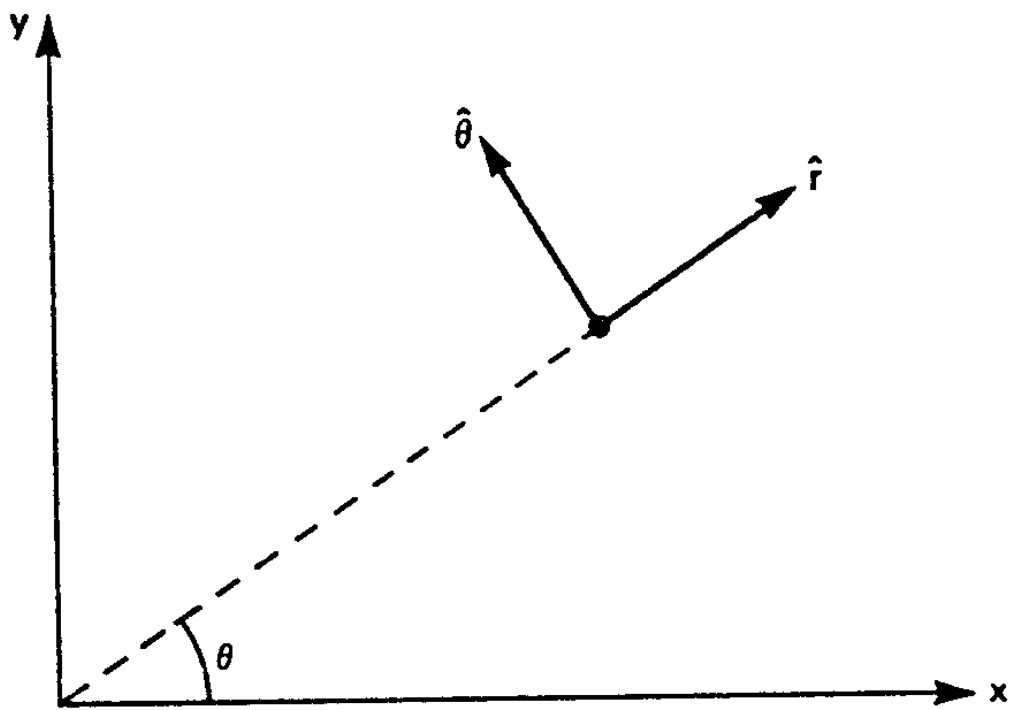}}
\vspace{1.53truein}
}
\caption{Polar Coordinate Unit Vectors}
\label{mcp}
\end{figure}

Part of the reason that this concept of coordinate and noncoordinate bases
is tricky is that there is no need for
such a distinction in Cartesian coordinates.  There, the coordinate basis
is identical to the usual orthonormal basis
\begin{equation}
\frac{\partial}{\partial x^\alpha}={\hat{\bf e}}_\alpha.
\end{equation}
A good example of a familiar situation for which the distinction {\it is}
important is that of polar coordinates in two dimensions.  The usual
polar unit vectors, $\hat{\bf r}$ and $\hat{\bf \theta},$ are not a
coordinate basis since
\begin{equation}
\left[ \hat{\bf r},\hat{\bf \theta} \right]=-\frac{\hat{\bf \theta}}{r}.
\end{equation}
On the other hand, $\partial/\partial r$ and $\partial/\partial \theta$ 
{\it do} constitute a valid coordinate basis, and these are related to
the above orthonormal basis by
\begin{eqnarray}
\frac{\partial}{\partial r}&=&\hat{\bf r}\nonumber\\
\noalign{\hbox{and}}
\frac{\partial}{\partial \theta}&=&r\hat{\bf \theta}.\label{mcm}
\end{eqnarray}
The important point is that there are {\it no} pair of coordinates,
$\xi$ and $\eta,$ such that $\hat{\bf r}=\partial/\partial\xi$ {\it and}
$\hat{\bf \theta}=\partial/\partial\eta .$  Geometrically, this is because
if we traverse an infinitesimal loop following first the $\hat{\bf r}$
vector field and then the $\hat{\bf \theta}$ vector field (and then
returning along them, respectively) we will not arrive at our starting
point (see Fig.~\ref{mcp}).  The factor of $r$ on the right hand side of the
second of Eqs.~(\ref{mcm}) corrects for this and gives us a coordinate basis.

As mentioned previously, the above distinction also holds for covectors.
To pursue the above example, the covector basis consisting of $dr$ and
$d\theta$ is dual to the vector basis consisting of $\partial/\partial r$
and $\partial/\partial\theta.$  It follows that the covector basis
\begin{eqnarray}
\tilde{\bf r}&=&dr\nonumber\\
\noalign{\hbox{and}}
\tilde{\bf \theta}&=&rd\theta \label{mdw}
\end{eqnarray}
is dual to the vector basis $\hat{\bf r}$ and $\hat{\bf \theta}.$  Once
again, there is no pair of coordinates, $\xi$ and $\eta,$ such that
$\tilde{\bf r}=d\xi$ {\it and} $\tilde{\bf \theta}=d\eta.$

\subsection{Differential Forms}
An $s$-{\it form} is defined as a tensor of type $(0,s)$ that is antisymmetric 
in every pair of its $s$ vector arguments.  In particular, a zero form is a
scalar and a one form is a covector; a two form, $\tilde{\Omega},$  obeys
\begin{equation}
\tilde{\Omega}({\bf U},{\bf V})=-\tilde{\Omega}({\bf V},{\bf U}),
\end{equation}
etc.  It follows that the components of an $s$-form are antisymmetric 
under interchange of any pair of indices.  In particular, this means that the
$s$ indices must all be different, or else the component will vanish.  Hence
the requirement of antisymmetry means that there are no longer $n^s$
independent components.  Instead, a standard combinatorial argument shows
that only
\begin{equation}
\left( \begin{array}{c}
                  n  \\
                  s 
               \end{array}\right)\equiv\frac{n!}{s!(n-s)!}
\end{equation}
of the components are truly independent.  This means, among other things, that 
there are no nontrivial $s$-forms in an $n$-dimensional space if $s>n,$ that 
an $n$-form has only one nontrivial component, etc.  The total number of
independent components of {\it all} forms in a space of dimension $n$ is thus
\begin{equation}
\sum_{s=0}^n \left( \begin{array}{c}
                           n  \\
                           s 
                    \end{array}\right) =(1+1)^n=2^n.
\end{equation}

Note that $s$-forms inherit some properties from the fact that they are
tensors of type $(0,s).$  In particular, two $s$-forms may be added to get a
third $s$-form.  Thus, the set of all forms at a point in an $n$-dimensional
manifold may be thought of as a $2^n$-dimensional vector space, although it
is somewhat strange in that its elements may be added if and only if they
are both $s$-forms for some $s.$  Note, however, that this space is {\it not}
an algebra under the tensor product operation because it is not closed under
that operation:  The tensor product of two forms is not necessarily a form.
If we take the tensor product of an $s_1$-form with an $s_2$-form, we get
a tensor of type $(0,s_1+s_2)$ that is clearly antisymmetric under interchange
of any two of its first $s_1$ or last $s_2$ arguments, but is not necessarily
antisymmetric under interchange of one of its first $s_1$ components with one
of its last $s_2$ components.

\subsection{The Wedge Product, the Interior Product, Dual Tensors}
It would thus be nice to define a product under which the set of all forms
becomes a closed algebra.  Such a product is called the {\it wedge product},
and is denoted by the symbol $\wedge .$  We motivate its definition as follows:
The wedge product of a scalar (zero form) with any $s$-form is the $s$-form
obtained by simple multiplication by the scalar.  The wedge product of two
one forms, ${\bf a}^1$ and ${\bf a}^2,$ is the two form given by
\begin{equation}
{\bf a}^1\wedge {\bf a}^2={\bf a}^1\otimes {\bf a}^2
   -{\bf a}^2\otimes {\bf a}^1.
\end{equation}
It is clear that the two form thus obtained is antisymmetric.  For three or
more one forms, we demand that the wedge product be associative, so, for
example
\begin{eqnarray}
{\bf a}^1\wedge {\bf a}^2\wedge {\bf a}^3
  &=&{\bf a}^1\wedge ({\bf a}^2\wedge {\bf a}^3)\nonumber\\
  &=&({\bf a}^1\wedge {\bf a}^2)\wedge {\bf a}^3\nonumber\\
  &=&\phantom{+}
    {\bf a}^1\otimes {\bf a}^2\otimes {\bf a}^3
   +{\bf a}^2\otimes {\bf a}^3\otimes {\bf a}^1
   +{\bf a}^3\otimes {\bf a}^1\otimes {\bf a}^2\nonumber\\ 
  &\phantom{=}&
   -{\bf a}^3\otimes {\bf a}^2\otimes {\bf a}^1
   -{\bf a}^1\otimes {\bf a}^3\otimes {\bf a}^2
   -{\bf a}^2\otimes {\bf a}^1\otimes {\bf a}^3.
  \label{mcr}
\end{eqnarray}

Recall that the total number of independent components of an $s$-form in a
space of dimension $n$ is given by 
$\left( \begin{array}{c} n \\ s \end{array}\right).$  A moment's thought 
convinces one that the vector space of all such $s$-forms is spanned by the
$\left( \begin{array}{c} n \\ s \end{array}\right)$ independent basis
$s$-forms
\begin{equation}
{\tilde{\bf \omega}}^{\beta_1}\wedge\ldots\wedge 
  {\tilde{\bf \omega}}^{\beta_s},
\end{equation}
where the $\beta$ indices range from $1$ to $n,$ and must all be different
(else the above expression will vanish).  It is then straightforward to
see that an arbitrary $s$-form, ${\bf \Omega},$ is given by
\begin{equation}
\Omega=\frac{1}{s!}\Omega_{\beta_1\ldots\beta_s}
  {\tilde{\bf \omega}}^{\beta_1}\wedge\ldots\wedge 
  {\tilde{\bf \omega}}^{\beta_s}.
\end{equation}
Note that the factor of $s!$ appears here because we did not put it into
the definition of the wedge product; we could have done it either way, and
authors differ in this convention.

Now that we know how the wedge product operates on scalars and one forms,
we can extend its definition to arbitrary forms by writing them in terms of
wedge products of basis one forms, as shown above.  This makes the 
$2^n$-dimensional vector space of all forms into an algebra, called a
{\it Grassmann algebra}.  Note that it is not a commutative algebra:
If $\Omega^1$ and $\Omega^2$ are $s_1$ and $s_2$-forms, respectively, then
\begin{equation}
\Omega_1\wedge\Omega_2=(-1)^{s_1s_2}\Omega_2\wedge\Omega_1.
\end{equation}

If we contract the first index of an $s$-form (where $s\geq 1$), $\Omega,$
with a vector, ${\bf V},$ then it is straightforward to see that we get
an $(s-1)$-form.  We call this new form the {\it interior product} of
$\Omega$ with ${\bf V},$ and we denote it by $i_V\Omega.$  Thus
\begin{equation}
i_V\Omega=\frac{1}{(s-1)!}V^{\beta_1}\Omega_{\beta_1\ldots\beta_s}
  {\tilde{\bf \omega}}^{\beta_2}\wedge\ldots\wedge 
  {\tilde{\bf \omega}}^{\beta_s}.
\label{tac}
\end{equation}
If $\Omega^1$ and $\Omega^2$ are $s_1$ and $s_2$-forms, respectively, then
it is straightforward to show
\begin{equation}
i_V(\Omega^1\wedge\Omega^2)=(i_V\Omega^1)\wedge\Omega^2
  +(-1)^{s_1}\Omega^1\wedge (i_V\Omega^2).
\end{equation}
Also, the antisymmetry of forms makes it clear that
\begin{equation}
i_V i_V \Omega=0
\end{equation}
for any $n$-form $\Omega$ with $n\geq 2.$

Recall that we defined an $s$-form as a completely antisymmetric tensor of type
$(0,s).$  Note that we could have done the same thing for completely 
antisymmetric tensors of type $(s,0).$  Next
note that a completely antisymmetric tensor of type $(0,s)$ has exactly the 
same number of components as a completely antisymmetric tensor of type
$(n-s,0)$ in a space of dimension $n\geq s.$  This is because
\begin{equation}
\left( \begin{array}{c} n \\ s \end{array}\right)
  =\left( \begin{array}{c} n \\ n-s \end{array}\right).
\end{equation}
This suggests that there may be a one-to-one correspondence between $s$-forms
and completely antisymmetric tensors of type $(n-s,0).$

For example, note that there is only one independent component of a
completely antisymmetric tensor of type $(n,0).$  This is because the
components of such a tensor must be proportional to those of the
Levi-Civita symbol, $\epsilon^{\beta_1\ldots\beta_n}.$  The proportionality
constant is a scalar (zero form).  Similarly, we can put any
scalar (zero form) in front of the Levi-Civita symbol, and obtain the 
components of a completely antisymmetric tensor of type $(n,0).$  Thus, there
is a one-to-one correspondence between scalars (zero forms) and completely
antisymmetric tensors of type $(n,0).$

More generally, we can use the Levi-Civita symbol to obtain a one-to-one
correspondence between $s$-forms and completely antisymmetric tensors of
type $(n-s,0)$ as follows:
\begin{eqnarray}
T^{\beta_1\ldots\beta_{n-s}}
 &=&\frac{1}{s!}\epsilon^{\beta_1\ldots\beta_n}
   \Omega_{\beta_{n-s+1}\ldots\beta_n},
 \label{mcs}\\
\noalign{\hbox{and}}
\Omega_{\beta_{n-s+1}\ldots\beta_n}
 &=&\frac{1}{(n-s)!}\epsilon_{\beta_1\ldots\beta_n}
   T^{\beta_1\ldots\beta_n}.
 \label{mct}
\end{eqnarray}
Here we have used the easily verified relation
\begin{equation}
\epsilon^{\beta_1\ldots\beta_n}\epsilon_{\beta_1\ldots\beta_n}=n!.
\end{equation}
Referring to Eq.~(\ref{mcs}), we say that ${\bf T}$ is {\it dual} to
${\bf \Omega}$ with respect to $\epsilon.$  This is often abbreviated
${\bf T}=\hbox{$^*{\bf \Omega}$}.$  Referring to Eq.~(\ref{mct}), we say that
${\bf \Omega}$ is {\it dual} to ${\bf T}$ with respect to $\epsilon,$ or
${\bf \Omega}=\hbox{$^*{\bf T}$}.$  Note that for any form, ${\bf \Omega},$ we
have $\hbox{$^{**}{\bf \Omega}$}=(-1)^{s(n-s)}{\bf \Omega}.$

\subsection{The Exterior Derivative and the Homotopy Formula}
We now define a differential operator, $d,$ that converts $s$-forms 
into $(s+1)$-forms.  This operator is defined as follows:  When applied
to a scalar (zero form), $f,$ it yields the one form, $df,$ such that
\begin{equation}
df({\bf V})=\hat{\bf V}f.
\end{equation}
Thus, in a coordinate basis, $z^\alpha,$ we have
\begin{equation}
df=\frac{\partial f}{\partial z^\alpha}dz^\alpha.
\end{equation}
Next, we demand that the operator be linear, so if ${\bf \Omega}$ and
${\bf \Lambda}$ are two $s$-forms then
\begin{equation}
d({\bf \Omega}+{\bf \Lambda})=d{\bf \Omega}+d{\bf \Lambda}.
\end{equation}
Next, we demand that if ${\bf \Omega}_1$ is an $s_1$-form and ${\bf \Omega}_2$
is an $s_2$-form,
\begin{equation}
d({\bf \Omega}_1\wedge {\bf \Omega}_2)=d{\bf \Omega}_1\wedge {\bf \Omega}_2
  +(-1)^{s_1}{\bf \Omega}_1\wedge d{\bf \Omega}_2.
\end{equation}
Finally, we demand that for any $s$-form, ${\bf \Omega},$ we have
\begin{equation}
dd{\bf \Omega}=0.
\end{equation}
The above demands define the operator $d$ uniquely and unambiguously.  We can
apply the exterior derivative to an arbitrary form by first expanding it
in terms of wedge products of basis one forms, and then applying the above 
rules.

In terms of components in a coordinate basis, the exterior derivative of a 
scalar is
\begin{equation}
(df)_\alpha=f_{,\alpha},
\end{equation}
and the exterior derivative of a one form is
\begin{equation}
(d{\bf a})_{\alpha\beta}=a_{\beta , \alpha}-a_{\alpha , \beta}.
\end{equation}
More generally,
\begin{equation}
(d{\bf \Omega})_{\alpha_1\ldots\alpha_{s+1}}
  =\frac{(-1)^s}{s!}\epsilon_{\beta_1\ldots\beta_{s+1}}
   \Omega_{\alpha_{\beta_1}\ldots\alpha_{\beta_s},\alpha_{\beta_{s+1}}}.
\end{equation}

A form whose exterior derivative vanishes is said to be {\it closed}.
A form that is the exterior derivative of another form is said to be 
{\it exact}.  Clearly, any exact form is closed.  The interesting question
is whether or not any closed form is exact.  The answer to this depends on
the global topology of the manifold on which the closed form lives.
Locally, it is always true.

There is a marvelous relationship between Lie derivatives, interior products,
and exterior derivatives.  It is possible to prove that
\begin{equation}
{\cal L}_V\Omega=i_Vd\Omega+di_V\Omega
\label{mby}
\end{equation}
for any $n$-form, $\Omega,$ with $n\geq 1,$ and any
vector field, ${\bf V}.$  This relationship is
called the {\it homotopy formula}.  The proof usually given (see for example
Section 4.20 of Schutz~\cite{zaa}) proceeds by induction:  It is first proved
for a one-form, and then it is shown that it works for an $n$-form if it works
for an $(n-1)$-form.

The {\it generalized homotopy formula},
\begin{equation}
{\cal L}_V^j\Omega=(i_Vd)^j\Omega+(di_V)^j\Omega
\label{mbz}
\end{equation}
for $j\geq 1,$ is proved by induction as follows:  First note that it reduces
to the ordinary homotopy formula when $j=1.$  Next, assume that it is true for
$j=l.$  Then
\begin{eqnarray}
{\cal L}_V^{l+1}\Omega&=&(i_Vd+di_V)[(i_Vd)^l\Omega+(di_V)^l\Omega]\nonumber\\
  &=&(i_Vd)^{l+1}\Omega+(di_V)^{l+1}\Omega, \label{mcu}
\end{eqnarray}
where we have used the fact that application of $dd$ or $i_Vi_V$ causes
any form to vanish.  Note that the generalized homotopy formula 
is not true for $j=0.$

Finally, we can show that Lie derivatives commute with exterior derivatives.
This is done as follows:
\begin{equation}
d{\cal L}_V=d(i_Vd+di_V)=di_Vd=(i_Vd+di_V)d={\cal L}_Vd,
\end{equation}
where we have used the homotopy formula.

\subsection{Integration on Manifolds}
Differential $s$-forms can also be introduced as integrands of $s$-dimensional
integrals.  See Flanders~\cite{zbc} for more on this approach.  Adopting this
point of view, it is possible to prove the generalized {\it Stokes' theorem}
\begin{equation}
\int_U d{\bf \Omega}=\int_{\partial U}{\bf \Omega},
\end{equation}
where $U$ is an $(s+1)$-dimensional volume, and $\partial U$ is the
$s$-dimensional surface that bounds it.

We shall not attempt to prove the generalized Stokes' theorem here (see
Schutz~\cite{zaa} for a good presentation), but we shall make it plausible
by showing how it reduces to the familiar Stokes' theorem and divergence 
theorem of three dimensional vector calculus.  In three dimensional Euclidean
space, with Cartesian coordinates, the gradient is given in our notation by
\begin{equation}
(\delr f)_i=(df)_i,
\end{equation}
the divergence is given by
\begin{equation}
\delr\cdot {\bf V}=\hbox{$^*d$}\hbox{$^*{\bf V}$},
\end{equation}
and the curl is given by
\begin{equation}
\delr\times {\bf V}=\hbox{$^*d$}{\bf v},
\end{equation}
where ${\bf v}$ is the one form whose Cartesian components are identical to
those of the vector ${\bf V}.$  Note that
\begin{equation}
\delr\times\delr f=\hbox{$^*d$}df=0
\end{equation}
and
\begin{equation}
\delr\cdot(\delr\times {\bf V})=\hbox{$^*d$}\hbox{$^{**}{d{\bf v}}$}
                               =\hbox{$^*d$}d{\bf v}=0
\end{equation}
both follow from $dd=0.$  Then
\begin{equation}
\int_U \delr\cdot {\bf V} dx^3=\int_U d\hbox{$^*{\bf V}$}
  =\int_{\partial U} \hbox{$^*{\bf V}$}
  =\int_{\partial U} {\bf V}\cdot d{\bf \sigma},
\end{equation}
and
\begin{equation}
\int_U \delr\times {\bf V}\cdot d{\bf \sigma}
  =\int_U d{\bf v}
  =\int_{\partial U} {\bf v}
  =\int_{\partial U}{\bf V}\cdot d{\bf \ell}.
\end{equation}
Thus we see that our formalism is the natural generalization of three
dimensional vector calculus to manifolds of arbitrary dimension.  

\subsection{Metric Spaces}
The usual {\it dot product} of linear algebra is a rule for taking two vectors,
say ${\bf U}$ and ${\bf V},$ and associating with them a real number, denoted
${\bf U}\cdot {\bf V}.$  The result depends bilinearly on the two vectors
involved, so we see that there is a tensor of type $(0,2)$ at work here.
Furthermore, the dot product is required to be commutative, so the tensor
must be symmetric.  Denoting this tensor by ${\bf g},$ we have
\begin{equation}
{\bf U}\cdot {\bf V}={\bf g}({\bf U},{\bf V}).
\end{equation}
This tensor is called the {\it metric tensor}.  If we also demand that it 
have an inverse, then we can find a basis for which it has diagonal form
with entries equal to $\pm 1$ (if all the diagonal entries can be made equal
to $+1,$ then we say that the metric is {\it definite}, otherwise we say
that it is {\it indefinite}).  The trace of the metric in this canonical
diagonal form is called its {\it signature}.

A {\it metric tensor field} is the association of 
such a type $(0,2)$ symmetric tensor
with every point of a manifold.  It must have an inverse at every point.
It follows that the signature is the same at every point of the manifold.

A manifold endowed with a metric has all sorts of new structure.  For the
purposes of our discussion, its most important role is to provide a
one-to-one correspondence between vectors and covectors.  For, given any
vector, say ${\bf V},$ we can form the covector, 
${\bf g}({\bf V},\phantom{{\bf V}}).$  The components of this new covector
are then $g_{\alpha\beta}V^\beta.$  Denote the inverse of $g_{\alpha\beta}$
by $g^{\beta\gamma},$ so
\begin{equation}
g_{\alpha\beta}g^{\beta\gamma}=\delta^\gamma_\alpha.
\end{equation}
Then, given any covector, say ${\bf a},$ we can form the vector with
components $g^{\alpha\beta}a_\beta.$  Note that this is a one to one
correspondence.

Frequently we shall use the same symbol to denote a vector and its
corresponding covector in a metric space.  That is, we may write
\begin{equation}
V_\alpha=g_{\alpha\beta}V^\beta,
\end{equation}
or
\begin{equation}
V^\alpha=g^{\alpha\beta}V_\beta.
\end{equation}
This process is called {\it index raising} or {\it index lowering}, as the
case may be.  It can be used to raise or lower the indices of any tensor
of any type.

We shall frequently abuse notation by using the dot product to denote the
interior product of a vector with a covector.  That is, we may write
\begin{equation}
{\bf a}\cdot {\bf V}={\bf a}({\bf V})=a_\alpha V^\alpha.
\end{equation}
When this is done, it will be obvious from context, so no confusion should
arise.  We shall occasionally further abuse notation by using a ``double dot''
notation for two contracted indices.  That is, given two tensors of type
$(0,2)$ and $(2,0),$ respectively, we may write
\begin{equation}
{\bf F}:{\bf G}=F_{\alpha\beta}G^{\alpha\beta}.
\end{equation}
Once again, things should be clear from context.

A metric tensor field does far more than provide an invertible map from
vectors to covectors.  It also induces on the manifold something called
an {\it affine connection}.  This makes it possible to compare vectors
in nearby tangent spaces.  Recall that Lie dragging gave us a way to
do this, but there had to be a vector field present in the first place
along which to drag, and we could drag only in the direction of that field.
An affine connection allows us to {\it parallel transport} vectors from
one tangent space to any other one nearby; that is, it gives us a notion of
parallelism between vectors in different tangent spaces.  Furthermore, it
does not require the presence of any vector field there to begin with.
One does not need a metric to have an affine connection, but the presence
of a metric induces an affine connection in a natural way.

Armed with an affine connection, it is possible to go on to define such things
as curvature and torsion.  While knowledge of this material is certainly
helpful in understanding the material presented in this thesis
(especially the curvature and polarization guiding-center drifts and
the intimate relationship between torsion and spin angular momentum), it is not
essential.  Thus we shall not go on to discuss these topics; the interested
reader is referred to Schutz~\cite{zaa} for a good introduction, and to
Misner, Thorne and Wheeler~\cite{zbd} or Chandrasekhar~\cite{zbe} for a more
detailed presentation.

\section{Noncanonical Hamiltonian and\\ Lagrangian Mechanics}
\subsection{Canonical Versus Noncanonical Coordinates}
In elementary classical mechanics courses, Hamiltonian mechanics is 
derived by application of a Legendre transformation to the system 
Lagrangian.  This process gives rise to canonical coordinates in a very 
natural way.  When it becomes necessary to change coordinates on phase 
space, the student is taught to restrict attention to the limited class of 
transformations that will  maintain this separation of the coordinates 
into canonically conjugate pairs; these are the so-called canonical 
transformations. 

The Poisson bracket of two scalar phase functions, $A$ and $B,$  is then 
introduced by {\it defining} it in terms of partial derivatives with respect 
to the canonically conjugate pairs of coordinates, $q^i$ and $p_i$ (the 
index $i$ ranges over all the degrees of freedom): 
\begin{equation} 
\{A,B\}=\frac{\partial A}{\partial q^i}\frac{\partial B}{\partial p_i}
       -\frac{\partial A}{\partial p_i}\frac{\partial B}{\partial q^i},
\label{mah} 
\end{equation} 
where we have adopted the convention of summation over repeated indices. 
It is then shown that this bracket is bilinear: 
\begin{equation} 
\{xA+yB,C\}=x\{A,C\}+y\{B,C\} 
\label{mai} 
\end{equation} 
where $x$ and $y$ are constants, that it is antisymmetric: 
\begin{equation} 
\{A,B\}=-\{B,A\}, 
\end{equation} 
that it obeys the Jacobi identity: 
\begin{equation} 
\{A,\{B,C\}\}+\{C,\{A,B\}\}+\{B,\{C,A\}\}=0, 
\end{equation} 
and that it obeys the chain rule: 
\begin{equation} 
\{f(A),C\}=f'(A)\{A,C\}
\label{maj}
\end{equation} 
or, equivalently, the Leibniz product rule: 
\begin{equation} 
\{AB,C\}=A\{B,C\}+B\{A,C\}. 
\label{mak} 
\end{equation} 

Mathematicians have a different way of looking at all of this.  In 
mathematics courses on Hamiltonian mechanics, one is more likely to 
{\it define} the Poisson bracket as any rule for taking a pair of scalar 
phase functions and associating with them a third scalar phase function 
consistent with the properties listed in Eqs.~(\ref{mai}) through 
(\ref{mak}) above.  Now it is manifest that any Poisson bracket 
given by the physicists' definition is also a Poisson bracket according to 
the mathematicians' definition.  The converse, however, is {\it not} true; 
that is, there exist Poisson brackets that obey all of the above-listed 
properties, but are not given by  Eq.~(\ref{mah}) for any set of 
canonical coordinates, $q$ and $p.$  Thus, by adopting the mathematicians' 
definition, we can generalize what is meant by a Poisson bracket in a very 
powerful way.

To see how this comes about, let us take the mathematicians' viewpoint
and suppose that we have a phase space with coordinates, $z^\alpha ,$ where
$\alpha$ ranges from $1$ to $N.$  For canonical coordinates, $N$ is twice the
number of degrees of freedom and the $z^\alpha$ are the $q$'s and $p$'s, but
let us not restrict ourselves to this special case in any way; in particular,
$N$ could be an odd number, and there need not be any natural pairing 
amongst the coordinates.

Denote the Poisson bracket of coordinate $z^\alpha$ with 
coordinate $z^\beta$ by:
\begin{equation}
J^{\alpha \beta} \equiv \{ z^\alpha , z^\beta \} .
\end{equation}
Suppose that we changed our phase space coordinates, $z\mapsto z'.$
Then, using the chain rule, Eq.~(\ref{maj}), we see that the Poisson
bracket of two of the new coordinates is given by:
\begin{equation}
{J'}^{\alpha \beta}\equiv \{ {z'}^\alpha , {z'}^\beta \} 
  = \frac{\partial {z'}^\alpha}{\partial z^\xi} \{ z^\xi , z^\eta \}
    \frac{\partial {z'}^\beta}{\partial z^\eta}
\end{equation}
or
\begin{equation}
{J'}^{\alpha \beta} = \frac{\partial {z'}^\alpha}{\partial z^\xi}
            \frac{\partial {z'}^\beta}{\partial z^\eta} J^{\xi \eta}.
\end{equation}
This makes it clear that the $J^{\alpha \beta}$ are 
the components of a second rank
contravariant tensor.  This tensor will henceforth be called the {\it Poisson
tensor}.  Using the chain rule once again, we see that the Poisson bracket 
of any two phase functions, $A$ and $B,$ may be written in terms of the 
Poisson tensor as follows:
\begin{equation}
\{ A,B \} = \frac{\partial A}{\partial z^\alpha} J^{\alpha \beta} 
            \frac{\partial B}{\partial z^\beta}
\label{mal}
\end{equation}

The general form of the bracket given by Eq.~(\ref{mal}) is
clearly bilinear and obeys the chain rule (or, equivalently, the Leibniz
product rule).  Now, the other two defining properties of the 
Poisson bracket may be expressed as properties of the Poisson tensor.
It is easily seen that antisymmetry of the bracket implies and is implied
by antisymmetry of the Poisson tensor:
\begin{equation}
J^{\alpha \beta}=-J^{\beta \alpha}.
\label{mam}
\end{equation}
Somewhat more algebra shows that the Jacobi property of the bracket 
implies and is implied by the following property of the Poisson tensor:
\begin{equation}
J^{\alpha \xi}J^{\beta \gamma}_{\phantom{\beta \gamma},\xi}+ 
J^{\gamma \xi}J^{\alpha \beta}_{\phantom{\alpha \beta},\xi}+ 
J^{\beta \xi}J^{\gamma \alpha}_{\phantom{\gamma \alpha},\xi}=0,
\label{man}
\end{equation}
where the commas denote partial differentiation.
Thus, our philosophy shall be that any tensor that has these two 
properties defines a perfectly legitimate Poisson bracket according to 
Eq.~(\ref{mal}).

Let us see how this works for canonical coordinates, $q^i$ and $p_i,$
where $i$ ranges from $1$ to the number of degrees of freedom, $I.$  
Write $z^\alpha=q^\alpha$ for $\alpha=1,\ldots,I,$ 
and $z^\alpha=p_{\alpha-I}$ for $\alpha=I+1,\ldots,N$
where $N = 2I.$  Now canonical coordinates have 
the bracket relations, $\{ q^i , q^j \} = \{ p_i , p_j \} = 0$ and 
$\{ q^i , p_j \} = -\{ p_j , q^i \} = \delta^i_j,$ so the matrix of components
of the Poisson tensor is:
\begin{equation}
{\bf J} \equiv \{ {\bf z} , {\bf z} \} 
   =\left(
      \begin{array}{lr}
         \phantom{-}{\bf 0} & {\bf 1} \\
                  - {\bf 1} & {\bf 0} 
      \end{array}
    \right),
\label{mao}
\end{equation}
where ${\bf 0}$ and ${\bf 1}$ are the $I \times I$ null and unit matrices,
respectively.

Using this Poisson tensor in Eq.~(\ref{mal}), we easily recover
the usual expression for the canonical bracket, Eq.~(\ref{mah}).
Furthermore, this Poisson tensor is obviously antisymmetric, and it
obeys Eq.~(\ref{man}) since its components are constants so their
derivatives are all zero.

If we start with canonical coordinates, then a {\it canonical transformation}
is any transformation that leaves the Poisson tensor unchanged.  If we denote
the Jacobian matrix of the transformation by:
\begin{equation}
{\bf M} \equiv \partial {\bf z'} / \partial {\bf z},
\end{equation} 
then this condition may be written as the matrix equation:
\begin{equation}
{\bf J}={\bf M}{\bf J}{\bf M}^T,
\end{equation}
where the superscript ``$T$'' denotes ``transpose,'' and ${\bf J}$
is the canonical Poisson tensor given by Eq.~(\ref{mao}).  In
what follows, we shall generalize the term {\it canonical transformation} to
mean any bracket-preserving transformation, regardless of whether or not
we started from canonical coordinates.

Thus far, we have said nothing about the equations of motion.  For canonical
coordinates these are well known to be:
\begin{eqnarray}
\dot{q^i}&=& \frac{\partial H}{\partial p_i}, \nonumber\\
\noalign{\hbox{and}}
\dot{p_i}&=&-\frac{\partial H}{\partial q^i}, 
\label{map}
\end{eqnarray}
where $H$ is the Hamiltonian.
These may be written in terms of the Poisson bracket as follows:
\begin{eqnarray}
\dot{q^i}&=&\{ q^i,H \}, \nonumber\\
\noalign{\hbox{and}}
\dot{p_i}&=&\{ p_i,H \}. 
\label{maq}
\end{eqnarray}
If we use $z$ to refer to the $q$'s and $p$'s, this becomes even simpler
to write:
\begin{equation}
\dot{z}^\alpha=\{ z^\alpha,H \}.
\end{equation}
Alternatively, this may be written in terms of the Poisson tensor:
\begin{equation}
\dot{z}^\alpha=J^{\alpha \beta}\frac{\partial H}{\partial z^\beta}.
\label{mcv}
\end{equation}
Since this last equation is in tensor form, and since it is known to hold
for canonical coordinates, it must be the correct generalization of the
equation of motion for noncanonical coordinates.  Thus, the complete
specification of a Hamiltonian system in this new generalized sense
requires the specification of both a Poisson tensor and a scalar 
Hamiltonian.

Any dynamical system on phase space can be expressed in the form
$\dot{z}^\alpha=V^\alpha,$ where $\bf{V}$ is some vector field on the
phase space.  Eq.~(\ref{mcv}) for a Hamiltonian dynamical system has this
form.  Note, however, that in order to qualify as ``Hamiltonian,'' the
vector field on the right cannot be just any vector field; it must be given
by the Poisson tensor contracted with the gradient of some scalar function.
A vector field on phase space is called a {\it Hamiltonian vector field} if
there exists some scalar field for which this is true.  Thus, if a manifold
is endowed with a Poisson tensor, then scalar fields generate Hamiltonian 
vector fields.

\subsection{An Example of a Noncanonical Poisson Structure}
There are several ways that noncanonical Poisson structures can arise in
a problem.  The first and most obvious way is to start with canonical 
coordinates and make a noncanonical transformation.  The canonical
Poisson tensor is known to obey Eqs.~(\ref{mam}) 
and (\ref{man}), and since these are tensorial equations 
they will hold in all frames if they hold in any one 
frame.  So the result of a noncanonical transformation will be a new
bracket that obeys all the required properties.

A particularly beautiful example of this has been given by 
Littlejohn~\cite{zas} for
the problem of a charged particle in a magnetic field.  For canonical
coordinates, ${\bf q}$ and ${\bf p},$ the Hamiltonian is well known to be:
\begin{equation}
H=\frac{1}{2m} \left( {\bf p}-\frac{e}{c} {\bf A}({\bf q}) \right)^2,
\end{equation}
where ${\bf A}({\bf q})$ is the vector potential.
Make the noncanonical transformation to new coordinates,${\bf r}$
and ${\bf v},$ where:
\begin{eqnarray}
   {\bf r} &\equiv& {\bf q} \nonumber\\
\noalign{\hbox{and}}
   {\bf v} &\equiv& \frac{1}{m} \left( {\bf p}
                  -\frac{e}{c} {\bf A}({\bf q}) \right).
  \label{mar}
\end{eqnarray}
The bracket relations among the new coordinates are easily calculated:
\begin{eqnarray}
   \{ {\bf r},{\bf r} \} &=& {\bf 0}, \nonumber\\
   \{ {\bf r},{\bf v} \} &=& \frac{1}{m}{\bf 1}, \nonumber\\
\noalign{\hbox{and}}
   \{ {\bf v},{\bf v} \} &=& \frac{1}{m}{\bf \Omega},
  \label{mas}
\end{eqnarray}
where we have defined the matrix ${\bf \Omega}$ with components:
\begin{equation}
\Omega_{ij} \equiv \frac{e}{mc} \left( A_{j,i}-A_{i,j} \right)
   = \frac{e}{mc} \epsilon_{ijk}B^k,
\end{equation}
and where the $B^k$ are the components of the ordinary magnetic 
field pseudovector.  Thus the bracket of any two scalar phase functions,
$R$ and $S,$ is given by:
\begin{equation}
\{ R,S \} = \frac{1}{m}\left(
  \frac{\partial R}{\partial {\bf r}}\cdot\frac{\partial S}{\partial {\bf v}}
 -\frac{\partial R}{\partial {\bf v}}\cdot\frac{\partial S}{\partial {\bf r}}
  \right)+\frac{e}{m^2c}{\bf B}\cdot\left(
  \frac{\partial R}{\partial {\bf v}}\times\frac{\partial S}{\partial {\bf v}}
  \right).
\label{mat}
\end{equation}
This bracket is easily seen to be antisymmetric.  That it satisfies the Jacobi
identity is less obvious; we know that it must from the arguments given above,
but a direct proof involves some tedious algebra.
The new Hamiltonian is simply:
\begin{equation}
H({\bf r},{\bf v})=\frac{m}{2}v^2,
\label{mau}
\end{equation}
and it is readily verified that this Hamiltonian, together with the bracket
given in Eq.~(\ref{mat}) yield the correct equations of motion.
Note that the vector potential is absent from the new formulation; this is
construed as an advantage, since the vector potential is a gauge-dependent
quantity.  The above Hamiltonian system was the starting point for
Littlejohn's work on guiding-center theory~\cite{zas}.

Now that we have seen how noncanonical Poisson structures can arise from
noncanonical transformations of a canonical system, it is natural to ask the
opposite question:  Given a noncanonical Hamiltonian system, is it always
possible to find a transformation to canonical coordinates?  For noncanonical
Hamiltonian systems with a nonsingular Poisson tensor (that is, systems 
for which the matrix of components of the Poisson tensor is nonsingular), 
there is an important theorem, called Darboux's theorem, that tells us that
the answer is ``yes.''  A proof of Darboux's theorem is given by 
Littlejohn~\cite{zas} and is constructive; that is, it gives a
prescription for actually finding the transformation to canonical coordinates.
For Hamiltonian systems with singular Poisson structures, the situation is
more complicated, and will be discussed shortly.


\subsection{Reduction}
\label{yaz}
\subsubsection{Reduction and Noether's Theorem}
Noncanonical transformations from canonical coordinates is only one 
of many ways that interesting Poisson structures can arise naturally.
The process of ``reduction'' of a Hamiltonian system with symmetry is
another.  Work in this area has been pioneered by 
Marsden and Weinstein (see, for example, reference~\cite{zaz}).

A detailed discussion of reduction 
would be out of place in this work, but the general idea is this:
Suppose that we have a canonical Hamiltonian system with a configuration
space symmetry (e.g. spatial translation, rotation, etc.).
Make the configuration space symmetry group parameter one of the
generalized coordinates.  Noether's theorem then tells us that the
corresponding momentum is conserved.  It is then possible to eliminate
this degree of freedom from the system, thus reducing the dimensionality
of the phase space by two.  This much is familiar from elementary courses
in classical mechanics.  Reduction is an important generalization of Noether's
theorem that allows us to similarly ``mod out'' by a
symmetry group that acts on all of phase space rather than just configuration
space.  After reduction is performed, the
resulting Hamiltonian system may very well be noncanonical.

The set of all phase functions together with the Poisson bracket operation
constitutes a Lie algebra.  From a computational point of view, in order 
to perform reduction we must find a representation for which this Lie algebra
has a closed Lie subalgebra.  Furthermore, the Hamiltonian must depend only on 
the elements of this subalgebra.  The elements of the subalgebra then 
constitute coordinates for a reduced description of the problem.  This is
best illustrated by example.

\subsubsection{The Free Rigid Body}
One of the most elementary (but nontrivial) examples of 
this process is the Hamiltonian system for a free rigid body.
The usual generalized coordinates for this problem
are the Eulerian angles, $\theta,$ $\phi,$ and $\psi,$
with respect to some fixed space frame.  By introducing
their canonically conjugate momenta, $p_\theta,$ $p_\phi,$ and $p_\psi,$
it is possible to write the equations
of motion in a canonical Hamiltonian format with a six-dimensional 
phase space.  If we choose a body frame for which the inertia tensor is
diagonalized, then the Hamiltonian for the free rigid body problem is
\begin{eqnarray}
H&=&\frac{p_\psi^2}{2I_3}+\frac{1}{2I_2}\left[ (p_\phi\csc\theta
  -p_\psi\cot\theta)\cos\psi-p_\theta\sin\psi\right]^2\nonumber\\
  &&\qquad               +\frac{1}{2I_1}\left[ (p_\phi\csc\theta
  -p_\psi\cot\theta)\sin\psi+p_\theta\cos\psi\right]^2,
  \label{mav}
\end{eqnarray}
where $I_1,$ $I_2,$ and $I_3$ are the three diagonal elements of the inertia
tensor.

Consider the three components of the angular momentum resolved in the body
frame.  These can be expressed in terms of our canonical phase space
coordinates as follows:
\begin{eqnarray}
m_1&=&(p_\phi\csc\theta-p_\psi\cot\theta)\sin\psi+p_\theta\cos\psi,\nonumber\\
m_2&=&(p_\phi\csc\theta-p_\psi\cot\theta)\cos\psi-p_\theta\sin\psi,\nonumber\\
\noalign{\hbox{and}}
m_3&=&p_\psi.\label{maw}
\end{eqnarray}
(See Goldstein~\cite{zbb} for details.  Only the result is needed here.)

By direct calculation with the canonical bracket, we can verify the
following relations
\begin{eqnarray}
\{ m_1 , m_2 \}&=&-m_3\nonumber\\
\{ m_2 , m_3 \}&=&-m_1\nonumber\\
\{ m_3 , m_1 \}&=&-m_2.\label{max}
\end{eqnarray}
Thus, the three components of the angular momentum in the body frame
constitute a closed Lie subalgebra under the operation of the canonical
Poisson bracket.  This means that the subset of functions on the
canonical phase space that are functions of the $m$'s alone 
(that is, those functions that depend on
$\theta,$ $\phi,$ $\psi,$ $p_\theta,$ $p_\phi,$ and $p_\psi$ {\it only} 
through their dependence on the $m$'s) constitutes a Lie subalgebra of
the Lie algebra of all canonical phase functions.  

We thus adopt the $m$'s as generalized coordinates on a {\it reduced} 
phase space of three dimensions.  The Poisson tensor on this reduced
phase space is then given by 
$J^{\alpha \beta}=-\epsilon^{\alpha \beta \gamma}m_\gamma,$ or:
\begin{equation}
{\bf J}=\left(
           \begin{array}{lcr}
              \phantom{-}  0 &          - m_3 & \phantom{-}m_2 \\
              \phantom{-}m_3 &              0 &          - m_1 \\
                       - m_2 & \phantom{-}m_1 &              0
           \end{array}
        \right),
\label{may}
\end{equation}
so that the Poisson bracket of any two functions of ${\bf m},$ say
$A$ and $B,$ is given by:
\begin{equation}
\{ A,B \}=-{\bf m}\cdot\left(\frac{\partial A}{\partial {\bf m}}
   \times \frac{\partial B}{\partial {\bf m}} \right).
\label{maz}
\end{equation}
This bracket {\it must} satisfy all the required properties of a 
Poisson bracket, since it was derived by specializing the domain of a
canonical bracket; nevertheless, it is straightforward and instructive to
verify this by direct calculation.

It is possible to perform reduction only if the Hamiltonian is
expressible in terms of the reduced coordinate set.  For the free rigid
body, we have
\begin{equation}
H({\bf m})=\frac{m_1^2}{2I_1}+\frac{m_2^2}{2I_2}+\frac{m_3^2}{2I_3}.
\end{equation}
As usual, the equations of motion 
are given by $\dot{{\bf m}}=\{ {\bf m},H \} ,$ or:
\begin{eqnarray}
{\dot{m}}_1&=&\left(\frac{1}{I_3}-\frac{1}{I_2}\right)m_2m_3 \nonumber\\
{\dot{m}}_2&=&\left(\frac{1}{I_1}-\frac{1}{I_3}\right)m_3m_1 \nonumber\\
{\dot{m}}_3&=&\left(\frac{1}{I_2}-\frac{1}{I_1}\right)m_1m_2.
  \label{mba}
\end{eqnarray}
As expected, these are indeed Euler's equations for the free rigid body.
If the rigid body were not free (say, if it were in a gravitational field),
then a potential energy term would have been present in the Hamiltonian,
and that term would {\it not} have been expressible in terms of the $m$'s.  
Thus, the reduction process would have failed.  This is because the 
gravitational field breaks the SO(3) symmetry that makes the reduction 
possible.

As we shall see later on in this thesis, the passage from particle
coordinates to guiding-center coordinates is another example of reduction.
The symmetry involved is the group of rotations by the gyroangle, SO(2), and
the reduction eliminates the corresponding degree of freedom from the system.
If this gyrosymmetry is somehow broken (say, by a variation in the
background field configuration whose length scale is on the order of a
gyroradius), then the guiding-center description is invalidated.

\subsubsection{Euler's Fluid Equations}
Our next example is a Hamiltonian field theory for Euler's equations
for the flow of an inviscid, incompressible fluid.  Let us adopt a
Lagrangian description for such a fluid wherein each fluid particle
is labelled by a reference position, ${\bf x}_0.$
Then the configuration of the fluid at time $t$ may be specified by
giving the particle's current position, ${\bf x}$ as a function of
${\bf x}_0$ and $t.$  Thus, our dynamical field variable is
${\bf x}({\bf x}_0,t).$  The system Lagrangian consists solely of the
kinetic energy
\begin{equation}
L=\int d^3x_0 \frac{\rho}{2} {\dot{x}}^2({\bf x}_0,t),
\end{equation}
where $\rho$ is the constant uniform mass density.
The canonical momentum field is then given by
\begin{equation}
{\bf p}({\bf x}_0,t)=\frac{\delta L}{\delta \dot{\bf x}({\bf x}_0,t)}
  =\rho\dot{\bf x}({\bf x}_0,t),
\end{equation}
where the $\delta$'s denote functional differentiation.  Performing the
Legendre transformation, we see that the system Hamiltonian is
\begin{equation}
H=\int d^3x_0 \frac{1}{2\rho} {\bf p}^2({\bf x}_0,t).
\label{mcy}
\end{equation}
The canonical bracket of two functionals of ${\bf x}$ and ${\bf p},$
say $A$ and $B,$ is then
\begin{equation}
\{ A,B \}=\int d^3x_0
  \left(\frac{\delta A}{\delta {\bf x}({\bf x}_0,t)}\cdot
        \frac{\delta B}{\delta {\bf p}({\bf x}_0,t)}
       -\frac{\delta A}{\delta {\bf p}({\bf x}_0,t)}\cdot
        \frac{\delta B}{\delta {\bf x}({\bf x}_0,t)}\right).
\end{equation}

Now suppose that the fluid particles are identical.  In that case, 
specification of ${\bf x}({\bf x}_0,t)$ is far more information than is
really necessary to determine the configuration of the fluid.  This is because
${\bf x}({\bf x}_0,t)$ effectively keeps track of particle labels; two
configurations that differ only by swapping identical particles will
actually have different ${\bf x}({\bf x}_0,t).$  For a fluid of identical
particles, an Eulerian description, wherein the flow velocity is given as
a function of spatial position and time, say ${\bf v}({\bf \xi},t),$
suffices to determine the fluid configuration.  The Lagrangian description
just keeps track of too much information.  Thus, in passing from the
Lagrangian to the Eulerian description, we are effectively reducing by
the group of identical particle interchanges.  The Eulerian description is
therefore the reduced description.  The reduced phase space is the (smaller,
though still infinite dimensional) space of all divergenceless vector fields,
${\bf v},$ that satisfy the boundary conditions (${\bf v}$ tangential to
the boundary).  The requirement that ${\rm div}{\bf v}=0$ stems from the fact 
that we are considering only incompressible flows.

So, from a computational point of view, how do we perform this reduction?  
Note that the Eulerian velocity field may be written in terms of the Lagrangian
fields as follows:
\begin{equation}
{\bf v}({\bf \xi},t)=\frac{1}{\rho} {\bf p}({\bf x}^{-1}({\bf \xi},t),t).
\end{equation}
This may be interpreted as follows:  If we want the Eulerian velocity at
spatial point ${\bf \xi},$ first take ${\bf x}^{-1}({\bf \xi},t)$ to
get the reference position of the fluid element currently at ${\bf \xi},$
then evaluate the momentum ${\bf p}$ of the fluid element with this reference
position, then divide the result by $\rho$ to get the desired answer.  Now
the above equation may be written
\begin{equation}
{\bf v}({\bf \xi},t)=\frac{1}{\rho} \int d^3x_0 {\bf p}({\bf x}_0,t)
  \delta ({\bf x}({\bf x}_0,t)-{\bf \xi}),
\label{mcw}
\end{equation}
where we have used the fact that the Jacobian,
$|\partial {\bf x}/\partial {\bf x}_0|,$ is equal to unity because the flow
is incompressible.  Thus we have succeeded in expressing the reduced
field variable, ${\bf v},$ in terms of the canonical field variables,
${\bf x}$ and ${\bf p}.$  In this respect, Eq.~(\ref{mcw}) is the exact
analog of Eqs.~(\ref{maw}) for the free rigid body problem.

Thus, we can take the Poisson bracket of the Eulerian field with itself
using the canonical bracket.  This is straightforward, and the result is
\begin{equation}
\{ {\bf v}({\bf \xi},t),{\bf v}({\bf \xi}',t) \}
  =\frac{1}{\rho}
   \left( {\bf v}({\bf \xi}',t)\delta' ({\bf \xi}'-{\bf \xi})
         -\delta' ({\bf \xi}-{\bf \xi}'){\bf v}({\bf \xi},t)\right),
\end{equation}
where $\delta'$ denotes the gradient of the delta function.  Note that we have
been able to express this bracket in terms of the Eulerian (reduced) field
variables alone.  This equation is thus the analog of Eqs.~(\ref{max}) for
the free rigid body problem.

So we see that the functionals of the Eulerian field variables constitute a 
closed Lie subalgebra of the Lie algebra of all phase functionals.  We thus 
adopt the Eulerian field variables as coordinates on a reduced phase space.
The Poisson bracket of any two functionals of
${\bf v},$ say $A$ and $B,$ is then calculated by the Leibniz rule
\begin{eqnarray}
\{ A,B \}
  &=&\int d^3\xi \int d^3\xi'
  \frac{\delta A}{\delta {\bf v}({\bf \xi},t)}\cdot
  \{ {\bf v}({\bf \xi},t),{\bf v}({\bf \xi'},t) \}\cdot
  \frac{\delta B}{\delta {\bf v}({\bf \xi'},t)}\nonumber\\
  &=&-\frac{1}{\rho} \int d^3\xi
  {\bf v}({\bf \xi},t)\cdot \left[
  \frac{\delta A}{\delta {\bf v}({\bf \xi},t)},
  \frac{\delta B}{\delta {\bf v}({\bf \xi},t)} \right],
  \label{mcx}
\end{eqnarray}
where the square brackets are Lie brackets, and where the functional
derivatives $\delta A/\delta {\bf v}({\bf \xi},t)$ and $\delta B/\delta
{\bf v}({\bf \xi},t)$ are regarded as vector fields.

Note that Eq.~(\ref{mcx}) is
the analog of Eq.~(\ref{maz}) for the free rigid body problem.

We must also check that the Hamiltonian may be expressed in terms of the
reduced variables.  Fortunately, this is not difficult.  A change of
variables in Eq.~(\ref{mcy}) gives
\begin{equation}
H=\frac{\rho}{2}\int d^3 \xi v^2({\bf \xi},t),
\label{mcz}
\end{equation}
where we have again made use of the fact that the Jacobian,
$|\partial {\bf x}/\partial {\bf x}_0|,$ is equal to unity.

It remains to check that the Hamiltonian in Eq.~(\ref{mcz}) together with the
bracket in Eq.~(\ref{mcx}) actually yield Euler's fluid equations.  This is
slightly tricky.  Consider a functional $A({\bf v}).$  Its equation of motion
is
\begin{equation}
\frac{\partial A}{\partial t}=\{ A,H \}.
\end{equation}
We insert Eq.~(\ref{mcz}) for the Hamiltonian.  After some straightforward
manipulation, including an integration by parts where the surface term 
vanishes due to the boundary condition, we get
\begin{equation}
0=\int d^3 \xi \frac{\delta A}{\delta {\bf v}}\cdot
  \left[ \frac{\partial {\bf v}}{\partial t}+{\bf v}\cdot\delr {\bf v}
  +\delr \left( \frac{v^2}{2}\right) \right].
\label{meb}
\end{equation}

At this point, we might be tempted to set the expression in square brackets
above equal to zero on the grounds that $A$ is an arbitrary functional.  This
would, however, be incorrect because $\delta A/\delta {\bf v}$ is not really
arbitrary.  Recall that our phase space consists only of those vector fields
that have zero divergence.  This causes an ambiguity in the usual definition of
the functional derivative which is such that the equation
\begin{equation}
A({\bf v}+\delta {\bf v})=A({\bf v})+\int d^3 \xi \delta {\bf v}\cdot
   \frac{\delta A}{\delta {\bf v}}+{\cal O}({\delta v}^2)
\end{equation}
is satisfied.
If ${\bf v}$ and ${\bf v}+\delta {\bf v}$ are both divergenceless, it follows
that $\delta {\bf v}$ is divergenceless.  This means that the gradient of an
arbitrary function, $\phi,$ may be added to $\delta A/\delta {\bf v},$ since
\begin{equation}
\int d^3 \xi \delta {\bf v} \cdot \delr \phi
  =-\int d^3 \xi \phi \delr \cdot (\delta {\bf v})=0.
\end{equation}
We can make the definition of the functional derivative unique by demanding
that $\delr\cdot (\delta A/\delta {\bf v})=0.$  This gives a well-posed
problem for the determination of $\phi.$

Now, in order to incorporate this constraint that
$\delr\cdot (\delta A/\delta {\bf v})=0,$ note that if we were to add the
gradient of {\it any} scalar function, $\psi,$ to the expression in square
brackets in Eq.~(\ref{meb}), the equation would still hold because
\begin{equation}
\int d^3 \xi \frac{\delta A}{\delta {\bf v}}\cdot\delr\psi
  =-\int d^3 \xi \psi\delr\cdot\left(\frac{\delta A}{\delta {\bf v}}\right)=0.
\end{equation}
So the most that we can write is
\begin{equation}
\frac{\partial {\bf v}}{\partial t}+{\bf v}\cdot\delr {\bf v}
  +\delr \left( \frac{v^2}{2}+\psi\right)=0.
\end{equation}

We now identify the pressure
\begin{equation}
p\equiv \rho\left(\frac{v^2}{2}+\psi\right),
\end{equation}
so we finally arrive at Euler's fluid equation
\begin{equation}
\frac{\partial {\bf v}}{\partial t}+{\bf v}\cdot\delr {\bf v}=
  -\frac{1}{\rho}\delr p.
\label{mda}
\end{equation}
Finally, we note that the pressure is not really arbitrary, but is rather
determined by taking the divergence of both sides of Eq.~(\ref{mda}) to get
\begin{equation}
\nabla^2 p=-\delr\cdot ({\bf v}\cdot\delr {\bf v}),
\end{equation}
and by dotting both sides of Eq.~(\ref{mda}) with the unit normal to the
boundary surface, $\hat{\bf n},$ to get
\begin{equation}
\frac{\partial p}{\partial n}=-\hat{\bf n}\cdot ({\bf v}\cdot\delr {\bf v}).
\end{equation}
This constitutes a well-posed Neumann problem for $p$ as a functional 
of ${\bf v}.$  Thus, Eq.~(\ref{mda}), coupled with the constraint of
incompressibility, determines both ${\bf v}$ and $p.$

It is intriguing that the equations of motion for both examples considered
thus far are named after Euler; one wonders if he knew about the beautiful
analogy between them.  In fact, the first published
reference to this analogy seems to be a 1966 paper of Arnold~\cite{zbf}.

\subsubsection{The Poisson-Vlasov System}
Our final example of reduction is also a Hamiltonian field theory, this time
for the Poisson-Vlasov equations of plasma physics.  For simplicity, we
consider a one-dimensional plasma (the methods are trivially generalized to
three dimensions).  Once again, we label
particles by their initial conditions.  This time, however, the flow is in
phase space, so the initial conditions are $r_0$ and $p_0,$ and the present
phase space position is $r$ and $p.$  The dynamical fields are thus
$r(r_0,p_0,t)$ and $p(r_0,p_0,t).$  We shall use $z$ to refer to the set
of coordinates, $r$ and $p,$ and $z_0$ to refer to the set of initial
conditions, $r_0$ and $p_0.$  The fields may thus be abbreviated $z(z_0,t).$

The Lagrangian for this system that includes the electrostatic potential energy
of interaction was first written down by Low~\cite{zbg}.  It is
\begin{equation}
L=\int dz_0 f(z_0)\left[ \frac{m}{2}{\dot{r}}^2(z_0)-\frac{e^2}{2}\int dz'_0
  f(z'_0)g(r(z_0),r(z'_0))\right].
\end{equation}
Here we have ignored species labels for simplicity.  Also, $f(z_0)$ is the
distribution of initial conditions on phase space, and $g(r,r')$ is the
Coulomb potential kernel.  The canonical momentum field is then
\begin{equation}
\pi(z_0)=\frac{\delta L}{\delta \dot{r} (z_0)}=f(z_0)m\dot{r}(z_0).
\end{equation}
The Hamiltonian is obtained by Legendre transformation
\begin{equation}
H=\int dz_0 \frac{\pi^2(z_0)}{2mf(z_0)}+\frac{e^2}{2}
  \int dz_0 \int dz'_0 f(z_0)f(z'_0) g(r(z_0),r(z'_0)).
\end{equation}
The bracket is canonical, with $r$ and $\pi$ canonically conjugate.

Now suppose that the particles are identical.  Just as with Euler's fluid
equations, it turns out that we can reduce to an Eulerian description.  This
time, the Eulerian field variable is the usual distribution function on phase
space, ${\sf f}(Z).$  This may be expressed in terms of the Lagrangian field
variables as follows:
\begin{equation}
{\sf f}(R,P,t)=\int dz_0 f(z_0) \delta \left(R-r(z_0,t)\right) 
  \delta \left(P-\frac{\pi(z_0,t)}{f(z_0)}\right).
\end{equation}
This is the analog of Eqs.~(\ref{maw}) and (\ref{mcw}).

Now we can take the canonical bracket of $f(Z)$ with $f(Z').$  We get
\begin{equation}
\{ {\sf f}(Z),{\sf f}(Z') \}
  =\int dZ'' {\sf f}(Z'') \{ \delta (Z-Z''),\delta (Z'-Z'') \}''_0,
\end{equation}
where $\{a,b\}''_0$ denotes the single-particle Poisson bracket of
$a(R'',P'')$ with $b(R'',P'').$  Note that we have been able to express the
bracket of the Eulerian field variables in terms of the canonical field
variables; thus we have achieved the desired reduction.  The bracket of
any two functionals of ${\sf f}$ is found by application of the Leibniz rule.
The result is
\begin{equation}
\{ A,B \}=\int dZ {\sf f}(Z) \left\{ \frac{\delta A}{\delta {\sf f}(Z)},
                          \frac{\delta B}{\delta {\sf f}(Z)} \right\}_0.
\end{equation}
This form for the bracket was first given by Iwinski and Turski~\cite{zbx},
by Morrison who credits it to Kaufman~\cite{zbh}, and by Gibbons~\cite{zgb}.
A derivation similiar to that above can be found in a paper by Kaufman and
Dewar~\cite{zgc}.

Finally, we see that the Hamiltonian can be expressed in terms of ${\sf f}$
as follows
\begin{equation}
H=\int dZ {\sf f}(Z)\frac{P^2}{2m}+\frac{e^2}{2}
  \int dZ \int dZ' {\sf f}(Z){\sf f}(Z') g(R,R').
\end{equation}
It is now readily verified that the above brackets and Hamiltonian yield
the Poisson-Vlasov equations of motion,
\begin{equation}
\frac{\partial {\sf f}}{\partial t}+\frac{P}{m}\frac{\partial {\sf f}}{\partial R}
  -e\frac{\partial \phi}{\partial R}\frac{\partial {\sf f}}{\partial P}=0,
\end{equation}
where
\begin{equation}
\phi (R)=e\int dZ' {\sf f}(Z') g(R,R')
\end{equation}
is the electrostatic potential.

Note the similarity in structure of the brackets for all three of the 
above examples.  For example, all three have a Poisson tensor that is
linear in the coordinates used.  All are examples of what are called
{\it Lie-Poisson brackets}, and there is a rich mathematical literature
on brackets of this sort (see, for example, Marsden~\cite{zaz}).

\subsection{Singular Poisson Structures}
There are a few very important observations to be made about the above
examples before we go on to talk about perturbation theory.
First consider the free rigid body problem.
Note that the matrix in Eq.~(\ref{may}) is singular with
rank two for ${\bf m} \neq 0,$ and rank zero for ${\bf m}=0.$
Indeed, any odd dimensional phase space {\it must} have a singular
Poisson structure, because antisymmetric matrices always have even rank.
For these systems, Darboux's theorem does not apply and it is not
possible to find a transformation to canonical coordinates; of course,
this should have been obvious because canonical coordinates always come
in pairs and you can't pair an odd number of things.

When a system has a singular Poisson structure, the Poisson tensor will have
at least one null eigenvector.  Let's say it has $n$ of them; note that $n$
is equal to the dimensionality of the phase space, $N,$ minus the rank of the
Poisson tensor, $r.$  In this case, it has been shown by 
Littlejohn~\cite{zat} that it is {\it always} possible to find
a set of $n=N-r$ scalar phase functions whose gradients are those null
eigenvectors.  This is not at all obvious and requires an application of the
Frobenius theorem of differential geometry, where use is made of the fact that
the Poisson tensor satisfies the Jacobi identity.

These $n$ scalar phase functions are very special in that their bracket
with {\it any} other scalar phase function must vanish.  This is obvious from
Eq.~(\ref{mal}).  Scalar phase functions with this property are
called {\it Casimir functions}.  In particular, their bracket with {\it any}
Hamiltonian is zero, so they are always conserved quantities; note that their
conservation follows directly from the bracket structure, independent
of the particular Hamiltonian under consideration.

For the free rigid body problem presented above, the null eigenvector of the
Poisson tensor is any multiple of {\bf m} itself.  The function:
\begin{equation}
C({\bf m})=m_1^2+m_2^2+m_3^2
\end{equation}
is then a Casimir function since its gradient is in the direction of
${\bf m},$ and we recognize it as the total angular momentum squared.
Of course, any other scalar phase function that is functionally
dependent upon $C$ could have been used equally well.  The pathology at
the point ${\bf m}=0$ where the rank of ${\bf J}$ changes
is called a {\it symplectic bone}, and is discussed at
length by Weinstein~\cite{zba}.

If we were to choose $C$ to be one of our generalized coordinates, say
the third coordinate in place of $m_3,$ then it is clear that the third
row and column of ${\bf J}$ would be zero.  The two by two submatrix
consisting of rows and columns one and two would be nonsingular, and
Darboux's theorem could be applied to that subsystem.  Thus, the correct
generalization of Darboux's theorem for singular Poisson structures
is to say that it is always possible to find a transformation to a
coordinate system for which the matrix of components of the Poisson 
tensor has an $r$ by $r$ submatrix in canonical form with the rest of
the entries vanishing.  For the free rigid body problem Poisson structure given
above, this has the form:
\begin{equation}
{\bf J}=\left(
           \begin{array}{lcr}
              \phantom{-} 0 & 1 & 0 \\
                       -  1 & 0 & 0 \\
              \phantom{-} 0 & 0 & 0
           \end{array}
        \right)
\end{equation}
for ${\bf m}\neq 0,$ and ${\bf J}=0$ for ${\bf m}=0.$

It is worth repeating that Casimir functions are conserved for {\it any}
Hamiltonian.  For example, the Hamiltonian:
\begin{equation}
H({\bf m})=\mu m_3,
\end{equation}
where $\mu$ is a constant, together with the same bracket used above for the
free rigid body problem, yields the equations of motion for a classical spin 
gyrating in a uniform magnetic field.  That is, $m_1$ and $m_2$ undergo
simple harmonic oscillations, while $m_3$ is conserved because it
commutes with the Hamiltonian.  Note that $C$ is a conserved quantity
for this system as well, because the bracket is the same.  In general,
the Poisson structure is considered to be a more fundamental entity than
the Hamiltonian.

The other two examples presented in the last subsection also have
singular Poisson structures.  It is readily verified that the bracket
for Euler's fluid equations has the Casimir functional
\begin{equation}
C=\int d^3 \xi {\bf v}({\bf \xi},t)\cdot
  \left[\delr\times {\bf v}({\bf \xi},t)\right]
\end{equation}
(the integrand here is called the {\it helicity}),
and that the bracket for the Poisson-Vlasov equations has the Casimir
functionals
\begin{equation}
C_\Phi=\int dZ \Phi ({\sf f}(Z))
\end{equation}
where $\Phi$ is an arbitrary function of its argument.

\subsection{Phase-Space Lagrangian Techniques}
In this section, we review the phase space Lagrangian formalism;
for more details on this subject see Littlejohn~\cite{zar} and
Littlejohn and Cary~\cite{zaf}.
For a system with canonical coordinates, $q$ and
$p,$ and time-independent Hamiltonian, $H(q,p),$ the phase space Lagrangian
is given by
\begin{equation}
L(q,p,\dot{q},\dot{p})=p\cdot \dot{q}-H(q,p),
\end{equation}
where a dot denotes differentiation with respect to
time, $t.$  Note that $L$ may depend upon {\it all} the phase
space coordinates and their time derivatives, unlike ordinary
configuration space Lagrangians, $L(q,\dot{q}).$  The
associated action is
\begin{equation}
A=\int dt L(q,p,\dot{q},\dot{p}),
\label{mdx}
\end{equation}
the variation of which yields the Euler-Lagrange equations
\begin{equation}
0=\frac{d}{dt}\left( \frac{\partial L}{\partial \dot{q}}\right)-
                      \frac{\partial L}{\partial      q }=
       \dot{p}-\left(-\frac{\partial H}{\partial q}\right)=
       \dot{p}+       \frac{\partial H}{\partial q},
\label{mbb}
\end{equation}
and
\begin{equation}
0=\frac{d}{dt}\left( \frac{\partial L}{\partial \dot{p}}\right)-
                      \frac{\partial L}{\partial      p }=
      0-\left(\dot{q}-\frac{\partial H}{\partial p}\right)=
             -\dot{q}+\frac{\partial H}{\partial p};
\label{mbc}
\end{equation}
these are recognized as the canonical equations of motion.

We denote by $z^\mu ,$ where $\mu =1,\ldots ,N,$ (where $N=2I$) the coordinates
of phase space.  The phase space Lagrangian may then be written
\begin{equation}
L(z,\dot{z})=\gamma_\mu {\dot{z}}^\mu-H(z),
\label{mbd}
\end{equation}
where the covector whose components are $\gamma_\mu$ will be
called the {\it action one-form}.  For the canonical 
coordinate system used above, these components are
\begin{equation}
\gamma_\mu =\left\{ 
               \begin{array}{ll}
                  p_\mu        & {\rm if } \; \mu=1,\ldots ,I \\
                  0            & {\rm if } \; \mu=I+1,\ldots ,N .
               \end{array}
            \right.
\label{mbe}
\end{equation}
The fact that $I$ of these components are zero is a manifestation
of the fact that the coordinate system is canonical.  For more
general coordinate systems this will not be true, as we shall see 
shortly.  Note that phase space Lagrangians are always linear
in $\dot{z}.$  Also note that knowledge of the action one-form and the
Hamiltonian is completely equivalent to knowledge of the phase space Lagrangian
by Eq.~(\ref{mbd}).  

The equations of motion may be written in 
this notation as follows:
\begin{eqnarray}
0&=&\frac{d}{dt}\left( \frac{\partial L}{\partial \dot{z^\mu}}\right)
  -\frac{\partial L}{\partial z^\mu }\nonumber\\
 &=&\frac{d\gamma_\mu}{dt}-\gamma_{\nu ,\mu}{\dot{z}}^\nu
  +\frac{\partial H}{\partial z^\mu }\nonumber\\
 &=&\left(\gamma_{\mu ,\nu}-\gamma_{\nu ,\mu}\right){\dot{z}}^\nu
  +\frac{\partial H}{\partial z^\mu }
 \label{mch}
\end{eqnarray}
or
\begin{equation}
\omega_{\mu \nu}{\dot{z}}^\nu=\frac{\partial H}{\partial z^\mu },
\label{mbf}
\end{equation}
where we have defined the Lagrangian two-form
\begin{equation}
\omega_{\mu \nu} \equiv \gamma_{\nu ,\mu}-\gamma_{\mu ,\nu},
\label{mbg}
\end{equation}
or
\begin{equation}
\omega \equiv d\gamma .
\end{equation}
For $z=(q,p),$ where $q$ and $p$ are canonically conjugate, it is easily
verified that Eq.~(\ref{mbf}) is equivalent to
Eqs.~(\ref{mbb}) and (\ref{mbc}).

We can recover the more familiar Hamiltonian formalism in the
following manner:  Assuming that $[\omega_{\mu\nu}]$ is a nonsingular matrix,
we denote its inverse by $J^{\mu\nu},$ so
\begin{equation}
J^{\mu\rho} \omega_{\rho\nu}=\delta^\mu_\nu.
\label{mbi}
\end{equation}
Then Eq.~(\ref{mbf}) becomes
\begin{equation}
{\dot{z}}^\mu=J^{\mu\nu}\frac{\partial H}{\partial z^\nu}.
\label{mbj}
\end{equation}
These are recognized as Hamilton's equations if we identify 
$J^{\mu\nu}$ as the Poisson tensor.
That the Poisson tensor is antisymmetric and obeys the Jacobi
identity is easily verified.  In particular, the Jacobi identity follows 
directly from $d\omega=dd\gamma=0.$

Under a (possibly noncanonical) transformation of phase space 
coordinates, $z\mapsto Z,$ the action one-form transforms in the 
usual fashion of a covariant vector to give
\begin{equation}
\Gamma_\mu=\frac{\partial z^\xi}{\partial Z^\mu}\gamma_\xi.
\label{mbk}
\end{equation}
Similarly, the Lagrangian two-form transforms like a second rank
covariant tensor
\begin{equation}
\Omega_{\mu \nu}=\frac{\partial z^\xi }{\partial Z^\mu}
                 \frac{\partial z^\eta}{\partial Z^\nu}
                 \omega_{\xi \eta}
                =\Gamma_{\nu ,\mu}-\Gamma_{\mu ,\nu},
\label{mbl}
\end{equation}
where the commas in Eq.~(\ref{mbl}) denote partial
differentiation with respect to $Z.$  The Hamiltonian, of course, transforms
as a scalar, $K(Z)=H(z).$  The new equation of motion is then
\begin{equation}
\Omega_{\mu \nu}{\dot{Z}}^\nu =\frac{\partial K}{\partial Z^\mu},
\end{equation}
which may be compared to Eq.~(\ref{mbf}).

Note that all of the above considerations assume a time-independent 
Hamiltonian.  This restriction is not important for two reasons:  First,
we could always work in extended phase space to treat a time-dependent
system; this is the approach taken by Littlejohn and Cary~\cite{zaf}.
Second, all of our relativistic equations of motion will have the 
single-particle proper time as the independent variable, and nothing depends
explicitly on this.

The transformation
\begin{equation}
\gamma_\xi \mapsto \gamma_\xi + \frac{\partial S}{\partial z^\xi},
\label{mbm}
\end{equation}
where $S$ is an arbitrary scalar field on extended phase space,
is called a {\it Lagrangian gauge transformation}.  Though it alters the action
one-form, it is easily seen to have no effect on the Lagrangian
two-form, and so it does not change the equation of 
motion, Eq.~(\ref{mbf}).

It is clear that if $L$ is independent of one of the extended
phase space coordinates, say $z^\mu,$ then the associated canonical
momentum, $\partial L/\partial {\dot{z}}^\mu,$ is conserved
by Noether's theorem.  Note, however, that a gauge transformation,
like Eq.~(\ref{mbm}),
using a scalar field, $S,$ that depends upon the ignorable
coordinate, could destroy the Noether symmetry, even though the
associated momentum would still be conserved.  The same is true
for coordinate transformations like Eq.~(\ref{mbk}).
Conversely, we see that it may be necessary to perform gauge 
or coordinate transformations in order to uncover
Noether symmetries and, hence, to discover conserved quantities.

The strategy for our treatment of the guiding-center problem will be to start
with the phase space Lagrangian for a single relativistic
charged particle in
an electromagnetic field, and, via a sequence of gauge and
coordinate transformations, find a representation in which the
gyroangle, $\theta,$ is ignorable.  This is the Noether symmetry
for the gyromomentum.  When this is achieved, the gyroangle
will no longer appear in the equations of motion for the other
variables, and the magnetic moment will appear only as a constant
parameter like the rest mass.  Thus, in this system of 
``gyrocoordinates,'' the rapid oscillatory motion is effectively
decoupled from the slower guiding-center motion, and the
dimensionality of our phase space is reduced by two.

\subsection{Constrained Systems}
\label{ybb}
Eqs.~(\ref{mbd}) and (\ref{mdx}) may be interpreted as follows:  The variation
of the action one form must vanish, subject to the constraint that the
Hamiltonian is constant.  By including other constraints, besides the fact that
the Hamiltonian is constant, we can discover new and interesting Poisson
structures that have those other constraints ``built in.''

For example, consider a particle that is constrained to move on the surface
of a sphere of radius $r.$  To model this system, we take the canonical action
one form,
\begin{equation}
\gamma={\bf p}\cdot d{\bf r}=p_xdx+p_ydy+p_zdz,
\end{equation}
and vary it subject to the constraints that the Hamiltonian, $H,$ be constant,
that the particle position be on the sphere
\begin{equation}
| {\bf r} |^2=x^2+y^2+z^2=r^2,
\label{mdb}
\end{equation}
and that the particle momentum be tangent to the sphere
\begin{equation}
{\bf r} \cdot {\bf p}=xp_x+yp_y+zp_z=0.
\label{mdc}
\end{equation}

The constrained variation may be done in any one of a number of ways; e.g. by
use of Lagrange multipliers.  Thus we write
\begin{equation}
L={\bf p}\cdot\dot{\bf r}-\frac{1}{2}\lambda_1 |{\bf r}|^2
  -\lambda_2 {\bf r}\cdot {\bf p}-H,
\end{equation}
and form the Euler-Lagrange equations
\begin{equation}
\dot{\bf p}=-\lambda_1 {\bf r}-\lambda_2 {\bf p}
            -\frac{\partial H}{\partial {\bf r}}
\label{mec}
\end{equation}
\begin{equation}
0=\dot{\bf r}-\lambda_2 {\bf r}-\frac{\partial H}{\partial {\bf p}}.
\label{med}
\end{equation}
Dot the first of these equations with ${\bf r}$ to get
\begin{equation}
{\bf r}\cdot\dot{\bf p}=-\lambda_1 |{\bf r}|^2
   -{\bf r}\cdot\frac{\partial H}{\partial {\bf r}},
\end{equation}
from which it follows that
\begin{equation}
\lambda_1=-\frac{1}{r^2} {\bf r}\cdot\left(\dot{\bf p}
   +\frac{\partial H}{\partial {\bf r}}\right).
\label{mee}
\end{equation}
Then dot the second with ${\bf r}$ to get
\begin{equation}
0={\bf r}\cdot\dot{\bf r}-\lambda_2 |{\bf r}|^2
   -{\bf r}\cdot\frac{\partial H}{\partial {\bf p}},
\end{equation}
from which it follows that
\begin{equation}
\lambda_2=\frac{1}{r^2} {\bf r}\cdot\left(\dot{\bf r}
   -\frac{\partial H}{\partial {\bf p}}\right).
\label{mef}
\end{equation}
Note that Eqs.~(\ref{mee}) and (\ref{mef}) may be written in the form
\begin{equation}
\left(
   \begin{array}{c}
      \lambda_1 \\
      \lambda_2
   \end{array}
\right)=\frac{1}{r^2}
\left(
   \begin{array}{cc}
      -{\bf r}\cdot (\{ {\bf p},{\bf r} \}+{\bf 1}) &
      -{\bf r}\cdot  \{ {\bf p},{\bf p} \} \\
       {\bf r}\cdot  \{ {\bf r},{\bf r} \} &
      -{\bf r}\cdot (\{ {\bf r},{\bf p} \}-{\bf 1})
   \end{array}
\right)\cdot
\left(
   \begin{array}{c}
      \partial H/\partial {\bf r} \\
      \partial H/\partial {\bf p}
   \end{array}
\right).
\label{meg}
\end{equation}

To get the Poisson brackets, first substitute the Lagrange multipliers,
(\ref{mee}) and (\ref{mef}), back into the equations of motion,
(\ref{mec}) and (\ref{med}).  We get
\begin{equation}
\left( {\bf 1}-\frac{{\bf r}{\bf r}}{r^2}\right)\cdot\dot{\bf r}
   =\left( {\bf 1}-\frac{{\bf r}{\bf r}}{r^2}\right)\cdot
    \frac{\partial H}{\partial {\bf p}}
\end{equation}
and
\begin{equation}
\left( {\bf 1}-\frac{{\bf r}{\bf r}}{r^2}\right)\cdot\dot{\bf p}
   =-\left( {\bf 1}-\frac{{\bf r}{\bf r}}{r^2}\right)\cdot
     \frac{\partial H}{\partial {\bf r}}
    -\frac{1}{r^2}{\bf p}{\bf r}\cdot\left(\dot{\bf r}
    -\frac{\partial H}{\partial {\bf p}}\right).
\end{equation}
Note that these two equations do not determine the motion completely;
they give only the projection of the motion on the sphere.  To fully
determine $\dot{\bf r}$ and $\dot{\bf p},$ we need to employ the
derivatives of the constraints,
\begin{equation}
{\bf r}\cdot\dot{\bf r}=0
\end{equation}
and
\begin{equation}
{\bf r}\cdot\dot{\bf p}+\dot{\bf r}\cdot {\bf p}=0.
\end{equation}
Using these, we finally get
\begin{equation}
\dot{\bf r}=\left( {\bf 1}-\frac{{\bf r}{\bf r}}{r^2}\right)
   \cdot\frac{\partial H}{\partial {\bf p}}
\end{equation}
and
\begin{equation}
\dot{\bf p}=-\left( {\bf 1}-\frac{{\bf r}{\bf r}}{r^2}\right)
   \cdot\frac{\partial H}{\partial {\bf r}}+\frac{1}{r^2}
   ({\bf p}{\bf r}-{\bf r}{\bf p})\cdot\frac{\partial H}{\partial {\bf p}}.
\end{equation}
These equations of motion are Hamiltonian with the quadratic Poisson structure
\begin{eqnarray}
\{ r^i,r^j \}&=&0\nonumber\\
\{ r^i,p_j \}&=&\delta^i_j-\frac{r^ir_j}{r^2}\nonumber\\
\{ p_i,p_j \}&=&\frac{r_jp_i-r_ip_j}{r^2}.\label{mdg}
\end{eqnarray}
Note that the constraints, Eqs.~(\ref{mdb}) and (\ref{mdc}), are Casimir
functions of this Poisson structure.  This means that the Hamiltonian equations
of motion will yield dynamics that respect these constraints for {\it any}
Hamiltonian whatsoever.

There is another approach to deriving the above set of brackets.  We could
have adopted the spherical coordinates,
\begin{equation}
r=\sqrt{x^2+y^2+z^2}
\end{equation}
\begin{equation}
\theta=\arctan (\sqrt{x^2+y^2} / z)
\end{equation}
\begin{equation}
\phi=\arctan (y/x),
\end{equation}
on $\Re^3.$  These have the canonically conjugate momenta
\begin{equation}
p_r=(xp_x+yp_y+zp_z)/\sqrt{x^2+y^2+z^2}
\end{equation}
\begin{equation}
p_\theta=z(xp_x+yp_y)/\sqrt{x^2+y^2}
\end{equation}
\begin{equation}
p_\phi=xp_y-yp_x,
\end{equation}
as is easily verified.  The advantage to using these spherical coordinates
is that the constraint surface in phase space is simply described by setting
$p_r$ equal to zero, and $r$ equal to a constant.  

Now we can write
\begin{eqnarray}
{\bf r}&=&x\hat{\bf x}+y\hat{\bf y}+z\hat{\bf z}\nonumber\\
       &=&\cos\theta\hat{\bf z}
        +\sin\theta\cos\phi\hat{\bf x}
        +\sin\theta\sin\phi\hat{\bf y}
       \label{meh}\\
\noalign{\hbox{and}}
{\bf p}&=&p_x\hat{\bf x}+p_y\hat{\bf y}+p_z\hat{\bf z}\nonumber\\
       &=&(p_r\cos\theta-\frac{p_\theta}{r}\sin\theta)\hat{\bf z}\nonumber\\
       &&\qquad +(p_r\sin\theta\cos\phi+\frac{p_\theta}{r}\cos\theta\cos\phi
         -\frac{p_\phi}{r}\csc\theta\sin\phi)\hat{\bf x}\nonumber\\
       &&\qquad +(p_r\sin\theta\sin\phi+\frac{p_\theta}{r}\cos\theta\sin\phi
         +\frac{p_\phi}{r}\csc\theta\cos\phi)\hat{\bf y}.
       \label{mei}
\end{eqnarray}
Eqs.~(\ref{meh}) and (\ref{mei}) and the Leibniz rule allow us to compute
the brackets for the system of coordinates $({\bf r},{\bf p})$ in terms of the
brackets for the system of coordinates $(r,\theta,\phi,p_r,p_\theta,p_\phi).$
If we ignore the constraint, then the latter system is canonical, and it
follows that the former system is also canonical.  If, on the other hand,
we incorporate the constraint by {\it dictating} that $r$ and $p_r$ are Casimir
functions and that $p_r=0,$ then the brackets (\ref{mdg}) follow immediately.

It is interesting to contrast these two methods for obtaining the brackets
(\ref{mdg}).  We shall use these methods
when we cast our guiding-center equations of motion in gyrogauge and boostgauge
invariant format, towards the end of the Chapter 3.  Our guiding-center
Poisson brackets will also have a quadratic Poisson structure, similar to
that of the above set of brackets.  Such quadratic Poisson structures seem
to arise naturally from this type of manipulation.  The reader who is
interested in pursuing this topic further is encouraged to read about
Dirac's theory of constraints~\cite{zbt}.

\section{Lie Transform Perturbation Theory}
\subsection{General Discussion of Lie Transforms}
\label{yab}
Recall that we first introduced coordinates on manifolds using the
concepts of charts and atlases.  A chart is a one-to-one map from a region of
$\Re^n$ to a region of an $n$-dimensional manifold.  Each coordinate,
$z^\alpha,$ may thus be thought of as a function on the manifold.
When we change coordinates, we are effectively transforming these functions.

Consider an {\it infinitesimal} transformation of coordinates given by
\begin{equation}
Z^\alpha=z^\alpha+h ({\cal L}_g z)^\alpha
  =z^\alpha+h g^\alpha (z),
\label{mde}
\end{equation}
where $h$ is an infinitesimal, $g$ is a vector field, and the Lie
derivative acts on the coordinates as though they were scalar functions.
From our geometrical interpretation of the Lie derivative, we see that we
are effectively taking the functions that define the coordinates, and sliding
them an infinitesimal parameter interval, $h,$ along the field lines of
$g.$  The inverse transformation is
\begin{equation}
z^\alpha=Z^\alpha-h ({\cal L}_g Z)^\alpha
  =Z^\alpha-h g^\alpha (Z).
\end{equation}
Of course, since $h$ is an infinitesimal, we are scrupulously ignoring
anything of order $h^2.$

Now we ask how basis vector components behave under the above transformation.
Assume a coordinate basis for simplicity.  We have
\begin{equation}
\frac{\partial}{\partial Z^\alpha}
  =\frac{\partial z^\beta}{\partial Z^\alpha}
   \frac{\partial}{\partial z^\beta}
  =\frac{\partial}{\partial z^\alpha}
  -h\frac{\partial g^\beta}{\partial z^\alpha}
   \frac{\partial}{\partial z^\beta}.
\end{equation}
Similarly, basis covector components transform as follows:
\begin{equation}
dZ^\alpha
  =\frac{\partial Z^\alpha}{\partial z^\beta} dz^\beta
  =dz^\alpha
  +h\frac{\partial g^\alpha}{\partial z^\beta} dz^\beta.
\end{equation}

Now suppose that ${\bf t}$ is some tensor field on the manifold.  We can ask
how the components of ${\bf t}$ behave under the above transformation.  Use
a prime to distinguish the components of ${\bf t}$ in the new coordinate system.
We demand
\begin{eqnarray}
{t'}^{\alpha_1\ldots\alpha_r}_{\beta_1\ldots\beta_s}(Z)&&
  \frac{\partial}{\partial Z^{\alpha_1}}\otimes\cdots\otimes
  \frac{\partial}{\partial Z^{\alpha_r}}\otimes
  dZ^{\beta_1}\otimes\cdots\otimes dZ^{\beta_s}\nonumber\\
&=&t^{\mu_1\ldots\mu_r}_{\nu_1\ldots\nu_s}(z)
  \frac{\partial}{\partial z^{\mu_1}}\otimes\cdots\otimes
  \frac{\partial}{\partial z^{\mu_r}}\otimes
  dz^{\nu_1}\otimes\cdots\otimes dz^{\nu_s}.
\label{mdd}
\end{eqnarray}
Now expand in $h,$ retaining only first order terms.  We find
\begin{eqnarray}
{t'}^{\mu_1\ldots\mu_r}_{\nu_1\ldots\nu_s}&=&
  t^{\mu_1\ldots\mu_r}_{\nu_1\ldots\nu_s}
   -h\bigl(
    t^{\mu_1\ldots\mu_r}_{\nu_1\ldots\nu_s,\alpha}
      g^\alpha\nonumber\\
  &&-t^{\alpha\mu_2\ldots\mu_r}_{\nu_1\ldots\nu_s}
      g^{\mu_1}_{\phantom{\mu_1},\alpha}-\cdots
   -t^{\mu_1\ldots\mu_{r-1}\alpha}_{\nu_1\ldots\nu_s}
      g^{\mu_r}_{\phantom{\mu_r},\alpha}\nonumber\\
  &&+t^{\mu_1\ldots\mu_r}_{\alpha\nu_2\ldots\nu_s}
      g^{\alpha}_{\phantom{\alpha},\nu_1}+\cdots
   +t^{\mu_1\ldots\mu_r}_{\nu_1\ldots\nu_{s-1}\alpha}
      g^{\alpha}_{\phantom{\alpha},\nu_s}\bigr).
  \label{mej}
\end{eqnarray}
Suppose that we define a new tensor field, ${\bf T},$ whose components in
the old system are the same as those of ${\bf t}$ in the new system.
Then, by comparison with Eq.~(\ref{mcn}), we may write
\begin{equation}
{\bf T}={\bf t}-h ({\cal L}_g {\bf t}),
\label{mdf}
\end{equation}
where comparison with Eq.~(\ref{mcn}) is helpful.  Furthermore, since this last
equation is in coordinate-free form, it is true for coordinate bases and
noncoordinate bases alike.

Compare the signs of the second terms on the right-hand sides of
Eqs.~(\ref{mdf}) and (\ref{mde}).  Despite the algebra that went into proving
the above result, it has a marvelously simple geometric interpretation.  If
we slide the values of the coordinates one way along a field line of $g,$
then we must slide the tensor field in the other direction.
In case this is not obvious, a trivial example is afforded by a scalar field
on $\Re,$ call it $f(x).$  If we transform coordinates to $X=x+h,$
then $F(X)=f'(X)=f(x)=f(X-h)=f(X)-h(df/dX)(X)=(f-h{\cal L}f)(X).$

Suppose that our tensor field is the tensor product of two tensor fields,
say ${\bf t}={\bf t}_1\otimes {\bf t}_2.$
Then, since Lie derivatives obey the Leibniz rule over the tensor product,
we have
\begin{eqnarray}
{\bf T}&=&{\bf t}-h({\cal L}_g{\bf t})\nonumber\\
       &=&{\bf t}_1\otimes {\bf t}_2-h{\cal L}_g({\bf t}_1\otimes {\bf t}_2)\nonumber\\
       &=&{\bf t}_1\otimes {\bf t}_2-h({\cal L}_g{\bf t}_1)\otimes {\bf t}_2
         -h{\bf t}_1\otimes ({\cal L}_g{\bf t}_2)\nonumber\\
       &=&[{\bf t}_1-h({\cal L}_g{\bf t}_1)]\otimes
         [{\bf t}_2-h({\cal L}_g{\bf t}_2)],
       \label{mdz}
\end{eqnarray}
where, as always, we neglect ${\cal O}(h^2).$
This result indicates that the infinitesimal transformation commutes with
the tensor product.

Next suppose that the tensor field is obtained by starting with a tensor of
higher rank and applying to it some number of vectors and/or covectors.
For example, say ${\bf t}={\bf s}({\bf a},{\bf U})$ where ${\bf a}$ is a
covector field and ${\bf U}$ is a vector field; we could have let ${\bf s}$ 
have more than one of each type of argument or other unfilled slots
without affecting the following reasoning in any way.  Apply the
transformation, and use Eq.~(\ref{mco}) to write
\begin{eqnarray}
{\bf T}&=&{\bf t}-h({\cal L}_g{\bf t})\nonumber\\
       &=&{\bf s}({\bf a},{\bf U})-h {\cal L}_g[{\bf s} ({\bf a},{\bf U})]\nonumber\\
       &=&{\bf s}({\bf a},{\bf U})-h({\cal L}_g {\bf s})({\bf a},{\bf U})
         -h{\bf s}({\cal L}_g{\bf a},{\bf U})
         -h{\bf s}({\bf a},{\cal L}_g{\bf U})\nonumber\\
       &=&({\bf s}-h{\cal L}_g{\bf s})({\bf a}-h{\cal L}_g{\bf a},
         {\bf U}-h{\cal L}_g{\bf U}).
       \label{mea}
\end{eqnarray}
This result indicates that the transformation commutes with the application
of the vectors and/or covectors.

Next suppose that the tensor field is an exact form.  That is, say
${\bf t}=d{\bf \Omega}.$  Since Lie derivatives commute with exterior
derivatives, it follows that the transformation commutes with the application
of the exterior derivative.

The above results indicate that {\it any} tensorial relationship, including
those with differential operators, retains its form under a transformation
of the form given in Eq.~(\ref{mde}).  This crucial point makes the Lie
transform method possible.

Now suppose that we wish to consider finite (rather than infinitesimal)
changes of coordinates.  That is, suppose we wish to slide the coordinate
values a finite parameter interval, $\epsilon,$ along the field lines of $g.$
The easiest approach is to divide the finite interval into a large number of
infinitesimal intervals by writing
\begin{equation}
Z=\lim_{N \rightarrow \infty}(1+\frac{\epsilon}{N}{\cal L}_g)^N z
  =\exp (\epsilon {\cal L}_g) z.
\label{mbn}   
\end{equation}
The finite transformation of the tensor, ${\bf t},$ is then
\begin{equation}
{\bf T}=\lim_{N \rightarrow \infty}(1-\frac{\epsilon}{N}{\cal L}_g)^N {\bf t}
  =\exp (-\epsilon {\cal L}_g) {\bf t}.
\end{equation}
The transformation given by the above equations is called a {\it Lie transform}
generated by the vector field, $g.$

Because the infinitesimal transformations of the form given in Eq.~(\ref{mde})
are known to preserve tensorial relationships, and because a Lie transform is
composed of nothing more than a large number of these infinitesimal
transformations, it follows that Lie transforms preserve tensorial
relationships.  That is
\begin{equation}
\exp (-\epsilon {\cal L}_g)({\bf t}_1\otimes {\bf t}_2)
  =(\exp (-\epsilon {\cal L}_g){\bf t}_1)\otimes
   (\exp (-\epsilon {\cal L}_g){\bf t}_2),
\end{equation}
and
\begin{equation}
\exp (-\epsilon {\cal L}_g)
   \bigl[ {\bf s}({\bf a},{\bf U}) \bigr]
  =\bigl[ \exp (-\epsilon {\cal L}_g){\bf s} \bigr]
   \bigl( \exp (-\epsilon {\cal L}_g){\bf a},
          \exp (-\epsilon {\cal L}_g){\bf U} \bigr),
\end{equation}
and
\begin{equation}
\exp (-\epsilon {\cal L}_g) (d{\bf \Omega})
  =d \bigl(\exp (-\epsilon {\cal L}_g) {\bf \Omega}\bigr).
\end{equation}
We now have a way of making finite coordinate transformations of any
tensorial equation that is guaranteed to preserve its tensorial form.

By Taylor expanding the exponential in Eq.~(\ref{mbn}) and   
using Eq.~(\ref{mbo}) for the Lie derivative, it is possible to   
develop the transformation to arbitrarily high order in $\epsilon.$     
In practice, we want to be able to control the transformation order by
order in $\epsilon.$  There are two ways to do this.  The first, due to
Deprit~\cite{zai}, is to order the generator, $g,$ in $\epsilon.$
The second, due to Dragt and Finn~\cite{zaj}, is
to make a succession of transformations like
Eq.~(\ref{mbn}), as follows:
\begin{equation}
Z= \exp (\epsilon {\cal L}_{g_1}) \exp (\epsilon^2 {\cal L}_{g_2})
                        \exp (\epsilon^3 {\cal L}_{g_3}) \cdots z.
\label{mbp}
\end{equation}
In this work, we adopt the second procedure, as it was shown by
Cary~\cite{zau} to involve fewer terms in the perturbation series at
each order.  Expanding the above   
equation in $\epsilon$ and using Eq.~(\ref{mbo}), we get   
\begin{equation}   
Z=z+\epsilon    {\cal L}_1 z +\epsilon^2 ({\cal L}_2+\frac{1}{2} 
       {\cal L}_1^2) z + \epsilon^3 ({\cal L}_3+{\cal L}_1 {\cal L}_2
       +\frac{1}{6} {\cal L}_1^3) z +\cdots,
\label{mbq}
\end{equation}    
Here we have used ${\cal L}_n$ to abbreviate ${\cal L}_{g_n}.$  The inverse
transformation is then   
\begin{equation}   
z=\cdots \exp (-\epsilon^3 {\cal L}_3) \exp (-\epsilon^2 {\cal L}_2)
              \exp (-\epsilon {\cal L}_1) Z.   
\end{equation}   
Developing this order by order, we get   
\begin{equation}   
z=Z-\epsilon {\cal L}_1 Z -\epsilon^2 ({\cal L}_2-\frac{1}{2} 
       {\cal L}_1^2) Z - \epsilon^3 ({\cal L}_3-{\cal L}_2 {\cal L}_1
       +\frac{1}{6} {\cal L}_1^3)Z -\cdots.
\label{mdi}
\end{equation}

The transformation of the tensor ${\bf t}$ is then
\begin{equation}
{\bf T}= \cdots \exp (-\epsilon^3 {\cal L}_3) 
                \exp (-\epsilon^2 {\cal L}_2)    
                \exp (-\epsilon   {\cal L}_1)  {\bf t}.
\label{mbr}
\end{equation}
Let us suppose that ${\bf t}$ is given as a power series in the
expansion parameter, $\epsilon,$ so
\begin{equation}
{\bf t}={\bf t}_0 + \epsilon {\bf t}_1 + \epsilon^2 {\bf t}_2 +
              \epsilon^3 {\bf t}_3 + \cdots.
\end{equation}  
Then Eq.~(\ref{mbr}) yields
\begin{equation}
{\bf T}={\bf T}_0 + \epsilon {\bf T}_1 + \epsilon^2 {\bf T}_2 +
              \epsilon^3 {\bf T}_3 + \cdots,
\end{equation}
where
\begin{equation}
{\bf T}_0={\bf t}_0,
\label{mbs}
\end{equation}
\begin{equation}
{\bf T}_1={\bf t}_1-{\cal L}_1 {\bf t}_0,
\label{mdh}
\end{equation}
\begin{equation}
{\bf T}_2={\bf t}_2-{\cal L}_2 {\bf t}_0-{\cal L}_1 {\bf t}_1+
                   \frac{1}{2}{\cal L}_1^2 {\bf t}_0,
\label{mbt}
\end{equation}
\begin{equation}
{\bf T}_3={\bf t}_3-{\cal L}_3 {\bf t}_0-{\cal L}_2 {\bf t}_1+
                  {\cal L}_2 {\cal L}_1 {\bf t}_0-{\cal L}_1 {\bf t}_2+
                  \frac{1}{2}{\cal L}_1^2 {\bf t}_1-\frac{1}{6}{\cal L}_1^3 
                  {\bf t}_0,
\label{mbu}
\end{equation}
etc.

Given any equation written in tensor form, we can now make near-identity
coordinate transformations to perform perturbation analyses.  That is, if
the equation has the form of a solvable equation plus a small perturbation,
we can make a Lie transform to coordinates for which the perturbation is
removed or at least simplified.  The form of the generator, $g,$ required to
achieve this simplification depends on the specific problem, and is chosen
order by order in the perturbation series.

Once this process has been carried out to first order, we could continue on
to second and higher order, or we could regard the first-order problem as
a new solvable problem and renormalize the perturbation series accordingly
before proceeding to higher order.  The latter strategy is called the
{\it superconvergent} Lie transform procedure; superconvergent perturbation
series were first investigated by
Kolmogorov~\cite{zbu}.  All this will be made clear by selected examples
in the next few subsections.

\subsection{Lie Transforming a Scalar Field}
Consider the scalar equation
\begin{equation}
f(x)=\epsilon x^2 +2x -2c=0,
\end{equation}
where $c$ is a constant and $\epsilon$ is our expansion parameter.  Let's
pretend for a moment that we do not know how to solve a quadratic equation.
The scalar field, $f,$ is ordered in $\epsilon$ as follows:
\begin{equation}
f_0(x)=2x-2c,
\end{equation}
\begin{equation}
f_1(x)=x^2,
\end{equation}
and $f_n(x)=0$ for $n\geq 2.$

We wish to perform a Lie transform to a new coordinate, $X,$ for which the
transformed scalar will be denoted by $F.$  Since we are working in $\Re,$
the generator, $g,$ has only one component.  At order zero, use 
Eq.~(\ref{mbs}),
\begin{equation}
F_0=f_0.
\end{equation}
At order one, use Eq.~(\ref{mdh}),
\begin{equation}
F_1=f_1-g_1 f'_0=x^2-2g_1.
\end{equation}
Thus, we see that we can make $F_1$ vanish by choosing $g_1=x^2/2.$
Moving on to second order, we use Eq.~(\ref{mbt}),
\begin{equation}
F_2=-2g_2-\frac{x^3}{2}.
\end{equation}
So we can make $F_2$ vanish by choosing $g_2=-x^3/4.$  Thus, to order
$\epsilon^2,$ we have the Lie transformed scalar equation
\begin{equation}
F(X)=2X-2c=0.
\end{equation}
This has solution, $X=c.$  Now $x$ is given in terms of $X$ by Eq.~(\ref{mdi})
which becomes
\begin{eqnarray}
x&=&X-\epsilon g_1 -\epsilon^2 (g_2-\frac{1}{2}g_1 g'_1)-\cdots\nonumber\\
 &=&X-\frac{\epsilon}{2}X^2+\frac{\epsilon^2}{2}X^3-\cdots\nonumber\\
 &=&c-\frac{\epsilon}{2}c^2+\frac{\epsilon^2}{2}c^3-\cdots.\label{mdj}
\end{eqnarray}
This matches the Taylor expansion of the exact solution to the quadratic
equation
\begin{equation}
x=\frac{1}{\epsilon}\left(-1+\sqrt{1+2\epsilon c}\right),
\end{equation}
to ${\cal O}(\epsilon^2),$ as is easily verified.

Note that there is another solution to the quadratic equation
\begin{equation}
x=\frac{1}{\epsilon}\left(-1-\sqrt{1+2\epsilon c}\right),
\end{equation}
of leading order $\epsilon^{-1}$ that our technique does not give.  This is
because it is not continuously connected to the solution of the unperturbed
problem as $\epsilon$ goes to zero.  Lie transforms are useful only for 
near-identity coordinate transformations.

\subsection{Lie Transforming a Vector Field}
Now consider the following dynamical system:
\begin{eqnarray}
\dot{x}&=&y\nonumber\\
\dot{y}&=&-x-\epsilon x^2.\label{mdu}
\end{eqnarray}
If we use ${\bf z}$ to denote $(x,y),$ then this may be written
\begin{equation}
\dot{\bf z}={\bf v}_0+\epsilon {\bf v}_1,
\end{equation}
where we have defined the vectors, ${\bf v}_0\equiv (y,-x)$ and
${\bf v}_1\equiv (0,-x^2).$  We now try to Lie transform to new coordinates,
$Z=(X,Y),$ in an attempt to get rid of the order $\epsilon$ term.  The
transformed vector field is ${\bf V}={\bf V}_0+\epsilon {\bf V}_1,$ where
${\bf V}_0={\bf v}_0,$ and ${\bf V}_1$ is given from Eq.~(\ref{mdh}),
\begin{equation}
{\bf V}_1={\bf v}_1-{\cal L}_g {\bf v}_0.
\end{equation}
Using the formula for the Lie derivative of a vector, the demand that
${\bf V}_1=0$ is seen to be equivalent to the following pair of equations:
\begin{eqnarray}
\left(y\frac{\partial}{\partial x}-x\frac{\partial}{\partial y}\right)
 g_1^x&=&g_1^y\nonumber\\
\noalign{\hbox{and}}
\left(y\frac{\partial}{\partial x}-x\frac{\partial}{\partial y}\right)
 g_1^y&=&-g_1^x+x^2.\label{mdk}
\end{eqnarray}
These may be solved by the method of characteristics to yield
\begin{eqnarray}
g_1^x&=&\frac{1}{3}(x^2+2y^2)\nonumber\\
\noalign{\hbox{and}}
g_1^y&=&-\frac{2}{3}xy.\label{mdl}
\end{eqnarray}
Note that the characteristic equations for this system are the unperturbed
equations of motion.  This ``integration along unperturbed orbits'' is a
generic feature of problems of this sort.

Now then, the new coordinates are given in terms of the old by
\begin{eqnarray}
X&=&x+\frac{\epsilon}{3}(x^2+2y^2)\nonumber\\
Y&=&y-\frac{2\epsilon}{3}xy.\label{mdm}
\end{eqnarray}
The inverse transformation is then
\begin{eqnarray}
x&=&X-\frac{\epsilon}{3}(X^2+2Y^2)\nonumber\\
y&=&Y+\frac{2\epsilon}{3}XY.\label{mdn}
\end{eqnarray}
Note that we are ignoring terms of order $\epsilon^2$ or higher.
Now the equations of motion for ${\bf Z}$ are
\begin{eqnarray}
\dot{X}&=&Y\nonumber\\
\dot{Y}&=&-X.\label{mdo}
\end{eqnarray}
These have solution
\begin{eqnarray}
X&=&X_0\cos t+Y_0\sin t\nonumber\\
Y&=&Y_0\cos t-X_0\sin t.\label{mdp}
\end{eqnarray}
Thus, the solution for $z(t)$ is given by Eqs.~(\ref{mdn}) and (\ref{mdp}).  If
desired, the initial conditions for ${\bf Z}$ can be expressed in terms of
the initial conditions for ${\bf z}$ using Eq.~(\ref{mdm}).

Frequently, in physical applications of this formalism, it happens that the new
coordinates have physical significance.  For example, in guiding-center theory,
we shall find a Lie transform that takes us from the phase space coordinates of
a particle to those of a guiding center.  In such a circumstance, very little
is gained by expressing the initial conditions of the transformed problem in
terms of those of the original problem.  Instead, the new coordinates acquire
their own physical significance, and we can speak of ``the equations of motion
of a guiding center'' and ``the initial conditions of a guiding center,'' and
forget all about the original single-particle coordinates.

For a less trivial example of the vector Lie transform technique, see
Appendix~\ref{ybd} where the method is used to calculate the gyrofrequency
shift for two-dimensional nonrelativistic guiding-center motion in a spatially
nonuniform electromagnetic field.

\subsection{Canonical Lie Transforms of a Hamiltonian System}
When using perturbation theory to study a Hamiltonian dynamical system,
the above technique of Lie transforming the dynamical vector field
could be used, but there is a serious problem with this approach:
There is no guarantee that the Lie transform of a Hamiltonian vector field
will be another Hamiltonian vector field.

Recall that a Hamiltonian vector field is given by contracting the Poisson
tensor with the gradient of a scalar function.  This suggests the following
solution to the above problem:  Instead of Lie transforming the Hamiltonian
vector field, Lie transform the Poisson tensor and Hamiltonian separately.
This will insure that the transformed equations of motion are still in
Hamiltonian form.

Let us examine a little more closely why this should work.
Hamiltonian equations of motion are given by Eq.~(\ref{mcv}).  If we write
\begin{eqnarray}
{\bf Z} &=&\exp ( \epsilon {\cal L}) {\bf z},\nonumber\\
{\bf J}'&=&\exp (-\epsilon {\cal L}) {\bf J},\nonumber\\
\noalign{\hbox{and}}
H'&=&\exp (-\epsilon {\cal L}) H,\label{mdq}
\end{eqnarray}
then since our equations of motion are in tensor form, we are guaranteed that
the new equations of motion will be
\begin{equation}
\dot{\bf Z}={\bf J}'\cdot \frac{\partial H'}{\partial {\bf Z}}.
\label{mdr}
\end{equation}
Furthermore, we are guaranteed that ${\bf J}'$ is antisymmetric and obeys the
Jacobi identity because these requirements can also be written as tensorial
equations (see Eqs.~(\ref{mam}) and (\ref{man}), respectively).  Thus,
Eq.~(\ref{mdr}) qualifies as a bona fide Hamiltonian system.

We can now prove a marvelous theorem that considerably simplifies the work
involved in making {\it canonical} (bracket-preserving) Lie transformations
of a Hamiltonian system, and is probably responsible for the popularity of
the Lie transform technique:  A Poisson tensor is a Lie-dragged tensor along
{\it any} vector field that is Hamiltonian with respect to it.  Suppose the
Poisson tensor is denoted by ${\bf J}.$  Let ${\bf V}$ be given by
\begin{equation}
V^\alpha=J^{\alpha\beta}\frac{\partial W}{\partial z^\beta}
\label{mdt}
\end{equation}
for some (any) scalar field, $W.$  Then the theorem states
\begin{equation}
{\cal L}_V {\bf J}=0.
\end{equation}

This is easily proved using the formula for the Lie derivative of a second
rank contravariant tensor.  We write
\begin{eqnarray}
({\cal L}_V {\bf J})^{\alpha\beta}&=&
 V^\xi J^{\alpha\beta}_{\phantom{\alpha\beta},\xi}-
 V^\alpha_{\phantom{\alpha},\xi}J^{\xi\beta}-
 V^\beta_{\phantom{\beta},\xi}J^{\alpha\xi}\nonumber\\
 &=&-(J^{\alpha \xi}J^{\beta \gamma}_{\phantom{\beta \gamma},\xi}+ 
     J^{\gamma \xi}J^{\alpha \beta}_{\phantom{\alpha \beta},\xi}+ 
     J^{\beta \xi}J^{\gamma \alpha}_{\phantom{\gamma \alpha},\xi})
     W_{,\gamma}\nonumber\\
 &&\qquad -J^{\alpha\gamma}W_{,\xi\gamma}(J^{\xi\beta}+J^{\beta\xi}),
 \label{mds}
\end{eqnarray}
where we have used Eq.~(\ref{mdt}) for ${\bf V}.$
The first term vanishes by the Jacobi identity, the second term vanishes
by antisymmetry, and the theorem is proved.

It immediately follows that a Lie transform along the vector field ${\bf V}$
leaves ${\bf J}$ unchanged.  This is because a Lie transform is the
exponentiation of a Lie derivative (set ${\cal L}_m {\bf t}_n=0$ in
Eqs.~(\ref{mbs}) through (\ref{mbu}) to recover ${\bf T}={\bf t}$).
Thus, Lie transforms generated by Hamiltonian vector fields are always
canonical.  Now Hamiltonian vector fields are in one-to-one correspondence
with scalar phase functions, $W,$ by Eq.~(\ref{mdt}), so we have found a way
to generate canonical transformations with scalars.

Thus, to perform a canonical Lie transform of a Hamiltonian system, we need
only to transform the Hamiltonian.  Now the Lie derivative of a scalar with
respect to a Hamiltonian vector field is given by
\begin{equation}
{\cal L}_V H=V^\alpha H_{,\alpha}=J^{\alpha\beta}W_{,\beta}H_{,\alpha}
 =-\{ W,H \}.
\end{equation}
Thus, for a canonical Lie transform of a Hamiltonian, we may rewrite
Eqs.~(\ref{mbs}) through (\ref{mbu}) as follows:
\begin{equation}
K_0=H_0,
\end{equation}
\begin{equation}
K_1=H_1+\{ W_1 H_0 \},
\label{mek}
\end{equation}
\begin{equation}
K_2=H_2+\{ W_2,H_0 \}+\{ W_1,H_1 \}+\frac{1}{2}\{ W_1,\{ W_1,H_0 \} \}
\label{mel}
\end{equation}
\begin{eqnarray}
K_3&=&H_3+\{ W_3,H_0 \}+\{ W_2,H_1 \}+\{ W_2,\{ W_1,H_0 \} \}+\{ W_1,H_2 \}\nonumber\\
   &&\qquad +\frac{1}{2}\{ W_1,\{ W_1,H_1 \} \}
           +\frac{1}{6}\{ W_1,\{ W_1,\{ W_1,H_0 \} \} \},
   \label{mdy}
\end{eqnarray}
etc.  Here we have denoted the new Hamiltonian by $K.$

To see how this is used, consider the following example:  We perturb a
harmonic oscillator Hamiltonian by the addition of a nonlinear term,
\begin{equation}
H=\frac{1}{2}(q^2+p^2)-\frac{\epsilon}{3}p^4.
\end{equation}
Note that the unperturbed motion oscillates with unit frequency.  We can
introduce action-angle variables for the unperturbed Hamiltonian,
\begin{equation}
J=\frac{1}{2}(q^2+p^2)
\end{equation}
\begin{equation}
\theta=\arctan (q/p),
\end{equation}
so that
\begin{equation}
H=J-\frac{\epsilon}{2}J^2\left(1+\frac{4}{3}\cos (2\theta)
  +\frac{1}{3}\cos (4\theta)\right).
\end{equation}
Thus we have
\begin{equation}
H_0=J
\end{equation}
and
\begin{equation}
H_1=-\frac{1}{2}J^2\left(1+\frac{4}{3}\cos (2\theta)
  +\frac{1}{3}\cos (4\theta)\right).
\end{equation}
We now try to remove $H_1$ by a canonical Lie transform generated by the
scalar, $W_1$ (we shall work only to order one in $\epsilon$).  We have
$K_0=H_0,$ and
\begin{equation}
K_1=H_1+\{ W_1,H_0 \}=H_1+\{ W_1,J \}=H_1+\frac{\partial W_1}{\partial \theta}.
\end{equation}
Note that we cannot demand that $K_1=0$ since that would cause $W_1$ to be
multivalued (that is, secular terms would appear in $W_1$).  The best that
we can hope for is to make $K_1$ equal to the $\theta$-average of $H_1.$  That is,
\begin{equation}
K_1=-\frac{1}{2}J^2.
\end{equation}
Then
\begin{equation}
\frac{\partial W_1}{\partial \theta}
  =\frac{1}{6}J^2\left[4\cos (2\theta)+\cos (4\theta)\right],
\end{equation}
and this integrates to give
\begin{equation}
W_1=\frac{1}{24}J^2\left[8\sin (2\theta)+\sin (4\theta)\right].
\end{equation}
Using this generator we can work out the transformation equations, and
hence completely solve the problem (to order $\epsilon$).  For now we note
that the perturbed frequency is given by
\begin{equation}
\Omega\equiv\frac{\partial K}{\partial J}=1-\epsilon J.
\end{equation}

Note how the Lie transform has taken us to a new set of coordinates in
which the perturbation is averaged; that is, independent of the angle
variable.  Since the resulting Hamiltonian depends only on the action
variable, it is integrable by definition.  Furthermore, secular terms were
avoided by this absorbing of the averaged part of the perturbation into the
new Hamiltonian.

Aforementioned problems of resonant perturbations occur
when the unperturbed motion has characteristic frequencies that vary with
the action (this is true generically, but not in our above example).
When this happens, $\partial W_1/\partial \theta$ can equal a quantity that
is oscillatory but whose frequency passes through zero on some set of
measure zero in phase space.  Thus, in some neighborhood of this region,
problems of secular behavior can develop.  Various techniques exist for
dealing with this problem, but we shall not consider such problematic regions
of phase space in this thesis.

\subsection{Noncanonical Lie Transforms of a Hamiltonian System}
It sometimes happens that a canonical transformation is not the best way
to solve a particular problem in perturbation theory.  This may be because
it is best to express the unperturbed problem in noncanonical coordinates
for which the perturbation alters not only the Hamiltonian but also the
Poisson structure.  This is the case for both the guiding-center and
oscillation-center problems whose solution forms the core of this thesis.
In this case, we must resort to noncanonical transformations, but we demand
that they preserve the Hamiltonian nature of the equations of motion.  As has
already been pointed out, this can be accomplished by Lie transforming the
Poisson tensor along with the Hamiltonian; this means that the vector
generator of the Lie transform should not be a Hamiltonian vector field.

Consider once again the harmonic oscillator Hamiltonian,
\begin{equation}
H=\frac{1}{2}(q^2+p^2).
\end{equation}
This time, we introduce a perturbation not in the Hamiltonian but rather in the
Poisson structure.  Suppose that the perturbed brackets are
\begin{equation}
\{ q,p \}=1-\epsilon p^2.
\end{equation}
Thus we have ${\bf J}={\bf J}_0+\epsilon {\bf J}_1,$ where ${\bf J}_0$ is
the canonical Poisson tensor.  We wish to perform a Lie transform that will
restore the bracket to its canonical form.  We demand
\begin{equation}
0={\bf J}'_1={\bf J}_1-{\cal L}_g {\bf J}_0.
\end{equation}
Straightforward computation shows that this imposes only one independent
requirement on the generating vector field, ${\bf g},$ namely
\begin{equation}
\frac{\partial g^q}{\partial q}+\frac{\partial g^p}{\partial p}=p^2.
\end{equation}
It is easy enough to solve this equation; for example, we could take
\begin{equation}
g^p=\frac{p^3}{3}
\end{equation}
and
\begin{equation}
g^q=0.
\end{equation}
This effectively restores the bracket to canonical form, but it alters the
Hamiltonian as follows:
\begin{equation}
K=H-\epsilon {\cal L}_g H=\frac{1}{2}(q^2+p^2)-\frac{\epsilon}{3}p^4.
\end{equation}
Note that this transformed problem is coincidentally the same one that we
treated in the last subsection.  Thus, we could now apply a second (this time
canonical) Lie transform to finally solve it.  Once again, we would find
the perturbed frequency, $\Omega=1-\epsilon J.$

The important thing to note here is that $g$ is not a Hamiltonian vector field.
If it were, there would have to exist a scalar function $W$ such that
$0=\partial W/\partial p$ and $p^3/3=-\partial W/\partial q.$  Examination of
the mixed second derivatives shows these to be incompatible requirements.

\subsection{Lie Transforming the Phase-Space Lagrangian}
There is another way to go about making noncanonical transformations of a
Hamiltonian system that is guaranteed to keep it Hamiltonian.  Recall that
specifying the action one form is equivalent to specifying the Poisson
tensor (assuming that everything is nonsingular).  We can simply take the
exterior derivative of $\gamma$ to get ${\bf \omega},$ and then invert
${\bf \omega}$ to get ${\bf J}.$  These are all tensorial relationships, so we
could just as well Lie transform $\gamma$ and $H$ instead of ${\bf J}$ and $H.$

Indeed, there are several advantages to this approach.  First, it is easier
to take Lie derivatives of one forms than of second rank contravariant tensors;
there is one less term to worry about, and, more importantly, we can use the
homotopy formula to help us Lie differentiate one forms.  Second, when we Lie
transform the Poisson tensor, we are guaranteed that the resulting tensor will
be a valid Poisson structure only to the order we are keeping.  When we Lie
transform the action one form on the other hand,
its exterior derivative is still going to be closed even if we truncate it.
Thus ${\bf \omega}$ is exactly closed, so ${\bf J}={\bf \omega}^{-1}$ will 
obey the Jacobi identity {\it exactly}.

Consider a Lie transformation of the original action one form, $\gamma,$ into
a new action one form, $\Gamma.$
Using the homotopy formula, Eqs.~(\ref{mbs}) through (\ref{mbu}) become
\begin{equation}
\Gamma_0=\gamma_0,
\label{mca}
\end{equation}
\begin{equation}
\Gamma_1=\gamma_1-i_1 \omega_0+dS_1,
\label{mcb}
\end{equation}
\begin{equation}
\Gamma_2=\gamma_2-i_2 \omega_0-\frac{1}{2}i_1
                 (\omega_1+ \Omega_1)+dS_2,
\label{mcc}
\end{equation}
\begin{equation}
\Gamma_3=\gamma_3-i_3 \omega_0-i_2 \Omega_1-i_1[\omega_2 -
                   \frac{1}{3}di_1(\omega_1+\frac{1}{2}\Omega_1)]+dS_3,
\label{mcd}
\end{equation}
etc.  Here, we have defined $\omega_n \equiv d\gamma_n,$ and
$\Omega_n \equiv d\Gamma_n.$  
Note that in these equations, we have also made near-identity
gauge transformations by adding $dS_n$ at order $n$ for all $n\ge 1.$  
In fact, any other one-forms
in these equations that were given by the exterior derivative of a scalar
(typically arising from the second term on the right of 
Eqs.~(\ref{mby}) and (\ref{mbz})),
were absorbed in the definitions of the $S_n.$

Thus, these last transformation equations are capable of dealing with
any near-identity coordinate or gauge transformations, and so
it is these that we shall use in the sections to follow.  The vectors
$g_n$ and the scalars $S_n$ will be determined by certain
desiderata:  We want the transformation to average away the
rapidly oscillating terms of the Hamiltonian and action one-form, and
we want to avoid secular terms.  For the guiding-center problem, we shall 
also want the action
one-form to be invariant with respect to certain transformations
called {\it gyrogauge} and {\it boostgauge} transformations.  This will be
explained in more detail later.

For now, we consider another simple example.  Consider once again the
harmonic oscillator Hamiltonian, and perturb the canonical action one form
as follows:
\begin{equation}
\gamma=pdq+\frac{\epsilon}{3}p^3dq.
\end{equation}
We have
\begin{equation}
\omega=d\gamma=(1+\epsilon p^2)dp\wedge dq.
\end{equation}
This inverts to give $(1+\epsilon p^2)^{-1}$ times the canonical Poisson
tensor, and to order $\epsilon$ this is the same as the perturbation that
was examined in the last subsection (which is why we chose it).  We can now
compare the two methods of doing the problem.

Demand that $\Gamma_1=0,$ so Eq.~(\ref{mcb}) gives
\begin{equation}
0=\Gamma_1=\gamma_1-i_1 \omega_0+dS_1=(\frac{1}{3}p^3-g^p)dq-g^qdp+dS_1.
\end{equation}
Thus we can take $S_1=0,$ and
\begin{equation}
g^p=\frac{p^3}{3}
\end{equation}
and
\begin{equation}
g^q=0.
\end{equation}
These are precisely the same generators that we discovered in the last
subsection, they have precisely the same effect on the Hamiltonian, and the
rest of the problem follows in identical fashion.  That is, a second
canonical Lie transformation is necessary to get to averaged coordinates.

%% file: relgc.tex
\chapter{Relativistic Guiding-Center Theory}
\textheight=8.30truein
\setcounter{equation}{280}
\label{yac}
\pagestyle{myheadings}
\markboth{Relativistic Guiding-Center Theory}{Relativistic Guiding-Center Theory}
\section{Discussion}
Relativistic guiding-center motion occurs in many applications of
plasma physics, including controlled fusion, free-electron lasers,
and astrophysics.  The tandem mirror and bumpy torus plasma confinement
devices, for example, utilize populations of magnetized electrons
at relativistic energies in complicated field-line geometries.
In free-electron lasers, relativistic electron beams travel along 
strong magnetic fields with superposed wiggler fields.  Near a
neutron star, relativistic plasma can be confined in strong
electromagnetic and gravitational fields.

All these examples point out the need for a formalism that
is able to treat general electromagnetic field geometries.  Particle
simulation codes used for studying the properties of guiding-center plasmas
in controlled fusion confinement devices sometimes require the guiding-center
equations of motion to one order higher than the usual drifts; 
this indicates the need for a simplified
and systematic perturbative treatment, such as that afforded by the use of
Lie transforms.  The free-electron
laser problem has no obvious preferred frame of reference, and this suggests
that a manifestly covariant description would best reveal the essence of the
physical processes involved.  The neutron star problem involves coupling to
a general relativistic gravitational field, and this absolutely requires
a manifestly covariant formulation.  All these desiderata will be satisfied
by our theory.

Nonrelativistic theories of guiding-center motion in arbitrary magnetic
geometry frequently make use of orthonormal triads of unit vectors at
each point of three-dimensional physical space.  One member of each
such triad is required to lie in the direction of the magnetic field at
that point.  Such a basis affords great clarity and relative ease in the
computation and exposition of the results of guiding-center theory.

One of the first problems to be addressed in any relativistic formulation
of guiding-center theory is thus that of finding the relativistic analogs
of these basis triads.  Fortunately, this problem has been solved by
Fradkin~\cite{zae}, who gives a straightforward method for finding
orthonormal tetrads of unit vectors at each point of four-dimensional
spacetime.  In a frame for which the perpendicular electric field vanishes,
one pair of unit vectors in these
tetrads lies perpendicular to the magnetic field, while the other pair
spans the two-dimensional subspace determined by the direction of the
magnetic field and the direction of time.

Fradkin shows that these two two-dimensional subspaces are covariantly
defined, and that the rapid gyration takes place in the
first of these, while the slower parallel motion takes place in
the second.  This formalism is therefore useful for isolating the
oscillatory motion so that it can be effectively averaged to obtain
the guiding-center equations of motion.  It is described from first
principles in Sections~\ref{yad}, \ref{yae}, and \ref{yaf}.

Lie transform perturbation theory is used to perform the averaging.
Though this technique has been known for some time~\cite{zao}, its
use for the guiding-center problem poses special difficulties which
were first overcome by Littlejohn~\cite{zas}.  The
difficulties are due to the fact that the Poisson structure as well as
the Hamiltonian depends upon the rapidly gyrating variables, so that the 
transformation required to gyroaverage the system of equations is not 
canonical.

A Lie transform in its most general sense is 
a coordinate transformation generated by a vector 
field on phase space.  If this vector field generator is a Hamiltonian
vector field (that is, a vector field that is the flow generated by some
scalar Hamiltonian-like function) then the transformation it induces
is canonical; in this case one often simply speaks of the
transformation as being generated by the corresponding scalar function.  
For the guiding-center problem, however, the vector generator of the 
averaging transformation cannot be a Hamiltonian vector field, since it 
must generate a noncanonical transformation.

In the nonrelativistic
guiding-center problem, it was found by Littlejohn~\cite{zaq}
to be easiest to apply the general Lie transform to the action one form.
This is the approach that is followed here; it was
described from first principles in Chapter~\ref{yaa}.

In any calculation that goes beyond the lowest order drifts, it was
found by Littlejohn~\cite{zak} to be necessary to worry about
maintaining a certain gauge invariance property of the action
one form which for the nonrelativistic case is known as
{\it gyrogauge} invariance.  If the averaging transformation does not preserve
this invariance property, then the final guiding-center equations of motion 
will depend unavoidably on the arbitrarily chosen basis vectors used to set
up the problem, as was noted by Hagan and Frieman~\cite{zan}.
In Section~\ref{yag},
we work out the relativistic generalization of this invariance property,
and we find that the relativistic case admits another similar gauge
invariance property which we call {\it boostgauge} invariance.

The Lie transforms are carried out in Sections~\ref{yah}, \ref{yai} and
\ref{yaj}, and the guiding-center Lagrangian and Hamiltonian are presented.
The Poisson bracket structure is then given in Section~\ref{yak} and the
equations of motion are presented and discussed in Section~\ref{yal}.  In 
Section~\ref{yam}, a complete summary of the transformation equations is
given for reference and the correction to the gyromomentum is
derived.  In Section~\ref{yan}, we show how to write our results in ``$1+3$''
notation, and we compare our results to those of Northrop~\cite{zap}.
In Section~\ref{yao} we cast all our results in 
{\it manifestly} gyrogauge and boostgauge invariant format.  

\section{Conventions and Notation}
\label{yad}
In this work, we adopt the following conventions:  The particle space-time
coordinate will be denoted by $r^\mu,$ where $\mu=0,\ldots ,3.$
The Minkowski metric, $g_{\mu\nu}={\rm diag}(-1,+1,+1,+1),$ is used
throughout our derivation of the guiding-center equations, but the results
will be written in manifestly covariant form so that this assumption can be 
relaxed.  The four potential is given by $A^\mu = (\phi , {\bf A}),$ 
so the antisymmetric field tensor is $F=dA,$ or
\begin{equation}
F_{\mu\nu}=A_{\nu , \mu}-A_{\mu , \nu}
          =\left(
               \begin{array}{cccc}
                  0  & -E_x & -E_y & -E_z \\
                 E_x &   0  &  B_z & -B_y \\
                 E_y & -B_z &   0  &  B_x \\
                 E_z &  B_y & -B_x &   0
               \end{array}
            \right).
\label{gaa}
\end{equation}
The dual field tensor, ${\cal F}=\hbox{$^*F$},$ is given by
\begin{equation}
{\cal F}^{\mu\nu}=\frac{1}{2} \epsilon^{\mu\nu\alpha\beta} F_{\alpha\beta}
          =\left(
               \begin{array}{cccc}
                  0  & -B_x & -B_y & -B_z \\
                 B_x &   0  &  E_z & -E_y \\
                 B_y & -E_z &   0  &  E_x \\
                 B_z &  E_y & -E_x &   0
               \end{array}
            \right).
\label{gab}
\end{equation}
where $\epsilon_{\mu \nu \alpha \beta}$ is the completely
antisymmetric fourth rank Levi-Civita tensor with 
$\epsilon_{0123}=+1.$  Note carefully that $\epsilon^{0123}=-1,$ thanks to
the Minkowski metric.

It is often convenient to use ``$1+3$'' notation.  Then, the matrix of
components of the mixed field tensor, $F^\mu_{\phantom{\mu}\nu},$ may
be written
\begin{equation}
   F = \left(
         \begin{array}{cc}
                 0 & {\bf E}               \\
           {\bf E} & {\bf 1} \times {\bf B} 
         \end{array}
       \right),
\end{equation}
and that of the mixed dual field tensor, ${\cal F}^\mu_{\phantom{\mu}\nu},$ may
be written
\begin{equation}
{\cal F} = \left(
             \begin{array}{cc}
                      0 & -{\bf B}               \\
               -{\bf B} &  {\bf 1} \times {\bf E} 
             \end{array}
           \right).
\end{equation}
Note that we have used the notation
$({\bf 1}\times {\bf B})_{ij}=\epsilon_{ikl}\delta_{jk}B_l=\epsilon_{ijl}B_l.$
Also note that the mixed field tensors are neither symmetric nor antisymmetric.
The advantage to dealing with the mixed tensors is that one may contract them
with other tensors using ordinary matrix multiplication.  Of course, we could
equally well do this with the completely covariant or contravariant forms, but
we would have to remember to use the Minkowski metric when multiplying a row
by a column.

Thus, when the field tensor is applied to an arbitrary four-vector, the result 
may be written
\begin{equation}
   F \cdot \left(
                  \begin{array}{c}
                     a \\
                     {\bf a}
                  \end{array}
                \right)=
                \left(
                   \begin{array}{c}
                      {\bf E} \cdot {\bf a} \\
                      a {\bf E} + {\bf a} \times {\bf B}
                   \end{array}
                \right).
\label{gan}
\end{equation}
The analogous equation for the dual field tensor is
\begin{equation}
   {\cal F} \cdot \left(
                  \begin{array}{c}
                     a \\
                     {\bf a}
                  \end{array}
                \right)=
                \left(
                   \begin{array}{c}
                      -{\bf B} \cdot {\bf a} \\
                      -a {\bf B} + {\bf a} \times {\bf E}
                   \end{array}
                \right).
\label{gao}
\end{equation}
This ``$1+3$'' notation will prove to be useful and convenient throughout the
remainder of this thesis.

The two familiar Lorentz scalars can be expressed in terms of these tensors by
\begin{eqnarray}
\lambda_{1} &\equiv& \frac{1}{2} F_{\mu\nu} F^{\mu\nu}
   = \frac{1}{2} F:F = B^2-E^2,
\label{gac}\\
\noalign{\hbox{and}}
\lambda_{2} &\equiv& \frac{1}{4} {\cal F}_{\mu\nu} F^{\mu\nu}
   = \frac{1}{4}{\cal F}:F = {\bf E \cdot B}.
\label{gad}
\end{eqnarray}
Note carefully that
$F:F\equiv F_{\mu\nu}F^{\mu\nu}=-F_{\mu\nu}F^{\nu\mu}=-{\rm Tr}(F\cdot F).$

The Lorentz equation of motion may then be written
\begin{equation}
m\frac{du}{d\tau}=\frac{e}{c}F(r)\cdot u,
\label{gae}
\end{equation}
where
\begin{equation}
u=\frac{dr}{d\tau}
\end{equation}
is the four-velocity, $\tau$ is the proper time, $m$ is the rest
mass and $e$ is the charge. 

Equation~(\ref{gae}) makes it clear that if the field is 
independent of space-time position, then the
frequencies of the motion are the eigenvalues of $F$ times
$-ie/mc.$  Now the characteristic equation for the matrix $F$ is
\begin{equation}
{\rm det}(F-\lambda {\bf 1})=\lambda^4+\lambda_1 \lambda^2 -
                             \lambda_2^2 = 0.
\label{gaf}
\end{equation}
This biquadratic in $\lambda$ is easily solved to give $\lambda
=\pm \lambda_E,$ or $\lambda=\pm i\lambda_B,$ where we have
defined the Lorentz scalars
\begin{eqnarray}
\lambda_E &\equiv& {\rm sgn}(\lambda_2)
                 \sqrt{\frac{1}{2}
                 (\sqrt{\lambda_1^2 + 4\lambda_2^2} - \lambda_1)},
\label{gag}\\
\noalign{\hbox{and}}
\lambda_B &\equiv& \sqrt{\frac{1}{2}
                 (\sqrt{\lambda_1^2 + 4\lambda_2^2} + \lambda_1)}.
\label{gah}
\end{eqnarray}
We can write $\lambda_1$ and $\lambda_2$ in terms of $\lambda_E$
and $\lambda_B$ as follows:
\begin{eqnarray}
\lambda_1 &=&\lambda_B^2 - \lambda_E^2,
\label{gai}\\
\noalign{\hbox{and}}
\lambda_2 &=&\lambda_B \lambda_E.
\label{gaj}
\end{eqnarray}
We can now define the two Lorentz scalars
\begin{equation}
\Omega_E \equiv \frac{e\lambda_E}{mc},
\end{equation}
and
\begin{equation}
\Omega_B \equiv \frac{e\lambda_B}{mc}.
\label{gkf}
\end{equation}
The first of these is the inverse of the characteristic proper time 
required to accelerate to relativistic velocities along field lines, 
while the second is the gyrofrequency with respect to proper time.

\section{The Electromagnetic Projection Operators}
\label{yae}
In this section, we summarize the work of Fradkin~\cite{zae}
that is relevent to this study.
It is straightforward to verify the following identities:
\begin{equation}
F^2-{\cal F}^2=-\lambda_1 {\bf 1},
\end{equation}
and
\begin{equation}
F\cdot {\cal F}={\cal F}\cdot F=-\lambda_2 {\bf 1}.
\end{equation}
Premultiplying the first of these by $F,$ and employing the second
gives
\begin{equation}
F^3=-\lambda_2 {\cal F}-\lambda_1 F.
\end{equation}
Premultiplying by $F$ once again gives
\begin{equation}
F^4+\lambda_1 F^2 - \lambda_2^2 {\bf 1}=0.
\label{gak}
\end{equation}
Comparing this with Eq.~(\ref{gaf}), we see that we have
proven that $F$ obeys its own characteristic equation, as it must
by the Hamilton-Cayley theorem.  Now it is clear that 
Eq.~(\ref{gak}) may be written as follows:
\begin{equation}
(F-\lambda_E {\bf 1})\cdot (F+\lambda_E {\bf 1})\cdot
   (F-i\lambda_B {\bf 1})\cdot (F+i\lambda_B {\bf 1})=0,
\end{equation}
and the four factors in this expression commute, so any of them
could have been written first.  Thus, if $\Psi$ is an arbitrary
column four-vector, then
\begin{equation}
(F-\lambda_E {\bf 1})\cdot \biggl[(F+\lambda_E {\bf 1})\cdot
   (F-i\lambda_B {\bf 1})\cdot (F+i\lambda_B {\bf 1})\cdot \Psi \biggr]=0,
\end{equation}
so that $(F+\lambda_E {\bf 1})\cdot (F-i\lambda_B {\bf 1})\cdot
(F+i\lambda_B {\bf 1})\cdot \Psi $ is an (unnormalized) eigenvector
of $F$ with eigenvalue $\lambda_E.$  Thus, the operator
$(F+\lambda_E {\bf 1})\cdot (F-i\lambda_B {\bf 1})\cdot (F+i\lambda_B {\bf 1})$
is a (unnormalized) projection operator that projects arbitrary
four-vectors onto the vector subspace spanned 
by the zeroth eigenvector of
$F.$  Proceeding in this manner, it is easy to see that the
projection operator
\begin{equation}
P_\parallel = \frac{F^2 + \lambda_B^2 {\bf 1}}
                   {\lambda_B^2+\lambda_E^2}
\label{gal}
\end{equation}
projects arbitrary four-vectors onto the vector subspace spanned
by the eigenvectors of F with eigenvalues $\pm \lambda_E,$ while the 
projection operator
\begin{equation}
P_\perp     = \frac{-F^2 + \lambda_E^2 {\bf 1}}
                   {\lambda_B^2+\lambda_E^2}
\label{gam}
\end{equation}
projects arbitrary four-vectors onto the vector subspace spanned 
by the eigenvectors of F with eigenvalues $\pm i\lambda_B.$  The
normalization constants were chosen to make the projection
operators idempotent; that is
\begin{equation}
P_\parallel \cdot P_\parallel =P_\parallel,
\end{equation}
\begin{equation}
P_\perp \cdot P_\perp =P_\perp,
\end{equation}
\begin{equation}
P_\parallel \cdot P_\perp = P_\perp \cdot P_\parallel = 0,
\end{equation}
and
\begin{equation}
P_\parallel + P_\perp = {\bf 1}.
\end{equation}
We have thus decomposed the tangent space at each point of space-time into the
Cartesian product of two two-dimensional ``two-flats.''  The
rapid gyromotion takes place in the perpendicular two-flat since it
is spanned by the eigenvectors corresponding to the imaginary
eigenvalues, while the parallel motion takes place in the parallel
two-flat since it is spanned by the eigenvectors corresponding to
the real eigenvalues.
These two-flats will play an indispensible role in our theory.  We shall
use them to isolate the gyrational components of the particle velocity
in preparation for the guiding-center Lie transform.

In Section~\ref{yah}, we shall order the fields in an expansion parameter
and, for reasons that will be explained at that time, we shall demand that
our lowest-order field have $\lambda_E=0.$  Furthermore, the two-flats that
we shall use will always be defined in terms of the zero-order field; that
is, the field tensor that appears on the right hand side of Eqs.~(\ref{gal})
and (\ref{gam}) is always the lowest-order field tensor with $\lambda_E=0.$
Thus, these equations can be simplified to read
\begin{equation}
P_\parallel = {\bf 1}+\frac{F^2}{\lambda_B^2} = \frac{{\cal F}^2}{\lambda_B^2}
\label{gig}
\end{equation}
and
\begin{equation}
P_\perp     =-\frac{F^2}{\lambda_B^2} = {\bf 1}-\frac{{\cal F}^2}{\lambda_B^2}.
\label{gih}
\end{equation}
In ``$1+3$'' notation, Eqs.~(\ref{gig}) and (\ref{gih}) become
\begin{equation}
P_\parallel = \frac{1}{B^2-E^2}\left(
   \begin{array}{cc}
      B^2 & -{\bf E} \times {\bf B} \\
      {\bf E} \times {\bf B} & {\bf B}{\bf B}+{\bf E}{\bf E}-E^2 {\bf 1}
   \end{array}\right),
\end{equation}
and
\begin{equation}
P_\perp     = \frac{1}{B^2-E^2}
   \left(
   \begin{array}{cc}
      -E^2 & {\bf E} \times {\bf B} \\
      -{\bf E} \times {\bf B} & -{\bf B}{\bf B}-{\bf E}{\bf E}+B^2 {\bf 1}
   \end{array}
   \right).
\end{equation}
Henceforth, all our results concerning the nature of the two-flats and the
unit vectors that span them will contain this assumption that the underlying
field tensor has $\lambda_E=0.$

\section{The Orthonormal Basis Tetrad}
\label{yaf}
We wish to show how to construct a tetrad of unit vectors such that one pair
spans the parallel two-flat while the other pair spans the perpendicular
two-flat.  Clearly such a tetrad is not unique; it is defined only
to within an arbitrary rotation in the perpendicular two-flat,
and an arbitrary hyperbolic rotation (boost) in the parallel
two-flat.  We shall have much more to
say about this nonuniqueness later; for now we are simply looking for a way
to construct {\it any} such tetrad.

From the arguments presented in the last section, we know that one way to do
this is to examine the eigenvectors of the field tensor.  Here we shall take
a different approach that is perhaps more physically motivated.  Recall that
we are dealing with fields for which $E_\parallel=0$ (if this is true in any
one frame, it will be true in all frames because ${\bf E}\cdot {\bf B}$ is a
Lorentz scalar).  There exist a set of local
``preferred'' reference frames for which
${\bf E}_\perp$ also vanishes; hence there is no electric field at all in these
preferred frames.  Thus, in a preferred frame, the field tensors may be written
in ``$1+3$'' notation as follows:
\begin{equation}
   F = \left( \begin{array}{cc}
                         0 &  {\bf 0}               \\
                   {\bf 0} &  {\bf 1}\times {\bf B}
              \end{array} \right)
\label{gke}
\end{equation}
and
\begin{equation}
{\cal F} = \left( \begin{array}{cc}
                         0 & -{\bf B}               \\
                  -{\bf B} &  {\bf 0}
                  \end{array} \right).
\end{equation}
Also, in a preferred frame, the projection operators have the form
\begin{equation}
P_\parallel= \left( \begin{array}{cc}
                           1 & {\bf 0}                \\
                     {\bf 0} & {\bf b}{\bf b}
                    \end{array} \right)
\end{equation}
and
\begin{equation}
P_\perp   = \left( \begin{array}{cc}
                          0 & {\bf 0}                 \\
                    {\bf 0} & {\bf 1}-{\bf b}{\bf b}
                   \end{array} \right),
\end{equation}
where
\begin{equation}
{\bf b}\equiv {\bf B}/\left| {\bf B} \right|.
\end{equation}

The above forms for the projection operators in a preferred frame make it
clear that we can choose the following orthonormal basis tetrad for a preferred
frame:
\begin{equation}
   \begin{array}{cc}
      {\hat{\bf e}}_0=\left( 
                          \begin{array}{c}
                             1 \\
                             {\bf 0}
                          \end{array}
                       \right), &
      {\hat{\bf e}}_1=\left( 
                          \begin{array}{c}
                             0 \\
                             {\bf b}
                          \end{array}
                       \right),
   \end{array}
\end{equation}
and
\begin{equation}
   \begin{array}{cc}
      {\hat{\bf e}}_2=\left( 
                          \begin{array}{c}
                             0 \\
                             {\bftau}_1
                          \end{array}
                       \right), &
      {\hat{\bf e}}_3=\left( 
                          \begin{array}{c}
                             0 \\
                             {\bftau}_2
                          \end{array}
                       \right),
   \end{array}
\end{equation}
where ${\bftau}_1$ and ${\bftau}_2$ are unit three-vectors perpendicular
to ${\bf b},$ such that $\{ {\bf b},{\bftau}_1,{\bftau}_2 \}$ constitutes
an orthonormal triad in three-dimensional space.
We reiterate that the above choice is not unique.

Of course, we would like to be able to construct an orthonormal basis tetrad
in an {\it arbitrary} Lorentz frame.  To see how to do this, we consider a
Lorentz boost from the above-described preferred frame to a new frame.
The Lorentz transformation matrix for a boost is
\begin{equation}
\Lambda
  =\left(
     \begin{array}{cc}
       \gamma & -\gamma {\bfbet} \\
      -\gamma {\bfbet} & {\bf 1}+(\gamma-1)\beta^{-2}{\bfbet}{\bfbet}
     \end{array}
   \right),
\end{equation}
where the three-vector ${\bfbet}$ is the generator of the Lorentz boost
(it is the relative velocity of the two reference frames divided by $c$),
and where $\gamma\equiv (1-\beta^2)^{-1/2}.$  This matrix is an element of the
Lorentz group because it satisfies
$\Lambda^{-1}=g\cdot\Lambda^T\cdot g$ (here we have used a superscripted ``T''
to denote the transpose operation).  See
Jackson~\cite{zag} for more details on the Lorentz group and its generators.

The new field tensor components are then
\begin{equation}
F'=\Lambda \cdot F \cdot \Lambda^{-1}
  =\left(
     \begin{array}{cc}
       0 & \gamma {\bfbet}\times {\bf B} \\
       \gamma {\bfbet}\times {\bf B} & {\bf 1}\times [\gamma {\bf B}
       -(\gamma-1)\beta^{-2}{\bfbet}{\bfbet}\cdot {\bf B}]
     \end{array}
   \right).
\end{equation}
In writing this result, we have made use of the vector identity,
\begin{equation}
{\bfbet}\times {\bf B}{\bfbet}-{\bfbet}{\bfbet}\times {\bf B}
  ={\bf 1}\times (\beta^2 {\bf B}-{\bfbet}{\bfbet}\cdot {\bf B}).
\end{equation}
From this result for the field tensor,
we see that we can identify the electric and magnetic fields in the
new frame as
\begin{equation}
{\bf E}'=\gamma {\bfbet} \times {\bf B}
\end{equation}
and
\begin{equation}
{\bf B}'
   =\gamma {\bf B}-(\gamma-1)\beta^{-2} {\bfbet}{\bfbet}\cdot {\bf B}.
\end{equation}

At this point, there are a number of interesting observations to be made.
First note that if ${\bfbet}$ is parallel to ${\bf B}$ then ${\bf E}'=0,$
so the transformation takes us to another preferred frame.  Next note that if
${\bfbet}$ is perpendicular to ${\bf B}$ then ${\bf B}$ is parallel to
${\bf B}'.$  Next note that it
is possible to arrive at {\it any} desired ${\bf E}'$ by a transformation with
${\bfbet}$ perpendicular to ${\bf B}.$  Specifically, if we take
\begin{equation}
{\bfbet}=-{\bfbet}_E,
\end{equation}
where
\begin{equation}
{\bfbet}_E \equiv \frac{{\bf E}'\times {\bf B}'}{{B'}^2},
\end{equation}
then it is easy to see that the new electric field is ${\bf E}'.$  Conversely,
if we begin with a frame in which the (perpendicular) electric field is
${\bf E}',$ then a Lorentz boost with ${\bfbet}={\bfbet}_E$ gets us to
a preferred frame.

The orthonormal tetrad in the new frame is then
\begin{eqnarray}
      {\hat{\bf e}}'_0&=&\Lambda\cdot {\hat{\bf e}}_0=
                        \left(
                          \begin{array}{c}
                             \gamma_E \\
                             \gamma_E {\bfbet}_E
                          \end{array}
                        \right)\nonumber\\
      {\hat{\bf e}}'_1&=&\Lambda\cdot {\hat{\bf e}}_1=
                        \left( 
                          \begin{array}{c}
                             0 \\
                             {\bf b}
                          \end{array}
                        \right),
                      \label{gjk}
\end{eqnarray}
and
\begin{eqnarray}
      {\hat{\bf e}}'_2&=&\Lambda\cdot {\hat{\bf e}}_2=
                        \left( 
                          \begin{array}{c}
                             \gamma_E {\bfbet}_E \cdot {\bftau}_1 \\
                             {\bftau}_1+(\gamma_E-1)\beta_E^{-2}
                             {\bfbet}_E {\bfbet}_E \cdot {\bftau}_1
                          \end{array}
                        \right)\nonumber\\
      {\hat{\bf e}}'_3&=&\Lambda\cdot {\hat{\bf e}}_3=
                        \left(
                          \begin{array}{c}
                             \gamma_E {\bfbet}_E \cdot {\bftau}_2 \\
                             {\bftau}_2+(\gamma_E-1)\beta_E^{-2}
                             {\bfbet}_E {\bfbet}_E \cdot {\bftau}_2
                          \end{array}
                        \right),
                      \label{gjl}
\end{eqnarray}
where $\gamma_E \equiv (1-\beta_E^2)^{-1/2}.$

At this point we note that we can choose ${\bftau}_1$ to lie along the
direction of ${\bfbet}_E$ without any loss of generality.  We can now
write the results for the unit tetrad in the general frame, dropping the
primes which are no longer needed because {\it all} quantities will refer to
the general frame.  Thus
\begin{equation}
   \begin{array}{cc}
      {\hat{\bf e}}_0=\left(
                          \begin{array}{c}
                             \gamma_E \\
                             \gamma_E {\bfbet}_E
                          \end{array}
                      \right), &
      {\hat{\bf e}}_1=\left(
                          \begin{array}{c}
                             0 \\
                             {\bf b}
                          \end{array}
                      \right),
   \end{array}
\label{gij}
\end{equation}
and
\begin{equation}
   \begin{array}{cc}
      {\hat{\bf e}}_2=\left( 
                          \begin{array}{c}
                             \gamma_E \beta_E \\
                             \gamma_E {\hat{\bfbet}}_E
                          \end{array}
                      \right), &
      {\hat{\bf e}}_3=\left( 
                          \begin{array}{c}
                             0 \\
                             {\bf b}\times {\hat{\bfbet}}_E
                          \end{array}
                      \right),
   \end{array}
\label{gik}
\end{equation}
where
\begin{equation}
{\bfbet}_E \equiv \frac{{\bf E}\times {\bf B}}{{B}^2},
\end{equation}
and $\gamma_E \equiv (1-\beta_E^2)^{-1/2}.$ Here we have also introduced the
notation ${\hat{\bfbet}}_E$ for a unit vector in the
direction of ${\bfbet}_E$ if ${\bfbet}_E\neq 0.$  If ${\bfbet}_E=0,$
one may choose ${\hat{\bfbet}}_E$ to be any unit three-vector perpendicular
to ${\bf b}.$

Using Eqs.~(\ref{gan}) and (\ref{gao}), the following useful 
identities are readily demonstrated:
\begin{equation}
   \begin{array}{cc}
      F \cdot {\hat{\bf e}}_0=0, &
      F \cdot {\hat{\bf e}}_1=0, 
   \end{array}
\end{equation}
\begin{equation}
   \begin{array}{cc}
      F \cdot {\hat{\bf e}}_2=-\lambda_B {\hat{\bf e}}_3, &
      F \cdot {\hat{\bf e}}_3=+\lambda_B {\hat{\bf e}}_2, 
   \end{array}
\end{equation}
and
\begin{equation}
   \begin{array}{cc}
      {\cal F} \cdot {\hat{\bf e}}_0=-\lambda_B {\hat{\bf e}}_1, &
      {\cal F} \cdot {\hat{\bf e}}_1=-\lambda_B {\hat{\bf e}}_0, 
   \end{array}
\end{equation}
\begin{equation}
   \begin{array}{cc}
      {\cal F} \cdot {\hat{\bf e}}_2=0, &
      {\cal F} \cdot {\hat{\bf e}}_3=0.
   \end{array}
\end{equation}
Thus, the field tensor and its dual have the effect of rotating
these unit vectors {\it within} their respective two-flats.

\begin{figure}[p]
\center{
\vspace{1.70truein}
\mbox{\includegraphics[bbllx=0,bblly=0,bburx=221,bbury=156,width=5.82truein]{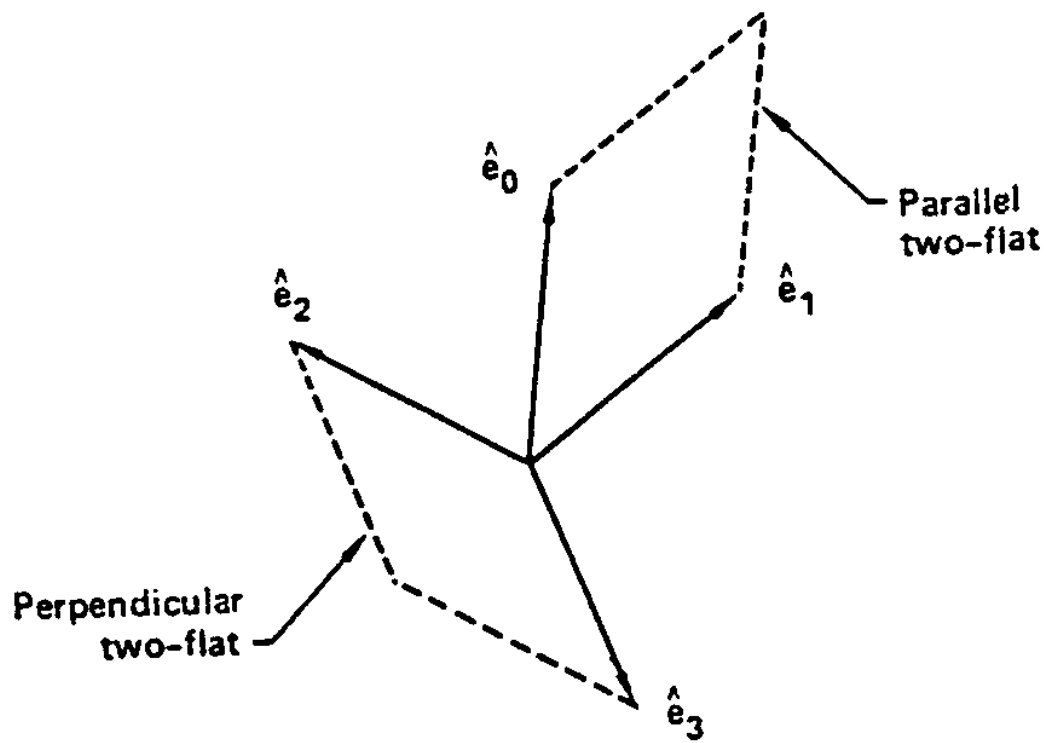}}
\vspace{1.70truein}
}
\caption{The Orthonormal Basis Tetrad}
\label{gic}
\end{figure}

Using Eqs.~(\ref{gig}) and (\ref{gih}),
it is easy to verify that $P_\parallel$ leaves ${\hat{\bf e}}_0$
and ${\hat{\bf e}}_1$ unchanged and annihilates ${\hat{\bf e}}_2$
and ${\hat{\bf e}}_3,$ while $P_\perp$ annihilates ${\hat{\bf e}}_0$
and ${\hat{\bf e}}_1$ and leaves ${\hat{\bf e}}_2$ and ${\hat{\bf e}}_3$
unchanged.  It is also easy to verify that this tetrad is orthonormal with 
respect to the Minkowski metric; that is, that
\begin{equation}
{\hat{\bf e}}_\mu \cdot {\hat{\bf e}}_\nu = g_{\mu \nu}.
\end{equation}
So ${\hat{\bf e}}_0$ and ${\hat{\bf e}}_1$ span the parallel two-flat,
and ${\hat{\bf e}}_2$ and ${\hat{\bf e}}_3$ span the perpendicular
two-flat, as asserted.  The geometrical situation is illustrated  
schematically in Fig.~\ref{gic}.

In terms of the ${\hat{\bf e}}_\alpha ,$ the projection operators may
be written
\begin{equation}
P_\parallel=-{\hat{\bf e}}_0 {\hat{\bf e}}_0
            +{\hat{\bf e}}_1 {\hat{\bf e}}_1
\end{equation}
and
\begin{equation}
 P_\perp      = {\hat{\bf e}}_2 {\hat{\bf e}}_2
               +{\hat{\bf e}}_3 {\hat{\bf e}}_3.
\end{equation}
This should be clear from the geometrical picture, but may also be verified
by direct algebra.

When applied to the particle four-velocity, these projection
operators will allow us to isolate the rapid gyrational motion in
the perpendicular two-flat from the nongyrational motion in the
parallel two flat.  Thus
\begin{equation}
u=u^\mu {\hat{\bf e}}_\mu,
\end{equation}
or, if we introduce polar coordinates $(w,\theta )$ for the 
perpendicular four-velocity components and hyperbolic polar 
coordinates $(k,\beta )$ for the parallel velocity components,
then we may write
\begin{equation}
u={\hat{\bf e}}_0 k \cosh \beta+
  {\hat{\bf e}}_1 k \sinh \beta-
  {\hat{\bf e}}_2 w \sin \theta-
  {\hat{\bf e}}_3 w \cos \theta
\label{gii}
\end{equation}
or
\begin{equation}
u=k {\hat {\bf t}}+w {\hat {\bf c}},
\label{gap}
\end{equation}
where we have defined
\begin{eqnarray}
{\hat{\bf t}} &\equiv&  {\hat{\bf e}}_0 \cosh \beta +
                       {\hat{\bf e}}_1 \sinh \beta ,
\label{gaq}\\
\noalign{\hbox{and}}
{\hat{\bf c}} &\equiv& -{\hat{\bf e}}_2 \sin \theta -
                       {\hat{\bf e}}_3 \cos \theta .
\label{gar}\\
\noalign{\hbox{If we also define}}
{\hat{\bf b}} &\equiv&  {\hat{\bf e}}_0 \sinh \beta +
                       {\hat{\bf e}}_1 \cosh \beta ,
\label{gas}\\
\noalign{\hbox{and}}
{\hat{\bf a}} &\equiv&  {\hat{\bf e}}_2 \cos \theta -
                       {\hat{\bf e}}_3 \sin \theta ,
\label{gat}
\end{eqnarray}
then $({\hat{\bf t}},{\hat{\bf b}},{\hat{\bf c}},{\hat{\bf a}})$
form a new velocity-dependent basis tetrad that is also orthonormal 
with respect to the Minkowski metric.  Please do not confuse the basis
four-vector $\hat{\bf b}$ with the basis three-vector ${\bf b},$
and do not confuse the hyperbolic polar coordinate $\beta$ with the
Lorentz transformation generator $\bfbet .$

Some useful relations among the elements of this new basis tetrad are
\begin{equation}
   \begin{array}{cc}
      \frac{\partial{\hat{\bf t}}}{\partial \beta} = {\hat{\bf b}}, &
      \frac{\partial{\hat{\bf b}}}{\partial \beta} = {\hat{\bf t}}, 
   \end{array}
\label{gau}
\end{equation}
\begin{equation}
   \begin{array}{cc}
      \frac{\partial{\hat{\bf c}}}{\partial \theta} = -{\hat{\bf a}}, &
      \frac{\partial{\hat{\bf a}}}{\partial \theta} = {\hat{\bf c}}, 
   \end{array}
\label{gav}
\end{equation}
and
\begin{equation}
   \begin{array}{cc}
      F \cdot {\hat{\bf t}}=0, &
      F \cdot {\hat{\bf b}}=0, 
   \end{array}
\label{gaw}
\end{equation}
\begin{equation}
   \begin{array}{cc}
      F \cdot {\hat{\bf c}}=-\lambda_B {\hat{\bf a}}, &
      F \cdot {\hat{\bf a}}=+\lambda_B {\hat{\bf c}}, 
   \end{array}
\label{gax}
\end{equation}
and
\begin{equation}
   \begin{array}{cc}
      {\cal F} \cdot {\hat{\bf t}}=-\lambda_B {\hat{\bf b}}, &
      {\cal F} \cdot {\hat{\bf b}}=-\lambda_B {\hat{\bf t}}, 
   \end{array}
\label{gay}
\end{equation}
\begin{equation}
   \begin{array}{cc}
      {\cal F} \cdot {\hat{\bf c}}=0, &
      {\cal F} \cdot {\hat{\bf a}}=0.
   \end{array}
\end{equation}
Also, the projection operators may now be written
\begin{equation}
P_\parallel=-{\hat{\bf t}} {\hat{\bf t}}
            +{\hat{\bf b}} {\hat{\bf b}}
\label{gaz}
\end{equation}
and
\begin{equation}
P_\perp      = {\hat{\bf c}} {\hat{\bf c}}
              +{\hat{\bf a}} {\hat{\bf a}}.
\label{gba}
\end{equation}

It is useful to compare the above description of the four-velocity
in terms of $(k,\beta,w,\theta )$
with the more conventional ``$1+3$'' representation,
$u=c(\gamma_v , \gamma_v {\bfbet}_v),$ where
${\bfbet}_v \equiv {\bf v}/c.$  We shall do this using the unit
tetrad that we constructed above.  Combining Eqs.~(\ref{gij}), (\ref{gik}) and
(\ref{gii}), we find
\begin{equation}
c\gamma_v = \gamma_E (k\cosh\beta-\beta_E w\sin\theta)
\end{equation}
and
\begin{equation}
c\gamma_v {\bfbet}_v = \gamma_E ({\bfbet}_E k\cosh\beta
   -{\hat{\bfbet}}_E w\sin\theta)
   -{\bf b}\times{\hat{\bfbet}}_E w\cos\theta+{\bf b}k\sinh\beta.
\end{equation}
From these equations, it follows that
\begin{eqnarray}
\beta_{v1} &\equiv& {\bfbet}_v\cdot {\bf b}
  =\frac{ k\sinh\beta}{\gamma_E(k\cosh\beta-\beta_E w\sin\theta)}
  \label{gil}\\
\beta_{v2} &\equiv& {\bfbet}_v\cdot {\hat{\bfbet}}_E
  =\frac{\beta_E k\cosh\beta-w\sin\theta}{k\cosh\beta
  -\beta_E w\sin\theta}
  \label{gim}\\
\beta_{v3} &\equiv& {\bfbet}_v\cdot ({\bf b}\times {\hat{\bfbet}}_E)
  =\frac{-w\cos\theta}{\gamma_E(k\cosh\beta-\beta_E w\sin\theta)}
  \label{gin}\\
\noalign{\hbox{and}}
k &=&c\gamma_E\gamma_v\sqrt{1-\beta_{v1}^2-2\beta_E\beta_{v2}
  +\beta_E^2\beta_{v1}^2+\beta_E^2\beta_{v2}^2}
  \label{gio}\\
\beta &=&\tanh^{-1}\left(\frac{\beta_{v1}}{\gamma_E(1-\beta_E\beta_{v2})}\right)
  \label{gip}\\
w &=&c\gamma_E\gamma_v\sqrt{\beta_{v2}^2+\beta_{v3}^2-2\beta_E\beta_{v2}
  +\beta_E^2-\beta_E^2\beta_{v3}^2}
  \label{giq}\\
\theta &=&{\rm arg}\left(-\beta_{v3}-i\gamma_E(\beta_{v2}-\beta_E)\right)
  \label{gir}
\end{eqnarray}
Note that the four coordinates $(k,\beta,w,\theta )$
obey the constraint $k^2-w^2=c^2,$ and this is why they can be determined by
the three components of ${\bfbet}_v.$
Naturally, the above transformation equations depend upon the choice we made
for the unit tetrad.  This arbitrariness will be discussed further in
Section~\ref{yag}.  These transformation equations will be most useful when
we want to compare our results to those of other authors who have used
``$1+3$'' notation; this will be done in Section~\ref{yan}.

\section{Phase Space Lagrangian for a Charged Particle
in an Electromagnetic Field}
For a relativistic charged particle in an electromagnetic field,
one possible choice for
the Hamiltonian, $H,$ in canonical coordinates, $(q,p),$ is given 
by~\cite{zag}
\begin{equation}
H(q,p)=\frac{1}{2m}\left(p-\frac{e}{c} A(q)\right)^2,
\label{gkb}
\end{equation}
and the action one form for canonical coordinates is, by Eq.~(\ref{mbe})
\begin{equation}
\gamma=p\cdot dq.
\end{equation}
Note that the independent variable is the particle's proper time; the
equations of motion are thus of the form of Eq.~(\ref{mbf}), but the dot
in that equation now denotes differentiation with respect to proper time.

We begin by making a noncanonical transformation to the new
coordinates $(r,u),$ where
\begin{equation}
   \left\{
      \begin{array}{l}
         r=q \\
         u=\frac{1}{m} \left( p-\frac{e}{c} A(q)\right).
      \end{array}
   \right.
\label{gjz}
\end{equation}
Thus we have eliminated the unphysical canonical momentum, $p,$ in
favor of the particle velocity, $u.$  The new Hamiltonian is
\begin{equation}
H'(r,u)=\frac{m}{2}u^2
\label{gka}
\end{equation}
and the new action one form is
\begin{equation}
\gamma'=\left( mu+\frac{e}{c} A(r)\right)\cdot dr.
\end{equation}

If we now use Eq.~(\ref{gap}) to eliminate the four 
components of $u$ in favor of $(k,\beta ,w,\theta ),$ then the new
Hamiltonian is
\begin{equation}
H''(r,k,\beta ,w,\theta)=\frac{m}{2}(-k^2+w^2)
\label{gie}
\end{equation}
and the new action one form is
\begin{equation}
\gamma''= \left(mk\hat{\bf t}+mw\hat{\bf c}+\frac{e}{c} A(r)\right) \cdot dr.
\label{gbn}
\end{equation}
It is important to remember that $\hat{\bf t}$ and $\hat{\bf b}$
are functions of $r$ and $\beta,$ and $\hat{\bf c}$ and
$\hat{\bf a}$ are functions of $r$ and $\theta.$  Thus, the second
term in the parenthesis on the right hand side of
Eq.~(\ref{gbn}) is rapidly oscillating due to its
dependence on $\theta$ (this will be made more precise shortly).
We are now ready to apply the Lie transform procedure that will
effectively average $H''$ and $\gamma''$ by transforming to gyrocoordinates
in which $\theta$ is ignorable.

\section{Gyrogauge and Boostgauge Transformations}
\label{yag}
We now discuss the afore-mentioned arbitrariness in choosing the
orthonormal unit vectors, ${\hat{\bf e}}_\alpha.$  
A {\it boostgauge} transformation replaces our choices for 
${{\hat{\bf e}}_0}$ and ${{\hat{\bf e}}_1}$ as follows:
\begin{eqnarray}
{{\hat{\bf e}}_0}'&=& {\hat{\bf e}}_0 \cosh \Phi (r) -
                     {\hat{\bf e}}_1 \sinh \Phi (r),
\label{gcg}\\
{{\hat{\bf e}}_1}'&=& {\hat{\bf e}}_1 \cosh \Phi (r) -
                     {\hat{\bf e}}_0 \sinh \Phi (r),
\label{gch}
\end{eqnarray}
while a {\it gyrogauge} transformation replaces our choices for
${{\hat{\bf e}}_2}$ and ${{\hat{\bf e}}_3}$ as follows:
\begin{eqnarray}
{{\hat{\bf e}}_2}'&=& {\hat{\bf e}}_2 \cos \Psi (r) +
                     {\hat{\bf e}}_3 \sin \Psi (r),
\label{gci}\\
{{\hat{\bf e}}_3}'&=& {\hat{\bf e}}_3 \cos \Psi (r) -
                     {\hat{\bf e}}_2 \sin \Psi (r).
\label{gcj}
\end{eqnarray}
Note that the new unit vectors are still orthonormal, that 
${\hat{\bf e}}_0$ and ${\hat{\bf e}}_1$ still span the parallel
two-flat, and that ${\hat{\bf e}}_2$ and ${\hat{\bf e}}_3$
still span the perpendicular two-flat.  The gyrogauge and boostgauge
transformations have simply given each of these two pairs of unit vectors a
rotation within its respective two-flat.  The amount of rotation is measured
by $\Phi$ in the parallel two-flat, and by $\Psi$ in the perpendicular
two-flat.  Note that these can be functions of the particle's spacetime
position, $r.$

Recall that we used the unit tetrad to decompose the particle
velocity into parallel and perpendicular parts, and to coordinatize
these by $(k,\beta )$ and $(w,\theta ),$ respectively.  It is fairly
easy to see that the transformation given by Eqs.~(\ref{gcg})
through (\ref{gcj}) will have no effect on
$k$ and $w,$ but will shift $\beta$ and $\theta.$  Hence, we add
\begin{equation}
\beta '= \beta + \Phi(r)
\label{gck}
\end{equation}
to our boostgauge transformation equations, and
\begin{equation}
\theta '= \theta +  \Psi(r)
\label{gcl}
\end{equation}
to our gyrogauge transformation equatons.
None of the other phase space coordinates are affected by the 
transformations.  

Equations~(\ref{gcg}) through (\ref{gcl})
constitute the full gyrogauge and boostgauge transformation equations.  A
quantity that is left unchanged by these transformation equations will be said
to be gyrogauge or boostgauge invariant, respectively.  The concept of
gyrogauge invariance has a
nonrelativistic analog which was first discussed by
Littlejohn~\cite{zak}.  In the remainder of this section, we
shall extend his methods to our relativistic problem.

To begin with, we note that
the unit vectors $({\hat{\bf t}}, {\hat{\bf b}}, {\hat{\bf c}},
{\hat{\bf a}})$ are all gyrogauge and boostgauge invariant.  This is
demonstrated for $ {\hat{\bf t}} $ as follows:
\begin{eqnarray}
{\hat{\bf t}}'& =&{{\hat{\bf e}}_0}' \cosh \beta '
                            +{{\hat{\bf e}}_1}' \sinh \beta '  \nonumber\\
                            & =&({\hat{\bf e}}_0 \cosh \Phi 
                            -{\hat{\bf e}}_1 \sinh \Phi )
                            \cosh (\beta +\Phi ) 
                            +({\hat{\bf e}}_1 \cosh \Phi 
                            -{\hat{\bf e}}_0  \sinh \Phi )
                            \sinh (\beta +\Phi ) \nonumber\\
                            & =& {\hat{\bf e}}_0 
                            [ \cosh \Phi \cosh (\beta +\Phi )
                            - \sinh \Phi \sinh (\beta +\Phi )] \nonumber\\
                            &&\qquad + {\hat{\bf e}}_1 
                            [ -\sinh \Phi \cosh (\beta +\Phi )
                            + \cosh \Phi \sinh (\beta +\Phi )] \nonumber\\
                            &=& {\hat{\bf e}}_0 \cosh \beta  
                            +{\hat{\bf e}}_1 \sinh \beta  \nonumber\\
                            &=& {\hat{\bf t}}; 
                            \label{gcm}
\end{eqnarray}
the demonstration for the other three unit vectors follows similarly.
Because the parallel and perpendicular projection operators may be
written in the form of Eqs.~(\ref{gaz}) and
(\ref{gba}), their gyrogauge and boostgauge invariance is manifest.

The fact that the quantities above are gyrogauge and boostgauge invariant means
that they may be expressed in terms of purely physical tensor quantities; more
precisely, they may be expressed in terms of quantities 
that are completely independent
of our choice of the orientation of the basis tetrad, ${\hat{\bf e}}_\alpha,$
at each point in spacetime.
For example, $P_\parallel$ and $P_\perp$ can be expressed in terms of
the field tensor, as was done in Eqs.~(\ref{gal}) and
(\ref{gam}).  The gyrogauge and boostgauge invariant quantities $k$ and $w$ can
be written in terms of the projection operators and the particle four-velocity
with the help of Eq.~(\ref{gap})
\begin{equation}
k=\sqrt{-u \cdot P_\parallel \cdot u},
\end{equation}
and
\begin{equation}
w=\sqrt{u \cdot P_\perp \cdot u }.
\end{equation}
Finally, the members of the tetrad $({\hat{\bf t}}, {\hat{\bf b}},
{\hat{\bf c}},{\hat{\bf a}})$ can all be expressed in terms of the field
tensor and the particle four-velocity, with the help of 
Eqs.~(\ref{gap}), (\ref{gax}), and (\ref{gay})
\begin{eqnarray}
{\hat{\bf t}}&=&\frac{1}{k} P_\parallel \cdot u,
\label{gcn}\\
{\hat{\bf b}}&=&-\frac{1}{\lambda_B} {\cal F} \cdot {\hat{\bf t}},
\label{gco}\\
{\hat{\bf c}}&=&\frac{1}{w} P_\perp \cdot u,
\label{gcp}\\
{\hat{\bf a}}&=&-\frac{1}{\lambda_B} F \cdot {\hat{\bf c}}.
\label{gcq}
\end{eqnarray}

Now consider the pair of one-forms:
\begin{equation}
{\cal Q} \equiv   (\delr {\hat{\bf e}}_1) \cdot {\hat{\bf e}}_0
           = -(\delr {\hat{\bf e}}_0) \cdot {\hat{\bf e}}_1
           =   (\delr {\hat{\bf b}}) \cdot {\hat{\bf t}}
           = -(\delr {\hat{\bf t}}) \cdot {\hat{\bf b}},
\label{gzq} 
\end{equation}
and
\begin{equation}
{\cal R} \equiv (\delr {\hat{\bf e}}_2) \cdot {\hat{\bf e}}_3
           = -(\delr {\hat{\bf e}}_3) \cdot {\hat{\bf e}}_2
           =   (\delr {\hat{\bf c}}) \cdot {\hat{\bf a}}
           = -(\delr {\hat{\bf a}}) \cdot {\hat{\bf c}},
\label{gzr}
\end{equation}
where $\delr$ is a shorthand for the spacetime gradient.
It is a straightforward exercise to show that ${\cal Q}$ is {\it not}
boostgauge invariant, and that ${\cal R}$ is {\it not} gyrogauge invariant;
this is essentially because the spacetime 
derivatives are taken at constant $\beta$ and $\theta,$ and these
latter two quantities are obviously not boostgauge and gyrogauge invariant,
respectively.  First note that $\delr$ transforms under a general boostgauge
and gyrogauge transformation as follows:
\begin{equation}
\delr ' = \delr - (\delr \Phi )\frac{\partial}{\partial \beta}
                  - (\delr \Psi )\frac{\partial}{\partial \theta},
\end{equation}
where we have made use of Eqs.~(\ref{gck}) and (\ref{gcl}).
Thus we have
\begin{equation}
{\cal Q}' = (\delr ' {\hat{\bf b}}') \cdot {\hat{\bf t}}'
    =[\delr {\hat{\bf b}}-(\delr \Phi) {\hat{\bf t}}] \cdot {\hat{\bf t}}
    ={\cal Q}+\delr \Phi,
\end{equation}
and
\begin{equation}
{\cal R}'  =(\delr ' {\hat{\bf c}}') \cdot {\hat{\bf a}}'
    =[\delr {\hat{\bf c}}+(\delr \Psi) {\hat{\bf a}}] \cdot {\hat{\bf a}}
    ={\cal R}+\delr \Psi.
\end{equation}
Here we have used Eqs.~(\ref{gau}) and (\ref{gav}).
The one-forms ${\cal Q}$ and ${\cal R}$ will be useful to us momentarily.
Furthermore, they have
great geometrical significance as will become clear later when we
discuss the guiding-center equations of motion.

We now ask what it means for a general one-form 
in our phase space to be boostgauge and gyrogauge invariant.
Using Eq.~(\ref{mbk}), we find that the $r$ component of the
one-form transforms as follows:
\begin{eqnarray}
\Gamma_r &=&\frac{\partial r}{\partial r'} \gamma_r+
                    \frac{\partial \beta}{\partial r'} \gamma_\beta+
                    \frac{\partial \theta}{\partial r'} \gamma_\theta \nonumber\\
                    &=&\gamma_r-
                    (\delr \Phi)\gamma_\beta-
                    (\delr \Psi)\gamma_\theta, 
                    \label{gcr}
\end{eqnarray}
while all of the other components ($k$, $\beta$, $w$, and $\theta$)
are unchanged.  Thus it is clear that the charged particle Hamiltonian and
action one form given by Eqs.~(\ref{gie}) and (\ref{gbn}) are boostgauge and
gyrogauge invariant, since they have no $\beta$ or $\theta$ components.

Now we demand that our Lie transformations, when applied to gauge
invariant quantities, preserve their gauge invariance.  This, coupled with
the established boostgauge and gyrogauge invariance of the particle 
action one-form, will guarantee the boostgauge and gyrogauge
invariance of the guiding-center action one-form.  Suppose that we
have a boostgauge and gyrogauge invariant scalar field, $f.$  Applying the Lie
derivative operator, ${\cal L}_g,$ we find from Eq.~(\ref{mbo})
\begin{equation}
{\cal L}_g f=g^r \cdot \delr f+ g^k \frac{\partial f}{\partial k}
                        + g^\beta \frac{\partial f}{\partial \beta}
                        + g^w \frac{\partial f}{\partial w}
                        + g^\theta \frac{\partial f}{\partial \theta}.
\end{equation}
If we now subject this to a general boostgauge and gyrogauge transformation, we
find
\begin{eqnarray}
({\cal L}_g f)' &=&g^{\prime r} \cdot \delr ' f
                    + g^{\prime k} \frac{\partial f}{\partial k}
                    + g^{\prime \beta} \frac{\partial f}{\partial \beta}
                    + g^{\prime w} \frac{\partial f}{\partial w}
                    + g^{\prime \theta} \frac{\partial f}{\partial \theta}\nonumber\\
                  & =& g^{\prime r} \cdot \delr f
                    + g^{\prime k} \frac{\partial f}{\partial k}
                    +(g^{\prime \beta}-\delr \Phi \cdot g^{\prime r})
                           \frac{\partial f}{\partial \beta}
                    + g^{\prime w} \frac{\partial f}{\partial w} \nonumber\\
            &&\qquad +(g^{\prime \theta}-\delr \Psi \cdot g^{\prime r})
                           \frac{\partial f}{\partial \theta},
                  \label{gcs}
\end{eqnarray}
where we have made use of the assumed gauge invariance of $f.$  Thus,
${\cal L}_g f$ will be gauge invariant if all the components of $g$ are gauge
invariant, with the exception of $g^\beta$ and $g^\theta$ which must
transform as follows:
\begin{equation}
g^{\prime \beta}=g^\beta + \delr \Phi \cdot g^{\prime r},
\end{equation}
and
\begin{equation}
g^{\prime \theta}=g^\theta + \delr \Psi \cdot g^{\prime r}.
\end{equation}
Thus, if we use a subscripted ``0'' to denote a gauge invariant quantity,
we see that the components of the vector $g$ must be of the form
\begin{eqnarray}
g^r            &=&(g^r)_0\nonumber\\ 
           g^k            &=&(g^k)_0\nonumber\\
           g^\beta        &=&(g^\beta)_0  + {\cal Q} \cdot (g^r)_0\nonumber\\ 
           g^w            &=&(g^w)_0\nonumber\\
           g^\theta       &=&(g^\theta)_0 + {\cal R} \cdot (g^r)_0
           \label{gct}
\end{eqnarray}
Using the homotopy formula,
it is a straightforward exercise to show that this result is valid not only
for gauge invariant scalars, but also for {\it any} gauge invariant $n$-form.
In particular, this restriction on the form of $g$ is necessary
to guarantee the gauge invariance of the Lie transformed action 
one-form, so we shall demand that it hold in the sections to follow.

\section{The Zero-Order Problem}
\label{yah}
We order the particle Hamiltonian and action one-form with the prescription
$e \mapsto e/ \epsilon;$  equivalently, we could say that we are ordering the
electromagnetic field at order ${\epsilon}^{-1}.$  The electromagnetic
contribution to the canonical momentum thus dominates the kinetic
contribution.  This ordering procedure has been discussed at length
by Kruskal~\cite{zal} and by Littlejohn~\cite{zak}.

We shall also order the four potential of the electromagnetic
field in the parameter $\epsilon,$ so
\begin{equation}
A=\sum_{i=0}^{\infty} \epsilon^i A_i.
\end{equation}
Clearly, this induces an ordering of the field itself
\begin{equation}
F=\sum_{i=0}^{\infty} \epsilon^i F_i,
\end{equation}
where
\begin{equation}
F_i=dA_i.
\end{equation}
Henceforth, when we refer to the Lorentz scalars ($\lambda_1,
\lambda_2, \lambda_E$ and $\lambda_B$) or to the unit basis tetrads
or to the projection operators,
it is to be understood that they are calculated on the basis of the
{\it zero order} field tensor, $F_0.$

The Hamiltonian, Eq.~(\ref{gie}), is thus an order unity scalar.
The particle action one-form, Eq.~(\ref{gbn}), may be written
\begin{equation}
\gamma=\frac{1}{\epsilon} \sum_{i=0}^{\infty} \epsilon^i \gamma_i,
\end{equation}
where $\gamma_0$ has the component
\begin{equation}
{\gamma_0}_r=\frac{e}{c}A_0(r),
\end{equation}
$\gamma_1$ has the component
\begin{equation}
{\gamma_1}_r=\frac{e}{c}A_1(r)+mk{\hat{\bf t}}+mw{\hat{\bf c}},
\label{gcu}
\end{equation}
and $\gamma_i$ has the component
\begin{equation}
{\gamma_i}_r=\frac{e}{c}A_i(r)
\end{equation}
for $i\geq 2.$  All components not listed above are zero.

Suppose that we now write the equations of motion to lowest order as
$\omega_0 \cdot \dot{z} =0,$ where $\omega_0 \equiv d\gamma_0.$
This turns out to be an instructive exercise even though, as we
shall see in a moment, it is somewhat misleading.
We see that the only surviving component of $\omega_0$ is
\begin{equation}
{\omega_0}_{rr}=\frac{e}{c} F_0,
\label{gcw}
\end{equation}
so we get the following equation of motion:
\begin{equation}
F_0 \cdot \dot{r} = 0.
\label{gcx}
\end{equation}
Now we know that $\dot{t}$ is never zero,
so $F_0$ must have at least one null eigenvector with nonzero time component.
In particular, this must be true in a preferred frame,
for which ${\bfbet}_E=0.$
Thus the parallel two-flat must be the nullspace of $F_0.$  So we demand that
\begin{equation}
\lambda_E=0,
\end{equation}
where we again emphasize that
$\lambda_E$ is computed from Eqs.~(\ref{gab}),
(\ref{gac}), (\ref{gad}) and (\ref{gag}) using 
$F_0$ in place of $F.$  This is a restriction on the allowed zero order
fields.  It is the relativistic analog of the usual nonrelativistic
restriction that $E_\parallel = 0$ to lowest order.  Recall that we used
this assumption in Section~\ref{yaf} when we first discussed the basis tetrads.

Thus, when we order the four potential in $\epsilon,$ we must keep in mind
that the field derived from $A_0$ should have no $E_\parallel.$  If we have
a problem in which there is nonzero $E_\parallel,$ then it must be
included in $A_n$ where $n\geq 1.$  In particular, it could all be put into
$A_1.$  The only reason for keeping $A_n$ where $n\geq 2$ in our theory is
that sometimes a problem admits another expansion parameter in the field
geometry (the stellarator expansion parameter and the long-thin parameter in
mirrors are examples), and in some asymptotic theories that other expansion
parameter may be taken to be equal to the guiding-center expansion parameter.
In such cases, one might want to expand the field in a general power series in
$\epsilon,$ rather than just restrict oneself to the use of $A_0$ and $A_1.$

Thus, Eq.~(\ref{gcx}) constitutes only two independent
conditions on the four components of $\dot{r}.$  Dotting it with
${\hat{\bf c}}$ and ${\hat{\bf a}}$ and using 
Eq.~(\ref{gax}) gives ${\hat{\bf c}} \cdot \dot{r}=
{\hat{\bf a}} \cdot \dot{r}=0,$ so $\dot{r}$ must lie in the
parallel two-flat; that is, the particle motion is constrained to
lie along the field lines like that of a bead sliding along a wire.
The rapid oscillatory motion is then considered to be a modification
to this motion along the field lines, to be transformed away except
for the residual perpendicular drifting motion.

What is perhaps most disturbing about Eq.~(\ref{gcx}) is
that it gives only two dynamical equations of motion when there are
really eight independent phase space coordinates.  It gives us no
description of the motion along the field lines, and no description
of the rate of change of the velocity components.  This is because
the matrix of components of the zero order Lagrangian two-form
is a eight by eight matrix whose rank is only two.  This is thus an
example of a problem in asymptotics with no well-defined limit 
problem; this phenomenon is by no means rare and has been discussed in
a general context by Kruskal~\cite{zam}.

To get a better idea of what is going on here, we should consider the
full particle equations of motion, retaining the lowest order nonzero
contributions to each component of $\omega = d\gamma,$ even if some
are higher order than others.  We find
\begin{eqnarray}
\omega_{rr}&=&\frac{e}{\epsilon c}F_0 + {\cal O}(1),
   \label{gcy}\\
\omega_{rk}&=&-m{\hat{\bf t}},
   \label{gcz}\\
\omega_{r\beta}&=&-mk{\hat{\bf b}},
   \label{gda}\\
\omega_{rw}&=&-m{\hat{\bf c}},
   \label{gdb}\\
\omega_{r\theta}&=&+mw{\hat{\bf a}},
   \label{gdc}
\end{eqnarray}
with all other components vanishing.  Forming the equations of
motion, $\omega \cdot \dot{z}=\partial H/\partial z,$ we find that
\begin{equation}
\dot{r}=k{\hat{\bf t}}+w{\hat{\bf c}},
\end{equation}
so there is no longer any ambiguity in the parallel motion.
Similarly we can now find the equations of motion for the
velocity components.  We get
\begin{eqnarray}
\dot{k}&=&{\cal O}(1),
   \label{gdf}\\
\dot{\beta}&=&{\cal O}(1),
   \label{gdg}\\
\dot{w}&=&{\cal O}(1),
   \label{gdh}\\
\noalign{\hbox{and}}
\dot{\theta}&=&\frac{1}{\epsilon} \Omega_B + {\cal O}(1).
   \label{gdi}
\end{eqnarray}
This makes it clear that the dominant motion at lowest order is
the gyration, in accordance with our intuition.
Thus, as $\epsilon \rightarrow 0,$ we have the rate of
change of $\theta$ dominating that of all the other
dynamical variables, including $r.$  Hence, averages over the 
unperturbed motion will simply be averages over $\theta.$

Note that in order to get this zero order equation of motion,
we needed $\gamma_r$ only to order $\epsilon^{-1},$ while all the other
components of $\gamma$ were needed to order unity.  This peculiar 
mixing of orders persists to higher order; so to obtain the $n$-th
order guiding-center equations of motion, we will need $\gamma_r$ only
to order $n-1,$ while all the other components of $\gamma$ will be
needed to order $n.$


\section{The Preparatory Lie Transform}
\label{yai}
All treatments of guiding-center motion share one feature in common:
In the transformation from particle position, $r,$ to guiding-center 
position, $R,$ they all include the term, $-w{\hat{\bf a}}/\Omega_B.$
This is the gyroradius vector, and
it is the most intuitive term in the entire guiding-center 
transformation (indeed, one might argue that it is the {\it only}
intuitive term in the entire guiding-center transformation).
We shall make this transformation before we do anything else, as
this was found to facilitate the remainder of the calculation
in Littlejohn's nonrelativistic treatment~\cite{zaq}.

From Eqs.~(\ref{mbo}) and (\ref{mbq}), we see
that, to first order, the difference between $z$ and $Z$ is simply
given by the components of the generator vector, $g.$  So since we want
to have $R=r-w{\hat{\bf a}}/\Omega_B,$ we see that we should choose
\begin{equation}
g_p^r = -\frac{w}{\Omega_B}{\hat{\bf a}},
\end{equation}
where the subscript ``p'' denotes ``preparatory.''

Now $g_p^r$ is clearly boostgauge and gyrogauge invariant, but
from Eq.~(\ref{gct}) we see that a Lie transform
generated by this vector alone would not preserve the
gauge invariance of the action one-form.  Consequently, we must append
the following additional components to $g_p$:
\begin{equation}
g_p^\beta = -\frac{w}{\Omega_B}{\hat{\bf a}} \cdot {\cal Q},
\end{equation}
and
\begin{equation}
g_p^\theta = -\frac{w}{\Omega_B}{\hat{\bf a}} \cdot {\cal R}.
\end{equation}

First note that the Hamiltonian, Eq.~(\ref{gie}), is unaffected by
the preparatory Lie transform because it is independent of $r$, $\beta$
and $\theta$ (so ${\cal L}_pH''=0$).  Next,
using Eqs.~(\ref{mca}) through (\ref{mcd}),
we calculate the new action one-form resulting from the 
transformation generated by this vector.  This transformation
takes place at first order only, so we may set $g_1=g_p$ and
$g_2=g_3=0$ in those equations.  Also, since we are interested in 
calculating the guiding-center equations 
of motion to third order (this turns out to
be one order higher than the usual perpendicular
drifts), we do not need ${\Gamma_3}_r.$

At zero order, we have the obvious
\begin{equation}
\Gamma_0=\gamma_0.
\end{equation}
This has the single nonzero component,
\begin{equation}
{\Gamma_0}_r=\frac{e}{c}A_0.
\end{equation}
The corresponding Lagrangian two-form, $\omega_0,$ was given in 
Eq.~(\ref{gcw}).

Moving on to first order, it is readily found that $i_p \omega_0$ 
(where, in keeping with past convention, $i_p\equiv i_{g_p}$) has only one
nonzero component,
\begin{equation}
(i_p \omega_0)_r=mw{\hat{\bf c}}.
\end{equation}
We take $S_1=0,$ so Eq.~(\ref{mcb}) gives the following
nonzero component for $\Gamma_1$:
\begin{equation}
{\Gamma_1}_r = \frac{e}{c}A_1+mk {\hat{\bf t}}.
\label{gdj}
\end{equation}
Note that the aforementioned rapidly oscillating term, $ mw{\hat{\bf c}},$
has been removed from ${\gamma_1}_r$ by the transformation.

Before proceeding to second order, we need to calculate
$\omega_1 \equiv d\gamma_1$ and $\Omega_1 \equiv d\Gamma_1.$ 
The first of these has the following nonzero components:
\begin{eqnarray}
{\omega_1}_{rr}&=&\frac{e}{c}F_1
   +mk(\delr \hat{\bf t}-\hat{\bf t} \dell) \nonumber\\
   &\qquad +mw (\delr \hat{\bf c}-\hat{\bf c} \dell),
   \label{gdl}\\
{\omega_1}_{rk}&=&-m \hat{\bf t},
   \label{gdm}\\
{\omega_1}_{r\beta}&=&-mk \hat{\bf b},
   \label{gdn}\\
{\omega_1}_{rw}&=&-m \hat{\bf c},
   \label{gdo}\\
{\omega_1}_{r\theta}&=&+mw \hat{\bf a}.
   \label{gdp}
\end{eqnarray}
The second has the following nonzero components:
\begin{eqnarray}
{\Omega_1}_{rr}&=&\frac{e}{c}F_1
   +mk(\delr \hat{\bf t}-\hat{\bf t} \dell),
   \label{gds}\\
{\Omega_1}_{rk}&=&-m \hat{\bf t},
   \label{gdt}\\
{\Omega_1}_{r\beta}&=&-mk \hat{\bf b}.
   \label{gdu}
\end{eqnarray}
Note that we have introduced the notation $\hat{\bf t} \dell$ for the
transpose of $\delr \hat{\bf t}.$

We are now ready to proceed to second order.  First note that 
$\frac{1}{2} i_p \omega_1$ has the following nonzero components:
\begin{eqnarray}
(\frac{1}{2} i_p \omega_1)_r &=& -\frac{1}{2} \frac{w}{\Omega_B}
      \hat{\bf a} \cdot \Bigl[ \frac{e}{c}F_1 + mk(\delr \hat{\bf t} \cdot
      P_\perp - \hat{\bf t} \dell) \nonumber\\ 
      &&\qquad + mw(\delr \hat{\bf c} \cdot
      P_\parallel - \hat{\bf c} \dell) \Bigr] ,
      \label{gdx}\\
\noalign{\hbox{and}}
(\frac{1}{2} i_p \omega_1)_\theta &=& -\frac{mw^2}{2\Omega_B}.
      \label{gdy}
\end{eqnarray}
Next note that $\frac{1}{2} i_p \Omega_1$ has the single nonzero 
component,
\begin{equation}
(\frac{1}{2} i_p \Omega_1)_r = -\frac{1}{2} \frac{w}{\Omega_B}
      \hat{\bf a} \cdot \left[ \frac{e}{c}F_1 + mk(\delr \hat{\bf t} \cdot
      P_\perp - \hat{\bf t} \dell) \right] .
\end{equation}
Now, using Eq.~(\ref{mcc}) and choosing $S_2=0,$
 we can write down the nonzero components of $\Gamma_2,$
\begin{eqnarray}
{\Gamma_2}_r&=&\frac{e}{c}A_2+\frac{w}{\Omega_B} \hat{\bf a} \cdot
      \Bigl[ \frac{e}{c}F_1 
      +mk (\delr \hat{ \bf t} \cdot
      P_\perp - \hat{\bf t} \dell) \nonumber\\ 
      &&\qquad + \frac{mw}{2} (\delr \hat{ \bf c} \cdot
      P_\parallel - \hat{\bf c} \dell) \Bigr],
      \label{gdz}\\
\noalign{\hbox{and}}
{\Gamma_2}_\theta &=& \frac{mw^2}{2 \Omega_B}.
      \label{gea}
\end{eqnarray}
Note that ${\Gamma_2}_r$ has rapidly oscillating terms; these will be
removed by subsequent Lie transforms.  Also note the appearance of the
gyromomentum as the $\theta$ component of $\Gamma_2.$

Moving on to third order, we recall that we do not need ${\Gamma_3}_r.$
Referring to Eq.~(\ref{mcd}),
it is easily seen that $\gamma_3$ and $i_p \omega_2$ both have 
only an $r$-component, so we do not bother with these terms.
Then $\frac{1}{3}i_p di_p \omega_1$ has a nonzero $r$-component
which we shall not calculate,
and it also has a nonzero $\theta$ component given by
\begin{equation}
(\frac{1}{3}i_p di_p \omega_1)_\theta = -\frac{mw^3}{3\Omega_B^3}
      \hat{\bf a} \cdot \delr \Omega_B -\frac{w^2}{3\Omega_B^2}
      \hat{\bf c} \cdot \left[ \frac{e}{c} F_1+mk(\delr \hat{\bf t} -
      \hat{\bf t} \dell) \right] \cdot \hat{\bf a}.
\end{equation}
Similarly, $\frac{1}{6}i_p di_p \Omega_1$ has a nonzero $r$-component
which we shall not calculate,
and it also has a nonzero $\theta$ component given by
\begin{equation}
(\frac{1}{6}i_p di_p \Omega_1)_\theta = -\frac{w^2}{6\Omega_B^2}
      \hat{\bf c} \cdot \left[ \frac{e}{c} F_1+mk(\delr \hat{\bf t} -
      \hat{\bf t} \dell) \right] \cdot \hat{\bf a}.
\end{equation}
Taking $S_3=0,$ we see that the nonzero components of $\Gamma_3$ are
${\Gamma_3}_r$ and
\begin{equation}
{\Gamma_3}_\theta=-\frac{mw^3}{3\Omega_B^3}
      \hat{\bf a} \cdot \delr \Omega_B -\frac{w^2}{2\Omega_B^2}
      \hat{\bf c} \cdot \left[ \frac{e}{c} F_1+mk(\delr \hat{\bf t} -
      \hat{\bf t} \dell) \right] \cdot \hat{\bf a}.
\end{equation}
Note that this has rapidly oscillating terms which will have to be removed
by subsequent Lie transforms.  This completes the preparatory 
transformation.

\section{The Averaging Lie Transforms}
\label{yaj}
We now perform the averaging Lie transformations that will take us to
the guiding-center action one-form.  These are somewhat more difficult
than the preparatory transformation, since we do not know the generators
in advance.  For economy of notation, we reset our variables as follows:
We shall henceforth refer to the Hamiltonian and action one-form that 
resulted from the preparatory transformation as $H''$ and $\gamma,$ 
respectively, and these new Lie transforms will take us to $H'''$ and $\Gamma.$

First consider the action one form.  Once again, nothing changes at order 
zero, so
\begin{equation}
\Gamma_0=\gamma_0,
\end{equation}
and the only nonzero component of this is
\begin{equation}
{\Gamma_0}_r=\frac{e}{c} A_0.
\end{equation}
The corresponding Lagrangian two-form, $\omega_0,$ was given in 
Eq.~(\ref{gcw}); its only nonzero component was
${\omega_0}_{rr}.$

At order one, we take $g_1^r=0$ and $S_1=0$ because we have already
succeeded in averaging ${\Gamma_1}_r$ by the preparatory 
transformation, and we don't want to ruin this.  It follows that
$i_1 \omega_0 = 0,$ and so $\Gamma_1=\gamma_1.$  The only nonvanishing
component of $\Gamma_1$ is then
\begin{equation}
{\Gamma_1}_r = \frac{e}{c}A_1+mk {\hat{\bf t}}.
\label{geb}
\end{equation}
Note that we have not yet had to specify 
$g_1^k, g_1^\beta, g_1^w,$ or $g_1^\theta,$ since it is clear 
that these have no effect on $\Gamma_1.$  These components of $g_1$ will be 
useful in the averaging of $\Gamma_2.$  Also note that 
$\Omega_1 = \omega_1$ is given by Eqs.~(\ref{gds}) 
through (\ref{gdu}).

A word of caution is in order concerning the coordinate $\tau.$
It is not altered in any way by the
transformation.  This means that after we complete the transformation 
to guiding-center coordinates, $\tau$ will still be the 
{\it single-particle} proper time; it will {\it not} be the guiding-center
proper time.  So $g_{\mu\nu}dr^\mu dr^\nu=-d\tau^2,$ but
$g_{\mu\nu}dR^\mu dR^\nu \neq -d\tau^2.$  Thus, throughout the remainder
of this calculation, it is best to regard $\tau$ as simply an orbit parameter,
devoid of relevant physical significance.

Now we proceed to second order.  Note that $i_2 \omega_0$ has only
an $r$-component,
\begin{equation}
(i_2 \omega_0)_r = \frac{e}{c} g_2^r \cdot F_0.
\end{equation}
Next note that $\frac{1}{2} i_1 \Omega_1 = \frac{1}{2} i_1 \omega_1$
has the following nonzero component:
\begin{equation}
(\frac{1}{2} i_1 \omega_1)_r = \frac{1}{2} (mkg_1^\beta \hat{\bf b}+
      mg_1^k \hat{\bf t}),
\label{ged}
\end{equation}
We then take $S_2=0$ because we have already
succeeded in averaging ${\Gamma_2}_\theta$ by the preparatory 
transformation, and we don't want to ruin this.
Equation~(\ref{mcc}) then gives the following nonzero components
for $\Gamma_2$:
\begin{eqnarray}
{\Gamma_2}_r &=& \frac{e}{c}A_2 +
      \frac{w}{\Omega_B} \hat{\bf a} \cdot
      \Bigl[ \frac{e}{c}F_1 
      +mk (\delr \hat{ \bf t} \cdot
      P_\perp - \hat{\bf t} \dell) \nonumber\\ 
      &&\qquad + \frac{mw}{2} (\delr \hat{ \bf c} \cdot
      P_\parallel - \hat{\bf c} \dell) \Bigr] 
      -\frac{e}{c} g_2^r \cdot F_0 - mkg_1^\beta \hat{\bf b}
      -mg_1^k \hat{\bf t}, 
      \label{gef}\\
      \noalign{\hbox{and}}
      {\Gamma_2}_\theta &=& \frac{mw^2}{2 \Omega_B}.
      \label{geg}
\end{eqnarray}

We now proceed to third order, and once again we do not need the
$r$-component of $\Gamma_3.$ 
Referring to Eq.~(\ref{mcd}),
it is easily seen that $i_3 \omega_0$ 
has only an $r$-component, so we do not bother with this term.
Then $i_2 \Omega_1 = i_2 \omega_1$ has
a nonzero $r$-component which we shall not calculate; its other
nonzero components are
\begin{eqnarray}
(i_2 \omega_1)_k &=& -mg_2^r \cdot \hat{\bf t},
   \label{gei}\\
\noalign{\hbox{and}}
(i_2 \omega_1)_\beta &=& -mkg_2^r \cdot \hat{\bf b}.
   \label{gej}
\end{eqnarray}
Next, $i_1 \omega_2$ has a nonzero $r$-component which we shall not 
calculate; its other nonzero components are
\begin{eqnarray}
(i_1 \omega_2)_w &=& -\frac{mw}{\Omega_B} g_1^\theta,
   \label{gel}\\
\noalign{\hbox{and}}
(i_1 \omega_2)_\theta &=& +\frac{mw}{\Omega_B} g_1^w.
   \label{gem}
\end{eqnarray}
Next, $\frac{1}{3} i_1 di_1 (\omega_1+\frac{1}{2}\Omega_1) = 
\frac{1}{2} i_1 di_1 \omega_1$ has 
a nonzero $r$-component which we shall not calculate; it has no other
nonzero components.  From Eq.~(\ref{mcd}) we see that the nonzero components of
$\Gamma_3$ are ${\Gamma_3}_r$ and the following:
\begin{eqnarray}
{\Gamma_3}_k &=& mg_2^r \cdot \hat{\bf t}
      +\frac{\partial S_3}{\partial k}, 
      \label{geo}\\
{\Gamma_3}_\beta &=& mkg_2^r \cdot \hat{\bf b}
      +\frac{\partial S_3}{\partial \beta}, 
      \label{gep}\\ 
{\Gamma_3}_w &=& \frac{mw}{\Omega_B} g_1^\theta
      +\frac{\partial S_3}{\partial w}, 
      \label{geq}\\ 
\noalign{\hbox{and}}
{\Gamma_3}_\theta &=& 
      -\frac{w^2}{2\Omega_B^2} \hat{\bf c} \cdot \left[ \frac{e}{c}F_1
      +mk(\delr \hat{\bf t}
      -\hat{\bf t} \dell)\right] \cdot \hat{\bf a} \nonumber\\
      &&\qquad -\frac{mw^3}{3\Omega_B^3} \hat{\bf a} \cdot \delr 
      \Omega_B
      -\frac{mw}{\Omega_B} g_1^w
      +\frac{\partial S_3}{\partial \theta}.
      \label{ger}
\end{eqnarray}

Now we apply the Lie transform to the Hamiltonian.  This is straightforward,
and we get
\begin{equation}
H'''=H'''_1+\epsilon H'''_2+{\cal O}(\epsilon^2),
\end{equation}
where 
\begin{equation}
H'''_1=H''=m(-k^2+w^2)/2,
\end{equation}
and
\begin{equation}
H'''_2=mkg_1^k-mwg_1^w.
\label{geh}
\end{equation}
Thus, the Hamiltonian, which emerged unscathed from the preparatory
Lie transform, may indeed be modified by the averaging Lie transform.

We must now choose the vector generator components,
$g_1^k, g_1^\beta, g_1^w, g_1^\theta,$ and $g_2^r,$ and the scalar
gauge transformation generator, $S_3,$ in order to average and maximally
simplify ${\Gamma_2}_r, H'''_2, {\Gamma_3}_k, {\Gamma_3}_\beta,
{\Gamma_3}_w,$ and ${\Gamma_3}_\theta.$  These are given by 
Eqs.~(\ref{gef}), (\ref{geh}),
(\ref{geo}), (\ref{gep}), (\ref{geq}), and
(\ref{ger}), respectively.  We proceed by taking the averaged
parts of these equations,
\begin{eqnarray}
   {\Gamma_2}_r &=&\frac{e}{c} A_2 - \frac{mw^2}{2\Omega_B}
   \left[ {\cal R}-\frac{1}{2}\left( \hat{\bf a} \cdot \delr \hat{\bf c}
   -\hat{\bf c} \cdot \delr \hat{\bf a}\right) \cdot P_\parallel \right]\nonumber\\
   &&\qquad -\frac{e}{c}\bar{g_2^r}\cdot F_0 - m\bar{g_1^k}\hat{\bf t}
   -mk\bar{g_1^\beta}\hat{\bf b},
   \label{get}\\
   H'''_2 &=&mk\bar{g_1^k}-mw\bar{g_1^w},
   \label{geu}\\
   {\Gamma_3}_k &=&m\bar{g_2^r}\cdot \hat{\bf t}
   +\frac{\partial \bar{S_3}}{\partial k},
   \label{gev}\\
   {\Gamma_3}_\beta &=&mk\bar{g_2^r}\cdot \hat{\bf b}
   +\frac{\partial \bar{S_3}}{\partial \beta},
   \label{gew}\\
   {\Gamma_3}_w &=&\frac{mw}{\Omega_B}\bar{g_1^\theta}
   +\frac{\partial \bar{S_3}}{\partial w},
   \label{gex}\\
   {\Gamma_3}_\theta &=&-\frac{mw^2}{4\Omega_B^3}\left( \frac{e}{mc}\right)
   F_0:\left[ \left( \frac{e}{mc}\right) F_1
   +mk\left( \delr \hat{\bf t} - \hat{\bf t} \dell \right)
   \right] \nonumber\\
   &&\qquad -\frac{mw}{\Omega_B}\bar{g_1^w}
   +\frac{\partial \bar{S_3}}{\partial \theta},
   \label{gey}\\
   \noalign{\hbox{and the fluctuating parts,}}
   0 &=&\frac{w}{\Omega_B}\hat{\bf a}\cdot\left[ \frac{e}{c}
   +mk\left( \delr \hat{\bf t}\cdot P_\perp -\hat{\bf t} \dell
   \right) \right] \nonumber\\ &&\qquad +\frac{mw^2}{4\Omega_B}
   \left( \hat{\bf a} \cdot \delr \hat{\bf c}
   +\hat{\bf c} \cdot \delr \hat{\bf a}\right) \cdot P_\parallel \nonumber\\
   &&\qquad -\frac{e}{c}\tilde{g_2^r}\cdot F_0 - m\tilde{g_1^k}\hat{\bf t}
   -mk\tilde{g_1^\beta}\hat{\bf b},
   \label{gez}\\
   0 &=&mk\tilde{g_1^k}-mw\tilde{g_1^w},
   \label{gfa}\\
   0 &=&m\tilde{g_2^r}\cdot \hat{\bf t}
   +\frac{\partial \tilde{S_3}}{\partial k},
   \label{gfb}\\
   0 &=&mk\tilde{g_2^r}\cdot \hat{\bf b}
   +\frac{\partial \tilde{S_3}}{\partial \beta},
   \label{gfc}\\
   0 &=&\frac{mw}{\Omega_B}\tilde{g_1^\theta}
   +\frac{\partial \tilde{S_3}}{\partial w},
   \label{gfd}\\
   0 &=&-\frac{mw^3}{3\Omega_B^3}-\frac{mw}{\Omega_B}\tilde{g_1^w}
   +\frac{\partial \tilde{S_3}}{\partial \theta},
   \label{gfe}
\end{eqnarray}
where we have demanded that the Hamiltonian and one-form components 
themselves be purely averaged.  In the above equations, an overbar denotes the 
averaged part of a quantity, while an overtilde denotes the fluctuating part.

Solve Eq.~(\ref{gfe}) for $\partial \tilde{S_3}/\partial \theta$
in terms of $\tilde{g_1^w}.$  Then use Eq.~(\ref{gfa}) to
get $\tilde{g_1^w}$ in terms of $\tilde{g_1^k}.$  Then dot 
Eq.~(\ref{gez}) with $\hat{\bf t}$ in order to get $\tilde{g_1^k}.$
The result is
\begin{eqnarray}
   \frac{\partial \tilde{S_3}}{\partial \theta}&=&\frac{mw^3}{3\Omega_B^3}
   \hat{\bf a}\cdot \delr \Omega_B - \frac{k}{\Omega_B}\Bigl\{
   \frac{w}{\Omega_B}\left[ \frac{e}{c}\hat{\bf a}\cdot F_1 \cdot \hat{\bf t}
   - mk(tta)\right] \nonumber\\
   &&\qquad +\frac{mw^2}{4\Omega_B}\left[ (act)+(cat)\right] \Bigr\},
   \label{gff}
\end{eqnarray}
where the abbreviation $(act)$ is shorthand for 
$\hat{\bf a}\cdot \delr \hat{\bf c} \cdot \hat{\bf t},$ etc.  Now this
equation is easily integrated to give
\begin{eqnarray}
   \tilde{S_3}&=&-\frac{mw^3}{3\Omega_B^3}
   \hat{\bf c}\cdot \delr \Omega_B + \frac{wk}{\Omega_B^2}
   \left[ \frac{e}{c}\hat{\bf c}\cdot F_1 \cdot \hat{\bf t}
   - mk(ttc)\right] \nonumber\\
   &&\qquad +\frac{mw^2k}{8\Omega_B^2}\left[ (ata)+(ctc)\right].
   \label{gfg}
\end{eqnarray}
We can now back substitute to get the oscillatory parts of the
vector generator components,
\begin{eqnarray}
   \tilde{g_1^k}&=&\frac{w}{\lambda_B}\hat{\bf t}\cdot F_1 \cdot \hat{\bf a}
   +\frac{kw}{\Omega_B}(tta)-\frac{w^2}{4\Omega_B}\left[ (act)+(cat)\right],
   \label{gfh}\\
   \tilde{g_1^\beta}&=&\frac{w}{k\lambda_B}\hat{\bf a}\cdot F_1 \cdot
   \hat{\bf b}
   -\frac{w}{\Omega_B}(bta)+\frac{w^2}{4k\Omega_B}\left[ (acb)+(cab)\right],
   \label{gfi}\\
   \tilde{g_1^w}&=&\frac{k}{\lambda_B}\hat{\bf t}\cdot F_1 \cdot \hat{\bf a}
   +\frac{k^2}{\Omega_B}(tta)-\frac{kw}{4\Omega_B}\left[ (act)+(cat)\right],
   \label{gfj}\\
   \tilde{g_1^\theta}&=&\frac{w}{\Omega_B^2}\hat{\bf c}\cdot \delr \Omega_B
   -\frac{k}{w\lambda_B}\hat{\bf c}\cdot F_1 \cdot \hat{\bf t} \nonumber\\
   &&\qquad +\frac{k^2}{w\Omega_B}(ttc)
   -\frac{k}{4\Omega_B}\left[ (ata)-(ctc)\right],
   \label{gfk}\\
   \noalign{\hbox{and}}
   \tilde{g_2^r}&=&\frac{w}{\lambda_B \Omega_B}\left( P_\parallel - P_\perp 
   \right) \cdot F_1 \cdot \hat{\bf c} + \frac{w^2}{8\Omega_B^2}
   \left( \hat{\bf a}\cdot \delr \hat{\bf a} -
   \hat{\bf c}\cdot \delr \hat{\bf c}\right) \cdot P_\parallel \nonumber\\
   &&\qquad +\frac{kw}{\Omega_B^2}\Bigl[ \left( \hat{\bf c} \cdot
   \delr \hat{\bf t}\cdot P_\perp - P_\perp \cdot \delr \hat{\bf t}
   \cdot \hat{\bf c}\right)\nonumber\\
   &&\qquad -\left( \hat{\bf t} \cdot
   \delr \hat{\bf c}\cdot P_\parallel - P_\parallel \cdot \delr \hat{\bf c}
   \cdot \hat{\bf t}\right) \Bigr].
   \label{gfl}
\end{eqnarray}

Next we consider the equations for the averaged parts of the generators, 
Eqs.~(\ref{get}) through (\ref{gey}).  These constitute
nine equations (Eq.~(\ref{get}) is really four equations) in
seventeen unknowns (the nine components of $\Gamma,$ and the eight
components of $\bar{g}$).  Thus, we can choose eight unknowns at will.
So we demand
\begin{eqnarray}
   {\Gamma_2}_r &=&\frac{e}{c}A_2-\frac{mw^2}{2\Omega_B}{\cal R},
   \label{gfm}\\
   {\Gamma_3}_k &=&0,
   \label{gfn}\\
   {\Gamma_3}_\beta &=&0,
   \label{gfo}\\
   {\Gamma_3}_w &=&0,
   \label{gfp}\\
   \noalign{\hbox{and}}
   {\Gamma_3}_\theta &=&0.
   \label{gfq}
\end{eqnarray}
Here we have retained the term involving ${\cal R}$ in ${\Gamma_2}_r$ in order
to preserve boostgauge and gyrogauge invariance, according to Eq.~(\ref{gcr}).
Taking $\bar{S_3}=0,$ we can now solve for $\bar{g}.$ We get
\begin{eqnarray}
   \bar{g_2^r} &=&0,
   \label{gfr}\\
   \bar{g_1^k} &=&\frac{w^2}{4\Omega_B}\left[ (cat)-(act)\right],
   \label{gfs}\\
   \bar{g_1^\beta} &=&-\frac{w^2}{4k\Omega_B}\left[ (cab)-(acb)\right],
   \label{gft}\\
   \bar{g_1^w} &=&\frac{w}{2\lambda_B}\hat{\bf a}\cdot F_1 \cdot \hat{\bf c}
   +\frac{kw}{2\Omega_B}\left[ (atc)-(cta)\right],
   \label{gfu}\\
   \bar{g_1^\theta} &=&0.
   \label{gfv}
\end{eqnarray}
We can now solve for $H'''_2$ using Eq.~(\ref{geu}) to get
\begin{eqnarray}
H'''_2 &=&-\frac{mw^2}{2\lambda_B}\hat{\bf a}\cdot F_1 \cdot \hat{\bf c}
   -\frac{mkw^2}{4\Omega_B}\left[ (atc)-(cta)\right] \nonumber\\
   &=&\frac{mw^2}{4\Omega_B^2}\left( \frac{e}{mc}\right) F_0 : \left[
   \left( \frac{e}{mc}\right) F_1 + \frac{k}{2}\left( \delr \hat{\bf t} -
   \hat{\bf t} \dell \right) \right].
   \label{gfw}
\end{eqnarray}
This completes the averaging transformation.

Henceforth, we shall write transformed quantities as functions of the
guiding-center variables ($R$, $K$, ${\cal B}$, $W$, $\theta$) instead of
their lower-case counterparts.  Note that this has no mathematical 
significance, and is done only to emphasize the {\it physical} interpretation
of the various quantities that emerge from the theory.  We
regard functions in the mathematicians' sense of the word:
functional arguments are nothing more than dummy placeholders.

We may now write out the full guiding-center Hamiltonian and action one form 
to the above-described order.  We have
\begin{eqnarray}
   H'''_2 &=& \frac{m}{2}(-K^2+W^2)+\epsilon 
   \frac{mW^2}{4\Omega_B^2}\left( \frac{e}{mc}\right) F_0 : \Bigl[
   \left( \frac{e}{mc}\right) F_1 \nonumber\\ 
   &&\qquad + \frac{K}{2}\left( \delr \hat{\bf t} -
   \hat{\bf t} \dell \right)
   \Bigr] +O(\epsilon^2). 
   \label{gif}\\
   \noalign{\hbox{and}}
   \Gamma &=&\left[ \frac{e}{\epsilon c} \left( A_0+\epsilon A_1+\epsilon^2 
   A_2 \right) +mK\hat{\bf t} -\frac{\epsilon mW^2}{2\Omega_B}{\cal R}
   +O(\epsilon^2 ) \right]\cdot dR\nonumber\\
   &&\qquad +\epsilon \frac{mW^2}{2\Omega_B}
   d\Theta +{\cal O}(\epsilon^3).
   \label{gfx}
\end{eqnarray}
Note that $\theta$ is an ignorable coordinate, so that its canonically
conjugate momentum, $\mu \equiv mW^2/2\Omega_B,$ is conserved.  This can
now be identified as the gyromomentum, and it is useful to eliminate
the coordinate $W$ in favor of $\mu.$  The results will be denoted
\begin{eqnarray}
H_{gc}&=&-\frac{m}{2} K^2+\mu \Omega_B \nonumber\\ &&\qquad +
   \frac{\epsilon \mu}{2\lambda_B} F_0 : \left[
   \left( \frac{e}{mc}\right) F_1 + \frac{K}{2}\left( \delr \hat{\bf t} -
   \hat{\bf t} \dell \right)
   \right] +O(\epsilon^2)
   \label{gfz}\\
\noalign{\hbox{and}}
\Gamma_{gc}&=&\left[ \frac{e}{\epsilon c} A
   +mK\hat{\bf t} -\epsilon \mu {\cal R} +O(\epsilon^2 )
   \right]\cdot dR +\epsilon \mu d\Theta .
   \label{gfy}
\end{eqnarray}
This is the form of the guiding-center Hamiltonian and action one form that 
will be used in subsequent sections.  Note that the order $\epsilon$ term
in the Hamiltonian may be neglected if only the classical drifts (usual
gradient, polarization and curvature drifts) are desired.


\section{The Guiding-Center Poisson Brackets}
\label{yak}
As a first step towards writing down the guiding-center equations of
motion, we form the guiding-center Lagrangian two-form.  The nonzero
components are
\begin{eqnarray}
\Omega_{RR}&=&\frac{e}{\epsilon c}(F_0+\epsilon F')+mK(\delr \hat{\bf t}-
   \hat{\bf t} \dell),
   \label{gga}\\
\Omega_{RK}&=&-m\hat{\bf t},
   \label{ggb}\\
\Omega_{R{\cal B}}&=&-mK\hat{\bf b},
   \label{ggc}\\
\Omega_{R\mu}&=&\left\{
                   \begin{array}{ll}
                      0& {\rm (classical \, order)}\\
                     \epsilon {\cal R}& {\rm (higher \, order),}
                   \end{array}
                \right.
   \label{ggd}\\
\Omega_{\mu \Theta}&=&\epsilon,
   \label{ggh}
\end{eqnarray}
where
\begin{equation}
F'\equiv \left\{
           \begin{array}{ll}
   F_1 &{\rm (classical \, order)}\\
   F_1+\epsilon F_2-\frac{\epsilon c}{e} \mu {\cal N} &{\rm (higher \, order),}
           \end{array}
        \right.
\label{gjs}
\end{equation}
and
\begin{equation}
{\cal N} \equiv d{\cal R}.
\end{equation}
Here we have drawn a distinction between two cases, just as we did with
the Hamiltonian.  Terms of {\it classical}
order are all that are necessary to retain if only the usual gradient,
curvature and polarization drifts are desired.  If one would like the equations
of motion to one order higher than that, one must also retain the terms
labelled {\it higher} order.  This makes a difference only in $\Omega_{R\mu}$
and in the definition of $F'.$

Now we can get the Poisson brackets using Eq.~(\ref{mbi}).
We do this by inverting the eight by eight matrix consisting of the
components of $\Omega.$  This is a tedious but straightforward exercise, and 
the nonvanishing results are presented below.  We have performed this matrix
inversion for both the classical-order and the higher-order cases separately.
\begin{eqnarray}
\{ R,R \}&=&-\frac{\epsilon F_0}{m\lambda_B \Omega_B \Upsilon},
   \label{ggj}\\
\{ R,K \}&=&-\frac{\hat{\bf t}}{m}\cdot \Xi,
   \label{ggk}\\
\{ R,{\cal B} \}&=&\frac{\hat{\bf b}}{mK}\cdot \Xi,
   \label{ggl}\\
\{ R,\Theta \}&=&\left\{
                   \begin{array}{ll}
                      0& {\rm (classical \, order)}\\
                     \epsilon \{ R,R \}\cdot {\cal R}& {\rm (higher \, order),}
                   \end{array}
                \right.
   \label{ggm}\\
\{ K,{\cal B} \}&=&-\frac{e}{m^2ck}\hat{\bf t}\cdot \Xi
   \cdot F'' \cdot \hat{\bf b},
   \label{ggn}\\
\{ K,\Theta \}&=&\left\{
                   \begin{array}{ll}
                      0& {\rm (classical \, order)}\\
                     \epsilon \{ K,R \}\cdot {\cal R}& {\rm (higher \, order),}
     \end{array}
                 \right.
   \label{ggo}\\
\{ {\cal B} , \Theta \}&=&\left\{
                   \begin{array}{ll}
               0& {\rm (classical \, order)}\\
             \epsilon \{ {\cal B} ,R \}\cdot {\cal R}& {\rm (higher \, order),}
                 \end{array}
                 \right.
   \label{ggp}\\
\noalign{\hbox{and}}
\{ \Theta ,\mu \}&=&\epsilon^{-1},
   \label{ggq}
\end{eqnarray}
where we have defined the scalar
\begin{equation}
\Upsilon \equiv 1+\frac{\epsilon F_0 : F''}{2 \lambda_B^2},
\label{gkg}
\end{equation}
and the tensors
\begin{equation}
\Xi \equiv  {\bf 1}+\frac{\epsilon F'' \cdot F_0}{\lambda_B^2 \Upsilon}
\label{gkh}
\end{equation}
and
\begin{equation}
F'' \equiv F'+\frac{mcK}{e}( \delr \hat{\bf t}-\hat{\bf t} \dell ),
\label{gkj}
\end{equation}
and where $F'$ is given by Eq.~(\ref{gjs}).
Note carefully that the bracket of $R$ with $R$ is nonzero because $R$ is
really four coordinates; thus $\{ R,R \}$ is a four by four antisymmetric
matrix and, consequently, its diagonal elements vanish but the rest of it
may be nonzero.

Note that $\Theta$ and $\mu$ are decoupled from the other dynamical variables
at the classical order, but that $\Theta$ is not decoupled at higher order.
The reason for this will be clarified shortly, but for now we note that this
coupling is not at all problematic.  The important point is that the set of
functions of $R,$ $K$ and ${\cal B}$ form a subset of the set of all phase
functions that is a closed Lie subalgebra under the operation of these Poisson 
brackets.  Then, since our Hamiltonian is independent of $\Theta,$ we can
eliminate that degree of freedom and still have a valid Hamiltonian system for
guiding centers.  This is an example of the
{\it reduction} of a Hamiltonian system, discussed in Chapter~\ref{yaa}.

Next note that we could have expanded all of the above expressions in
pure power series in $\epsilon.$  For example, $\Upsilon$ appears in the
denominators of several brackets, and consists of an order one term and
an order $\epsilon$ term.  One might argue that, since our expressions
are valid only to a certain power of $\epsilon$ anyway, we ought to
expand this in powers of $\epsilon.$  There is, however, a
compelling reason not to do this:  The above brackets are
guaranteed to obey the Jacobi identity {\it exactly} because they
are elements of the inverse matrix of the matrix of components of 
the Lagrange tensor which obeys $d\Omega=dd\Gamma=0.$  If we were to 
expand the brackets in $\epsilon,$
and retain $\epsilon$ only to a certain power, then the Jacobi identity
would be satisfied only to that power of $\epsilon.$
Now one might counter that in an asymptotic theory of this nature, 
that is all we
have a right to demand.  In practice, however, guiding-center equations of
motion are often integrated numerically, and violations of the Jacobi 
identity invalidate Liouville's theorem which guarantees
phase space area preservation.  This, in turn, can lead to an observed
``fuzziness'' of KAM tori which might cause one to draw erroneous
conclusions about the presence of stochasticity.

To elaborate on this last point, in studies of mirror-confined plasmas, for
example, one might integrate the guiding-center equations
numerically and produce a ``puncture plot'' of the places where the
trajectory of the guiding center intersects the midplane of the device.
If such a plot exhibits stochasticity, one might well 
expect the radial transport of the plasma
to be enhanced significantly as compared to a case for which the plot is a
smooth KAM surface.  Thus, in a study of mirror plasma radial transport, one
might vary some parameter to see for what value this transition from regular to
stochastic motion takes place.  The decision might be made by comparing the
numerically-generated puncture-plots for several different parameter values in
some range.  Yet if one uses guiding-center equations of motion that do not
satisfy Liouville's theorem {\it exactly}, one runs the risk of misinterpreting
``fuzziness'' in plots that is due only to violations of Liouville's theorem
(which is, after all, the only reason that KAM tori exist in the first place)
as the presence of true stochasticity.

This is why we inverted the Lagrange tensor for the classical and the
higher-order cases {\it separately}, rather than do a single inversion for the
higher-order case and truncate to get the classical case.  As things stand,
the brackets for both cases presented above are guaranteed to satisfy the
Jacobi identity {\it exactly}.

\section{Guiding-Center Equations of Motion}
\label{yal}
These brackets together with the Hamiltonian, Eq.~(\ref{gfz}), give the
guiding-center equations of motion according to Eq.~(\ref{mbj}).
First consider the equation for $\dot{R}.$  To the classical order, this may be
written
\begin{eqnarray}
\dot{R}&=&\{ R,R \} \cdot \mu \delr \Omega_B - \{ R,K \} mK \nonumber\\
       &=&K\hat{\bf t} + \frac{\epsilon}{\lambda_B^2 \Upsilon}
         \left( K\hat{\bf t} \cdot F'' + \frac{c}{e} \mu \delr \Omega_B
         \right) \cdot F_0.
       \label{ggr}
\end{eqnarray}
The first term contains the usual parallel motion and the
${\bf E}\times {\bf B}$ drift.  The order $\epsilon$ contribution
consists of two parts:  The first contains the relativistic analog of the
curvature and polarization drifts (they are in $F''$),
and the second is the relativistic analog 
of the grad-$B$ drift; these statements will be clarified when we cast these
results in ``$1+3$'' notation.  Of course, the above apparatus is sufficient to
get $\dot{R}$ to one order higher than this, but the expression itself is
rather unenlightening to look at, so we shall not bother to write it down.

The equations for $\dot{K}$ and $\dot{\cal B}$ are then
\begin{eqnarray}
\dot{K}&=&\{ K,R \} \cdot \mu \delr \Omega_B\nonumber\\
       &=&\frac{\mu}{m}\hat{\bf t}\cdot\Xi\cdot\delr\Omega_B
       \label{gia}
\end{eqnarray}
and
\begin{eqnarray}
\dot{\cal B}&=&\{ {\cal B},R \} \cdot \mu \delr \Omega_B 
         - \{ {\cal B},K \} mK \nonumber\\
       &=&-\frac{\mu}{mK}\hat{\bf b}\cdot\Xi\cdot\delr\Omega_B
         -\frac{e}{mc}\hat{\bf t}\cdot\Xi\cdot F''\cdot\hat{\bf b}.
       \label{gib}
\end{eqnarray}
The terms containing $\delr\Omega_B$ contain the mirroring force, and
the contribution of $F_1$ contains the force due to the parallel electric
field; once again, these statements will be clarified when we cast these
results in ``$1+3$'' notation.

Next note that $\dot{\mu}$ is exactly zero, even at the higher order; this,
of course, was our aim all along.  The higher order equation of motion 
for $\Theta$ is
\begin{equation}
\dot{\Theta}=\frac{1}{\epsilon}\Omega_B+\epsilon {\cal R} \cdot \dot{R}
   +\frac{\epsilon}{2\lambda_B} F_0 : \left[
   \left( \frac{e}{mc}\right) F_1 + \frac{K}{2}\left( \delr \hat{\bf t} -
   \hat{\bf t} \dell \right)
   \right].
\label{ggs}
\end{equation}
The first term is the lowest-order gyromotion.  The second term 
arises from the bracket structure, and corrects for the
possibility that as the guiding-center moves in $R,$ the perpendicular
unit vectors upon which the definition of $\Theta$ is based may rotate
within the perpendicular two-flat.  This term arose from our demand of
boostgauge and gyrogauge invariance, and it is the reason that the Poisson
bracket of $\Theta$ with $R,K,$ and ${\cal B}$ cannot vanish at higher
order.  The necessity of this has been discussed by Littlejohn~\cite{zak}
and by Hagan and Frieman~\cite{zan}.

The third term on the right side of Eq.~(\ref{ggs}) arises from the first-order
piece of the Hamiltonian and consists of two subterms in the square brackets.
The first of these subterms is the correction to the gyrofrequency
due to $F_1.$  To see this, define the {\it total gyrofrequency} due to both
$F_0$ and $F_1$ by $\Omega_{BT}\equiv e\lambda_{BT}/mc,$ where $\lambda_{BT}$
is given by Eq.~(\ref{gah}).  We quickly find
\begin{eqnarray}
\Omega_{BT}
  &=&\frac{e}{mc}\sqrt{\frac{1}{2}(F_0+\epsilon F_1):(F_0+\epsilon F_1)}
    +{\cal O}(\epsilon^2)\nonumber\\
  &=&\frac{e}{mc}\sqrt{\lambda_B^2+\epsilon F_0:F_1}
    +{\cal O}(\epsilon^2)\nonumber\\
  &=&\Omega_B+\frac{\epsilon}{2\lambda_B}\left(\frac{e}{mc}\right)F_0:F_1
    +{\cal O}(\epsilon^2).
  \label{gkc}
\end{eqnarray}
The second subterm of the third term on the right of Eq.~(\ref{ggs})
is the gyrofrequency shift due to gradients of
the perpendicular electric field.  This is not expected to be obvious, and
will be discussed further in Section~\ref{yan}, when we cast our results in
``$1+3$'' notation.

\begin{figure}[p]
\center{
\vspace{2.42truein}
\mbox{\includegraphics[bbllx=0,bblly=0,bburx=251,bbury=115,width=5.82truein]{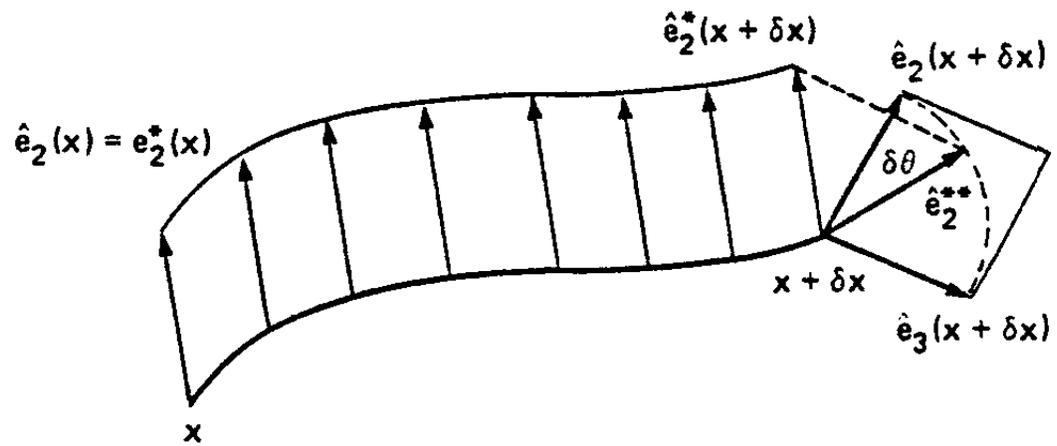}}
\vspace{2.42truein}
}
\caption{Change in Gyroangle due to Rotation of Basis Tetrad as Guiding Center
Moves in Spacetime}
\label{gid}
\end{figure}

The geometrical significance of the second
term in Eq.~(\ref{ggs}) is illustrated in
Fig.~\ref{gid} (here we temporarily revert to using lower-case $r$ and
$\theta$).  In order to compare the unit tetrad at one point
in spacetime, $r,$ with that at another point, $r+\delta r,$ (to see 
how much it rotated) we need some way of transporting the unit vectors
from one point to another.  The correct way of doing this was elucidated
by Littlejohn~\cite{zak}.  Since we have assumed flat spacetime
throughout this calculation, we can simply translate the unit vector
${\hat{\bf e}}_2$ from $r$ to $r+\delta r$ in the usual manner of
Euclidean geometry.  Of course, when we arrive at $r+\delta r,$ the
translated unit vector, called ${\hat{\bf e}}_2^*,$ will not be the 
same as the unit vector ${\hat{\bf e}}_2.$  Furthermore, it need not
even lie in the perpendicular two-flat.  To remedy this, we project
it onto the perpendicular two-flat and normalize the result to get a
new unit vector, called ${\hat{\bf e}}_2^{**}.$  The angle between
${\hat{\bf e}}_2$ and ${\hat{\bf e}}_2^{**}$ at the point $r+\delta r$
is defined to be $\delta \theta.$  The calculation goes as follows:
\begin{eqnarray}
{\hat{\bf e}}_2^*(r+\delta r)&=&{\hat{\bf e}}_2(r) \nonumber\\
   &=&{\hat{\bf e}}_2(r+\delta r-\delta r) \nonumber\\
   &=&{\hat{\bf e}}_2(r+\delta r)-\delta r \cdot \delr
     {\hat{\bf e}}_2(r+\delta r)+ \nonumber\\ 
   &&\qquad \frac{1}{2}\delta r \delta r :
     \delr \delr {\hat{\bf e}}_2(r+\delta r) + \cdots .
   \label{ggt}
\end{eqnarray}
Henceforth, all quantities are evaluated at the point $r+\delta r$
so this will not be noted explicitly.  Continuing,
\begin{eqnarray}
{\hat{\bf e}}_2^{**} &\equiv \frac{P_\perp \cdot {\hat{\bf e}}_2^*}
     {\left| P_\perp \cdot {\hat{\bf e}}_2^* \right| } \nonumber\\
   &=&{\hat{\bf e}}_2-\delta r \cdot \delr {\hat{\bf e}}_2 \cdot
     {\hat{\bf e}}_3 {\hat{\bf e}}_3 + \frac{1}{2}\delta r \delta r :
     \delr \delr {\hat{\bf e}}_2 \cdot {\hat{\bf e}}_3 {\hat{\bf e}}_3 \nonumber\\
   &&\qquad -\frac{1}{2}\left( \delta r \cdot \delr {\hat{\bf e}}_2 \cdot
     {\hat{\bf e}}_3 \right)^2 {\hat{\bf e}}_2 + \cdots .
   \label{ggu}
\end{eqnarray}
Thus
\begin{eqnarray}
\cos \delta \theta &=&1-\frac{\delta \theta^2}{2}+\cdots \nonumber\\
   &=&{\hat{\bf e}}_2 \cdot {\hat{\bf e}}_2^{**} \nonumber\\
   &=&1-\frac{1}{2}\left( \delta r \cdot \delr {\hat{\bf e}}_2
     \cdot {\hat{\bf e}}_3 \right)^2 + \cdots \nonumber\\
   &=&1-\frac{1}{2}\left( {\cal R} \cdot \delta r \right)^2 + \cdots ,
   \label{ggv}
\end{eqnarray}
so we identify
\begin{equation}
\delta \theta = {\cal R} \cdot \delta r.
\end{equation}
This is the change in $\theta$ due to the rotation of the unit vectors
alone, and it explains the second term on the right of 
Eq.~(\ref{ggs}).  A similar term, ${\cal Q} \cdot \dot{R},$ would
appear in the equation of motion of ${\cal B}$ if we went to higher order.

It was noted by Littlejohn~\cite{zak} that the one-form, ${\cal R}$ is
the potential for the gauge field ${\cal N}=d{\cal R}$ 
which obeys the field equation $d{\cal N}=dd{\cal R}=0.$  
In the relativistic problem, we also have the gauge field
${\cal M}=d{\cal Q},$ and this also obeys $d{\cal M}=dd{\cal Q}=0.$
These are the gauge fields corresponding to the boostgauge and gyrogauge 
gauge groups.  Note that ${\cal M}$ and ${\cal N}$ are gauge invariant 
even though ${\cal Q}$ and ${\cal R}$ are not.  Thus, they can be
expressed in terms of the field tensor directly; in index notation
\begin{equation}
{\cal M}_{\mu \nu}=\frac{1}{\lambda_B}
                   {\cal F}^\alpha_{\phantom{\alpha} \beta}
                   P^{\gamma \beta}_{\parallel \phantom{\beta},\mu}
                   P_{\parallel \gamma \alpha ,\nu},
\label{taa}
\end{equation}
and
\begin{equation}
{\cal N}_{\mu \nu}=\frac{1}{\lambda_B}
                   F^\alpha_{\phantom{\alpha} \beta}
                   P^{\gamma \beta}_{\perp \phantom{\beta},\mu}
                   P_{\perp \gamma \alpha ,\nu}.
\label{tab}
\end{equation}
The $\dot{R}$ term of the guiding-center Lagrangian, 
Eq.~(\ref{gfy}), thus couples the two gauge potentials, $A$ and ${\cal R},$
and the coupling constant is the gyromomentum.

\section{Summary of Guiding-Center Transformation}
\label{yam}
The entire transformation that we have made from the particle coordinates
may be written in the form of Eq.~(\ref{mbp}) as follows:
\begin{equation}
Z= \exp (\epsilon {\cal L}_{g_p})   \exp (\epsilon {\cal L}_{g_1}) 
   \exp (\epsilon^2 {\cal L}_{g_2}) \exp (\epsilon^3 {\cal L}_{g_3}) \cdots z.
\label{ggw}
\end{equation}   
It is possible to expand these equations in $\epsilon,$ and plug in our
expressions for the generators to get the coordinate transformation
equations.  For reference, we present these here:
\begin{eqnarray}
R&=&r-\frac{\epsilon w \hat{\bf a}}{\Omega_B}+\epsilon^2 \Bigl\{
   \frac{w^2}{2\Omega_B}\hat{\bf a}\cdot \delr \left(
   \frac{\hat{\bf a}}{\Omega_B}\right) \nonumber\\ 
   &&\qquad +\frac{w^2\hat{\bf a}\cdot {\cal R}\hat{\bf c}}{2\Omega_B^2}
   +\frac{w}{\lambda_B \Omega_B}(P_\parallel -P_\perp )\cdot F_1 \cdot
   \hat{\bf c} \nonumber\\
   &&\qquad -\frac{2kw}{\Omega_B^2}(ttc)\hat{\bf t}
   +\frac{w^2}{8\Omega_B^2} \left[ (ata)-(ctc) \right] \hat{\bf t} \nonumber\\
   &&\qquad +\frac{kw}{\Omega_B^2} \left[ (btc)+(tbc) \right] \hat{\bf b}
   -\frac{w^2}{8\Omega_B^2}\left[ (aba)-(cbc) \right] \hat{\bf b} \nonumber\\
   &&\qquad +\frac{kw}{\Omega_B^2}\left[ (cta)-(atc) \right] \hat{\bf a}
   \Bigr\}+O(\epsilon^3),
   \label{ggx}\\
K&=&k+\epsilon \Bigl[ \frac{kw}{\Omega_B}(tta)-\frac{w^2}{2\Omega_B}(act) \nonumber\\
   &&\qquad +\frac{w}{\lambda_B}\hat{\bf t}\cdot F_1 \cdot \hat{\bf a}
   \Bigr] +O(\epsilon^2),
   \label{ggy}\\
{\cal B}&=&\beta+\epsilon \Bigl[ \frac{w^2}{2k\Omega_B}(acb)-\frac{w}{\Omega_B}
   [(bta)-(atb)] \nonumber\\
   &&\qquad -\frac{w}{k\lambda_B}\hat{\bf b}\cdot F_1
   \cdot \hat{\bf a} \Bigr] +O(\epsilon^2),
   \label{ggz}\\
W&=&w+\epsilon \Bigl[ \frac{k^2}{\Omega_B}(tta)-\frac{kw}{4\Omega_B}
   [3(act)-(cat)] \nonumber\\
   &&\qquad +\frac{k}{\lambda_B}\hat{\bf t}\cdot
   F_1 \cdot \hat{\bf a}+\frac{w}{2\lambda_B}\hat{\bf a} \cdot F_1
   \cdot \hat{\bf c}\Bigr]+O(\epsilon^2),
   \label{gha}\\
\noalign{\hbox{and}}
\Theta &=&\theta+\epsilon \Bigl[\frac{k^2}{w\Omega_B}(ttc)-\frac{k}{4\Omega_B}
   [(ata)-(ctc)]-\frac{w}{\Omega_B}(aca) \nonumber\\
   && \qquad +\frac{w}{\Omega_B^2}
   \hat{\bf c}\cdot \delr \Omega_B + \frac{k}{w\lambda_B}\hat{\bf t}\cdot
   F_1 \cdot \hat{\bf c} \Bigr]+O(\epsilon^2).
   \label{ghb}
\end{eqnarray}
In the above equations, the capitalized variables are the guiding-center
coordinates and the lower case variables are the particle coordinates;
it is emphasized that {\it all} quantities on the right hand sides of
these equations (e.g. unit vectors, field tensor, etc.) 
are evaluated at the particle coordinates.  The inverse
transformation is given by
\begin{eqnarray}
r&=&R+\frac{\epsilon W \hat{\bf a}}{\Omega_B}+\epsilon^2 \Bigl\{
   \frac{W^2}{2\Omega_B}\hat{\bf a}\cdot \delr \left(
   \frac{ \hat{\bf a}}{\Omega_B}\right) 
   - \frac{W}{\lambda_B \Omega_B}(P_\parallel - P_\perp)\cdot F_1
   \cdot \hat{\bf c} \nonumber\\
   &&\qquad +\hat{\bf t}\Bigl[\frac{2KW}{\Omega_B^2}(ttc)
   - \frac{W^2}{8\Omega_B^2}[(ata)-(ctc)]\Bigr] \nonumber\\
   &&\qquad +\hat{\bf b}\Bigl[\frac{W^2}{8\Omega_B^2}[(aba)-(cbc)]
   - \frac{KW}{\Omega_B^2}[(btc)+(tbc)]\Bigr] \nonumber\\
   &&\qquad +\hat{\bf c}\Bigl[-\frac{W^2}{\Omega_B^3}\hat{\bf c}\cdot
   \delr \Omega_B + \frac{K}{\lambda_B \Omega_B}\hat{\bf t}\cdot F_1 \cdot
   \hat{\bf c}-\frac{K^2}{\Omega^2}(ttc)\nonumber\\ &&\qquad +\frac{KW}{4\Omega_B^2}
   [(ata)-(ctc)]+\frac{W^2}{2\Omega_B^2}(aca)\Bigr] \nonumber\\
   &&\qquad +\hat{\bf a}\Bigl[-\frac{W}{2\lambda_B \Omega_B}\hat{\bf a}\cdot
   F_1\cdot \hat{\bf c} - \frac{K}{\lambda_B \Omega_B}
   \hat{\bf t}\cdot F_1 \cdot
   \hat{\bf a} \nonumber\\ &&\qquad -\frac{K^2}{\Omega^2}(tta) +\frac{KW}{4\Omega_B^2}
   [3(cat)-(act)]\Bigr] \Bigr\} +O(\epsilon^3),
   \label{ghc}\\ 
k&=&K-\epsilon \Bigl[ \frac{KW}{\Omega_B}(tta)-\frac{W^2}{2\Omega_B}(act) \nonumber\\
   &&\qquad +\frac{W}{\lambda_B}\hat{\bf t}\cdot F_1 \cdot \hat{\bf a}
   \Bigr] +O(\epsilon^2),
   \label{ghd}\\
\beta &=&{\cal B}-\epsilon \Bigl[ \frac{W^2}{2K\Omega_B}(acb)-\frac{W}{\Omega_B}
   [(bta)-(atb)] \nonumber\\ &&\qquad -\frac{W}{K\lambda_B}\hat{\bf b} \cdot F_1
   \cdot \hat{\bf a} \Bigr] +O(\epsilon^2),
   \label{ghe}\\
w&=&W-\epsilon \Bigl[ \frac{K^2}{\Omega_B}(tta)-\frac{KW}{4\Omega_B}
   [3(act)-(cat)] \nonumber\\ &&\qquad +\frac{K}{\lambda_B}\hat{\bf t}\cdot F_1
   \cdot \hat{\bf a} + \frac{W}{2\lambda_B}\hat{\bf a}\cdot F_1 \cdot
   \hat{\bf c} \Bigr] +O(\epsilon^2),
   \label{ghf}\\
\noalign{\hbox{and}}
\theta &=&\Theta - \epsilon \Bigl[ \frac{K^2}{W\Omega_B}(ttc)-
   \frac{K}{4\Omega_B}[(ata)-(ctc)]-\frac{W}{\Omega_B}(aca) \nonumber\\
   &&\qquad +\frac{W}{\Omega_B^2}\hat{\bf c} \cdot \delr \Omega_B
   +\frac{K}{W\lambda_B}\hat{\bf t} \cdot F_1 \cdot \hat{\bf c}
   \Bigr] +O(\epsilon^2).
   \label{ghg}
\end{eqnarray}
In the above equations, everything on the right is evaluated
at the guiding-center position.

Recall that the gyromomentum in guiding-center coordinates is
given by $mW^2/2\Omega_B.$  In particle coordinates, this may be written
\begin{eqnarray}
\mu_{\rm part}&=&\frac{mw^2}{2\Omega_B} 
   + \epsilon \Bigl\{ \frac{mw^3}{2\Omega_B^3}
   \hat{\bf a} \cdot \delr \Omega_B + \frac{mw}{\Omega_B} \Bigl[
   \frac{w}{2\lambda_B}\hat{\bf a}\cdot F_1 \cdot \hat{\bf c} \nonumber\\
   &&\qquad +\frac{k}{\lambda_B}\hat{\bf t}\cdot F_1 \cdot \hat{\bf a}
   +\frac{k^2}{\Omega_B}(tta)-\frac{kw}{4\Omega_B}[3(act)-(cat)]
   \Bigr] \Bigr\} +O(\epsilon^2).
   \label{ghh}
\end{eqnarray}
This expression is useful because it gives the conserved quantity in
terms of particle coordinates.

\section{Comparison with Three-Vector Formulations}
\label{yan}
In order to compare our results with the three-vector formulation given by
Northrop~\cite{zap}, we must be able
to cast our results into ``$1+3$'' notation.
We learned how to do this for the particle coordinates back at the end of
Section~\ref{yaf} where we gave the explicit transformation equations,
Eqs.~(\ref{gil}) through (\ref{gir}).
These are scalar equations in phase space, and so
they will retain their form under the guiding-center Lie transform.  We need
only to replace $(k,\beta,w,\theta)$ by $(K,{\cal B},W,\Theta),$ and to
reinterpret ${\bfbet}_v$ as the {\it guiding-center} three-velocity
(divided by $c$).  Then we can write down the equations of motion for
${\bfbet}_v$ by differentiating Eqs.~(\ref{gil}) through (\ref{gin})
with respect to proper time, using the known equations of motion for the
guiding-center coordinates, and expressing the results back in terms of
${\bfbet}_v$ by using Eqs.~(\ref{gio}) through (\ref{gir}).

The above-described program seems rather tedious.  Fortunately, there are
two things that we can do to simplify the task.  First, we need only check
our results to the order of the classical drifts.  This is the order given
in the text by Northrop~\cite{zap}.  Second, we can check our
results in one of the ``preferred'' frames of reference, as were described back
in Section~\ref{yaf}.  If they hold there, they have to hold in all other
frames as well because our results are in manifestly covariant format.  These
two simplifications make the problem straightforward.

First note that in a preferred frame ${\bfbet}_E=0,$ so Eqs.~(\ref{gil})
through (\ref{gir}) become
\begin{eqnarray}
\beta_{v1} &=&\tanh {\cal B} \label{gis}\\
\beta_{v2} &=&-\frac{W\sin\Theta}{K\cosh {\cal B}} \label{git}\\
\beta_{v3} &=&-\frac{W\cos\Theta}{K\cosh {\cal B}} \label{giu}\\
\noalign{\hbox{and}}
K &=&c\gamma_v\sqrt{1-\beta_{v1}^2} \label{giv}\\
{\cal B} &=&\tanh^{-1}\beta_{v1} \label{giw}\\
W &=&c\gamma_v\sqrt{\beta_{v2}^2+\beta_{v3}^2} \label{gix}\\
\Theta &=&{\rm arg}(-\beta_{v3}-i\beta_{v2}), \label{giy}
\end{eqnarray}
where, as noted in the last paragraph, all variables are now
{\it guiding-center} variables.  In particular, the equations
\begin{equation}
K\cosh {\cal B}=c\gamma_v
\label{gjw}
\end{equation}
and
\begin{equation}
K\sinh {\cal B}=c\gamma_v\beta_{v1}=\gamma_v v_\parallel,
\end{equation}
where $v_\parallel\equiv c\beta_{v1},$ will turn out to be particularly useful.
The quantity ${\cal B}$ is sometimes called the {\it rapidity}.

Next note that, in a preferred frame, the unit vectors that we constructed in
Eqs.~(\ref{gij}) and (\ref{gik}) can be inserted into Eqs.~(\ref{gaq})
and (\ref{gas}) to yield
\begin{equation}
   \begin{array}{cc}
      {\hat{\bf t}}=\left(
                          \begin{array}{c}
                             \cosh {\cal B} \\
                             {\bf b}\sinh {\cal B}
                          \end{array}
                      \right), &
      {\hat{\bf b}}=\left( 
                          \begin{array}{c}
                             \sinh {\cal B} \\
                             {\bf b}\cosh {\cal B}
                          \end{array}
                      \right).
   \end{array}
\label{giz}
\end{equation}
These will also be useful in what follows.

Now examine Eq.~(\ref{ggr}).  We can consider the terms individually.  First
\begin{equation}
K\hat{\bf t}=\left(
               \begin{array}{c}
                 \gamma_v c \\
                 \gamma_v v_\parallel {\bf b}
               \end{array}
             \right)
\end{equation}
follows immediately.  Next
\begin{eqnarray}
\hat{\bf t}\cdot F'' 
   &=&\hat{\bf t}\cdot F_1+\frac{mcK}{e}\hat{\bf t}\cdot\delr\hat{\bf t}\nonumber\\
   &=&\hat{\bf t}\cdot F_1+\frac{mcK}{e}
     \left(\cosh {\cal B}\frac{1}{c}\frac{\partial}{\partial t}
     +\sinh {\cal B}{\bf b}\cdot {\bf \nabla}\right)\hat{\bf t}\nonumber\\
   &=&\hat{\bf t}\cdot F_1
     +\frac{mcv_\parallel}{eK}\gamma_v^2
     \left(
     \begin{array}{c}
     0 \\
     \frac{\partial {\bf b}}{\partial t}
      +v_\parallel {\bf b}\cdot {\bf \nabla}{\bf b}
     \end{array}
     \right)\nonumber\\
   &&\qquad +\frac{mc}{eK}\gamma_v^2
     \left(
     \begin{array}{c}
     0 \\
     \frac{\partial {\bf u}_E}{\partial t}
      +v_\parallel {\bf b}\cdot {\bf \nabla}{\bf u}_E
     \end{array}
     \right),
   \label{gja}
\end{eqnarray}
where
\begin{equation}
{\bf u}_E\equiv c\frac{{\bf E}\times {\bf B}}{B^2},
\end{equation}
also follows after a short computation.  Note that ${\bf u}_E$ vanishes in a
preferred frame, but its derivatives may not; thus we had to apply the
derivative to $\hat{\bf t}$ {\it before} specializing to a preferred frame.

Next we write the components of $F_1$ as follows
\begin{equation}
 F_1 = \left(
         \begin{array}{cc}
           0         & {\bf E}_1 \\
           {\bf E}_1 & {\bf 1}\times {\bf B}_1
         \end{array}
       \right).
\label{gjv}
\end{equation}
Recall that ${\bf E}_1$ must contain all of the parallel electric field.

It now follows from Eq.~(\ref{ggr}) that
\begin{equation}
c\dot{t}=c\gamma_v+{\cal O}(\epsilon)
\end{equation}
and
\begin{eqnarray}
\dot{\bf R}&=&\gamma_v v_\parallel {\bf b}_T+\frac{\epsilon}{B} {\bf b}_T\times
   \biggl\{ \frac{mc}{e}\gamma_v^2 \biggl[ v_\parallel\left(
   \frac{\partial {\bf b}_T}{\partial t}+v_\parallel {\bf b}_T\cdot
   {\bf \nabla}{\bf b}_T\right)\nonumber\\
   &&\qquad +\left(
   \frac{\partial {\bf u}_E}{\partial t}+v_\parallel {\bf b}_T\cdot
   {\bf \nabla}{\bf u}_E\right)\biggr]
   +\frac{\mu}{m}{\bf \nabla}B\biggr\}
   +{\cal O}(\epsilon^2),
   \label{gjt}
\end{eqnarray}
where
\begin{equation}
{\bf b}_T\equiv\frac{{\bf B}+\epsilon {\bf B}_1}{\left|{\bf B}
   +\epsilon {\bf B}_1\right| }.
\end{equation}
Now take the perpendicular part of $\dot{\bf R}$ by dotting it with
${\bf 1}-{\bf b}_T{\bf b}_T,$ then divide by $\dot{t}$ to get
\begin{eqnarray}
\frac{d {\bf R}_\perp}{dt}&=&\frac{\epsilon}{\Omega_B} {\bf b}_T\times
   \biggl\{ \gamma_v \biggl[ v_\parallel\left(
   \frac{\partial {\bf b}_T}{\partial t}+v_\parallel {\bf b}_T\cdot
   {\bf \nabla}{\bf b}_T\right)\nonumber\\
   &&\qquad +\left(
   \frac{\partial {\bf u}_E}{\partial t}+v_\parallel {\bf b}_T\cdot
   {\bf \nabla}{\bf u}_E\right)\biggr]
   +\frac{\mu}{m\gamma_v}{\bf \nabla}\Omega_B\biggr\}
   +{\cal O}(\epsilon^2).
   \label{gju}
\end{eqnarray}
This is identical to Eq.~(1.76) in the text by Northrop~\cite{zap} in a
preferred frame.  Recall that $\lambda_B=B$ in a preferred frame, so that
$\Omega_B$ in the above equation is simply $eB/mc.$  The classical curvature,
gradient and polarization drifts are readily visible in the above equation.  If
we had instead done the calculation for a general frame of reference, the
${\bf E}\times {\bf B}$ drift would appear as well.
The reader is referred to Northrop~\cite{zap} for a good discussion of these
results.

Next differentiate $\gamma_v v_\parallel=K\sinh {\cal B}$ to get
\begin{equation}
\frac{d}{dt}(\gamma_v v_\parallel)=\frac{1}{\gamma_v}(\dot{K}\sinh {\cal B}
    +K\dot{\cal B}\cosh {\cal B}).
\end{equation}
Insert Eqs.~(\ref{gia}) and (\ref{gib}) for $\dot{K}$ and $\dot{\cal B},$
respectively, and after a little algebra we find
\begin{equation}
\frac{d}{dt}(\gamma_v v_\parallel)=\frac{1}{\gamma_v}\left(-\frac{\mu}{m}
    {\bf b}_T\cdot {\bf \nabla}\Omega_B-\frac{e}{m}\gamma_v
    \hat{\bf t}\cdot F_1 \cdot\hat{\bf b}\right)
    +{\cal O}(\epsilon).
\end{equation}
Now it follows from Eq.~(\ref{gjv}) that
\begin{equation}
\hat{\bf t}\cdot F_1 \cdot\hat{\bf b}=-{\bf b}\cdot {\bf E}_1=-E_\parallel,
\label{gjb}
\end{equation}
So we finally have
\begin{equation}
\frac{d}{dt}(\gamma_v v_\parallel)=-\frac{\mu}{m\gamma_v} {\bf b}_T\cdot
    {\bf \nabla}\Omega_B+\frac{e}{m}E_\parallel+{\cal O}(\epsilon).
\end{equation}
This is identical to Eq.~(1.77) in the text by Northrop~\cite{zap} in a
preferred frame.  The terms on the right are the mirroring force and the
force due to the parallel electric field, respectively.

Northrop's Eq.~(1.78) is immediately seen to be equivalent to the fact that
our gyromomentum $\mu$ is a constant of the motion.  Note that Northrop's
magnetic moment $M_r$ is related to our $\mu$ as follows:  $M_r=e\mu/mc.$

Next, we know from Eq.~(\ref{gjw}) that $c\gamma_v=K\cosh {\cal B},$ so
\begin{equation}
\frac{d}{dt}(mc^2\gamma_v)
   =\frac{mc}{\gamma_v}(\dot{K}\cosh {\cal B}+K\sinh {\cal B}\dot{\cal B}).
\end{equation}
Now use Eqs.~(\ref{gia}), (\ref{gib}) and (\ref{gjb}) to get
\begin{equation}
\frac{d}{dt}(mc^2\gamma_v)=\frac{\mu}{\gamma_v}
   \frac{\partial \Omega_B}{\partial t}+ev_\parallel E_\parallel+{\cal O}(\epsilon)
\end{equation}
after a short calculation.  This is identical to Eq.~(1.79) in the text by
Northrop~\cite{zap} in a preferred frame.

Finally, as promised, we discuss the nonrelativistic limit of the second subterm
of the third term on the right side of Eq.~(\ref{ggs}).  This term is given by
$(K/4\lambda_B)F_0:(\delr\hat{\bf t}-\hat{\bf t}\dell).$  To simplify the
evaluation of this term, we specialize to a preferred frame where the perpendicular
electric field vanishes (though we shall be careful to retain its gradient).
We also specialize to the case of time-independent fields, spatially
uniform magnetic field, and zero parallel
velocity.  These assumptions are not at all necessary; they serve only to simplify
an otherwise tedious calculation, to aid the reader in seeing an effect that would
otherwise be masked by lots of other less interesting terms, and to facilitate
comparison with Appendix~\ref{ybd}.  Under these circumstances, we find that
\begin{equation}
\delr K\hat{\bf t}=c\gamma_v
  \left(
  \begin{array}{cc}
  0 & {\bf 0} \\
  {\bf 0} & {\bf \nabla}{\bfbet}_E
  \end{array}
  \right),
\end{equation}
and
\begin{equation}
F_0=\left(
    \begin{array}{cc}
    0 & {\bf 0} \\
    {\bf 0} & {\bf 1}\times {\bf b}
    \end{array}
    \right).
\end{equation}
It then follows after a short calculation that
\begin{equation}
\frac{K}{4\lambda_B} F_0:(\delr\hat{\bf t}-\hat{\bf t}\dell)
  =\frac{c\gamma_v}{2\lambda_B} {\bf B}\cdot\left({\bf \nabla}\times {\bfbet}_E\right)
  =-\frac{e\gamma_v}{2m\Omega_B} {\bf \nabla}_\perp\cdot {\bf E}_\perp.
\end{equation}
Except for the factor $\gamma_v,$ which is clearly a relativistic effect,
this is identical to the gyrofrequency shift due to perpendicular electric fields
that is derived in Appendix~\ref{ybd}.  This shift
was discovered by Kaufman~\cite{zcc} in 1960, who also showed that it gives rise
to the phenomenon of {\it gyroviscosity}.

The reader is urged to consult the text by Northrop~\cite{zap} as well as
a paper by Vandervoort~\cite{zby} for a further discussion and alternative
presentation of the above results.

\section{Manifestly Boostgauge and Gyrogauge Invariant Format}
\label{yao}
The guiding-center equations of motion presented above
contain expressions, such as $\delr \hat{\bf t},$ that are not boostgauge or
gyrogauge invariant.  Of course, the equations as a whole are guaranteed to be
gauge invariant by our method of derivation; but they are not {\it manifestly}
so.  This is due to the fact that our chosen coordinates, namely 
$(R,K,{\cal B},\mu,\Theta),$ are themselves not gauge invariant, thanks to
the inclusion of ${\cal B}$ and $\Theta.$  This observation suggests that if we
were to transform to a new set of gauge invariant coordinates, we could write
our results in manifestly gauge invariant format; that is, without any mention
of the unit vectors, $\hat{{\bf e}}_\alpha.$  In this section, we shall
derive two new versions of the Poisson brackets:  The first will be manifestly
boostgauge invariant, but it will not be manifestly gyrogauge invariant.  The
second will be both manifestly boostgauge invariant and manifestly gyrogauge
invariant.

\subsection{Manifest Boostgauge Invariance}
To get manifestly boostgauge invariant results,
we would like to replace $K$ and ${\cal B}$ by
the new boostgauge invariant coordinate
\begin{equation}
U\equiv K\hat{\bf t}.
\end{equation}
The inverse transformation would then be
\begin{equation}
K=\sqrt{-U^2}
\end{equation}
and
\begin{equation}
{\cal B}=\tanh^{-1} \left( -\frac{U\cdot\hat{\bf e}_1(R)}
                              {U\cdot\hat{\bf e}_0(R)}\right).
\end{equation}

Alas, there is a problem with this approach.
Since the new coordinate $U$ is a four vector, it contains 
four degrees of freedom, whereas $K$ and ${\cal B}$ represent only two 
degrees of freedom.  This discrepency stems from the fact that $U$ is not an 
arbitrary four vector because it is constrained to lie in the parallel two 
flat; that is, it obeys the constraint equation
\begin{equation}
P_\perp (R)\cdot U=0.
\label{ghi}
\end{equation}
This constraint restricts $U$ to two degrees of freedom, but it also means that
the coordinates $R$ and $U$ are no longer {\it independent} variables.  The
coordinate transformation is not a diffeomorphism (it is injective rather than
bijective) and so we cannot proceed in the usual manner.

We can remedy this difficulty by temporarily relaxing the constraint
in Eq.~(\ref{ghi}).  We make the following coordinate transformation (where,
for clarity, we use primes to distinguish the new coordinates):
\begin{eqnarray}
R'&=&R\nonumber\\
U'&=&K\hat{\bf t}(R,{\cal B})+\frac{C_{1a}}{\lambda_B(R)}\hat{\bf c}(R,\Theta)
                            +C_{1b}\hat{\bf a}(R,\Theta)\nonumber\\
\mu'&=&\mu\nonumber\\
\Theta'&=&\Theta . \label{gjc}
\end{eqnarray}
The reason for including $\lambda_B$ in the second term on the right hand side
of the equation for $U'$ will become clear in the next subsection.
The inverse transformation is then
\begin{eqnarray}
R&=&R'\nonumber\\
K&=&\sqrt{-U'\cdot P_\parallel(R')\cdot U'}\nonumber\\
{\cal B}&=&\tanh^{-1}\left(
   -\frac{U'\cdot {\hat{\bf e}}_1(R')}{U'\cdot {\hat{\bf e}}_0(R')}\right)\nonumber\\
\mu&=&\mu'\nonumber\\
\Theta&=&\Theta'\nonumber\\
C_{1a}&=&\lambda_B(R') U'\cdot \hat{\bf c} (R',\Theta')\nonumber\\
C_{1b}&=&U'\cdot \hat{\bf a} (R',\Theta').\label{gjd}
\end{eqnarray}
Here, $U'$ is no longer constrained to lie in the parallel two-flat, and its
perpendicular components are called $C_{1a}/\lambda_B$ and $C_{1b}.$  In order to have
the same number of variables before and after the transformation, we have
appended $C_{1a}$ and $C_{1b}$ to our usual set of variables before making the
transformation.

We now have a diffeomorphism, but we still have to decide how
to deal with these two new variables in the unprimed system.  Our strategy
will be to demand that they are Casimir functions.
That way, the dynamics is constrained to lie on hypersurfaces for which they
both are constant.  If we start the phase space trajectory  on the
hypersurface for which they are both zero, it will remain on that hypersurface.
Of course, the equations of motion that we end up with will also be capable of
describing dynamics on other hypersurfaces for which they are nonzero, but we
ignore these other orbits as physically irrelevent.

So our phase space coordinates before this transformation are now taken to be
$(R,K,{\cal B},\mu,\Theta,C_{1a},C_{1b}).$  The bracket relations among these
coordinates are given by Eqs.~(\ref{ggj}) through (\ref{ggq}) for the brackets
not involving $C_{1a}$ and $C_{1b}.$  Then, following the strategy discussed in
the last paragraph,  we simply say that the bracket of
$C_{1a}$ or $C_{1b}$ with {\it any} of the other coordinates is zero.  We now
have dynamics in a ten dimensional phase space, but we are interested in
what is going on only in the eight dimensional subspace defined by
$C_{1a}=C_{1b}=0.$  We have simply imbedded the guiding-center dynamics in a
higher dimensional phase space.  It is clear that the Poisson bracket still
obeys antisymmetry and the Jacobi identity.

It is now straightforward to write the Poisson bracket relations among the
new set of coordinates, $(R',U',\mu',\Theta').$  Once we are finished doing
this, it will be alright to set $C_{1a}$ and $C_{1b}$ equal to zero, but not
until we have taken {\it every} derivative that needs to be taken in the
process; derivatives get messed up by coordinate transformations that are not
diffeomorphisms.

We illustrate this calculation for the $\{ R',U' \}$ bracket as follows:
\begin{eqnarray}
\{ R',U' \}&=&\{ R,K\hat{\bf t} \}
            +\{ R,C_{1a}\hat{\bf c}/\lambda_B(R) \} 
            +\{ R,C_{1b}\hat{\bf a} \} \nonumber\\
           &=&\{ R,K\hat{\bf t} \}
            +\{ R,C_{1a} \} \hat{\bf c}/\lambda_B(R)
            +\{ R,C_{1b} \} \hat{\bf a}\nonumber\\
           &&\qquad +\{ R,\hat{\bf c}/\lambda_B(R) \} C_{1a}
            +\{ R,\hat{\bf a} \} C_{1b}\nonumber\\
           &=&\{ R,K \} \hat{\bf t}+K\{ R,R \} \cdot \delr \hat{\bf t}
            +K\{ R,{\cal B} \} \hat{\bf b}.\label{ghj}
\end{eqnarray}
Note that all quantities on the right hand side in the above equation are
expressed in the old coordinate system.  Note also that all terms involving
$C_{1a}$ or $C_{1b}$ have vanished, either because they are bracketed with
something (recall that they are Casimir functions), or because they appear in
a term outside of all derivatives and so we have set them to zero.

Eqs.~(\ref{ggj}) through (\ref{ggl}) can now be substituted into
the right hand side of Eq.~(\ref{ghj}).  The result will still contain
objects such as $\delr \hat{\bf t}$ and $\delr \hat{\bf b}.$  Eliminate
these by means of the easily verified relations
\begin{equation}
\delr \hat{\bf t}=(\delr P_\parallel )\cdot\hat{\bf t}-{\cal Q}\hat{\bf b}
\end{equation}
\begin{equation}
\delr \hat{\bf b}=(\delr P_\parallel )\cdot\hat{\bf b}-{\cal Q}\hat{\bf t}.
\end{equation}
Because our results are guaranteed to be boostgauge invariant, all terms
involving ${\cal Q}$ will cancel, leaving a {\it manifestly} boostgauge
invariant result.  This being the case, the result can be expressed in terms
of the new coordinates.

Before presenting these results, a word of warning is in order.  When the
term $K(\delr P_\parallel )\cdot\hat{\bf t}$ is expressed in the new
coordinates, the result is easily found to be
\begin{equation}
K(\delr P_\parallel )\cdot\hat{\bf t}=
   \left( \delr' P_\parallel (R') \right) \cdot P_\parallel (R') \cdot U'.
\end{equation}
Upon applying the constraint, $P_\parallel (R') \cdot U'$ can be replaced by
simply $U'.$  One might thus be tempted to pull the following dubious maneuver:
\begin{equation}
\left( \delr' P_\parallel (R') \right) \cdot P_\parallel (R') \cdot U'
  =\left( \delr' P_\parallel (R') \right) \cdot U'
  =\delr' \left( P_\parallel (R') \cdot U' \right)
  =\delr' U'
  =0.
\end{equation}
This is incorrect because {\it after the constraint is applied}, $R'$
{\it and} $U'$ {\it are no longer independent variables}.  We thus had no
right to pull $U'$ inside the $\delr'$ operator, nor did we have a right to
say that $\delr' U'=0.$  This is subtle but important, as the brackets below
are full of things that look like
$\left( \delr' P_\parallel (R') \right) \cdot  U',$
and they are definitely {\it not} zero.

We now present the full set of brackets in the new coordinate system
(omitting the primes since ambiguity should no longer result from doing so).
We find
\begin{eqnarray}
\{ R,R \}&=&-\frac{\epsilon F_0}{m\lambda_B \Omega_B \Upsilon'},
   \label{ghk}\\
\{ R,U \}&=&\frac{1}{m}P_\parallel+\frac{\epsilon}{m\lambda_B^2\Upsilon'}
   F_0\cdot \left[ F'''\cdot P_\parallel
   -\frac{mc}{e}(\delr P_\parallel \cdot U)\right],
   \label{ghl}\\
\{ R,\Theta \}&=&\left\{
                   \begin{array}{ll}
                      0& {\rm (classical \, order)}\\
                     \epsilon \{ R,R \}\cdot {\cal R}& {\rm (higher \, order),}
                   \end{array}
                \right.
   \label{ghm}\\
\{ U,U \}&=&-\frac{\Omega_B}{2m\lambda_B^3} {\cal F}_0 :(F'''\cdot\Xi'^T)
   {\cal F}_0\nonumber\\
   &&\qquad -\frac{1}{m}\left[\left( P_\parallel \cdot\Xi'\cdot 
   \left(\delr P_\parallel \cdot U\right)\right)
                            -\left( P_\parallel \cdot\Xi'\cdot 
   \left(\delr P_\parallel \cdot U\right)\right)^T\right]\nonumber\\
   &-\frac{\epsilon}{m\lambda_B\Omega_B\Upsilon'}
   (\delr P_\parallel\cdot U)^T\cdot F_0\cdot\left(\delr P_\parallel\cdot U\right)
   \label{ghn}\\
\{ U,\Theta \}&=&\left\{
                   \begin{array}{ll}
                      0& {\rm (classical \, order)}\\
                     \epsilon \{ U,R \}\cdot {\cal R}& {\rm (higher \, order),}
                   \end{array}
                 \right.
   \label{gho}\\
\noalign{\hbox{and}}
\{ \Theta ,\mu \}&=&\epsilon^{-1},
   \label{ghp}
\end{eqnarray}
where we have defined
\begin{equation}
\Upsilon'\equiv 1+\frac{\epsilon F_0 : F'''}{2\lambda_B^2},
\label{gkd}
\end{equation}
\begin{equation}
\Xi'\equiv {\bf 1}+\frac{\epsilon F'''\cdot F_0}{\lambda_B^2 \Upsilon'},
\label{gki}
\end{equation}
\begin{equation}
F'''\equiv F'+\frac{mc}{e}\left( (\delr P_\parallel\cdot U)
   -(\delr P_\parallel\cdot U)^T \right),
\label{gkk}
\end{equation}
where $F'$ was defined in Eq.~(\ref{gjs}),
and where the superscripted $T$ means ``transpose.''  Note that $\Upsilon',$
$\Xi'$ and $F'''$ are the boostgauge invariant portions of $\Upsilon,$
$\Xi$ and $F''$; that is, they are related by
\begin{equation}
\Upsilon=\Upsilon',
\end{equation}
\begin{equation}
\Xi=\Xi'+\frac{\epsilon k}{\lambda_B \Omega_B \Upsilon'}\hat{\bf b}
   {\cal Q}\cdot F_0,
\end{equation}
and
\begin{equation}
F''=F'''+\frac{mcK}{e}(\hat{\bf b}{\cal Q}-{\cal Q}\hat{\bf b}).
\end{equation}
These new brackets may be compared to those for the old coordinates, given in
Eqs.~(\ref{ggj}) to (\ref{ggq}).

This Poisson structure has the Casimir function, $P_\perp \cdot U,$
so the constraint Eq.~(\ref{ghi}) is guaranteed to hold for all times if it
holds initially.  The physical motion takes place on the hypersurface for which
this Casimir function has the value zero.

The guiding-center Hamiltonian, Eq.~(\ref{gfz}), can now be expressed in the new
boostgauge invariant coordinates:
\begin{eqnarray}
H_{gc}(R,U,\mu)&=&\mu\Omega_B+\frac{m}{2}U^2+\frac{\epsilon\mu}{2\lambda_B}\nonumber\\
   &&\qquad \times\left[ \left(\frac{e}{mc}\right) F_0:F_1
   +P_\perp:\left( (\delr P_\perp\cdot U)\cdot F_0\right)\right].
   \label{ghq}
\end{eqnarray}
Note that this Hamiltonian is also gyrogauge invariant, since it does not
involve $\Theta.$

There is another way to derive the above manifestly boostgauge invariant
Poisson brackets.  We can write the phase space Lagrangian
corresponding to Eq.~(\ref{gfy}) in manifestly boostgauge invariant form as
follows:
\begin{eqnarray}
L_{gc}(R,U,\mu,\Theta,\dot{R},\dot{\Theta})&=&\left[ \frac{e}{\epsilon c} A
   +mU -\epsilon \mu {\cal R} +O(\epsilon^2 )
   \right]\cdot\dot{R} +\epsilon \mu \dot{\Theta}\nonumber\\
      &&\qquad -\lambda_{1a} U\cdot\hat{\bf c}(R,\Theta)
   -\lambda_{1b} U\cdot\hat{\bf a}(R,\Theta)\nonumber\\
      &&\qquad -H_{gc}(R,U,\mu).
      \label{gjm}
\end{eqnarray}
The action associated with this Lagrangian may be varied to yield the
same equations of motion given by the manifestly boostgauge invariant brackets
and Hamiltonian, but the variation of the action must be performed subject
to the constraint, Eq.~(\ref{ghi}).  Hence we have introduced the Lagrange
multipliers, $\lambda_{1a}$ and $\lambda_{1b}.$
Note that varying an action subject to
a constraint causes the constraint to appear as a Casimir of the resulting
Poisson structure; recall the example of this phenomenon given in
Subsection~\ref{ybb}.

The equations of motion in this coordinate system are then easily found either
by using the Poisson brackets given in Eqs.~(\ref{ghk}) through (\ref{ghp}) with the
Hamiltonian given in Eq.~(\ref{ghq}), or by finding the Euler-Lagrange equations
from the phase space Lagrangian given in Eq.~(\ref{gjm}).  The results are
\begin{eqnarray}
\dot{R} &=&
  U-\frac{\epsilon\mu F_0\cdot\delr\Omega_B}{m\lambda_B\Omega_B\Upsilon'}\nonumber\\
  &&\qquad +\frac{\epsilon}{\lambda_B^2\Upsilon'} F_0
  \cdot\left( F'''\cdot P_\parallel
  -\frac{mc}{e}(\delr P_\parallel\cdot U)\right)\cdot U
  \label{gjo}
\end{eqnarray}
\begin{eqnarray}
\dot{U} &=&
  -\frac{\mu}{m}P_\parallel\cdot\delr\Omega_B
  +\frac{\epsilon\mu}{m\lambda_B^2\Upsilon'}\left( F'''\cdot P_\parallel
  -\frac{mc}{e}(\delr P_\parallel\cdot U)\right)^T\cdot
  F_0\cdot\delr\Omega_B\nonumber\\
  &&\qquad -\frac{\Omega_B}{2\lambda_B^3} {\cal F}_0:(F'''\cdot\Xi'^T)
  {\cal F}_0\cdot U\nonumber\\
  &&\qquad -\left[ P_\parallel\cdot\Xi'\cdot (\delr P_\parallel\cdot U)
  -(\delr P_\parallel\cdot U)^T\cdot\Xi'^T\cdot P_\parallel\right]\cdot U\nonumber\\
  &&\qquad -\frac{\epsilon}{\lambda_B\Omega_B\Upsilon'}
  \left[ (\delr P_\parallel\cdot U)^T\cdot F_0\cdot
  (\delr P_\parallel\cdot U)\right]\cdot U
  \label{gjp}
\end{eqnarray}
\begin{equation}
\dot{\mu}=0
\label{gjy}
\end{equation}
\begin{equation}
\dot{\Theta}=\frac{\Omega_B}{\epsilon}+\epsilon {\cal R}\cdot\dot{R}
   +\frac{\epsilon}{2\lambda_B}
   \left[ \left(\frac{e}{mc}\right) F_0:F_1
   +P_\perp:\left( (\delr P_\perp\cdot U)\cdot F_0\right)\right].
\end{equation}
These equations of motion may be compared term for term with Eqs.~(\ref{ggr})
through (\ref{ggs}).  In the equation for $\dot{R},$ note that the parallel
motion is given simply by $U.$  The second term contains the grad-$B$ drift,
and the third term contains the curvature and polarization drifts.  The first
term of $\dot{U}$ contains the mirroring force, and the force due to the
parallel electric field arises from the terms that contain $F_1$ (via their
dependence on $F'''$).  Of course, $\dot{\mu}$ still vanishes, and the
equation for $\dot{\Theta}$ compares term for term with Eq.~(\ref{ggs}) in an
obvious way.

\subsection{Manifest Boostgauge and Gyrogauge Invariance}
Now we can use the same techniques to make our results gyrogauge invariant
as well.  To do this, we would like to replace the coordinate $\Theta$ by the
new coordinate
\begin{equation}
\hat{\alpha}\equiv\hat{\bf a}(R,\Theta).
\end{equation}
The inverse transformation would then be
\begin{equation}
\Theta=\arctan \left( -\frac{\hat{\alpha}\cdot\hat{\bf e}_3(R)}
                            {\hat{\alpha}\cdot\hat{\bf e}_2(R)}\right).
\end{equation}
Note that $\hat{\alpha},$ like $\Theta,$ has only one degree
of freedom, even though it is a four vector.  This is because it is subject 
to the constraints
\begin{equation}
P_\parallel (R)\cdot\hat{\alpha}=0,
\label{ghr}
\end{equation}
and
\begin{equation}
\hat{\alpha}\cdot\hat{\alpha}=1.
\label{ghs}
\end{equation}

In order to deal with this in a proper fashion, we have to use the same
techniques that we used above to get boostgauge invariant brackets.
Write the coordinate transformation
\begin{eqnarray}
R' &=&R\nonumber\\
U' &=&K\hat{\bf t}(R,{\cal B})+\frac{1}{\sqrt{C_3}}\left(
     \frac{C_{1a}}{\lambda_B(R)} \hat{\bf c}(R,\Theta)
     +C_{1b}\hat{\bf a}(R,\Theta)\right)\nonumber\\
\mu' &=&\mu\nonumber\\
{\hat{\alpha}}' &=&\sqrt{C_3}\hat{\bf a}(R,\Theta)+\frac{1}{K}\left(
     -C_{2a}\hat{\bf t}(R,{\cal B})+\frac{C_{2b}}{\lambda_B(R)}
     \hat{\bf b}(R,{\cal B})\right)\label{gje}
\end{eqnarray}
The inverse transformation is then
\begin{eqnarray}
R &=&R'\nonumber\\
K &=&\sqrt{-U' \cdot P_\parallel (R') \cdot U'}\nonumber\\
{\cal B} &=&\tanh^{-1}\left(
           -\frac{U' \cdot {\hat{\bf e}}_1(R')}{U' \cdot {\hat{\bf e}}_0(R')}
           \right)\nonumber\\
\mu &=&\mu'\nonumber\\
\Theta &
  =\arctan \left( -\frac{{\hat{\alpha}}'\cdot {\hat{\bf e}}_3(R')}
                        {{\hat{\alpha}}'\cdot {\hat{\bf e}}_2(R')} \right)\nonumber\\
C_{1a}&=&U'\cdot F_0(R')\cdot {\hat{\alpha}}'\nonumber\\
C_{1b}&=&U'\cdot P_\perp (R')\cdot {\hat{\alpha}}'\nonumber\\
C_{2a}&=&U'\cdot P_\parallel (R')\cdot {\hat{\alpha}}'\nonumber\\
C_{2b}&=&U'\cdot {\cal F}(R')\cdot {\hat{\alpha}}'\nonumber\\
C_{3 }&=&{\hat{\alpha}}'\cdot P_\perp (R')\cdot {\hat{\alpha}}'
      \label{gjf}
\end{eqnarray}
We demand that $C_{1a},$ $C_{1b},$ $C_{2a},$ $C_{2b}$ and $C_3$ are
Casimir functions, and that
the physical motion takes place on the submanifold defined by
$C_{1a}=C_{1b}=C_{2a}=C_{2b}=0$ and $C_3=1.$

We can now write the Poisson bracket relations among the new coordinates.
We use the easily verified relations
\begin{equation}
\delr \hat{\bf t}=(\delr P_\parallel )\cdot\hat{\bf t}+{\cal Q}\hat{\bf b}
\end{equation}
\begin{equation}
\delr \hat{\bf b}=(\delr P_\parallel )\cdot\hat{\bf b}-{\cal Q}\hat{\bf t}
\end{equation}
\begin{equation}
\delr \hat{\bf c}=(\delr P_\perp )\cdot\hat{\bf c}+{\cal R}\hat{\bf a}
\end{equation}
\begin{equation}
\delr \hat{\bf a}=(\delr P_\perp )\cdot\hat{\bf a}-{\cal R}\hat{\bf c}.
\end{equation}
Note that, because our results are guaranteed to be both boostgauge and
gyrogauge invariant, all terms involving ${\cal Q}$ and ${\cal R}$ will
cancel, leaving a {\it manifestly} boostgauge and gyrogauge invariant result.
Also note that the Hamiltonian $H_{gc}(R,U,\mu),$ given by Eq.~(\ref{ghq}),
is already manifestly gyrogauge invariant (this is because it is
$\Theta$-independent).  The new manifestly boostgauge and gyrogauge invariant
brackets are then
\begin{eqnarray}
\{ R,R \}&=&-\frac{\epsilon F_0}{m\lambda_B \Omega_B \Upsilon'},
   \label{gjg}\\
\{ R,U \}&=&\frac{1}{m}P_\parallel+\frac{\epsilon}{m\lambda_B^2\Upsilon'}
   F_0\cdot \left[ F'''\cdot P_\parallel
   -\frac{mc}{e}(\delr P_\parallel \cdot U)\right],
   \label{gjh}\\
\{ U,U \}&=&-\frac{\Omega_B}{2m\lambda_B^3} {\cal F}_0 :(F'''\cdot\Xi'^T)
   {\cal F}_0\nonumber\\
   &&\qquad -\frac{1}{m}\left( P_\parallel \cdot\Xi'\cdot 
   (\delr P_\parallel \cdot U)-(\delr P_\parallel \cdot U)^T\cdot
   \Xi'^T\cdot P_\parallel \right)\nonumber\\
   &&\qquad -\frac{\epsilon}{m\lambda_B\Omega_B\Upsilon'}
   (\delr P_\parallel\cdot U)^T\cdot F_0\cdot (\delr P_\parallel\cdot U)
   \label{gji}\\
\{ R,\hat{\alpha} \}&=&\left\{
                   \begin{array}{ll}
                      0
                         & {\rm (classical \, order)}\\
                      \epsilon \{ R,R \}\cdot \delr P_\perp\cdot\hat{\alpha}
                         & {\rm (higher \, order),}
                   \end{array}
                \right.
   \label{ght}\\
\{ U,\hat{\alpha} \}&=&\left\{
                   \begin{array}{ll}
                      0
                         & {\rm (classical \, order)}\\
                      \epsilon \{ U,R \}\cdot \delr P_\perp\cdot\hat{\alpha}
                         & {\rm (higher \, order),}
                   \end{array}
                \right.
   \label{ghu}\\
\{ \hat{\alpha} ,\mu \}&=&\frac{1}{\epsilon\lambda_B}F_0\cdot\hat{\alpha},
   \label{ghv}\\
\noalign{\hbox{and}}
\{ \hat{\alpha} ,\hat{\alpha} \}&=&-\frac{\epsilon}{m\lambda_B\Omega_B\Upsilon}
   (\delr P_\perp\cdot\hat{\alpha})^T\cdot F_0\cdot(\delr P_\perp\cdot
   \hat{\alpha}). 
   \label{ghw}
\end{eqnarray}
This Poisson structure has the Casimir functions, $P_\perp\cdot U,$
$P_\parallel\cdot\hat{\alpha},$ and $\hat{\alpha}\cdot\hat{\alpha}.$  This
insures that the constraint Eqs.~(\ref{ghi}), (\ref{ghr}) and (\ref{ghs})
will hold at all times if they hold initially.
The physical motion takes place on
the hypersurface for which the first two of these Casimir functions have the 
value zero and the third has the value one.

Note that $\hat{\alpha},$ like $\Theta,$ has nonvanishing brackets with
$R$ and $U$ at higher order.  Once again, however, the set of functions of
$R$ and $U$ form a subset of the set of all possible phase functions that
is closed under the operation of these Poisson brackets; also, $H_{gc}$ is
independent of $\hat{\alpha}.$  So we can still reduce to the guiding-center
description.

Next, we note that these results could have been derived by varying the 
action corresponding to the phase space Lagrangian obtained by rewriting
Eq.~(\ref{gfy}) in manifestly boostgauge and gyrogauge invariant format,
\begin{eqnarray}
L_{gc}&=&\left[ \frac{e}{\epsilon c} A +mU +O(\epsilon^2 )
   \right]\cdot\dot{R} -\frac{\epsilon\mu}{\lambda_B}\hat{\alpha}\cdot
   F_0\cdot \dot{\hat{\alpha}}\nonumber\\
      &&\qquad -\lambda_{1a} U\cdot F_0\cdot\hat{\alpha}
              -\lambda_{1b} U\cdot P_\perp\cdot\hat{\alpha}\nonumber\\
      &&\qquad -\lambda_{2a} U\cdot P_\parallel\cdot\hat{\alpha}
              -\lambda_{2b} U\cdot {\cal F}_0\cdot\hat{\alpha}\nonumber\\
      &&\qquad -\lambda_3    \hat{\alpha}\cdot P_\perp\cdot\hat{\alpha}
              -H_{gc}(R,U,\mu).
      \label{gjj}
\end{eqnarray}
This must be varied subject to the constraints, Eqs.~(\ref{ghi}), (\ref{ghr})
and (\ref{ghs}).  We have enforced these constraints by introducing the scalar
Lagrange multipliers, $\lambda_{1a},$ $\lambda_{1b},$ $\lambda_{2a},$
$\lambda_{2b},$ and $\lambda_{3}.$  Note that the term involving $\cal{R}$ has
disappeared from $\Gamma_{gc}$ when written in these coordinates, because
$-\mu {\cal R}\cdot\dot{R}+\mu\dot{\Theta}=-\mu\hat{\alpha}\cdot F_0
\cdot\dot{\hat{\alpha}}/\lambda_B.$

We are going to need these Lagrange multipliers in Chapter~\ref{yaq}, so
we compute them here for reference.  They are rather easy to calculate,
especially since we already know the Poisson brackets.  The Euler-Lagrange
equations for coordinates $U$ and $\hat{\alpha}$ are
\begin{equation}
0=m\dot{R}-\lambda_{1a}F_0\cdot\hat{\alpha}
  -\lambda_{1b}P_\perp\cdot\hat{\alpha}
  -\frac{\partial H_{gc}}{\partial U}
\end{equation}
and
\begin{equation}
\frac{d}{d\tau}
\left(\frac{\epsilon\mu}{\lambda_B}\hat{\alpha}\cdot F_0\right)
  =\frac{\epsilon\mu}{\lambda_B} F_0\cdot\dot{\hat{\alpha}}
  +\lambda_{2a}U\cdot P_\parallel+\lambda_{2b}U\cdot {\cal F}_0
  +2\lambda_3 P_\perp\cdot\hat{\alpha}
  +\frac{\partial H_{gc}}{\partial \hat{\alpha}},
\end{equation}
respectively.  Upon multiplication by $\hat{\alpha}\cdot F_0$ and
$\hat{\alpha},$ the first of these yields
\begin{equation}
\lambda_{1a}=-\frac{1}{\lambda_B^2}\hat{\alpha}\cdot F_0\cdot
  \left(m\dot{R}-\frac{\partial H_{gc}}{\partial U}\right)
\end{equation}
and
\begin{equation}
\lambda_{1b}=\hat{\alpha}\cdot
  \left(m\dot{R}-\frac{\partial H_{gc}}{\partial U}\right),
\end{equation}
respectively.  Upon multiplication by $U,$ $U\cdot {\cal F}_0$ and
$\hat{\alpha},$ the second yields
\begin{equation}
\lambda_{2a}=\frac{1}{U^2}U\cdot\left[\epsilon\mu\hat{\alpha}\cdot
  \left(\frac{F_0}{\lambda_B}\right)\dell\cdot\dot{R}
  -\frac{\partial H_{gc}}{\partial \hat{\alpha}}\right]
\end{equation}
and
\begin{equation}
\lambda_{2b}=\frac{-1}{\lambda_B^2 U^2}U\cdot {\cal F}_0\cdot
  \left[\epsilon\mu\hat{\alpha}\cdot
  \left(\frac{F_0}{\lambda_B}\right)\dell\cdot\dot{R}
  -\frac{\partial H_{gc}}{\partial \hat{\alpha}}\right]
\end{equation}
and
\begin{equation}
\lambda_3=\frac{\epsilon\mu}{\lambda_B}\dot{\hat{\alpha}}\cdot F_0\cdot
  \hat{\alpha}+\frac{1}{2}\hat{\alpha}\cdot\left[\epsilon\mu\hat{\alpha}\cdot
  \left(\frac{F_0}{\lambda_B}\right)\dell\cdot\dot{R}
  -\frac{\partial H_{gc}}{\partial \hat{\alpha}}\right],
\end{equation}
respectively.  Note that, in perfect analogy with Eq.~(\ref{meg}), these
results can be cast in the form
\begin{equation}
\lambda_\nu=\xi^\alpha_\nu\frac{\partial H_{gc}}{\partial Z^\alpha},
\end{equation}
where the label $\nu$ runs over all the constraints present (1a, 1b, 2a, 2b, 3),
and where
\begin{eqnarray}
\xi_{1a}^R &=&-\frac{m}{\lambda_B^2}\hat{\alpha}\cdot F_0\cdot \{ R,R \}\nonumber\\
\xi_{1a}^U &=&-\frac{m}{\lambda_B^2}\hat{\alpha}\cdot F_0\cdot \left(
  \{ R,U \}-\frac{1}{m} {\bf 1}\right)\nonumber\\
\xi_{1a}^\mu &=&0\nonumber\\
\xi_{1a}^{\hat{\alpha}} &=&-\frac{m}{\lambda_B^2}\hat{\alpha}\cdot F_0
  \cdot \{ R,\hat{\alpha} \}\nonumber\\
\noalign{\hbox{and}}
\xi_{1b}^R &=&m\hat{\alpha}\cdot \{ R,R \}\nonumber\\
\xi_{1b}^U &=&m\hat{\alpha}\cdot \left(
  \{ R,U \}-\frac{1}{m} {\bf 1}\right)\nonumber\\
\xi_{1b}^\mu &=&0\nonumber\\
\xi_{1b}^{\hat{\alpha}} &=&m\hat{\alpha}\cdot \{ R,\hat{\alpha} \}\nonumber\\
\noalign{\hbox{and}}
\xi_{2a}^R &=&\frac{1}{U^2}U\cdot\left[\epsilon\mu\hat{\alpha}\cdot\left(
  \frac{F_0}{\lambda_B}\right)\dell\cdot \{ R,R \}\right]\nonumber\\
\xi_{2a}^U &=&\frac{1}{U^2}U\cdot\left[\epsilon\mu\hat{\alpha}\cdot\left(
  \frac{F_0}{\lambda_B}\right)\dell\cdot \{ R,U \}\right]\nonumber\\
\xi_{2a}^\mu &=&0\nonumber\\
\xi_{2a}^{\hat{\alpha}} &=&\frac{1}{U^2}U\cdot
  \left[\epsilon\mu\hat{\alpha}\cdot\left(
  \frac{F_0}{\lambda_B}\right)\dell\cdot \{ R,\hat{\alpha} \}-{\bf 1}\right]\nonumber\\
\noalign{\hbox{and}}
\xi_{2b}^R &=&\frac{-1}{U^2}U\cdot {\cal F}_0\cdot
  \left[\epsilon\mu\hat{\alpha}\cdot\left(
  \frac{F_0}{\lambda_B}\right)\dell\cdot \{ R,R \}\right]\nonumber\\
\xi_{2b}^U &=&\frac{-1}{U^2}U\cdot {\cal F}_0\cdot
  \left[\epsilon\mu\hat{\alpha}\cdot\left(
  \frac{F_0}{\lambda_B}\right)\dell\cdot \{ R,U \}\right]\nonumber\\
\xi_{2b}^\mu &=&0\nonumber\\
\xi_{2b}^{\hat{\alpha}} &=&\frac{-1}{\lambda_B^2 U^2}U\cdot {\cal F}_0\cdot
  \left[\epsilon\mu\hat{\alpha}\cdot\left(
  \frac{F_0}{\lambda_B}\right)\dell\cdot \{ R,\hat{\alpha} \}-{\bf 1}\right]\nonumber\\
\noalign{\hbox{and}}
\xi_3^R &=&-\frac{\epsilon\mu}{\lambda_B}\hat{\alpha}\cdot F_0 \cdot
  \{ \hat{\alpha},R \}+\frac{1}{2}\hat{\alpha}\cdot \left[\epsilon\mu
  \hat{\alpha}\cdot\left(\frac{F_0}{\lambda_B}\right)\dell\cdot
  \{ R,R \}\right]\nonumber\\
\xi_3^U &=&-\frac{\epsilon\mu}{\lambda_B}\hat{\alpha}\cdot F_0 \cdot
  \{ \hat{\alpha},U \}+\frac{1}{2}\hat{\alpha}\cdot \left[\epsilon\mu
  \hat{\alpha}\cdot\left(\frac{F_0}{\lambda_B}\right)\dell\cdot
  \{ R,U \}\right]\nonumber\\
\xi_3^\mu &=&-\frac{\epsilon\mu}{\lambda_B}\hat{\alpha}\cdot F_0 \cdot
  \{ \hat{\alpha},\mu \}\nonumber\\
\xi_3^{\hat{\alpha}}&=&-\frac{\epsilon\mu}{\lambda_B}\hat{\alpha}\cdot F_0 \cdot
  \{ \hat{\alpha},\hat{\alpha} \}+\frac{1}{2}\hat{\alpha}\cdot
  \left[\epsilon\mu
  \hat{\alpha}\cdot\left(\frac{F_0}{\lambda_B}\right)\dell\cdot
  \{ R,\hat{\alpha} \}-{\bf 1}\right].
  \label{gjn}
\end{eqnarray}

Finally, we note that the equations of motion in these
coordinates are easily found either by using the Poisson brackets
given in Eqs.~(\ref{gjg}) through (\ref{ghw}) with the
Hamiltonian given in Eq.~(\ref{ghq}), or by finding the Euler-Lagrange equations
from the phase space Lagrangian given in Eq.~(\ref{gjj}).  The results are
\begin{eqnarray}
\dot{R} &=&
  U-\frac{\epsilon\mu F_0\cdot\delr\Omega_B}{m\lambda_B\Omega_B\Upsilon'}\nonumber\\
  &&\qquad +\frac{\epsilon}{\lambda_B^2\Upsilon'} F_0
  \cdot\left( F'''\cdot P_\parallel
  -\frac{mc}{e}(\delr P_\parallel\cdot U)\right)\cdot U
  \label{gjq}
\end{eqnarray}
\begin{eqnarray}
\dot{U} &=&
  -\frac{\mu}{m}P_\parallel\cdot\delr\Omega_B
  +\frac{\epsilon\mu}{m\lambda_B^2\Upsilon'}\left( F'''\cdot P_\parallel
  -\frac{mc}{e}(\delr P_\parallel\cdot U)\right)^T\cdot
  F_0\cdot\delr\Omega_B\nonumber\\
  &&\qquad -\frac{\Omega_B}{2\lambda_B^3} {\cal F}_0:(F'''\cdot\Xi'^T)
  {\cal F}_0\cdot U\nonumber\\
  &&\qquad -\left[ P_\parallel\cdot\Xi'\cdot (\delr P_\parallel\cdot U)
  -(\delr P_\parallel\cdot U)^T\cdot\Xi'^T\cdot P_\parallel\right]\cdot U\nonumber\\
  &&\qquad -\frac{\epsilon}{\lambda_B\Omega_B\Upsilon'}
  \left[ (\delr P_\parallel\cdot U)^T\cdot F_0\cdot
  (\delr P_\parallel\cdot U)\right]\cdot U
  \label{gjr}
\end{eqnarray}
\begin{equation}
\dot{\mu}=0
\label{gjx}
\end{equation}
\begin{equation}
\dot{\hat{\alpha}}=\frac{e}{\epsilon mc} F_0\cdot\hat{\alpha}
  +\hat{\alpha}\cdot (P_\perp\dell)\cdot\dot{R}.
  -\frac{\epsilon}{2\lambda_B^2}\hat{\alpha}\cdot F_0\cdot
  \left[ \left(\frac{e}{mc}\right) F_0:F_1
  +P_\perp:\left( (\delr P_\perp\cdot U)\cdot F_0\right)\right].
\end{equation}
Note that Eqs.~(\ref{gjq}) through (\ref{gjx}) are identical to
the corresponding equations in the last subsection.  These were gyrogauge
invariant anyway, and so were unaffected by the manipulations carried out in
this subsection.  The equation for $\dot{\Theta}$ has been replaced by an
equation for $\dot{\hat{\alpha}}$; the two may, however, be compared term for
term in an obvious way.

%% file: reloc.tex
\chapter{Relativistic Oscillation-Center Theory}
\textheight=8.4truein
\setcounter{equation}{617}
\pagestyle{myheadings}
\markboth{Relativistic Oscillation-Center Theory}{Relativistic Oscillation-Center Theory}
\label{yap}
\section{Discussion}
In this chapter, we shall consider the perturbation of a guiding center
due to the presence of an electromagnetic wave of eikonal form.  In
doing so, we shall take as our unperturbed problem the guiding-center 
equations of motion, as derived in Chapter~\ref{yac}.  Thus we are effectively
using the superconvergent Lie transform procedure as described
in Subsection~\ref{yab}.

We are interested in understanding the response of
the guiding center to the presence of the wave.  Towards this end, we seek a
transformation to a new system of coordinates in which the wave perturbation
is removed.  Neglecting resonant phenomena, it turns out that it is possible
to do this to first order, but not to second order.  At second order, there
remains an averaged residual perturbation to the Hamiltonian that gives rise
to the ponderomotive force exerted by the wave on the guiding center.  
Thus, after we transform away the rapid fluctuations in the guiding-center
motion, we are left with the slower ponderomotive effects.

An analogy with the guiding-center problem may be helpful here.  In that
calculation, we averaged over the rapid gyromotion to find the slower drift
motion.  The thing that is drifting is then called a ``guiding center.''
A guiding center is a fictitious object whose position and momentum are
the gyroaverage of the particle position and momentum, respectively.  
Furthermore, a guiding center may be thought of as having an intrinsic or
{\it spin} angular momentum equal to the {\it orbital} angular momentum of
the underlying gyrating particle.  Thus, 
by finding the averaging transformation that
eliminates the fast degree of freedom, we have discovered a new 
``macroparticle'' that lives on the slow time scale, but whose properties 
derive from those of the original charged particle gyrating on the fast
time scale.

Similarly, when a perturbing wave is present and we transform away the
associated rapid fluctuations, the residual ponderomotive forces may be
thought of as acting on a new ``macroparticle'' that is averaged over a
wave oscillation time scale.  We call this new object an ``oscillation
center.''  Whereas an individual charged particle feels wave
fluctuations on a rapid time scale, an oscillation center feels only the
slower ponderomotive effects; it also feels resonant effects (since
these are also slow and do not average away), but we shall ignore these
in our treatment.  Thus, a kinetic equation for a plasma of oscillation
centers would contain only ponderomotive forces and resonant effects.

The averaged $n$th-order part of the ponderomotive Hamiltonian is called $K_n,$
and we shall derive this for a relativistic guiding center.  As has already 
been noted, $K_1$ vanishes if we neglect resonant effects.
It was discovered by Cary and Kaufman that there exists an intimate connection
between the ponderomotive Hamiltonian and the plasma's response to a wave.
Specifically, $K_2$ is a quadratic form in the
amplitude of the perturbing wave, and the kernel of this quadratic form is
the functional derivative of the linear susceptibility with respect to the
distribution function.  Subsequently, it was found by Kaufman that this
relationship persists to higher order;  that is, nonlinear corrections to the
susceptibility are related to $K_3,$ etc.

In the traditional approach to studying plasma response to a wave, one
begins with the field equations and the kinetic equation, and studies
perturbations in the fields and the distribution function about an 
equilibrium.  Though this approach is not as systematic as ours, it has at
least one advantage:  The vector potential never appears, so all results
obtained by such an analysis are guaranteed to be {\it manifestly} gauge
invariant.  In contrast, Hamiltonian or Lagrangian approaches to
ponderomotive theory seem to require the use of the vector potential, so
past attempts along these lines have produced results whose gauge invariance
was either not established, or established only by laborious calculation
after the fact.

In this chapter, we shall find that eikonal wave perturbations to the 
Lagrangian action for a relativistic charged particle in the guiding-center 
representation can be written in {\it manifestly} gauge-invariant form.  To 
do this, it is necessary to abandon the usual approach of expanding the
eikonal wave perturbation in a series of Bessel functions of $k_\perp\rho.$
Instead, we first perform a Lagrangian gauge transformation, and then we
expand in a series of functions that are related to indefinite integrals of
Bessel functions.  This allows us to develop an oscillation-center theory
to arbitrarily high order in the wave amplitude expansion parameter, and be
guaranteed of {\it manifest} gauge invariance at every step of the way.
Thus, we can enjoy the benefits of the systematic Lie transform approach to
ponderomotive theory without fear of losing manifest gauge invariance.

\section{Eikonal Wave Perturbation}
\label{yba}
In single-particle phase space coordinates, an eikonal wave has a four 
potential of the form
\begin{equation}
A_w(r)=\tilde{A}(r)\exp\left( \frac{i}{\epsilon}\psi (r)\right)+{\rm c.c.},
\label{oay}
\end{equation}
where $\tilde{A}$ is the amplitude and $\psi$ is the phase, and where c.c.
denotes the expression's complex conjugate.  The derivative
of $\psi$ with respect to spacetime position is the four wavevector, {\bf k}:
\begin{equation}
{\bf k}=\delr\psi (r).
\end{equation}
Both $\tilde{A}$ and {\bf k} are slowly varying functions of $r.$  That is, an
eikonal wave is {\it locally} a plane wave.  To reflect this,
we have placed $1/\epsilon$ in front of the phase.  Thus, the derivative of
$A_w$ with respect to $r$ is $i{\bf k}A_w/\epsilon$ plus terms of order unity
that involve derivatives of $\tilde{A}$ or of {\bf k}.  

Furthermore, in this work, we shall take this eikonal expansion parameter to 
be equal to the guiding-center expansion parameter (hence, it is no coincidence
that we are calling it $\epsilon$).  This means that we are considering waves
whose characteristic wavelengths are on the order of a gyroradius, and
whose characteristic frequencies are on the order of a gyrofrequency.

We shall now consider the effect of such a wave on the single particle
action one form in Eq.~(\ref{gbn}).  Replacing $A$ in that equation by
$A+\lambda A_w,$ we write
\begin{equation}
\gamma=\gamma''+\lambda\gamma_w,
\label{obx}
\end{equation}
where $\gamma_w$ is the perturbation in the action one form due to the 
wave, or
\begin{equation}
\gamma_w=\frac{e}{c}\tilde{A}(r)\cdot dr 
   \exp\left( \frac{i}{\epsilon}\psi (r)\right)+{\rm c.c.}
\label{oaa}
\end{equation}
Note that we have introduced a new expansion parameter, $\lambda,$ to
order the wave amplitude.  For the time being, we shall not compare
$\lambda$ and $\epsilon,$ though more will be said about this later.

As was remarked earlier, our starting point for the oscillation center
Lie transform will be the guiding-center equations of motion.  Hence, it
is necessary to write $\gamma_w$ in guiding-center coordinates (the
above form for $\gamma_w$ is in particle coordinates).  We apply the 
guiding-center Lie transform to the above equation for $\gamma$ to get
\begin{equation}
\Gamma=\Gamma_{gc}+\lambda\Gamma_w,
\end{equation}
where $\Gamma_{gc}$ is the guiding-center action one form, calculated in
Chapter~\ref{yac}.  Then, $\Gamma_w$ is given by
\begin{equation}
\Gamma_w=\exp (-\epsilon {\cal L}_g) \gamma_w,
\label{oab}
\end{equation}
where $g$ is the generator for the guiding-center transformation.  

Note that we are working only to first order in $\epsilon.$  To this order
we can take $g^r=-\rho \hat{\bf a},$ where $\rho\equiv w/\Omega_B.$  All
other components of $g$ are unnecessary, and may be ignored.
We shall use the boostgauge invariant set of coordinates ($R, U, \mu, \theta$)
described in Section~\ref{yao}.

\section{Manifest Gauge Invariance}
At this point in the calculation, the usual approach is to apply the Lie
transform in Eq.~(\ref{oab}) by simply substituting $R+\rho \hat{\bf a}$
for $r$ in Eq.~(\ref{oaa}).  This is straightforward, and the result is
\begin{equation}
\Gamma_w=\frac{e}{c}\left(\tilde{A}\cdot dR
         +\frac{\epsilon\tilde{A}\cdot\hat{\bf a}d\mu}{m\rho\Omega_B}
         +\epsilon\rho\tilde{A}\cdot\hat{\bf c}d\theta\right)
          \exp\left(\frac{i}{\epsilon}\psi\right)
          \exp (i\rho\hat{\bf a}\cdot {\bf k})+{\rm c.c.},
\label{oac}
\end{equation}
where we have retained the leading nonvanishing order for each component of
the one form, and where it is understood that all quantities on the right
(such as $\tilde{A}$ and $\hat{\bf c}$) are now evaluated at $R.$
Since $\hat{\bf a}\cdot {\bf k}$ is oscillatory, the second exponential in the
above expression gives rise to a series of Bessel functions of 
${\rm k}_\perp\rho.$

Unfortunately, the above expression for $\Gamma_w$ does {\it not} possess
manifest gauge invariance.  To understand why this is, we must qualify what
we mean by ``manifest gauge invariance.''  A term in the action one form is 
gauge invariant if it is unchanged to within a Lagrangian gauge transformation
when $\tilde{A}$ is replaced by $\tilde{A}+i{\bf k}\Lambda,$ where $\Lambda$
is any slowly varying scalar function of position.  Thus, the quantity
\begin{equation}
\tilde{F}\equiv i({\bf k}\tilde{A}-\tilde{A}{\bf k})
\end{equation}
is gauge invariant since it is unchanged by this transformation.  The
quantity $\tilde{A}\cdot dR\exp (i\psi/\epsilon)$ is also gauge invariant 
since it transforms to itself plus the term
\begin{equation}
i\Lambda{\bf k}\cdot dR\exp\left(\frac{i}{\epsilon}\psi\right)
     =d\left[\epsilon\Lambda\exp\left(\frac{i}{\epsilon}\psi\right)\right]
\end{equation}
(where we have neglected higher-order terms in $\epsilon$), and this can be
removed by a Lagrangian gauge transformation.  We shall say that a term is
{\it manifestly} gauge invariant if it has the form 
$\tilde{A}\cdot dR\exp (i\psi/\epsilon),$ or if it depends on $\tilde{A}$
only through its dependence on $\tilde{F}.$

Thus the first term on the right hand side of Eq.~(\ref{oac}) is manifestly
gauge invariant, but the other two terms are not.  They are gauge invariant
(as they must be), since to leading order in $\epsilon$ we have
\begin{eqnarray}
i\left(\frac{\epsilon\Lambda {\bf k}\cdot\hat{\bf a}d\mu}{m\rho\Omega_B}
+\epsilon\rho\Lambda {\bf k}\cdot\hat{\bf c}d\theta\right)
&&\exp\left(\frac{i}{\epsilon}\psi\right)
\exp (i\rho\hat{\bf a}\cdot {\bf k})\nonumber\\
   &=&d\left[\epsilon^2\rho\Lambda {\bf k}\cdot\hat{\bf a}
    \exp\left(\frac{i}{\epsilon}\psi\right)
    \exp (i\rho\hat{\bf a}\cdot {\bf k})\right]
   \label{oaw}
\end{eqnarray}
and this can be removed by a Lagrangian gauge transformation, but they are 
not {\it manifestly} gauge invariant.

If we were to use Eq.~(\ref{oac}) as the starting point for our ponderomotive
theory, we would obtain results for $K_n$ that are not manifestly gauge
invariant.  We could get around this problem if there were some way of
manipulating Eq.~(\ref{oac}) into manifestly gauge-invariant form.  It turns
out that this can be done by making a particular Lagrangian gauge
transformation, but this transformation is far from obvious and needs to be
motivated.  As we shall now see, this motivation comes from the homotopy
formula.

Return to Eq.~(\ref{oab}), and expand the exponential in a series of
Lie derivatives
\begin{equation}
\Gamma_w=\sum_{j=0}^{\infty}\frac{(-\epsilon)^j}{j!}{\cal L}_g^j\gamma_w.
\end{equation}
Applying the generalized homotopy formula, Eq.~(\ref{mbz}), we get
\begin{equation}
\Gamma_w=\gamma_w+\sum_{j=1}^{\infty}\frac{(-\epsilon)^j}{j!}
         \left[ (i_gd)^j+(di_g)^j \right] \gamma_w.
\end{equation}
Note that we have split off the $j=0$ term from the sum because Eq.~(\ref{mbz})
is valid only for $j\geq 1.$  The above may now be written in the suggestive
form
\begin{equation}
\Gamma_w=\left[ \gamma_w+\sum_{j=1}^{\infty}\frac{(-\epsilon)^j}{j!}
         (i_gd)^j\gamma_w\right]+d\left[\sum_{j=1}^{\infty}
         \frac{(-\epsilon)^j}{j!}i_g(di_g)^{j-1}\gamma_w\right].
\label{oad}
\end{equation}
Note that the second term in square brackets is an exact one form, and may
therefore be removed by a Lagrangian gauge transformation.  The first term in
square brackets has two pieces:  The first is $\gamma_w$ itself, which we know
is manifestly gauge invariant.  The second is a series of terms all of which
have the operator $i_gd,$ raised to some power, operating on $\gamma_w.$
Thus, in all these terms, the very first operator to be applied to $\gamma_w$
is the exterior derivative.  Now
\begin{equation}
d\gamma_w=\frac{e}{2\epsilon c}\tilde{F}:dr\wedge dr
          \exp\left( \frac{i}{\epsilon}\psi (r)\right)+{\rm c.c.}
\end{equation}
(plus higher-order terms), and this is manifestly gauge invariant.  Subsequent
applications of $i_g$ and $d$ preserve this manifest gauge invariance.  Thus
the term in the first square brackets on the right hand side of Eq.~(\ref{oad})
is manifestly gauge invariant.  Thus, Eq.~(\ref{oad}) gives us the Lagrangian
gauge transformation that leaves $\Gamma_w$ in {\it manifestly} gauge invariant
form.

At this point, one may wonder why we have bothered to keep all the terms in the
above series when we have said that we are interested in only the lowest
nonvanishing order in $\epsilon.$  Note that when we apply differential
operators to $\gamma_w,$ as given by Eq.~(\ref{oaa}), we pull out factors 
of $1/\epsilon.$  This means that even terms with very high $j$ can 
make order unity contributions.  Thus, it is important to keep {\it all} the 
terms of the series as given above.  This situation arises as a consequence of 
the nonanalyticity of $\gamma_w$ in $\epsilon.$  It will become more clear 
momentarily.

To proceed, we need expressions for $(i_gd)^j\gamma_w$ and
$i_g(di_g)^j\gamma_w,$ for $j\ge 1.$  To get such expressions, we simply 
evaluate them for the first few values of $j,$ notice the pattern, and prove it
by mathematical induction.  The results are
\begin{eqnarray}
\left[ (i_gd)^j\gamma_w\right]_r
   &=&-\frac{ie}{c}\left(\frac{i}{\epsilon}\right)^j
   (g^r\cdot {\bf k})^{j-1}g^r\cdot\tilde{F}\exp\left(\frac{i}{\epsilon}\psi
   \right)+{\cal O}\left(\frac{1}{\epsilon^{j-1}}\right)+{\rm c.c.}\nonumber\\
\left[ (i_gd)^j\gamma_w\right]_\mu
   &=&\frac{-e}{\epsilon c}(j-1)\left(\frac{i g^r\cdot {\bf k}}{\epsilon}\right)^{j-2}
   \frac{\partial g^r}{\partial\mu   }\cdot\tilde{F}\cdot 
   g^r\exp\left(\frac{i}{\epsilon}\psi\right)
   +{\cal O}\left(\frac{1}{\epsilon^{j-2}}\right)+{\rm c.c.}\nonumber\\
\left[ (i_gd)^j\gamma_w\right]_\theta
   &=&\frac{-e}{\epsilon c}(j-1)\left(\frac{i g^r\cdot {\bf k}}{\epsilon}\right)^{j-2}
   \frac{\partial g^r}{\partial\theta}\cdot\tilde{F}\cdot
   g^r\exp\left(\frac{i}{\epsilon}\psi\right)
   +{\cal O}\left(\frac{1}{\epsilon^{j-2}}\right)+{\rm c.c.}\nonumber\\
\noalign{\hbox{and}}
i_g(di_g)^j\gamma_w
   &=&\frac{e}{c}\left(\frac{i}{\epsilon}\right)^j(g^r\cdot {\bf k})^j
   g^r\cdot\tilde{A}\exp\left(\frac{i}{\epsilon}\psi\right)
   +{\cal O}\left(\frac{1}{\epsilon^{j-1}}\right)+{\rm c.c.}\label{oae}
\end{eqnarray}
Note that the components of $(i_gd)^j\gamma_w$ are manifestly gauge invariant,
as promised.  Then $i_g(di_g)^j\gamma_w$ is not manifestly gauge invariant, but
this is the term that will be removed by the Lagrangian gauge transformation.
Thus, everything is going as planned.

Now we must plug the above results into Eq.~(\ref{oad}), and sum the series 
over $j.$  This is straightforward, and the result is
\begin{eqnarray}
\Gamma_w&=&\gamma_w
   +\frac{e}{c}\left[-ig^r\cdot\tilde{F}\left(\frac{\exp (-ig^r\cdot
   {\bf k})-1}{g^r\cdot {\bf k}}\right)\exp\left(\frac{i}{\epsilon}\psi\right)+
   {\cal O}(\epsilon )\right]\cdot dR\nonumber\\
   && +\frac{e}{c}\left[
   -\epsilon\frac{\partial g^r}{\partial\mu}\cdot\tilde{F}
   \cdot g^r\left(\frac{(1+ig^r\cdot{\bf k})\exp (-ig^r\cdot {\bf k})-1}
   {(g^r\cdot {\bf k})^2}\right)\exp\left(\frac{i}{\epsilon}\psi\right)+
   {\cal O}(\epsilon^2 )\right] d\mu\nonumber\\
   && +\frac{e}{c}\left[
   -\epsilon\frac{\partial g^r}{\partial\theta}\cdot\tilde{F}
   \cdot g^r\left(\frac{(1+ig^r\cdot{\bf k})\exp (-ig^r\cdot {\bf k})-1}
   {(g^r\cdot {\bf k})^2}\right)\exp\left(\frac{i}{\epsilon}\psi\right)+
   {\cal O}(\epsilon^2 )\right] d\theta\nonumber\\
   && -d\left[\frac{i\epsilon e}{c}
   g^r\cdot\tilde{A}\left(\frac{\exp (-ig^r\cdot
   {\bf k})-1}{g^r\cdot {\bf k}}\right)\exp\left(\frac{i}{\epsilon}\psi\right)+
   {\cal O}(\epsilon^2 )\right]+{\rm c.c.}\label{oaf}
\end{eqnarray}
At this point, we can check the above result by actually applying the exterior
derivative to the last term in square brackets.  There is extensive
cancellation, and we are left with Eq.~(\ref{oac}), as expected.
We can now make the Lagrangian gauge transformation,
\begin{equation}
\Gamma'_w\equiv\Gamma_w+dS_T
\end{equation}
where
\begin{equation}
S_T\equiv\frac{i\epsilon e}{c}
   g^r\cdot\tilde{A}\left(\frac{\exp (-ig^r\cdot
   {\bf k})-1}{g^r\cdot {\bf k}}\right)\exp\left(\frac{i}{\epsilon}\psi\right)
   +{\rm c.c.},
\label{obe}
\end{equation}
thereby removing the last term of Eq.~(\ref{oaf}) to get a
{\it manifestly} gauge invariant one form, as desired.

Now $g^r=-\rho\hat{\bf a},$ and we can substitute this into Eq.~(\ref{oaf}).
Note that the $\mu$ component of $\Gamma'_w$ vanishes because $g^r$ and
$\partial g^r/\partial\mu$ are both in the $\hat{\bf a}$ direction, and they
are both dotted into the antisymmetric two form, $\tilde{F}.$  The $\theta$
component does not vanish, however, because $\partial g^r/\partial\theta$ is
in the $\hat{\bf c}$ direction.  We finally have
\begin{eqnarray}
\Gamma'_w&=&\frac{e}{c}\left[ \tilde{A}
   +\rho\hat{\bf a}\cdot\tilde{F}\left(\frac{\exp 
   (i\rho\hat{\bf a}\cdot {\bf k})-1}{i\rho\hat{\bf a}\cdot {\bf k}}\right)+
   {\cal O}(\epsilon)\right]\cdot dR\exp\left(\frac{i}{\epsilon}\psi\right)\nonumber\\
   &&\qquad -\left[\frac{\epsilon e\rho^2}{2c\lambda_B}
   F_0:\tilde{F}\left(\frac{(1-i\rho
   \hat{\bf a}\cdot {\bf k})\exp (i\rho\hat{\bf a}\cdot {\bf k})-1}{(\rho
   \hat{\bf a}\cdot {\bf k})^2}\right)+{\cal O}(\epsilon^2)\right] d\theta
   \exp\left(\frac{i}{\epsilon}\psi\right)\nonumber\\
   &&\qquad+{\cal O}(\epsilon^2)d\mu+{\cal O}(\epsilon^2)\cdot dU+{\rm c.c.}
   \label{oag}
\end{eqnarray}
To proceed, we must Fourier analyze the above expression in preparation for
the oscillation-center Lie transformation.

\section{Fourier Expansion in Gyroangle}
\label{yax}
\begin{figure}[p]
\center{
\vspace{2.04truein}
\mbox{\includegraphics[bbllx=0,bblly=0,bburx=230,bbury=161,width=5.82truein]{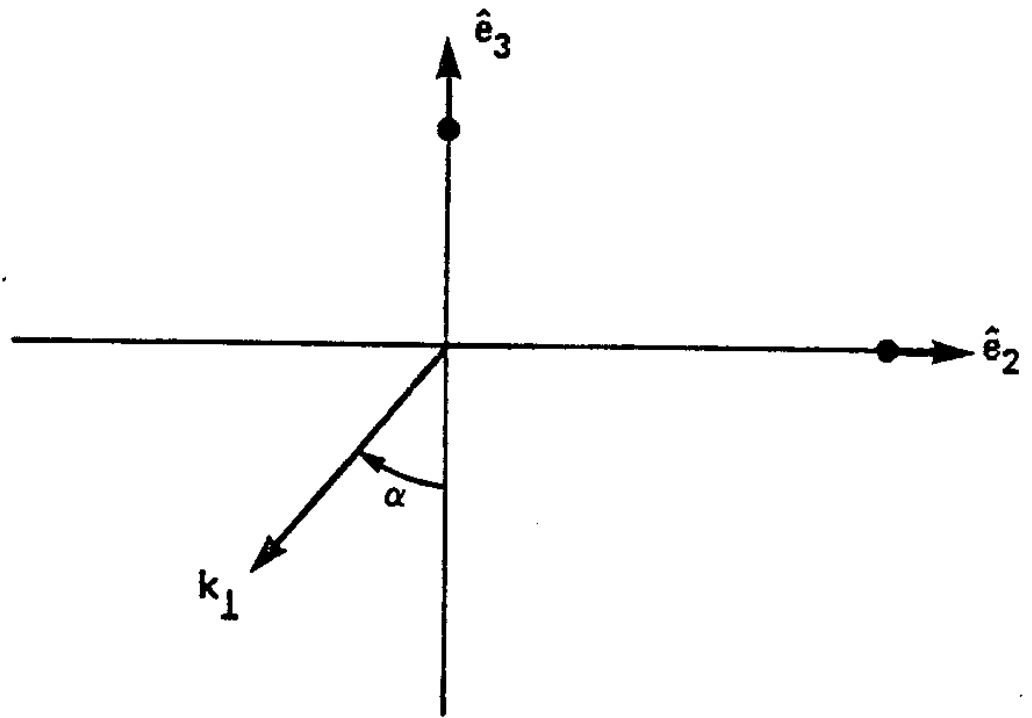}}
\vspace{2.04truein}
}
\caption{Components of the Four Wavevector}
\label{oax}
\end{figure}

We now write the components of ${\bf k}$ in the ${\hat{\bf e}}_\mu$ basis,
introduced back in Chapter~\ref{yac}, as follows:
\begin{equation}
{\bf k}={\bf k}_\parallel
       -{\rm k}_\perp    ({\hat{\bf e}}_2\sin\alpha+{\hat{\bf e}}_3\cos\alpha),
\end{equation}
where ${\bf k}_\parallel$ lies entirely within the parallel two-flat.
The geometrical situation is illustrated schematically in Fig.~\ref{oax}.
Then, using Eq.~(\ref{gat}), we find
\begin{equation}
\hat{\bf a}\cdot {\bf k}={\rm k}_\perp\sin (\theta-\alpha).
\end{equation}
Now we may Fourier Expand the quantities
\begin{eqnarray}
\exp (i\rho\hat{\bf a}\cdot {\bf k})
   &=&e^{i{\rm k}_\perp\rho\sin (\theta-\alpha)} \nonumber\\
   &=&\sum_\ell J_\ell({\rm k}_\perp\rho)e^{i\ell(\theta-\alpha)},
   \label{oaz}
\end{eqnarray}
\begin{eqnarray}
\frac{\exp (i\rho\hat{\bf a}\cdot {\bf k})-1}{i\rho\hat{\bf a}\cdot {\bf k}}
   &=&\frac{e^{i{\rm k}_\perp\rho\sin (\theta-\alpha)}-1}{i{\rm k}_\perp\rho\sin
   (\theta-\alpha)} \nonumber\\
   &=&\sum_\ell Q_\ell({\rm k}_\perp\rho)e^{i\ell(\theta-\alpha)}
   \label{oah}
\end{eqnarray}
and
\begin{eqnarray}
\frac{(1-i\rho\hat{\bf a}\cdot {\bf k})\exp (i\rho\hat{\bf a}\cdot {\bf k})
   -1}{(\rho\hat{\bf a}\cdot {\bf k})^2}
   &=&\frac{\left(1-i{\rm k}_\perp\rho\sin (\theta-\alpha)\right)
   e^{i{\rm k}_\perp\rho\sin (\theta-\alpha)}-1}{{\rm k}_\perp^2\rho^2
   \sin^2(\theta-\alpha)} \nonumber\\
   &=&\frac{1}{2}\sum_\ell R_\ell({\rm k}_\perp\rho)
   e^{i\ell(\theta-\alpha)},
   \label{oai}
\end{eqnarray}
where the $J_\ell$ are Bessel functions,
\begin{equation}
J_\ell(x)\equiv\frac{1}{2\pi}\int_0^{2\pi}d\xi e^{ix\sin\xi-i\ell\xi},
\end{equation}
where we have defined the special functions
\begin{equation}
Q_\ell(x)\equiv\frac{1}{2\pi}\int_0^{2\pi}d\xi\left(\frac{e^{ix\sin\xi}-1}{ix
   \sin\xi}\right)e^{-i\ell\xi}
\label{oqdf}
\end{equation}
and
\begin{equation}
R_\ell(x)\equiv\frac{1}{\pi}\int_0^{2\pi}d\xi\left(\frac{(1-ix\sin\xi)
   e^{ix\sin\xi}-1}{x^2\sin^2\xi}\right)e^{-i\ell\xi},
\label{ordf}
\end{equation}
and where the summations over $\ell$ extend from minus infinity to infinity.
The properties of the $Q$ and $R$ functions will be explored in detail in
Appendix~\ref{yaw}.

Now, along with the expressions in Eqs.~(\ref{oaz}), (\ref{oah}) and
(\ref{oai}), $\Gamma_w$ and $\Gamma'_w$ also contains the $\theta$-dependent
(and hence oscillatory) quantities, $\hat{\bf c}$ and $\hat{\bf a}.$  Thus
we need to know how to Fourier expand these as well.  Using Eq.~(\ref{gat}),
we may write
\begin{equation}
\hat{\bf c}=\frac{i}{\sqrt{2}}({\hat{\bf e}}_{+}e^{ i\theta}
                              -{\hat{\bf e}}_{-}e^{-i\theta})
\end{equation}
and
\begin{equation}
\hat{\bf a}=\frac{1}{\sqrt{2}}({\hat{\bf e}}_{+}e^{ i\theta}
                              +{\hat{\bf e}}_{-}e^{-i\theta}),
\end{equation}
where we have defined
\begin{equation}
{\hat{\bf e}}_{\pm}\equiv\frac{1}{\sqrt{2}}({\hat{\bf e}}_2
                                       \pm i{\hat{\bf e}}_3).
\end{equation}
Note that these are complex unit vectors that obey
${\hat{\bf e}}_{\pm}^*={\hat{\bf e}}_{\mp},$
${\hat{\bf e}}_{\pm}  \cdot {\hat{\bf e}}_{\pm}=0,$ and
${\hat{\bf e}}_{\pm}^*\cdot {\hat{\bf e}}_{\pm}=1.$  Because they contain
$e^{\pm i\theta},$ when we multiply them by the series in Eqs.~(\ref{oaz}),
(\ref{oah}) and (\ref{oai}), they will generate terms with
$e^{i(\ell\pm 1)\theta}.$  By defining new
summation variables we can restore these to the form $e^{i\ell\theta},$ but
then these terms will be left with
special functions that have indices $\ell\pm 1.$

Now then, we may write $\Gamma_w$ as follows:
\begin{equation}
\Gamma_w=\sum_\ell (\Gamma_{\ell R}\cdot dR+\epsilon\Gamma_{\ell\mu}d\mu
   +\epsilon\Gamma_{\ell\theta}d\theta)
   \exp\left(\frac{i}{\epsilon}\Psi_\ell\right)+{\cal O}(\epsilon^2)
   +{\rm c.c.},
\label{oba}
\end{equation}
where
\begin{equation}
\Gamma_{\ell R}\equiv\frac{e}{c}J_\ell\tilde{A},
\end{equation}
\begin{equation}
\Gamma_{\ell\mu}\equiv\frac{1}{\sqrt{2}\rho\lambda_B}
   {\cal J}_\ell^{+}\cdot\tilde{A}
\end{equation}
and
\begin{equation}
\Gamma_{\ell\theta}\equiv\frac{ie\rho}{\sqrt{2}c}
   {\cal J}_\ell^{-}\cdot\tilde{A},
\end{equation}
and where we have defined
\begin{equation}
\Psi_\ell ( R,\theta ) \equiv \psi (R) + \epsilon \ell (\theta-\alpha (R))
\label{oca}
\end{equation}
and
\begin{equation}
{\cal J}_\ell^\pm\equiv
       {\hat{\bf e}}_{+}e^{ i\alpha}J_{\ell-1}
   \pm {\hat{\bf e}}_{-}e^{-i\alpha}J_{\ell+1}.
\label{ojpm}
\end{equation}

Similarly, we may write $\Gamma'_w$ as follows:
\begin{equation}
\Gamma'_w=\sum_\ell (\Gamma'_{\ell R}\cdot dR+\epsilon\Gamma'_{\ell\theta}d\theta)
   \exp\left(\frac{i}{\epsilon}\Psi_\ell\right)+{\cal O}(\epsilon^2)
   +{\rm c.c.},
\label{oal}
\end{equation}
where
\begin{equation}
\Gamma'_{\ell R}\equiv\frac{e}{c}\left[\delta_{\ell 0}\tilde{A}
   +\frac{\rho}{\sqrt{2}} {\cal Q}_\ell^{+}\cdot\tilde{F}
   +{\cal O}(\epsilon)\right],
\end{equation}
and
\begin{equation}
\Gamma'_{\ell\theta}\equiv -\frac{e\rho^2}{4c\lambda_B}R_\ell F_0:\tilde{F}
   +{\cal O}(\epsilon^2),
\end{equation}
and where we have defined
\begin{equation}
{\cal Q}_\ell^\pm\equiv
       {\hat{\bf e}}_{+}e^{ i\alpha}Q_{\ell-1}
   \pm {\hat{\bf e}}_{-}e^{-i\alpha}Q_{\ell+1}.
\label{oqpm}
\end{equation}
In the above expressions, it is understood that $J_\ell,$ $Q_\ell$ and
$R_\ell$ are evaluated at ${\rm k}_\perp\rho.$

Finally, note that $S_T,$ as defined by Eq.~(\ref{obe}), has the Fourier
decomposition,
\begin{equation}
S_T=\epsilon\sum_\ell S_{T\ell}\exp\left(\frac{i}{\epsilon}\Psi_\ell\right)
    +{\rm c.c.}
\label{obf}
\end{equation}
where
\begin{equation}
S_{T\ell}\equiv -\frac{e\rho}{\sqrt{2}c}
  {\cal Q}_\ell^{+}\cdot\tilde{A}.
\end{equation}
Using Eqs.~(\ref{oba}), (\ref{oal}) and (\ref{obf}), it is possible to check
that $\Gamma'_w=\Gamma_w+dS_T.$

\section{The Oscillation-Center Lie Transform}
Our aim is to perform a Lie transform that will remove all the effects of the
wave from the Poisson structure, and put them into the Hamiltonian.  Thus, when
we have completed this task, our Poisson brackets will be identical to those
for a guiding center with no wave present (through order $\lambda^2$).  The
effect of the wave will be pushed into a term of order $\lambda^2$ in the
Hamiltonian.  We shall do this both for $\Gamma_w$ and for $\Gamma'_w,$ in
order to verify that we get the same answer either way.

We now reset our variables, so that $\Gamma_w$ (as given by Eq.~(\ref{oba}))
and $\Gamma'_w$ (as given by Eq.~(\ref{oal})) will henceforth be called
$\gamma_w$ and $\gamma'_w,$ respectively. The oscillation-center transform will
take us to $\Gamma_w$ and $\Gamma'_w,$ but we
want these to vanish by the above argument.
Thus, in Eqs.~(\ref{mcb}) and (\ref{mcc}) we demand that $\Gamma_1$ and
$\Gamma_2$ vanish.  This is the step at which we are neglecting resonant
effects.  Furthermore, in Eq.~(\ref{mcc}) we have $\gamma_2=0$ because our
wave perturbation is at first order in $\lambda$ only, and $\Omega_1=0$
because $\Gamma_1=0.$

First consider the oscillation-center transform of $\gamma_w.$  We have
\begin{equation}
0=\gamma_w-i_1 \omega_{gc}+dS_1,
\label{oam}
\end{equation}
and
\begin{equation}
0=-i_2 \omega_{gc}-\frac{1}{2}i_1 \omega_w+dS_2.
\label{oan}
\end{equation}
Meanwhile, the Hamiltonian transforms according to Eqs.~(\ref{mbs}) through
(\ref{mbu}) to give
\begin{equation}
K_0=H_{gc},
\end{equation}
\begin{equation}
K_1=-{\cal L}_1 H_{gc}=-i_1dH_{gc},
\end{equation}
and
\begin{equation}
K_2=-{\cal L}_2 H_{gc}+\frac{1}{2}{\cal L}_1^2 H_{gc}.
\label{oao}
\end{equation}

Now we demand that $K_1=-i_1dH_{gc}=0.$  Let $i_0$ denote interior
multiplication by $\dot{z}$ (the unperturbed flow), so
$i_0\omega_{gc}=-dH_{gc}$ (our unperturbed problem is the guiding-center
problem).  Then, applying $i_0$ to Eq.~(\ref{oam}) gives
\begin{equation}
{\dot{S}}_1=-i_0\gamma_w+i_1dH_{gc}=-i_0\gamma_w,
\label{oar}
\end{equation}
where the last step follows as a result of our demand that $K_1=0.$  We
can integrate this last equation along unperturbed orbits to get $S_1.$
Then $g_1$ is given by Eq.~(\ref{oam})
\begin{equation}
g_1=(\gamma_w+dS_1)\cdot J_{gc}.
\label{oas}
\end{equation}

At second order, we can solve Eq.~(\ref{oan}) for $g_2$ as follows:
\begin{equation}
g_2=(-\frac{1}{2}i_1\omega_w+dS_2)\cdot J_{gc}.
\end{equation}
Now then, we can insert these generators into Eq.~(\ref{oao}) to get $K_2,$
as follows:
\begin{eqnarray}
K_2&=&-{\cal L}_2 H_{gc}+\frac{1}{2}{\cal L}_1^2 H_{gc}\nonumber\\
   &=&-{\cal L}_2 H_{gc}\nonumber\\
   &=&(\frac{1}{2}i_1\omega_w-dS_2)\cdot J_{gc}\cdot dH_{gc}\nonumber\\
   &=&\frac{1}{2}i_0i_1\omega_w-{\dot{S}}_2.\label{oap}
\end{eqnarray}
Now we can choose $S_2$ to remove the oscillatory part of the first term.
Note that we cannot remove the averaged part of the first term, because
that would introduce secular terms in $S_2.$  So the best that we can do is
to take
\begin{equation}
K_2=\left\langle \frac{1}{2}i_0i_1\omega_w \right\rangle .
\label{obd}
\end{equation}
This is the ponderomotive Hamiltonian.

Now suppose that we had started with $\gamma'_w=\gamma_w+dS_T$ instead of
$\gamma_w.$  Instead of Eqs.~(\ref{oam}) and (\ref{oan}), we would have written
\begin{equation}
0=\gamma'_w-i_{1'} \omega_{gc}+dS'_1,
\label{obb}
\end{equation}
and
\begin{equation}
0=-i_{2'} \omega_{gc}-\frac{1}{2}i_{1'} \omega_w+dS'_2,
\label{obc}
\end{equation}
where $i_{n'}$ is an obvious shorthand for $i_{g'_n},$ and where we are adhering
to the convention of using primes to denote quantities arising from the Lagrangian
gauge transformed action one form.  Of course, we still would
have taken $K'_0=K_0=H_{gc}$ and we still would have demanded that
$K'_1=-i_{1'}dH_{gc}=0=K_1.$  From this it follows that $K'_2=-i_{2'}dH_{gc}.$
Thus, if we could show that $g_2=g'_2,$ it would immediately
follow that $K'_2=K_2$; that is, it would follow that the ponderomotive
Hamiltonian is invariant under the Lagrangian gauge transformation.

From Eq.~(\ref{obb}), we have
\begin{equation}
{\dot{S}}'_1=i_0dS'_1=-i_0\gamma'_w+i_{1'}dH_{gc}
   =-i_o\gamma'_w=-i_0(\gamma_w+dS_T)={\dot{S}}_1-{\dot{S}}_T,
\label{obg}
\end{equation}
so
\begin{equation}
S'_1=S_1-S_T.
\label{obj}
\end{equation}
Then
\begin{equation}
g'_1=(\gamma'_w+dS'_1)\cdot J_{gc}=(\gamma_w+dS_T+dS_1-dS_T)\cdot J_{gc}
   =(\gamma_w+dS_1)\cdot J_{gc}=g_1,
\label{obk}
\end{equation}
So $g_1$ is invariant under the Lagrangian gauge transformation.
Next, from Eq.~(\ref{obc}) we have
\begin{equation}
g'_2=\left(-\frac{1}{2}i_{1'}\omega_w+dS'_2\right)\cdot J_{gc},
\end{equation}
so
\begin{equation}
K'_2=\left(\frac{1}{2}i_{1'}\omega_w-dS'_2\right)\cdot J_{gc}\cdot dH_{gc}
   =\frac{1}{2}i_0i_{1'}\omega_w-{\dot{S}}'_2
   =\frac{1}{2}i_0i_1\omega_w-{\dot{S}}'_2.
\label{obv}
\end{equation}
Thus we have
\begin{equation}
{\dot{S}}'_2=\frac{1}{2}i_0i_1\omega_w
   -\left\langle\frac{1}{2}i_0i_1\omega_w\right\rangle
   ={\dot{S}}_2,
\end{equation}
so
\begin{equation}
S'_2=S_2,
\end{equation}
and so
\begin{equation}
g'_2=g_2.
\end{equation}
It immediately follows that
\begin{equation}
K'_2=\left\langle \frac{1}{2}i_0i_1\omega_w \right\rangle =K_2,
\end{equation}
so the ponderomotive Hamiltonian is indeed invariant under the Lagrangian gauge
transformation.  Note that $g_1,$ $g_2,$ and $S_2$ are also thus invariant,
but that $\gamma_w$ and $S_1$ are {\it not}.  The latter two quantities transform
under the Lagrangian gauge transformation as follows:
\begin{equation}
\gamma'_w=\gamma_w+dS_T
\end{equation}
and
\begin{equation}
S'_1=S_1-S_T,
\end{equation}
so that the combination $\gamma_w+dS_1$ {\it is} invariant.

Though we have just shown that we would get the same answer for the ponderomotive
Hamiltonian either way, it bears repeating that the advantage of starting with
$\gamma'_1$ is its {\it manifest} gauge invariance.  In the next section, we shall
further discuss the relative merits of each of the two ways of calculating $K_2.$

While the above expression, Eq.~(\ref{obd}), for the ponderomotive Hamiltonian is
wonderfully compact, it is also very formal.  We need to plug in Eq.~(\ref{oba})
and/or Eq.~(\ref{oal}), and work it out in detail.  This is done in the next section.

\section{The Ponderomotive Hamiltonian}
Our unperturbed equations of motion are
\begin{eqnarray}
\dot{R}&=&U+\epsilon U_d\nonumber\\
\dot{U}&=&{\cal O}(1)\nonumber\\
\dot{\mu}&=&0\nonumber\\
\noalign{\hbox{and}}
\dot{\theta}&=&\frac{1}{\epsilon}\Omega_B, \label{oaq}
\end{eqnarray}
where $U_d$ denotes the guiding-center drift motion, and where we do not
need to know anything about $\dot{U}$ other than the fact that it is order
unity in $\epsilon.$  Then Eq.~(\ref{oar}) for $S_1$ becomes
\begin{equation}
{\dot{S}}_1=-\sum_\ell \left[ \gamma_{\ell R}\cdot (U+\epsilon U_d)
  +\gamma_{\ell\theta} \Omega_B \right]
  \exp\left(\frac{i}{\epsilon}\Psi_\ell\right)+{\rm c.c.},
\end{equation}
and Eq.~(\ref{obg}) for $S'_1$  becomes
\begin{equation}
{\dot{S}}'_1=-\sum_\ell\left[ \gamma'_{\ell R}\cdot (U+\epsilon U_d)
  +\gamma'_{\ell\theta} \Omega_B \right]
  \exp\left(\frac{i}{\epsilon}\Psi_\ell\right)+{\rm c.c.}
\end{equation}
Integrating over unperturbed orbits, we get
\begin{equation}
S_1=i\epsilon\sum_\ell \left[ \gamma_{\ell R}\cdot (U+\epsilon U_d)
  +\gamma_{\ell\theta} \Omega_B \right]
  \frac{\exp\left(\frac{i}{\epsilon}\Psi_\ell\right)}{D_\ell}+{\rm c.c.},
\label{obh}
\end{equation}
and
\begin{equation}
S'_1=i\epsilon\sum_\ell\left[ \gamma'_{\ell R}\cdot (U+\epsilon U_d)
  +\gamma'_{\ell\theta} \Omega_B \right]
  \frac{\exp\left(\frac{i}{\epsilon}\Psi_\ell\right)}{D_\ell}+{\rm c.c.},
\label{obi}
\end{equation}
respectively.  Here we have defined the resonant denominator
\begin{equation}
D_\ell\equiv {\dot{\Psi}}_\ell={\bf k}\cdot (U+\epsilon U_d)+\ell\Omega_B.
\label{ord}
\end{equation}
Using Eqs.(\ref{obf}), (\ref{obh}) and (\ref{obi}), it is possible to verify
Eq.~(\ref{obj}); that is, it is possible to show explicitly that
$S'_1=S_1-S_T.$

Now we use Eq.~(\ref{oas}) to get the components of the generator $g_1$,
\begin{eqnarray}
g_1^R&=&-\frac{\epsilon}{m} \sum_\ell
  \left(H_{1\ell} {\bf k}+\frac{e}{c}J_\ell D_\ell\tilde{A}\right)\cdot
  \left(\frac{F_0D_\ell}{\lambda_B \Omega_B}+iP_\parallel\right)
  \frac{\exp\left(\frac{i}{\epsilon}\Psi_\ell\right)}{D_\ell^2}\nonumber\\
  &&\qquad +{\rm c.c.}
  +{\cal O}(\epsilon^2),
  \label{obm}\\
g_1^U&=&\frac{1}{m} \sum_\ell
  \left(H_{1\ell} {\bf k}+\frac{e}{c}J_\ell D_\ell\tilde{A}\right)\cdot
  P_\parallel
  \frac{\exp\left(\frac{i}{\epsilon}\Psi_\ell\right)}{D_\ell}+{\rm c.c.}
  +{\cal O}(\epsilon),
  \label{obn}\\
g_1^\mu&=&-\frac{1}{\Omega_B}\sum_\ell
  \left(H_{1\ell} {\bf k}+\frac{e}{c}J_\ell D_\ell\tilde{A}\right)
  \cdot U
  \frac{\exp\left(\frac{i}{\epsilon}\Psi_\ell\right)}{D_\ell}+{\rm c.c.}
  +{\cal O}(\epsilon),
  \label{obo}\\
\noalign{\hbox{and}}
g_1^\theta&=&\sum_\ell\Bigl[ -\frac{e}{mc\lambda_B} {\bf k}\cdot F_0\cdot
  \tilde{A} J_\ell +\frac{i{\rm k}_\perp}{2\rho\lambda_B}(J_{\ell+1}-J_{\ell-1})
  U\cdot\tilde{A}\nonumber\\
  &&\qquad -\frac{1}{\sqrt{2}\rho\lambda_B}\tilde{A}\cdot {\cal J}_\ell^{+}
  {\bf k}\cdot U \Bigr]
  \frac{\exp\left(\frac{i}{\epsilon}\Psi_\ell\right)}{D_\ell}+{\rm c.c.}
  +{\cal O}(\epsilon),
  \label{obp}
\end{eqnarray}
where we have defined
\begin{equation}
H_{1\ell}\equiv -\frac{e}{c}\left( UJ_\ell+\frac{i\rho\Omega_B}{\sqrt{2}}
  {\cal J}_\ell^{-}\right)\cdot\tilde{A}.
\label{obw}
\end{equation}
If we had instead used the the first of Eqs.~(\ref{obk}), we would have
obtained the following results for the components of $g'_1$:
\begin{eqnarray}
g_1^{\prime R}&=&\frac{i\epsilon e}{mc} \sum_\ell
  \left( U J_\ell +\frac{i\rho\Omega_B}{\sqrt{2}} {\cal J}_\ell^{-}
  \right)
  \cdot\tilde{F}\nonumber\\
  &&\qquad\cdot
  \left(\frac{F_0D_\ell}{\lambda_B \Omega_B}+iP_\parallel\right)
  \frac{\exp\left(\frac{i}{\epsilon}\Psi_\ell\right)}{D_\ell^2}+{\rm c.c.}
  +{\cal O}(\epsilon^2),
  \label{obq}\\
g_1^{\prime U}&=&-\frac{ie}{mc} \sum_\ell
  \left( U J_\ell +\frac{i\rho\Omega_B}{\sqrt{2}} {\cal J}_\ell^{-}
  \right)
  \cdot\tilde{F}\nonumber\\
  &&\qquad\cdot
  P_\parallel
  \frac{\exp\left(\frac{i}{\epsilon}\Psi_\ell\right)}{D_\ell}+{\rm c.c.}
  +{\cal O}(\epsilon),
  \label{obr}\\
g_1^{\prime\mu}&=&\frac{e\rho}{\sqrt{2}c}\tilde{F}:\sum_\ell
  \left( U {\cal J}_\ell^{-}\right)
  \frac{\exp\left(\frac{i}{\epsilon}\Psi_\ell\right)}{D_\ell}+{\rm c.c.}
  +{\cal O}(\epsilon),
  \label{obs}\\
\noalign{\hbox{and}}
g_1^{\prime\theta}&=&\frac{ie}{2mc\lambda_B}\tilde{F}:\sum_\ell
  \Bigl[ F_0\cdot\Bigl({\bf 1}J_\ell
  +\frac{\sqrt{2}i}{\rho\Omega_B} {\cal J}_\ell^{-}
  U \Bigr) \Bigr]
  \frac{\exp\left(\frac{i}{\epsilon}\Psi_\ell\right)}{D_\ell}\nonumber\\
  &&\qquad +{\rm c.c.}
  +{\cal O}(\epsilon).
  \label{obt}
\end{eqnarray}
By straightforward calculation, it is possible to directly verify that
$g'_1=g_1,$ as required by Eq.~(\ref{obk}).  To do this, simply substitute
$\tilde{F}\equiv i({\bf k}\tilde{A}-\tilde{A}{\bf k})$ into Eqs.~(\ref{obq})
through (\ref{obt}); upon simplification, the results will be Eqs.~(\ref{obm})
through (\ref{obp}).  If we had not made the Lagrangian gauge transformation,
and had instead started with only $\gamma_w$ and $S_1,$ we might have had
difficulty casting Eqs.~(\ref{obm}) through (\ref{obp}) in the manifestly
gauge-invariant form of Eqs.~(\ref{obq}) through (\ref{obt}).

Next we compute the components of $\omega_w=d\gamma_w.$  Direct calculation
gives
\begin{equation}
\omega_w=\sum_\ell \omega_{w\ell}
  \exp\left(\frac{i}{\epsilon}\Psi_\ell\right)+{\rm c.c.},
\end{equation}
where
\begin{equation}
\omega_{w\ell RR}=\frac{ie}{\epsilon c} J_\ell
  ({\bf k}\tilde{A}-\tilde{A} {\bf k}),
\end{equation}
\begin{equation}
\omega_{w\ell RU}={\cal O}(\epsilon^2),
\end{equation}
\begin{equation}
\omega_{w\ell R\mu}=\frac{i}{\sqrt{2}\rho\lambda_B}
  \left[{\bf k} {\cal J}_\ell^{+}
  -\frac{i{\rm k}_\perp}{2}
  \left(J_{\ell+1}-J_{\ell-1}\right) {\bf 1}\right]\cdot\tilde{A},
\end{equation}
\begin{equation}
\omega_{w\ell R\theta}=-\frac{e\rho}{\sqrt{2}c}
  \left({\bf k} {\cal J}_\ell^{-}
  +\frac{\sqrt{2} i\ell}{\rho}J_\ell {\bf 1}\right)\cdot\tilde{A},
\end{equation}
\begin{equation}
\omega_{w\ell UU}={\cal O}(\epsilon^2),
\end{equation}
\begin{equation}
\omega_{w\ell U\mu}={\cal O}(\epsilon^2),
\end{equation}
\begin{equation}
\omega_{w\ell U\theta}={\cal O}(\epsilon^2),
\end{equation}
and
\begin{equation}
\omega_{w\ell\mu\theta}=-\frac{i\epsilon}{\lambda_B^2}
  {\bf k}\cdot F_0\cdot\tilde{A} J_\ell.
\end{equation}
If we had instead used $\omega'_w=d\gamma'_w,$ we would have
obtained the following results:
\begin{equation}
\omega'_w=\sum_\ell \omega'_{w\ell}
  \exp\left(\frac{i}{\epsilon}\Psi_\ell\right)+{\rm c.c.},
\end{equation}
where
\begin{equation}
\omega'_{w\ell RR}=\frac{e}{\epsilon c} J_\ell \tilde{F},
\end{equation}
\begin{equation}
\omega'_{w\ell RU}={\cal O}(\epsilon^2),
\end{equation}
\begin{equation}
\omega'_{w\ell R\mu}=\frac{1}{\sqrt{2}\rho\lambda_B}\tilde{F}\cdot
  {\cal J}_\ell^{+},
\end{equation}
\begin{equation}
\omega'_{w\ell R\theta}=\frac{ie\rho}{\sqrt{2}c}\tilde{F}\cdot
  {\cal J}_\ell^{-},
\end{equation}
\begin{equation}
\omega'_{w\ell UU}={\cal O}(\epsilon^2),
\end{equation}
\begin{equation}
\omega'_{w\ell U\mu}={\cal O}(\epsilon^2),
\end{equation}
\begin{equation}
\omega'_{w\ell U\theta}={\cal O}(\epsilon^2),
\end{equation}
and
\begin{equation}
\omega'_{w\ell\mu\theta}=-\frac{\epsilon}{2\lambda_B^2}F_0:\tilde{F} J_\ell.
\end{equation}
By direct calculation, it is once again possible to verify that
$\omega'_w=d\gamma'_w=d(\gamma_w+dS_T)=d\gamma_w=\omega_w$
by simply substituting
$\tilde{F}\equiv i({\bf k}\tilde{A}-\tilde{A}{\bf k})$ into the results
for the components of $\omega'_w$ and simplifying to get the
components of $\omega_w.$

Before using the above results to calculate $K_2,$ we digress for one last
discussion about the relative merits of starting with $\gamma_w$ and
$\gamma'_w.$  First note that all of the components of $g_1$ and $\omega_w$
are indeed manifestly gauge invariant.  If we had started the calculation with
$\gamma'_w,$ this would not be a surprise since $\gamma'_w$ is itself manifestly
gauge invariant; if however we had started the calculation with $\gamma_w,$ the
manifest gauge invariance of the result would seem fortuitous.  In the latter
event, we would have had results in terms of $\tilde{A},$ and only through some
tedious algebraic manipulations would we have discovered that their dependence on
$\tilde{A}$ arose only through a dependence on $\tilde{F}.$
On the other hand, note that the only special functions that appear in
the components of $g_1$ and $\omega_w$ are the Bessel functions,
$J_\ell.$  The $Q_\ell$ and $R_\ell$ functions have all disappeared in favor
of the $J_\ell.$  If we had started the calculation with $\gamma_w,$ this would
not be a surprise since $\gamma_w$ itself depends only on the $J_\ell,$ and not
on the $Q_\ell$ and $R_\ell$; if however we had started the calculation with
$\gamma'_w,$ the disappearance of the $Q_\ell$ and $R_\ell$ functions would
seem fortuitous.  In the latter event, we would have had results in terms
of the $Q_\ell$ and $R_\ell$ functions, and only through some tedious algebraic
manipulations would we have discovered that the recursion relations and
derivative formulas could be used to cast them in terms of $J_\ell$ alone.
There is thus a peculiar duality between the presence of special functions and
of manifest gauge invariance.

We now insert the above formulas into our expression for $K_2.$  The averaging is
carried out as follows:
\begin{equation}
\left\langle
  \exp\left(\frac{i}{\epsilon}\Psi_{\ell }\right)^*
  \exp\left(\frac{i}{\epsilon}\Psi_{\ell'}\right)
\right\rangle = \delta_{\ell\ell'}.
\end{equation}
We get
\begin{equation}
K_2=\frac{e^2}{2mc^2}({\tilde{A}}^*\cdot P_\parallel\cdot\tilde{A}
  +\frac{i}{\lambda_B\Omega_B} {\bf k}\cdot U {\tilde{A}}^*\cdot F_0
   \cdot\tilde{A}+{\rm c.c.})
  +\sum_\ell K_{2\ell},
\label{obz}
\end{equation}
where
\begin{eqnarray}
K_{2\ell}&=&\frac{eH_{1\ell}^*}{mcD_\ell}
  \Bigl\{ {\bf k}\cdot P_\parallel\cdot\tilde{A} J_\ell
  -\frac{i{\bf k}\cdot U}{\rho\Omega_B}
  \Bigl[-\frac{i{\rm k}_\perp U}{2\Omega_B} (J_{\ell-1}-J_{\ell+1})\nonumber\\
  &&\qquad -\frac{{\bf k}\cdot U}{\sqrt{2}\Omega_B} {\cal J}_\ell^{+}
  +\frac{\rho}{\lambda_B} F_0\cdot {\bf k} J_\ell\Bigr]\cdot\tilde{A}\Bigr\}\nonumber\\
  &&\qquad +\frac{|H_{1\ell}|^2}{2mD_\ell^2}
  {\bf k}\cdot P_\parallel\cdot {\bf k}+{\rm c.c.}
  +{\cal O}(\epsilon),
  \label{obu}
\end{eqnarray}
where c.c. denotes the complex conjugate, and where $H_{1\ell}$ is defined
in Eq.~(\ref{obw}).  If we had instead computed $K'_2$
according to Eq.~(\ref{obv}), we would have obtained the result,
\begin{equation}
K'_2=\sum_\ell K'_{2\ell},
\label{oav}
\end{equation}
where
\begin{eqnarray}
K'_{2\ell}&=&\frac{ie^2}{2mc^2D_\ell^2}
  \left(J_\ell U+\frac{i\rho\Omega_B}{\sqrt{2}} {\cal J}_\ell^{-}\right)
  \cdot\tilde{F}\cdot
  \left(\frac{F_0D_\ell}{\lambda_B \Omega_B}
  +iP_\parallel\right)\nonumber\\
  &&\qquad\cdot{\tilde{F}}^*\cdot
  \left(J_\ell U+\frac{i\rho\Omega_B}{\sqrt{2}} {\cal J}_\ell^{-}\right)^*
  +\frac{e^2\rho}{2\sqrt{2} mc^2D_\ell}\tilde{F}:
  \Bigl( {\cal J}_\ell^{-} U\Bigr)\nonumber\\
  &&\qquad {\tilde{F}}^*:
  \left[\frac{1}{\lambda_B}F_0\cdot\left({\bf 1}J_\ell-\frac{\sqrt{2}i}{\rho\Omega_B}
  {\cal J}_\ell^{-} U\right)^*\right]
  +{\rm c.c.}
  +{\cal O}(\epsilon).
  \label{obl}
\end{eqnarray}
Once again, by substituting $\tilde{F}\equiv i({\bf k}\tilde{A}-\tilde{A}{\bf k})$
into Eq.~(\ref{obl}) and simplifying, it is possible to reduce the expression to
Eq.~(\ref{obu}), thus directly verifying that $K'_2=K_2.$  In the course of this
calculation, some of the sum rules of Appendix~\ref{ybe} are useful.
Henceforth we shall drop the prime in our notation, and refer to the
ponderomotive Hamiltonian only as $K_2,$ whether or not it is in manifestly
gauge-invariant form.

Note that $K_2$ is a function of
the phase space coordinates, $R, U, \mu$ and $\theta$; in particular, it
depends on $R$ through its dependence on the background fields,
$F_0(R)$ and $F_1(R),$ and through its dependence on the eikonal wave
field parameters, $\tilde{F}(R)$ and ${\bf k}(R).$  Thus we write
$K_2(Z;F_i(R),\tilde{F}(R),{\bf k}(R)),$ where $i=0,1.$

The ponderomotive Hamiltonian will be used extensively in the next chapter
where we shall study the self-consistent dynamics of magnetized relativistic
plasma in an eikonal wave field.

\section{Obtaining the Ponderomotive Hamiltonian Using Canonical Lie Transforms}

Grebogi and Littlejohn~\cite{zbn} have obtained the ponderomotive Hamiltonian
by first performing a single noncanonical coordinate transformation to remove
the perturbation from the action one form, and then using {\it canonical} Lie
transforms on the Hamiltonian.  We shall use that procedure in this section in
order to check our above result for $K_2.$

Let us return to the point at which the wave perturbation was first added to
the single-particle action one form in Eq.~(\ref{obx}).  Recall the definition
of the single-particle velocity $u$ in Eq.~(\ref{gjz}).  Suppose that we change
this definition to absorb the wave perturbation; that is, we adopt the following
new definition for $u$:
\begin{equation}
u=\frac{1}{m} \left( p-\frac{e}{c} A(q)\right)
  +\frac{\lambda e}{mc}\tilde{A}(q)
  \exp\left(\frac{i}{\epsilon}\psi\left(q\right)\right)+{\rm c.c.}
\end{equation}
This has the effect of returning the action one form to the functional form
that it had before the wave was introduced.  Of course, the {\it definitions}
of the quantities that appear in the one form will be different; that is, $u$
and anything that depends on $u$ (e.g. $k,$ $\beta,$ $w,$ and $\theta$) will
be defined differently in terms of the single-particle position and velocity.
Nevertheless, the action one form is returned to the form that it had when no
wave was present, and now we can apply the usual guiding-center transformation
to take it to the guiding-center action one form $\Gamma_w,$ given implicitly
in Eq.~(\ref{gjm}), with no remaining perturbation due to the wave.

Whereas the action one form has thus been simplified by this transformation,
the Hamiltonian, Eq.~(\ref{gka}), now becomes considerably more complicated.
Using the new definition of $u$ in Eq.~(\ref{gkb}), we have
\begin{equation}
H'(r,u)=H'_0(r,u)+\lambda H'_1(r,u)+\lambda^2 H'_2(r,u),
\end{equation}
where
\begin{equation}
H'_0(r,u)=\frac{m}{2}u^2
\end{equation}
\begin{equation}
H'_1(r,u)=-\frac{e}{c}u\cdot\tilde{A}(r)
  \exp\left(\frac{i}{\epsilon}\psi\left(r\right)\right)+{\rm c.c.}
\end{equation}
and
\begin{equation}
H'_2(r,u)=\frac{e^2}{2mc^2}\tilde{A}(r)\cdot {\tilde{A}}^*(r)
  +\frac{e^2}{2mc^2}\tilde{A}(r)\cdot\tilde{A}(r)
    \exp\left(\frac{2i}{\epsilon}\psi\left(r\right)\right)+{\rm c.c.}
\end{equation}
At this point we can apply the guiding-center transformation,
$(r,u)\mapsto (R,U,\mu,\Theta),$ which may be taken to be simply
$R=r-\epsilon\rho$ to the order to which we are working.  The result may be
Fourier expanded in the gyroangle using the usual Bessel function identities.
The result is
\begin{equation}
H=H_0+\lambda H_1+\lambda^2 H_2,
\end{equation}
where
\begin{equation}
H_0=\frac{m}{2} U^2+\mu\Omega_B
\end{equation}
is the usual guiding-center Hamiltonian (to lowest order), where
\begin{equation}
H_1=\sum_\ell H_{1\ell}
  \exp\left(\frac{i}{\epsilon}\Psi_\ell\right)+{\rm c.c.}
\end{equation}
with $H_{1\ell}$ given by Eq.~(\ref{obw}), and where
\begin{equation}
H_2=\frac{e^2}{mc^2}\tilde{A}\cdot {\tilde{A}}^*+{\rm oscillatory \, terms}.
\end{equation}

To recap, we have applied a noncanonical transformation to remove the
perturbation from the Poisson structure and deposit it in the Hamiltonian.
We can now use a {\it canonical} Lie transform to remove $H_1$ (neglecting
resonances) and average $H_2$ to get $K_2.$  Note that this method does not
preserve manifest gauge invariance; that was lost in the very first step
when we redefined $u$ in a gauge-dependent way.

Applying canonical Lie transform perturbation theory, at first order we have
from Eq.~(\ref{mek})
\begin{equation}
0=K_1=H_1+\{ W_1,H_0 \},
\end{equation}
so
\begin{equation}
\{ W_1,H_0 \}=-H_1=-\sum_\ell H_{1\ell}
  \exp\left(\frac{i}{\epsilon}\Psi_\ell\right)+{\rm c.c.}
\end{equation}
Integrate this along unperturbed orbits to get the scalar generator
\begin{equation}
W_1=i\epsilon\sum_\ell\frac{H_{1\ell}}{D_\ell}
  \exp\left(\frac{i}{\epsilon}\Psi_\ell\right)+{\rm c.c.}
\end{equation}

Proceeding to second order, we have from Eq.~(\ref{mel})
\begin{equation}
K_2=H_2+\{ W_2,H_0 \}+\frac{1}{2} \{ W_1,H_1 \}.
\end{equation}
Now $W_2$ is chosen to average the result, so without having to explicitly
calculate it, we can write
\begin{equation}
K_2=\left\langle H_2+\frac{1}{2} \{ W_1,H_1 \}\right\rangle.
\end{equation}
After a short calculation, this reduces to the result
\begin{eqnarray}
K_2&=&\frac{e^2}{2mc^2}\tilde{A}\cdot {\tilde{A}}^*
  -\frac{1}{2}\sum_\ell \{ \Psi_\ell , \frac{|H_{1\ell}|^2}{D_\ell} \}
  + {\rm c.c.}\nonumber\\
   &=&\frac{e^2}{2mc^2}\tilde{A}\cdot {\tilde{A}}^*
  -\frac{1}{2}\sum_\ell \left(\frac{1}{m} {\bf k}_\parallel\cdot
  \frac{\partial}{\partial U}+\ell\frac{\partial}{\partial\mu}\right)
  \frac{|H_{1\ell}|^2}{D_\ell}+ {\rm c.c.}
   \label{oby}
\end{eqnarray}
That this answer is equal to our previous result for $K_2$ may be proved
by expanding the derivatives in Eq.~(\ref{oby}), replacing $\ell$ by
$[D_\ell-{\bf k}\cdot (U+\epsilon U_d)]/\Omega_B,$ and using the sum rules
of Appendix~\ref{ybe} to sum the terms with no resonant denominator.
The result is Eqs.~(\ref{obz}) and (\ref{obu}).

Note that this is by far the easiest way to get $K_2.$  Furthermore, it
yields the result in a considerably more compact form than the Lagrangian
Lie transform approach does.  On the other hand, as has already been noted,
it does not yield the result in manifestly gauge invariant form.

This result may be compared with that of Grebogi and
Littlejohn~\cite{zbn} who used ``$1+3$'' notation and whose result was
gauge invariant but not manifestly so.  To make this comparison, use the
technique for translating our results into ``$1+3$'' notation that was
introduced back in Section~\ref{yan}.  It is then a straightforward
exercise to show that our ponderomotive Hamiltonian gives rise to the
same equations of motion as that of Littlejohn and Grebogi, though the
two are {\it not} numerically equal.  The reason that the two results
for $K_2$ are not numerically equal can be traced back to the fact that
the corresponding {\it unperturbed} Hamiltonians are not numerically
equal.  This is because Littlejohn and Grebogi started with the
Hamiltonian (written in terms of three-vector coordinates and
velocities),
\begin{equation}
H_{LG}=(\gamma_v-1)mc^2+e\phi,
\end{equation}
which is not numerically equal to the Hamiltonian that we started with,
though it does yield the same equations of motion.

It is easier to compare our result with that of Achterberg~\cite{zcd} who used
a four-vector approach, but who did not worry about manifest gauge invariance
and who used essentially the same method outlined in this section.  His result
is identical to our Eq.~(\ref{oby}), outside of some minor notational differences.

%% file: relgcp.tex
\chapter{The Relativistic Guiding-Center Plasma}
\textheight=8.64truein
\setcounter{equation}{736}
\pagestyle{myheadings}
\markboth{The Relativistic Guiding-Center Plasma}{The Relativistic Guiding-Center Plasma}
\label{yaq}
\section{Discussion}
The reason that a Vlasov plasma is a nonlinear medium is that the plasma
currents generate fields which in turn drive the motion of the plasma.
Up until now in this thesis, we have dealt only with single particles
(or single guiding centers or single guiding/oscillation centers)
moving in fields that are known in advance as fixed functions of spacetime.
In this final chapter,
we show how to pass from this single particle description to a
{\it self-consistent} description of the dynamics of the guiding-center
plasma; this includes the dynamics of the fields as well as that of the
particles.  We shall do this by imbedding the single particle action in a
system action, and coupling it to the Maxwell field.

In Section~\ref{yar}, we prove Liouville's theorem, and show how to write the
Vlasov equation in any desired coordinate system.
In Section~\ref{yas}, we sum the guiding-center Lagrangian action over a full
distribution of guiding centers and couple to the Maxwell field in order
to obtain the Lagrangian action of the full guiding-center Vlasov plasma.
The variation of this with respect to the guiding-center coordinates yields the
relativistic kinetic equation for guiding centers, while the variation with
respect to the four potential yields the self-consistent field equation
including the guiding-center magnetization and current densities.

In Section~\ref{yat}, Noether's theorem is applied constructively to 
obtain covariant conservation laws for the momentum-energy and the angular 
momentum of a guiding-center plasma.  That is, we obtain the
stress-energy and angular momentum tensors of the guiding-center plasma,
including the contribution to the angular momentum due to guiding-center
spin.

Finally, in Section~\ref{yau}, we employ
the results of Chapter~\ref{yap} to generalize the results of
Sections~\ref{yas} and \ref{yat} to the case of a guiding-center plasma in an
eikonal wave field.  We begin by forming a system action, this time including
the Maxwell action of the eikonal wave field, and the ponderomotive Hamiltonian
of the guiding/oscillation centers.  Variation with respect to the
coordinates again yields the kinetic equation, which now includes a term due
to the ponderomotive effects caused by the wave field.  Variation with respect to
the four potential of the {\it background} field again yields the self-consistent
field equation, which now includes a modification in the magnetization density
due to the presence of the wave.  There are then two new additional variations:
Variation with respect to the eikonal wave field amplitude yields the linear
dispersion relation for the wave, and variation with respect to the eikonal
wave phase yields the conservation law for wave action.  Constructive
application of Noether's theorem to this new system action yields the laws of
conservation of energy-momentum and angular momentum for the combined system
of plasma, background field, and wave field.  Specifically, the modification
to the stress-energy  and angular momentum tensors due to the presence of the
wave field is presented and discussed.

\section{Liouville's Theorem}
\label{yar}
\subsection{Lagrangian and Eulerian Descriptions of Relativistic Plasma}
In this section, we present a version of Liouville's theorem that is valid
for relativistic Hamiltonian systems with noncanonical
coordinates.  We begin by examining the difference between the Lagrangian and
Eulerian descriptions of relativistic kinetic theory.

Recall that a Lagrangian description
keeps track of the trajectory of each particle of the system, whereas an
Eulerian description uses a distribution function to specify the phase-space
density of particles (we discussed this briefly in Section~\ref{yaz}).
Thus, a Lagrangian description for a system of relativistic particles might be
the specification of $z(\eta,\tau),$ where $z$ denotes
a set of $n$-dimensional phase space coordinates, $\eta$ is a continuous
particle label, and $\tau (\eta)$ is an orbit parameter
along the world line of the particle with label $\eta.$
Specifying $z$ as a function of $\eta$ and
$\tau$ is equivalent to specifying the phase space orbit of every
particle in the system.  The corresponding Eulerian distribution is
\begin{equation}
f_n(Z)=\int dN(\eta) \int d\tau (\eta)\delta^n \bigl(Z-z(\eta,\tau)\bigr).
\label{pbu}
\end{equation}
Here $dN(\eta)$ is some measure describing the number of particles with
labels between $\eta$ and $\eta+d\eta.$  This measure appears when we pass
from the discrete to the continuum description; that is
\begin{equation}
\sum_{\rm particles}\rightarrow \int dN(\eta).
\end{equation}
In what follows, we shall frequently not bother to write the explicit
$\eta$ dependence of $\tau,$ but it should be kept in mind that each particle
has its own proper time.

Note that $f(Z)$ has support only on a space of dimension smaller than
that of the full $n$-dimensional phase space.  This is because there are
constraints that must be satisfied by the various coordinates involved.
For example, single-particle dynamics must remain on the mass shell, since
$u\cdot u=-c^2.$  Upon making the guiding-center transformation, this
requirement is easily seen to become $H_{gc}=-mc^2/2$ (the guiding-center
transformation is a diffeomorphism, so the mass shell is distorted but not
topologically altered).  So, when using the $(R,K,{\cal B},\mu,\Theta)$
coordinates, $f$ has support on a seven dimensional submanifold in an
eight dimensional phase space.  When we use the $(R,U,\mu,\Theta)$
coordinates the phase space is ten dimensional, and when we use the
$(R,U,\mu,\hat{\alpha})$ coordinates the phase space is thirteen dimensional;
in all cases, however, $f$ has support only on a manifold of seven dimensions
thanks to the constraints on these coordinates.

The Lagrangian description keeps track of the dynamics of all the particles in
the system as though they were distinguishable, and so it includes more degrees
of freedom than the Eulerian description.  That is why it is possible to write
the Eulerian distribution $f(Z)$ in terms of the Lagrangian description
$z(\eta,\tau),$ but it is impossible to do the reverse.  There are many
different functional forms for $z(\eta,\tau)$ that yield the same $f(Z).$
Nevertheless, for a plasma of indistinguishable particles (we are not going to
bother about species labelling in this thesis) it is clear that any
{\it physically relevant} quantity can be expressed in terms of the
Eulerian distribution, $f(Z).$  This is because any physically relevant
quantity should not depend on the identity of the individual particles in the
system.

This is really a gauge invariance issue.  The gauge group is the group of
identical particle interchanges.  The Lagrangian description
keeps track of extra nonphysical {\it gauge degrees of freedom}.  A physically
relevant quantity can be written in terms of the Eulerian distribution since it
is {\it gauge invariant} in this regard.

Consider for example the value of some phase function, $\Phi (z),$ summed over
all the particles in the system and integrated along world lines
\begin{equation}
N_\Phi = \int dN(\eta) \int d\tau \Phi \bigl( z(\eta,\tau) \bigr).
\end{equation}
This object is invariant under the gauge group of identical particle
interchanges because it can be written in terms of the Eulerian distribution
as follows:
\begin{eqnarray}
N_\Phi&=&\int d^nZ \int dN(\eta) \int d\tau \delta^n\bigl( Z-z(\eta,\tau) \bigr)
        \Phi (Z)\nonumber\\
      &=&\int d^nZ f_n(Z) \Phi (Z).
      \label{paa}\nonumber\\
\end{eqnarray}
Though we shall frequently work with the Lagrangian description of things,
we must be able to show that our results can be expressed in terms of the
Eulerian distribution.  Fortunately, this will pose no problem.

The Lagrangian description of the {\it dynamics} of the system is then given
by
\begin{equation}
\dot{z}(\eta,\tau)=V\bigl( [z],z(\eta,\tau) \bigr),
\end{equation}
where the dot denotes differentiation with respect to $\tau$, and where
$V$ is the dynamical vector field expressed as a function of $z(\eta,\tau)$
and as a functional of $z$ (since the dynamics of one particle may depend on
the phase space positions of all the other particles in the system).  The
corresponding Eulerian description of the dynamics is then found as follows:
\begin{eqnarray}
0&=&-\int dN(\eta) \int d\tau \frac{d}{d\tau}
   \delta^n \bigl( Z-z(\eta,\tau) \bigr)\nonumber\\
 &=&\int dN(\eta) \int d\tau \dot{z}(\eta,\tau) \cdot
   \frac{\partial}{\partial Z} \delta^n\bigl( Z-z(\eta,\tau)\bigr)\nonumber\\
 &=&\frac{\partial}{\partial Z}\cdot\left[\int dN(\eta)\int d\tau
    \dot{z}(\eta,\tau)\delta^n\bigl( Z-z(\eta,\tau)\bigr)\right]\nonumber\\
 &=&\frac{\partial}{\partial Z}\cdot\left[\int dN(\eta)\int d\tau
    V\bigl( [z],z(\eta,\tau) \bigr)
   \delta^n\bigl( Z-z(\eta,\tau)\bigr)\right]\nonumber\\
 &=&\frac{\partial}{\partial Z}\cdot\left[
   V( [f_n],Z )\int dN(\eta)\int d\tau
    \delta^n\bigl( Z-z(\eta,\tau)\bigr)\right]\nonumber\\
 &=&\frac{\partial}{\partial Z} \cdot \left[ V( [f_n],Z ) f_n(Z) \right].
 \label{pab}\nonumber\\
\end{eqnarray}
The first line above follows from the fact that at any finite time $\tau$ is
finite, so the delta function vanishes at the limits of integration
$\tau\rightarrow\pm\infty.$  Note that we had to assume that the
functional dependence of $V$ on $z$ could be replaced by a functional
dependence on $f_n$; this is just a statement of the very reasonable condition
that the dynamics cannot depend on particle labels.  The resulting kinetic
equation for $f_n(Z)$ is called the {\it continuity equation}, and it
expresses conservation of particles.  It is true for
any relativistic system of particles, regardless of the nature of the forces
involved (they could even be dissipative in nature).

\subsection{Conservation of Phase Space Volume}
One thing that distinguishes Hamiltonian systems from other dynamical systems
is the property that phase space volume is conserved by a Hamiltonian flow.
This means that if we take a volume element in phase space and drag each point
of its boundary surface along a Hamiltonian vector
field for some parameter increment,
the volume enclosed will be unchanged.  As we shall now see, this property
follows from the Jacobi identity; this fact was used in Section~\ref{yak}
as an argument for using brackets that satisfy the Jacobi identity
{\it exactly} (as opposed to satisfying it only to some order in an
expansion parameter).

Suppose that we have a set of canonical coordinates $Z_c,$ and that the
Eulerian distribution function in these coordinates is $f_c(Z_c).$  Now
under a (possibly noncanonical) coordinate transformation, $Z_c\mapsto Z,$
a distribution function transforms in such a way as to keep the number of
particles in a fixed phase space volume element constant.  That is
\begin{equation}
f(Z)d^nZ=f_c(Z_c)d^nZ_c,
\end{equation}
where $n$ is the number of dimensions in phase space.  Thus, $f$ transforms
like a pseudoscalar,
\begin{equation}
f(Z)=f_c(Z_c)D,
\end{equation}
where we have defined the Jacobian of the transformation
\begin{equation}
D(Z)=\frac{\partial^n Z_c}{\partial^n Z}.
\end{equation}
Alternatively, we can define a scalar distribution function,
${\sf f}(Z),$ which transforms as follows:
\begin{equation}
{\sf f}(Z)=f_c(Z_c).
\end{equation}
It follows that in {\it any} coordinate system we have
\begin{equation}
f(Z)={\sf f}(Z)D(Z).
\label{pac}
\end{equation}
Note that $f(Z)={\sf f}(Z)$ in any {\it canonical} coordinate system,
since the Jacobian of a canonical transformation is unity.  In noncanonical
coordinates, however, $f(Z)$ and ${\sf f}(Z)$ are different.

The Lagrangian two-form in coordinate system $Z$ is given by
\begin{equation}
\Omega_{\mu\nu}=\frac{\partial Z^\alpha_c}{\partial Z^\mu}
                \frac{\partial Z^\beta_c }{\partial Z^\nu}
                \Omega^c_{\alpha\beta},
\end{equation}
where $\Omega^c$ is the canonical Lagrangian two-form.  Taking the
determinant of both sides, we find
\begin{equation}
{\rm det}\Omega=D^2.
\label{pad}
\end{equation}
We now no longer need to make reference to the canonical coordinate system,
$Z_c.$  Eqs.~(\ref{pac}) and (\ref{pad}) tell us all we need to know, and
they are written entirely in the general coordinates, $Z.$

Take the gradient of both sides of Eq.~(\ref{pad}) to get
\begin{eqnarray}
2DD_{,\alpha}&=&({\rm det}\Omega)_{,\alpha}\nonumber\\
             &=&D^2 J^{\beta\gamma} \Omega_{\gamma\beta ,\alpha},
             \label{pae}\nonumber\\
\end{eqnarray}
where we used the formula for the derivative of a determinant,
\begin{equation}
({\rm det}A)_{,\alpha}
  =({\rm det}A) (A^{-1})^{\beta\gamma} A_{\gamma\beta ,\alpha}.
\end{equation}

We are now ready to prove Liouville's theorem.  We have
\begin{eqnarray}
D({\dot{Z}}^\alpha D)_{,\alpha}
  &=&D(J^{\alpha\beta}H_{,\beta}D)_{,\alpha}\nonumber\\
  &=&D^2 H_{,\beta} (J^{\alpha\beta}_{\phantom{\alpha\beta},\alpha}
   +\frac{1}{2} J^{\alpha\beta} J^{\mu\nu} \Omega_{\nu\mu ,\alpha})\nonumber\\
  &=&\frac{1}{2} D^2 H_{,\beta} J^{\beta\mu} J^{\alpha\nu}
   (\Omega_{\nu\mu ,\alpha}+\Omega_{\mu\alpha ,\nu}+\Omega_{\alpha\nu ,\mu})\nonumber\\
  &=&0,
  \label{paf}\nonumber\\
\end{eqnarray}
where we used the above formula for $DD_{,\alpha},$ and
where we used the Jacobi identity in the last step.  Thus, since $D$ is
never zero, we have proved Liouville's theorem,
\begin{equation}
\frac{\partial}{\partial Z} \cdot (\dot{Z}D)=0.
\end{equation}

Now Eq.~(\ref{pab}) may be written for a Hamiltonian system as follows:
\begin{eqnarray}
0&=&\frac{\partial}{\partial Z} \cdot (\dot{Z}f)\nonumber\\
 &=&\frac{\partial}{\partial Z} \cdot (\dot{Z}D {\sf f}).
 \label{pag}\nonumber\\
\end{eqnarray}
Applying Liouville's theorem, we get the Vlasov equation,
\begin{equation}
0=\dot{Z}\cdot\frac{\partial{\sf f}}{\partial Z}.
\label{pah}
\end{equation}
Our proof of this result has been quite general, and so in the future we can
simply write down the Vlasov equation for any Hamiltonian equations of motion.

The careful reader will have noticed that we assumed invertibility of the
Poisson tensor in the above proof, whereas our Poisson tensors in the
$(R,U,\mu,\Theta)$ and $(R,U,\mu,\hat{\alpha})$ coordinate systems are
definitely singular.  Recall, however, that we showed in Section~\ref{yao}
how these constrained coordinate systems could be imbedded in larger
unconstrained coordinate systems.  That is, we can obtain the
$(R,U,\mu,\Theta)$ coordinates by a smooth coordinate transformation from
the $(R,K,{\cal B},\mu,\Theta, C_{1a},C_{1b})$ coordinates, and we can obtain
the $(R,U,\mu,\hat{\alpha})$ coordinates by a smooth coordinate transformation
from the $(R,K,{\cal B},\mu,\Theta, C_{1a},C_{1b},C_{2a},C_{2b},C_3)$
coordinates.  In both cases, the physical motion takes place on the subspace
for which $C_{1a}=C_{1b}=C_{2a}=C_{2b}=0$ and $C_3=1$; if the initial conditions
are on this subspace, the dynamics will keep them there.  From this point of view,
there is nothing singular about the transformation that led to these coordinate
systems, and the only reason that their Poisson tensors are singular is that we
enforced the constraints by setting $C_{1a}=C_{1b}=C_{2a}=C_{2b}=0$ and $C_3=1$
at the very end of the calculation that led to them.

Armed with this insight, it is easy to compute the Jacobian $D$ for these
coordinate systems.  First we consider the guiding-center transformation that
led to the $(R,K,{\cal B},\mu,\Theta)$ coordinates from canonical coordinates.
The Jacobian of this transformation is
\begin{equation}
D_1=\sqrt{{\rm det}\Omega_{gc}},
\end{equation}
where $\Omega_{gc}$ is the Lagrangian two-form
given in Eqs.~(\ref{gga}) through (\ref{ggh}).  The result is
\begin{equation}
D_1=\frac{m^3}{\epsilon}K\Omega_B\Upsilon.
\end{equation}
The coordinates $(C_{1a},C_{1b},C_{2a},C_{2b},C_3),$ which can be thought
of as describing directions transverse to those described by the
$(R,K,{\cal B},\mu,\Theta)$ coordinates, are unaffected by the above
transformation.

We now transform to either the $(R,U,\mu,\Theta)$ system or the
$(R,U,\mu,\hat{\alpha})$ system.  This transformation will involve the
coordinates $(C_{1a},C_{1b},C_{2a},C_{2b},C_3).$  Its Jacobian is given by
\begin{equation}
D_2=\frac{\partial (R,U,\mu,\Theta)}
    {\partial (R,K,{\cal B},\mu,\Theta, C_{1a},C_{1b})}
\end{equation}
or
\begin{equation}
D_2=\frac{\partial (R,U,\mu,\hat{\alpha})}
    {\partial (R,K,{\cal B},\mu,\Theta, C_{1a},C_{1b},C_{2a},C_{2b},C_3)},
\end{equation}
respectively.  We can use the transformation equations, Eqs.~(\ref{gjc})
or (\ref{gje}), to calculate the above expressions.  The important thing
is that we take {\it all} of the derivatives involved in calculating the Jacobian
{\it before} enforcing the constraints by setting
$C_{1a}=C_{1b}=C_{2a}=C_{2b}=0$ and $C_3=1.$  The calculation is straightforward,
and we find that for either the $(R,U,\mu,\Theta)$ or the $(R,U,\mu,\hat{\alpha})$
coordinates we get
\begin{equation}
D_2=\frac{1}{2K\lambda_B^2}.
\end{equation}
The {\it overall} Jacobian of the above transformation is thus
\begin{equation}
D=D_1D_2=\frac{em^2}{2\epsilon c\lambda_B (R)}\Upsilon' (R),
\label{pbt}
\end{equation}
where $\Upsilon'$ is given by Eq.~(\ref{gkd}).
Note that this same expression may be used for the
guiding/oscillation-center problem, since it has exactly the same brackets as
the guiding-center problem with no wave present.  This is because our
oscillation-center Lie transform took the wave perturbation out of the
brackets and put it into the Hamiltonian (which is how we got $K_2$).

Thus by imbedding our singular coordinate systems in larger nonsingular ones,
we are able to validate the above derivation of the Vlasov equation for our
coordinates.  Because we had to introduce the coordinates
$(C_{1a},C_{1b},C_{2a},C_{2b},C_3),$ however, we should ask what the distribution
function looks like, and whether or not the kinetic equation that we have started
with makes sense.  Consider Eq.~(\ref{pbu}), written for the coordinate system
$Z=(R,K,{\cal B},\mu,\Theta, C_{1a},C_{1b},C_{2a},C_{2b},C_3).$  We adopt the
shorthand notation $Z=(Y,C)$ where $Y=(R,K,{\cal B},\mu,\Theta)$ and
$C=(C_{1a},C_{1b},C_{2a},C_{2b},C_3).$  Then we have
\begin{equation}
f_{13}(Z)=\int dN(\eta)\int d\tau \delta^8\left(Y-y\left(\eta,\tau\right)\right)
       \delta^5\left(C-c\left(\eta,\tau\right)\right),
\end{equation}
where $y(\eta,\tau)$ and $c(\eta,\tau)$ give the dynamics of $Y$ and $C,$
respectively.  Note, however, that since the integral, $\int dN(\eta),$
includes only particles that obey the constraints
$C_{1a}=C_{1b}=C_{2a}=C_{2b}=0$ and $C_3=1,$ and since the dynamics is known
to keep such particles on the constraint surface, it must be that
$c(\eta,\tau)=(0,0,0,0,1).$  Thus the delta functions involving $C$ can be
pulled out of the integral to finally yield
\begin{equation}
f_{13}(Z)=\delta (C_{1a})\delta (C_{1b})\delta (C_{2a})\delta (C_{2b})
       \delta (C_{3}-1)
       \int dN(\eta)\int d\tau \delta^8\left(Y-y\left(\eta,\tau\right)\right).
\end{equation}
The proportionality of $f_{13}$ to delta functions in the $C$ is simply a
mathematical restatement of our earlier observation that it has support only on a
space of dimension less than that coordinatized by $Z.$  In fact, it has support
only on a space of seven dimensions (there is another delta function still
hiding in the integral on the right hand side of the above equation due to the
fact that the Hamiltonian is a constant of the motion).  The Vlasov equation written
in these coordinates is then
\begin{equation}
0=\dot{Y}\cdot\frac{\partial {\sf f}_{13}}{\partial Y}, 
\end{equation}
where ${\sf f}_{13}=f_{13}/D_1$ and where the terms
$\dot{C}\cdot\partial {\sf f}_{13}/\partial C$ are not present because
$\dot{C}=0.$  We can now integrate the above Vlasov equation over the $C$
coordinates to get
\begin{equation}
0=\dot{Y}\cdot\frac{\partial {\sf f}_{8}(Y)}{\partial Y},
\end{equation}
where
\begin{eqnarray}
f_8(Y)
  &=&\int d^5C f_{13}(Z)\nonumber\\
  &=&\int dN(\eta)\int d\tau \delta^8\left(Y-y\left(\eta,\tau\right)\right),
  \label{pbv}\nonumber\\
\end{eqnarray}
and ${\sf f}_8=f_8/D_1.$  This is obviously the same Vlasov equation that we would
have obtained if we had used only the clearly nonpathological
$(R,K,{\cal B},\mu,\Theta)$ coordinates from the start.

It turns out to be easier (for reasons that will become clear shortly) to
write the Vlasov equation in terms of ${\sf f}$ and easier to write the
field equation in terms of $f.$  Since we know what $D$ is, however, there
is clearly no problem involved in writing both equations in terms of either
$f$ or ${\sf f}$ (recall that $f$ and ${\sf f}$ are related by Eq.~(\ref{pac})
with $D$ given by Eq.~(\ref{pbt})).

\section{Self-Consistent Kinetic and Field Equations}
\label{yas}
\subsection{Constructing the System Action}
We begin by considering the case in which there is no
eikonal wave field present.  Our action one-form and Hamiltonian
for a single guiding-center are
thus given by Eqs.~(\ref{gjj}) and (\ref{ghq}), respectively.  In
Section~\ref{yau}, we generalize our results to the case in which the plasma
is bathed in an eikonal wave field.  For now we construct the action for the
coupled system of guiding-center plasma and Maxwell field.  This has the form
\begin{equation}
S=S_{gc}+S_m,
\end{equation}
where $S_{gc}$ is the total action of the guiding centers, and where $S_m$
is the action of the Maxwell field.

Now the action of the guiding centers
is found by simply summing that for a single guiding center over the
full distribution.  Thus we write
\begin{eqnarray}
S_{gc}[Z,A_i]
 &=&\int dN(\eta) \int d\tau \Bigl[ \Gamma_{gc}\bigl( Z(\eta,\tau);
   A_i(R(\eta,\tau)),F_i(R(\eta,\tau))\bigr)\cdot\dot{Z}(\eta,\tau)\nonumber\\
 &&\qquad -\sum_\nu \lambda_\nu (\eta,\tau)C_\nu\bigl(
   Z(\eta,\tau);F_i(R(\eta,\tau))\bigr)\nonumber\\
 &&\qquad - H_{gc}\bigl(
   Z(\eta,\tau);F_i(R(\eta,\tau))\bigr)\Bigr].
 \label{pai}\nonumber\\
\end{eqnarray}
Here we have written $Z$ for the full set of boostgauge and gyrogauge
invariant guiding-center coordinates, $(R,U,\mu,\hat{\alpha}).$
We have enforced the constraints by means of Lagrange
multipliers, using $\lambda_\nu$ to denote the multiplier for constraint
$C_\nu,$ where the index $\nu$ runs over all the constraints present as usual.
Finally, we have indicated separately the functional dependence of the various
terms on the four potential $A_i$ and the background field $F_i$ (here $i$
denotes the ordering of the field as discussed in Section~\ref{yah}).

Now Eq.~(\ref{pai}) may be written in the form
\begin{equation}
S_{gc}=\int d^4x {\cal L}_{gc},
\end{equation}
where $x$ denotes spacetime position, and
where we have defined the Lagrangian density for the guiding centers,
\begin{eqnarray}
{\cal L}_{gc}(x)
 &=&\int dN(\eta) \int d\tau \delta^4 \bigl(x-R(\eta,\tau)\bigr)
   \Bigl[ \Gamma_{gc}\bigl( Z(\eta,\tau);
   A_i,F_i\bigr)\cdot\dot{Z}(\eta,\tau)\nonumber\\
 &&\qquad -\sum_\nu \lambda_\nu (\eta,\tau)C_\nu\bigl(
   Z(\eta,\tau);F_i\bigr)- H_{gc}\bigl(
   Z(\eta,\tau);F_i\bigr)\Bigr].
 \label{paj}\nonumber\\
\end{eqnarray}
Here we have adopted the convention that $A_i$ and $F_i$ denote $A_i(x)$
and $F_i(x),$ respectively.

The Maxwell action is well known to be (see, for example, Jackson~\cite{zag})
\begin{equation}
S_m=\int d^4x {\cal L}_m(x),
\label{pbb}
\end{equation}
where the Lagrangian density for the Maxwell field is
\begin{equation}
{\cal L}_m=-\frac{1}{16\pi}(F_0+\epsilon F_1+\cdots):(F_0+\epsilon F_1+\cdots).
\end{equation}
In this study, we shall retain terms in ${\cal L}_m$ only to order
$\epsilon$; thus we write
\begin{equation}
{\cal L}_m=-\frac{1}{16\pi}(F_0:F_0+2\epsilon F_0:F_1).
\label{pbc}
\end{equation}

\subsection{The Vlasov Equation for Guiding Centers}
We first vary the system action with respect to the particle field,
$Z(\eta,\tau).$  After a short calculation, we find
\begin{eqnarray}
0&=&\frac{\delta S}{\delta Z(\eta,\tau)}\nonumber\\
 &=&\Omega_{gc}\bigl( Z(\eta,\tau);
   A_i(R(\eta,\tau)),F_i(R(\eta,\tau))\bigr)\cdot\dot{Z}(\eta,\tau)\nonumber\\
 &&\qquad -\sum_\nu \lambda_\nu (\eta,\tau)
   \frac{\partial C_\nu}{\partial Z}\bigl(
   Z(\eta,\tau);F_i(R(\eta,\tau))\bigr)\nonumber\\
 &&\qquad -\frac{\partial H_{gc}}{\partial Z}\bigl(
   Z(\eta,\tau);F_i(R(\eta,\tau))\bigr).
 \label{pak}\nonumber\\
\end{eqnarray}
where $\Omega_{gc}=d\Gamma_{gc}.$  This equation, coupled with the constraints
\begin{equation}
C_\nu\bigl(Z(\eta,\tau);F_i(R(\eta,\tau))\bigr)=\delta_{\nu 3}
\end{equation}
(which are needed to determine the Lagrange multipliers), shows clearly that
the fields, $Z(\eta,\tau),$ obey the usual equations of motion for a
single guiding center.  Knowing this, and using the ideas developed in the
previous section, it is now possible to write down the Vlasov equation,
\begin{equation}
0=\dot{Z}\cdot\frac{\partial{\sf f}}{\partial Z},
\end{equation}
using the equations of motion for a single guiding center.

In particular, if we use the $(R,U,\mu,\Theta)$ coordinates, this becomes
\begin{equation}
0=\dot{R}\cdot\frac{\partial {\sf f}_{10}}{\partial R}
 +\dot{U}\cdot\frac{\partial {\sf f}_{10}}{\partial U}
 +\dot{\mu}   \frac{\partial {\sf f}_{10}}{\partial \mu}
 +\dot{\Theta}\frac{\partial {\sf f}_{10}}{\partial \Theta}.
\end{equation}
We can now define the {\it guiding-center distribution function},
\begin{equation}
\overline{{\sf f}}_9(R,U,\mu)
  \equiv\int_0^{2\pi}d\Theta {\sf f}_{10}(R,U,\mu,\Theta).
\end{equation}
This is nothing more than $2\pi$ times the $\Theta$-average of the full distribution
function ${\sf f}_{10}.$  Now because $\dot{Z}$ is independent of $\Theta$ (thanks
to our guiding-center transformation) and because $\dot{\mu}=0,$ taking the
$\Theta$-average of the above kinetic equation yields
\begin{equation}
0=\dot{R}\cdot\frac{\partial {\overline{\sf f}}_9}{\partial R}
 +\dot{U}\cdot\frac{\partial {\overline{\sf f}}_9}{\partial U}
\end{equation}
This is the {\it reduced} kinetic equation for the guiding-center distribution
function.

\subsection{The Field Equations}
Generally speaking, the idea is now to
vary the above action with respect to the four potential to get the dynamical
equations for the fields.  This must be done carefully, however, as there
are two additional constraints that such variation must respect.  Recall
that in our derivation of the guiding-center action we assumed that the
background field scale lengths were large in comparison to the gyroradius,
and we assumed that the zero-order fields have $\lambda_E=0.$  We must make
certain that the dynamics of the fields do not evolve them into a configuration
for which either of these assumptions are violated.  In order to get dynamical
equations for the fields that respect these constraints, our variation of
the action with respect to the four potential must be a {\it constrained
variation}; that is, arbitrary variations of the four potential are not
allowed.  Only those variations of the four potential that preserve the
vanishing of $\lambda_E$ to lowest order and the smallness of the ratio of
gyroradius to scale length are allowed.

We thus begin our derivation of the field equations by examining the variation
of the action due to variations of the $A_i,$ without assuming in any way that
the variations of the $A_i$ are arbitrary.  Recall that we have indicated
separately the functional dependence of the various terms in the action on the
four potential $A_i$ and the background field $F_i.$  Of course, $F_i=dA_i,$
so when we vary with respect to the $A_i$ we must take into account the $F_i$
dependence.  To do this, it is convenient to distinguish between {\it total}
and {\it partial} functional derivatives with respect to $A_i.$  We use the
chain rule to write
\begin{equation}
{\left.\frac{\delta S}{\delta {A_i}_\rho (x)}\right|}_{\rm total}
  =\frac{\delta S}{\delta {A_i}_\rho (x)}
  +\int d^4x' \frac{\delta S}{\delta {F_i}_{\mu\nu} (x')}
   \frac{\delta {F_i}_{\mu\nu} (x')}{\delta {A_i}_\rho (x)}.
\label{pam}
\end{equation}
To proceed, note that
\begin{eqnarray}
{F_i}_{\mu\nu}(x')
  &=&{A_i}_{\nu,\mu}(x')-{A_i}_{\mu,\nu}(x')\nonumber\\
  &=&\int d^4x \delta^4 (x-x') {A_i}_{\nu,\mu}(x)-{A_i}_{\mu,\nu}(x)\nonumber\\
  &=&\int d^4x \left\{ {A_i}_\mu [\delta^4 (x-x')]_{,\nu}
                   -{A_i}_\nu [\delta^4 (x-x')]_{,\mu} \right\},
  \label{pal}\nonumber\\
\end{eqnarray}
so that
\begin{equation}
\frac{\delta {F_i}_{\mu\nu}(x')}{\delta {A_i}_{\rho}(x)}
  =\delta_{\mu\rho} [\delta^4 (x-x')]_{,\nu}
  -\delta_{\nu\rho} [\delta^4 (x-x')]_{,\mu}.
\end{equation}
Using this in Eq.~(\ref{pam}), we get
\begin{equation}
{\left.\frac{\delta S}{\delta A_i}\right|}_{\rm total}
  =\frac{\delta S}{\delta A_i}
  -2\delr\cdot\left(\frac{\delta S}{\delta F_i}\right).
\label{pan}
\end{equation}
This formula is very useful in what follows.

Using Eq.~(\ref{pan}) to vary the action with respect to the four potential,
we arrive straightforwardly at the following result:
\begin{equation}
\delta S=\int d^4x
         \left[{\cal J}_0\left( x\right)\cdot\delta A_0\left( x\right)
              +{\cal J}_1\left( x\right)\cdot\delta A_1\left( x\right)\right],
\end{equation}
where we have defined
\begin{equation}
{\cal J}_0(x)\equiv\frac{1}{c}J(x)+\frac{1}{4\pi}\delr\cdot G_0(x)
\label{pcg}
\end{equation}
and
\begin{equation}
{\cal J}_1(x)\equiv\frac{\epsilon}{c}J(x)+\frac{\epsilon}{4\pi}\delr\cdot G_1(x),
\label{pch}
\end{equation}
where in turn we have defined the {\it guiding-center current density}
\begin{eqnarray}
J(x)&\equiv& c\int dN(\eta) \int d\tau \delta^4 (x-R(\eta,\tau))
  \frac{\partial \Gamma_{gc}}{\partial A_0}(Z(\eta,\tau);A_i,F_i)
  \cdot\dot{Z}(\eta,\tau)\nonumber\\
    &=&\frac{c}{\epsilon}\int dN(\eta) \int d\tau \delta^4 (x-R(\eta,\tau))
  \frac{\partial \Gamma_{gc}}{\partial A_1}(Z(\eta,\tau);A_i,F_i)
  \cdot\dot{Z}(\eta,\tau)\nonumber\\
    &=&\frac{e}{\epsilon}\int dN(\eta) \int d\tau \delta^4 (x-R(\eta,\tau))
  \dot{R}(\eta,\tau)\nonumber\\
    &=&\frac{e}{\epsilon}\int dR\int dU\int d\mu \int d\Theta
  {f}_{10}(R,U,\mu,\Theta) \delta^4 (x-R)
  \dot{R}(R,U,\mu)\nonumber\\
    &=&\frac{e}{\epsilon}\int dR\int dU\int d\mu
  {\overline{f}}_9(R,U,\mu) \delta^4 (x-R)
  \dot{R}(R,U,\mu),
    \label{pao}\nonumber\\
\end{eqnarray}
and the {\it macroscopic field tensors}
\begin{equation}
G_0(x)\equiv F_0(x)+\epsilon F_1(x)-4\pi M_0(x)
\label{pcb}
\end{equation}
\begin{equation}
G_1(x)\equiv F_0(x)-4\pi M_1(x),
\label{pcc}
\end{equation}
and where in turn we have defined the
{\it guiding-center magnetization densities}
\begin{eqnarray}
M_0(x)&\equiv& 2\int dN(\eta)
  \int d\tau \delta^4 (x-R(\eta,\tau)) \Bigl[
  \frac{\partial \Gamma}{\partial F_0}(Z(\eta,\tau);A_i,F_i)
  \cdot\dot{Z}(\eta,\tau)\nonumber\\
      &&\qquad -\sum_\nu \lambda_\nu(\eta,\tau)
  \frac{\partial C_\nu}{\partial F_0}(Z(\eta,\tau);F_i)-
  \frac{\partial H_{gc}}{\partial F_0}(Z(\eta,\tau);F_i) \Bigr]\nonumber\\
      &=&2\int dR\int dU\int d\mu \int d\Theta
  {f}_{10}(R,U,\mu,\Theta) \delta^4 (x-R) \Bigl[
  \frac{\partial \Gamma}{\partial F_0}(Z;A_i,F_i)\nonumber\\
  &&\qquad\cdot\dot{Z}(R,U,\mu)
      -\sum_\nu \lambda_\nu
  \frac{\partial C_\nu}{\partial F_0}(Z;F_i)-
  \frac{\partial H_{gc}}{\partial F_0}(Z;F_i) \Bigr]\nonumber\\
      &=&2\int dR\int dU\int d\mu
  {\overline{f}}_9(R,U,\mu) \delta^4 (x-R) \Bigl[
  \frac{\partial \Gamma}{\partial F_0}(Z;A_i,F_i)
  \cdot\dot{Z}(R,U,\mu)\nonumber\\
      &&\qquad -\sum_\nu \lambda_\nu
  \frac{\partial C_\nu}{\partial F_0}(Z;F_i)-
  \frac{\partial H_{gc}}{\partial F_0}(Z;F_i) \Bigr]
      \label{pap}\\
M_1(x)&\equiv& \frac{2}{\epsilon}
  \int dN(\eta) \int d\tau \delta^4 (x-R(\eta,\tau)) \Bigl[
  \frac{\partial \Gamma}{\partial F_1}(Z(\eta,\tau);A_i,F_i)
  \cdot\dot{Z}(\eta,\tau)\nonumber\\
      &&\qquad -\sum_\nu \lambda_\nu(\eta,\tau)
  \frac{\partial C_\nu}{\partial F_1}(Z(\eta,\tau);F_i)-
  \frac{\partial H_{gc}}{\partial F_1}(Z(\eta,\tau);F_i) \Bigr]\nonumber\\
      &=&2\int dR\int dU\int d\mu \int d\Theta
  {f}_{10}(R,U,\mu,\Theta) \delta^4 (x-R) \Bigl[
  \frac{\partial \Gamma}{\partial F_1}(Z;A_i,F_i)\nonumber\\
  &&\qquad\cdot\dot{Z}(R,U,\mu)
      -\sum_\nu \lambda_\nu
  \frac{\partial C_\nu}{\partial F_1}(Z;F_i)-
  \frac{\partial H_{gc}}{\partial F_1}(Z;F_i) \Bigr]\nonumber\\
      &=&2\int dR\int dU\int d\mu
  {\overline{f}}_9(R,U,\mu) \delta^4 (x-R) \Bigl[
  \frac{\partial \Gamma}{\partial F_1}(Z;A_i,F_i)
  \cdot\dot{Z}(R,U,\mu)\nonumber\\
      &&\qquad -\sum_\nu \lambda_\nu
  \frac{\partial C_\nu}{\partial F_1}(Z;F_i)-
  \frac{\partial H_{gc}}{\partial F_1}(Z;F_i) \Bigr].
      \label{pbw}
\end{eqnarray}
Note that the magnetization came from the second term
on the right of Eq.~(\ref{pan}).  Also note that
the only thing that depends explicitly on $F_1$ is the first-order
piece of the Hamiltonian, so that only the last term in square brackets
in the above expression for $M_1$ survives;
of course, $F_1$ also appears in the brackets due to the $A_1$ dependence of
$\Gamma_{gc}.$  Finally note that we were able to write the current and the
magnetizations in terms of the reduced Eulerian distribution function,
${\overline{\sf f}}_9.$

Now because the $\delta A_i$ are not arbitrary, we cannot simply set
${\cal J}_0={\cal J}_1=0.$  Instead, as discussed above, we must restrict the
variation so that it respects the constraints that $\lambda_E=0$ to lowest
order and that the ratio of gyroradius to scale length is small.  To deal
with the first of these constraints, let us temporarily introduce Clebsch
variables for the fields.  We define four scalar fields,
$\alpha (x), \beta (x), \kappa (x), \sigma (x),$ such that in terms of these
fields the four potential is given by
\begin{equation}
A_0=\alpha d\beta
\end{equation}
\begin{equation}
A_1=\kappa d\sigma,
\end{equation}
and consequently the field tensor is given by
\begin{equation}
F_0=dA_0=d(\alpha d\beta)=d\alpha\wedge d\beta
\label{pce}
\end{equation}
\begin{equation}
F_1=dA_1=d(\kappa d\sigma)=d\kappa\wedge d\sigma.
\label{pcf}
\end{equation}
That such scalar fields exist is guaranteed by the Darboux theorem.  That is,
because $F$ is a closed two-form, it can be written in the form
$F=d\alpha\wedge d\beta+\epsilon d\kappa\wedge d\sigma,$ where we are
guaranteed enough freedom to choose $\alpha$ and $\beta$ such that
$P_\parallel\cdot (d\alpha\wedge d\beta)=0.$

It is clear that the above construction insures that
\begin{equation}
P_\parallel\cdot F_0=0.
\end{equation}
Note that we are ignoring $F_i$ for $i\ge 2,$ and that the parallel electric
field must lie entirely within $F_1.$  Thus, the specification of the four functions
$\alpha (x), \beta (x), \kappa (x),$ and $\sigma (x)$ is a coordinatization
of the function space of all electromagnetic fields that automatically
ensures the satisfaction of the constraint that $\lambda_E=0$ to lowest order.

The variation of the action with respect to the four potentials may now be
written
\begin{eqnarray}
\delta S
  &=&\int d^4x \left[{\cal J}_0\cdot\delta (\alpha\delr\beta )
                 +{\cal J}_1\cdot\delta (\kappa\delr\sigma  )\right]\nonumber\\
  &=&\int d^4x (\delta\alpha {\cal J}_0\cdot\delr\beta
                  +\alpha {\cal J}_0\cdot\delr\delta\beta
            +\delta\kappa {\cal J}_1\cdot\delr\sigma
                  +\kappa {\cal J}_1\cdot\delr\delta\sigma )\nonumber\\
  &=&\int d^4x \left[ \delta\alpha {\cal J}_0\cdot\delr\beta
                  -\delta\beta\delr\cdot (\alpha {\cal J}_0)
                  +\delta\kappa {\cal J}_1\cdot\delr\sigma
                  -\delta\sigma \delr\cdot (\kappa {\cal J}_1) \right].
  \label{paq}\nonumber\\
\end{eqnarray}
We still cannot set the coefficients of the variations equal to zero, however,
because of the remaining constraint that the fields remain sufficiently
slowly varying for the guiding-center approximation to remain valid.  This
point requires some discussion.

Consider a general Fourier decomposition of the electromagnetic field in and
around a plasma.  We can divide the Fourier space into three regions.  The
first consists of slowly varying fields for which the guiding-center
approximation is clearly valid; we call these {\it background fields}.
The second consists of rapidly varying fields that are due to collective
motion of the plasma; we call these {\it wave fields}, and their effect on
a single guiding center was the subject of Chapter~\ref{yap}.  Note that
wave fields violate the guiding-center approximation, and the only reason that
we were able to treat them perturbatively was our assumption that their
amplitudes are small.  The third consists of the extremely rapid fluctuations
associated with collisions and higher correlations.

Now fields belonging to the third region of Fourier space are clearly
outside of the scope of this thesis; our Vlasov kinetic description of
the plasma neglects correlations.  Wave fields were studied
in a single particle context in
Chapter~\ref{yap}, and their self-consistent evolution will be studied in
Section~\ref{yau}.  For now we are interested in the dynamics of the
background fields.  We thus define a projection operator, ${\cal P},$
that, when applied to an arbitrary field, projects out the part that is
slowly varying.  We shall not be specific about the nature of this operator
except to say that, since it is a projection operator, we expect it to be
idempotent.  A moment's thought convinces one that this means that it must
be a convolution of the field with a filter function whose Fourier transform
is piecewise constant, having a value of either zero or one everywhere in
Fourier space.  Specifically, it has a value of one in the first of the
above-described three regions of Fourier space, and a value of zero in the
other two regions.  Exactly how one draws these boundaries is what we are
leaving unspecified.

Thus, although we cannot set the coefficients of
$\delta\alpha (x), \delta\beta (x), \delta\kappa (x),$ and
$\delta\sigma (x)$ equal to zero in Eq.~(\ref{paq}), we can enforce the
constraint that the fields are slowly varying by requiring that their
{\it variations} be slowly varying; thus
\begin{equation}
\delta\alpha (x)={\cal P}\delta\alpha (x)
\label{pcd}
\end{equation}
(and similarly for the other three variations).  We can also decompose the
coefficients of the variations into slowly varying and rapidly varying parts;
thus
\begin{equation}
{\cal J}_0\cdot\delr\beta={\cal P}({\cal J}_0\cdot\delr\beta)
  +({\bf 1}-{\cal P})({\cal J}_0\cdot\delr\beta)
\end{equation}
(and similarly for the other three coefficients).  Thus, upon multiplying
$\delta\alpha (x)$ and ${\cal J}_0\cdot\delr\beta,$ we get the product of
the slowly varying terms and a
cross term.  Now the cross term is clearly oscillatory and
vanishes upon integration over $x.$  It is then legal to
set the coefficients of the {\it slowly varying} parts of the variations
equal to zero.  This essentially means that we can set the projection of the
coefficients of the variations in Eq.~(\ref{paq}) equal to zero.

Thus, we get
\begin{equation}
{\cal P}[{\cal J}_0\cdot\delr\beta]=0
\end{equation}
\begin{equation}
{\cal P}[\delr\cdot (\alpha {\cal J}_0)]=0
\end{equation}
\begin{equation}
{\cal P}[{\cal J}_1\cdot\delr\sigma]=0
\end{equation}
\begin{equation}
{\cal P}[\delr\cdot (\kappa {\cal J}_1)]=0.
\end{equation}
Now note that from Eq.~(\ref{pao}), we have
\begin{eqnarray}
J^\mu_{\phantom{\mu} ,\mu}
  &=& \frac{e}{\epsilon}\int dN(\eta) \int d\tau
    \frac{\partial \delta^4 \left(x-R\left(\eta,\tau\right)\right)}{\partial x^\mu}
    \dot{R}^\mu(\eta,\tau)\nonumber\\
  &=&-\frac{e}{\epsilon}\int dN(\eta) \int d\tau
    \frac{\partial \delta^4 \left(x-R\left(\eta,\tau\right)\right)}{\partial R^\mu}
    \dot{R}^\mu(\eta,\tau)\nonumber\\
  &=&-\frac{e}{\epsilon}\int dN(\eta) \int dR^\mu
    \frac{\partial \delta^4 \left(x-R\left(\eta,\tau\right)\right)}{\partial R^\mu}
    \nonumber\\
  &=&0,
  \label{pbx}\nonumber\\
\end{eqnarray}
where the last step follows from the fact that the delta function vanishes at the
limits of integration for finite $x.$  This result expresses conservation of
particles.  From this it follows that
\begin{equation}
\delr\cdot {\cal J}_0=\frac{1}{\epsilon}\delr\cdot {\cal J}_1
  =\frac{1}{c}\delr\cdot J=0.
\end{equation}
So our field equations become
\begin{equation}
{\cal P}[{\cal J}_0\cdot\delr\alpha]={\cal P}[{\cal J}_0\cdot\delr\beta]=0
\end{equation}
\begin{equation}
{\cal P}[{\cal J}_1\cdot\delr\kappa]={\cal P}[{\cal J}_1\cdot\delr\sigma]=0.
\end{equation}
Thus it follows that
\begin{equation}
{\cal P}[{\cal J}_0\cdot (\delr\alpha\delr\beta-\delr\beta\delr\alpha)]=0
\end{equation}
\begin{equation}
{\cal P}[{\cal J}_1\cdot (\delr\kappa\delr\sigma-\delr\sigma\delr\kappa)]=0,
\end{equation}
or
\begin{equation}
{\cal P}[F_0\cdot {\cal J}_0]=0
\end{equation}
\begin{equation}
{\cal P}[F_1\cdot {\cal J}_1]=0.
\end{equation}
Note that the Clebsch potentials have disappeared from our final result; this
was essential since they have a gauge freedom and we expect our result to be
gauge invariant.  We simply used the Clebsch potentials to enforce our
constraints, and then we got rid of them.

The final results for the field equations are thus
\begin{equation}
{\cal P}[F_0\cdot (\frac{1}{4\pi}\delr\cdot G_0 +\frac{1}{c}J)]=0
\label{pbf}
\end{equation}
\begin{equation}
{\cal P}[F_1\cdot (\frac{1}{4\pi}\delr\cdot G_1 +\frac{1}{c}J)]=0.
\label{pbg}
\end{equation}
Note that the first describes field evolution due to perpendicular four current,
while the second describes field evolution due to parallel four current.

\subsection{Summary of Self-Consistent Kinetic and Field Equations}
To summarize the results of this section, we present the complete set
of kinetic and field equations for the guiding-center plasma.  The
kinetic equation is
\begin{equation}
0=\dot{R}\cdot\frac{\partial {\overline{\sf f}}_9}{\partial R}
 +\dot{U}\cdot\frac{\partial {\overline{\sf f}}_9}{\partial U},
\end{equation}
where $\dot{R}=\{ R,H_{gc} \}$ and $\dot{U}=\{ U,H_{gc} \},$ and where in turn
the Poisson brackets are given in Eqs.~(\ref{ghk}) through (\ref{ghp}) and
the Hamiltonian is given in Eq.~(\ref{ghq}).  The field equations are then
\begin{equation}
{\cal P}[F_0\cdot (\frac{1}{4\pi}\delr\cdot G_0 +\frac{1}{c}J)]=0
\end{equation}
\begin{equation}
{\cal P}[F_1\cdot (\frac{1}{4\pi}\delr\cdot G_1 +\frac{1}{c}J)]=0,
\end{equation}
where the current is given by
\begin{equation}
J(x)=\frac{e}{\epsilon}\int dR\int dU\int d\mu
  {\overline{f}}_9(R,U,\mu) \delta^4 (x-R)
  \dot{R}(R,U,\mu)
\end{equation}
and the macroscopic field tensors are given by
\begin{equation}
G_0(x)\equiv F_0(x)+\epsilon F_1(x)-4\pi M_0(x)
\end{equation}
\begin{equation}
G_1(x)\equiv F_0(x)-4\pi M_1(x),
\end{equation}
and where in turn the magnetization densities are given by
\begin{eqnarray}
M_0(x)&=&2\int dR\int dU\int d\mu
  {\overline{f}}_9(R,U,\mu) \delta^4 (x-R) \Bigl[
  \frac{\partial \Gamma}{\partial F_0}(Z;A_i,F_i)
  \cdot\dot{Z}(R,U,\mu)\nonumber\\
      &&\qquad -\sum_\nu \lambda_\nu
  \frac{\partial C_\nu}{\partial F_0}(Z;F_i)-
  \frac{\partial H_{gc}}{\partial F_0}(Z;F_i) \Bigr]
      \label{pby}\\
M_1(x)&=&2\int dR\int dU\int d\mu
  {\overline{f}}_9(R,U,\mu) \delta^4 (x-R) \Bigl[
  \frac{\partial \Gamma}{\partial F_1}(Z;A_i,F_i)
  \cdot\dot{Z}(R,U,\mu)\nonumber\\
      &&\qquad -\sum_\nu \lambda_\nu
  \frac{\partial C_\nu}{\partial F_1}(Z;F_i)-
  \frac{\partial H_{gc}}{\partial F_1}(Z;F_i) \Bigr].
      \label{pbz}
\end{eqnarray}
Of course, these must be supplemented by the homogeneous field equations,
\begin{equation}
\delr\cdot {\cal F}_0=0
\end{equation}
\begin{equation}
\delr\cdot {\cal F}_1=0.
\end{equation}
Note that ${\overline{f}}_9(R,U,\mu)$ and ${\overline{\sf f}}_9(R,U,\mu)$
are related by
\begin{equation}
{\overline{f}}_9(R,U,\mu)=D(R){\overline{\sf f}}_9(R,U,\mu),
\end{equation}
where the Jacobian $D$ is given by
\begin{equation}
D=\frac{em^2}{2\epsilon c\lambda_B (R)}\Upsilon' (R),
\end{equation}
and where in turn $\Upsilon' (R)$ is given by Eq.~(\ref{gkd}).

\section{Conservation Laws for the Guiding-Center Plasma}
\label{yat}
\subsection{The Noether Method}
We now employ Noether's theorem to deduce conservation laws for the
energy-momentum and the angular momentum of the guiding-center plasma.  The
technique has been described by Similon~\cite{zbr}, and we shall compare
our results to his.  We begin by considering the variation in the
Lagrangian density due to the variation of all the fields.  We start with
${\cal L}\equiv {\cal L}_{gc}+{\cal L}_{m},$ and apply the variation.
Whenever terms involving the derivative of a variation appear, we replace them
by a pure divergence minus a term for which the variation is not
differentiated; this is almost like integration by parts, but since there is no
integral sign, we must keep the pure divergence terms.  When we are done,
we shall find that $\delta {\cal L}$ is equal to a pure divergence minus
terms, for each field present, that consist of the variation of that field
times the corresponding equation of motion.
Thus, if we then use the equations of
motion, we can reduce $\delta {\cal L}$ to a pure divergence.  The algebra
is tedious but very straightforward, and we get
\begin{eqnarray}
\delta {\cal L} (x)
 &=&\delr\cdot \Bigl\{\frac{1}{4\pi}\delta A_0\cdot G_0+\frac{\epsilon}{4\pi}
  \delta A_1\cdot G_1+{\cal J}_0\alpha\delta\beta+{\cal J}_1\kappa\delta\sigma\nonumber\\
 &&\qquad +\int dN(\eta) \int d\tau \delta^4(x-R(\eta,\tau))\bigl[
  \dot{R}(\eta,\tau)\Gamma_{gc}(Z(\eta,\tau);A_i,F_i)
  \cdot\delta Z(\eta,\tau)\nonumber\\
 &&\qquad -\delta R(\eta,\tau)\bigl(\Gamma_{gc}(Z(\eta,\tau);A_i,F_i)
  \cdot\dot{Z}(\eta,\tau)\nonumber\\
 &&\qquad-\sum_\nu\lambda_\nu (\eta,\tau)
  C_\nu (Z(\eta,\tau);A_i,F_i)-H(Z(\eta,\tau);F_i)\bigr)\bigr]\Bigr\}.
 \label{par} \nonumber\\
\end{eqnarray}

\subsection{Conservation of Energy-Momentum}
To derive the conservation law for energy-momentum, we consider variations
in the coordinates that effectively translate in spacetime all the particles
of the plasma, the fields in the plasma, the external coils that generate the
fields, etc.  Following Similon~\cite{zbr}, we write these as follows:
\begin{equation}
\delta R=\xi
\label{pcj}
\end{equation}
\begin{equation}
\delta U=0
\end{equation}
\begin{equation}
\delta \mu=0
\end{equation}
\begin{equation}
\delta\hat{\alpha}=0,
\end{equation}
where $\xi$ is a constant vector.
Thus, the particles' position coordinates are pushed forward without
altering any of their other phase space coordinates.  The fields translate
according to the prescription
\begin{equation}
\delta\alpha=-\xi\cdot\delr\alpha
\end{equation}
\begin{equation}
\delta\beta=-\xi\cdot\delr\beta
\end{equation}
\begin{equation}
\delta\kappa=-\xi\cdot\delr\kappa
\end{equation}
\begin{equation}
\delta\sigma=-\xi\cdot\delr\sigma,
\end{equation}
so
\begin{eqnarray}
\delta A_0
  &=&\delta (\alpha\delr\beta)\nonumber\\
  &=&\delta\alpha\delr\beta+\alpha\delr\delta\beta\nonumber\\
  &=&-\xi\cdot\delr\alpha\delr\beta-\alpha\xi\cdot\delr\delr\beta\nonumber\\
  &=&-\xi\cdot (\delr\alpha\delr\beta+\alpha\delr\delr\beta)\nonumber\\
  &=&-\xi\cdot\delr (\alpha\delr\beta)\nonumber\\
  &=&-\xi\cdot\delr A_0,
  \label{pas}\nonumber\\
\end{eqnarray}
and similarly
\begin{equation}
\delta A_1=-\xi\cdot\delr A_1.
\end{equation}
Finally note that the Lagrangian densities transform like scalar fields so
\begin{equation}
\delta {\cal L}_{gc}=-\xi\cdot\delr {\cal L}_{gc}
\end{equation}
\begin{equation}
\delta {\cal L}_{m }=-\xi\cdot\delr {\cal L}_{m }.
\end{equation}

Inserting these into Eq.~(\ref{par}), a short manipulation yields
\begin{equation}
\delr\cdot T=0,
\label{pau}
\end{equation}
where we have introduced the {\it stress-energy tensor}
\begin{eqnarray}
T(x)&\equiv& -\frac{1}{4\pi}G_0(x)\cdot F_0(x)
     -\frac{\epsilon}{4\pi}G_1(x)\cdot F_1(x)
     +{\cal L}_m{\bf 1}\nonumber\\
    &&\qquad +\int dN(\eta) \int d\tau \delta^4 (x-R(\eta,\tau))
     \dot{R}(\eta,\tau) mU(\eta,\tau)\nonumber\\
    &=&      -\frac{1}{4\pi}G_0(x)\cdot F_0(x)
     -\frac{\epsilon}{4\pi}G_1(x)\cdot F_1(x)
     +{\cal L}_m{\bf 1}\nonumber\\
    &&\qquad +\int dR\int dU\int d\mu\int d\Theta f_{10} (R,U,\mu,\Theta)
     \delta^4 (x-R)\dot{R} (mU)\nonumber\\
    &=&      -\frac{1}{4\pi}G_0(x)\cdot F_0(x)
     -\frac{\epsilon}{4\pi}G_1(x)\cdot F_1(x)
     +{\cal L}_m{\bf 1}\nonumber\\
    &&\qquad +\int dR\int dU\int d\mu \overline{f}_9 (R,U,\mu)
     \delta^4 (x-R)\dot{R} (mU)
    \label{pat}\nonumber\\
\end{eqnarray}
Eq.~(\ref{pau}) expresses conservation of energy-momentum in the guiding-center
plasma.  Note that the last form for the stress-energy tensor given in
Eq.~(\ref{pat}) expresses the result in terms of the reduced Eulerian distribution
function, $\overline{f}_9.$

\subsection{Conservation of Angular Momentum}
To derive the conservation law for angular momentum, we consider variations
in the coordinates that effectively rotate about the origin of spacetime all the
particles of the plasma, the fields in the plasma, the external coils that generate
the fields, etc.  Following Similon~\cite{zbr}, we write these as follows:
\begin{equation}
\delta R=\Omega\cdot R
\label{pci}
\end{equation}
\begin{equation}
\delta U=\Omega\cdot U
\end{equation}
\begin{equation}
\delta \mu=0
\end{equation}
\begin{equation}
\delta\hat{\alpha}=\Omega\cdot\hat{\alpha},
\end{equation}
where $\Omega$ is a constant antisymmetric second rank tensor.
Thus, the particles' coordinates, $R, U,$ and $\hat{\alpha},$ transform like
vectors undergoing an infinitesimal rotation.  The fields rotate
according to the prescription
\begin{equation}
\delta\alpha=-(\Omega\cdot x)\cdot\delr\alpha
\end{equation}
\begin{equation}
\delta\beta=-(\Omega\cdot x)\cdot\delr\beta
\end{equation}
\begin{equation}
\delta\kappa=-(\Omega\cdot x)\cdot\delr\kappa
\end{equation}
\begin{equation}
\delta\sigma=-(\Omega\cdot x)\cdot\delr\sigma,
\end{equation}
so
\begin{eqnarray}
\delta A_0
  &=&\delta (\alpha\delr\beta)\nonumber\\
  &=&\delta\alpha\delr\beta+\alpha\delr\delta\beta\nonumber\\
  &=&-(\Omega\cdot x)\cdot\delr\alpha\delr\beta
    +\alpha\delr [-(\Omega\cdot x)\cdot\delr\beta ]\nonumber\\
  &=&-(\Omega\cdot x)\cdot (\delr\alpha\delr\beta+\alpha\delr\delr\beta)
    +\alpha\Omega\cdot\delr\beta\nonumber\\
  &=&-(\Omega\cdot x)\cdot\delr (\alpha\delr\beta)
    +\Omega\cdot (\alpha\delr\beta)\nonumber\\
  &=&-(\Omega\cdot x)\cdot\delr A_0+\Omega\cdot A_0,
  \label{pav}\nonumber\\
\end{eqnarray}
and similarly
\begin{equation}
\delta A_1=-(\Omega\cdot x)\cdot\delr A_1+\Omega\cdot A_1.
\end{equation}
Finally note that the Lagrangian densities transform like scalar fields so
\begin{equation}
\delta {\cal L}_{gc}=-(\Omega\cdot x)\cdot\delr {\cal L}_{gc}
\end{equation}
\begin{equation}
\delta {\cal L}_{m }=-(\Omega\cdot x)\cdot\delr {\cal L}_{m }.
\end{equation}

Inserting these into Eq.~(\ref{par}), a short manipulation yields
\begin{eqnarray}
\delr\cdot &&\left[ T\cdot\Omega\cdot x+\int dN(\eta)\int d\tau
   \delta^4 (x-R(\eta,\tau))\dot{R}(\eta,\tau)
  {\Gamma_{gc}}_{\hat{\alpha}}(Z(\eta,\tau);A_i,F_i)\cdot\Omega\cdot
  \hat{\alpha}\right]\nonumber\\
  &&\qquad =0,
  \label{paw}\nonumber\\
\end{eqnarray}
where $T$ is the stress-energy tensor given by Eq.~(\ref{pat}).  Since
$\Omega$ is the generator of an {\it arbitrary} rotation, this becomes
\begin{equation}
\delr\cdot (L+S)=0.
\label{pay}
\end{equation}
Here we have defined the third rank {\it orbital angular momentum tensor}
\begin{equation}
L^{\alpha\beta\gamma}\equiv T^{\alpha\beta}x^\gamma-T^{\alpha\gamma}x^\beta,
\label{pck}
\end{equation}
and the third rank {\it spin angular momentum tensor}
\begin{eqnarray}
S^{\alpha\beta\gamma}&\equiv&
  \int dN(\eta) \int d\tau \delta^4 (x-R(\eta,\tau))\nonumber\\
  &&\qquad {\dot{R}}^\alpha (\eta,\tau)
  [\Gamma_{\hat{\alpha}}^\beta (Z(\eta,\tau);A_i,F_i){\hat{\alpha}}^\gamma
  -\Gamma_{\hat{\alpha}}^\gamma (Z(\eta,\tau);A_i,F_i){\hat{\alpha}}^\beta]\nonumber\\
  &=&\int dN(\eta) \int d\tau \delta^4 (x-R(\eta,\tau))
  \frac{\epsilon\mu (\eta,\tau)}{\lambda_B (R(\eta,\tau))}
  \dot{R}^\alpha (\eta,\tau) F_0^{\beta\gamma}(R(\eta,\tau))\nonumber\\
  &=&\int dR\int dU\int d\mu\int d\Theta f_{10}(R,U,\mu,\Theta)\delta^4 (x-R)
  \frac{\epsilon\mu}{\lambda_B}\dot{R}^\alpha (R,U,\mu) F_0^{\beta\gamma}\nonumber\\
  &=&\int dR\int dU\int d\mu {\overline{f}}_9 (R,U,\mu)\delta^4 (x-R)
  \frac{\epsilon\mu}{\lambda_B}\dot{R}^\alpha (R,U,\mu) F_0^{\beta\gamma}.
  \label{pax}\nonumber\\
\end{eqnarray}
Eq.~(\ref{pay}) expresses conservation of angular
momentum in the guiding-center plasma.

We pause to interpret our result for the guiding-center spin, Eq.~(\ref{pax}).
In a preferred frame, $F_0^{\beta\gamma}=0$ if either $\beta=0$ or $\gamma=0,$
so we need consider only those components of $S^{\alpha\beta\gamma}$ for
which neither $\beta$ nor $\gamma$ is zero, as all the rest vanish.  Using
Eq.~(\ref{gke}) for $F_0$ in a preferred frame, we quickly find that
\begin{equation}
S^{\alpha ij}=\int dR\int dU\int d\mu {\overline{f}}_9 (R,U,\mu)\delta^4 (x-R)
  \epsilon\mu\dot{R}^\alpha (R,U,\mu) \epsilon^{ijk} {\bf b}_k,
\end{equation}
where Latin indices run from one to three, as usual.
Now in three dimensions one must take the three-dual of the angular momentum
tensor to get the angular momentum vector.  We can now do this for the last
two indices of $S^{\alpha ij}.$  The first index is present because the
relativistically covariant object is not the angular momentum itself, but rather
its four flux.  Taking the three dual, we find
\begin{equation}
\frac{1}{2}\epsilon_{kij}S^{\alpha ij}
  =\epsilon\int dR\int dU\int d\mu {\overline{f}}_9 (R,U,\mu)\delta^4 (x-R)
  \dot{R}^\alpha (R,U,\mu) \mu {\bf b}_k.
\end{equation}
Thus, to lowest order in $\epsilon,$ when $\alpha=0$ we get $c$ times the
spin density, which is the sum over the distribution
of guiding centers of the vector with magnitude
$\gamma_v\mu$ that points in the direction of ${\bf b}.$  Thus the spin angular
momentum for a single guiding center in a preferred frame may be thought of as
having magnitude $\gamma_v\mu$ and pointing in the direction of the magnetic field.
For $\alpha=l\neq 0,$ it is clear that we get the flux of this quantity, as the
integrand has an additonal factor of $v_\parallel^l$ (to lowest order).
This makes plausible our interpretation of $S$ as the spin.

Note that
\begin{eqnarray}
L^{\alpha\beta\gamma}_{\phantom{\alpha\beta\gamma} ,\alpha}&=&
   (T^{\alpha\beta}x^\gamma-T^{\alpha\gamma}x^\beta)_{,\alpha}\nonumber\\
  &=&T^{\alpha\beta}_{\phantom{\alpha\beta} ,\alpha}x^\gamma
   -T^{\alpha\gamma}_{\phantom{\alpha\gamma} ,\alpha}x^\beta
   +T^{\gamma\beta}-T^{\beta\gamma}\nonumber\\
  &=&T^{\gamma\beta}-T^{\beta\gamma},
  \label{paz}\nonumber\\
\end{eqnarray}
where we have used Eq.~(\ref{pau}).  Using this result, we can write the
angular momentum conservation law in the following form:
\begin{equation}
T-T^T+\delr\cdot S=0,
\label{pba}
\end{equation}
where the superscripted $T$ means ``transpose.''  Note that the antisymmetric
part of the stress-energy tensor is equal to the divergence of the spin tensor.

\section{The Guiding-center Plasma in the Presence of an Eikonal Wave Field}
\label{yau}
\subsection{Constructing the System Action}
We are now ready to extend the above analysis to the situation for which the
plasma is bathed in an eikonal wave field.  The full four potential is now
\begin{equation}
A(x)=A_0(x)+\epsilon A_1(x)+\lambda A_w(x),
\end{equation}
where the eikonal wave four potential
\begin{equation}
A_w(x)=\tilde{A}(x)\exp\left(\frac{i}{\epsilon}\psi (x)\right)+{\rm c.c.} 
\end{equation}
was introduced back in Eq.~(\ref{oay}) of Section~\ref{yba}.
The corresponding field is then
\begin{equation}
F(x)=F_0(x)+\epsilon F_1(x)+\lambda F_w(x),
\end{equation}
where
\begin{equation}
F_w(x)=\frac{1}{\epsilon}\tilde{F}(x)
    \exp\left(\frac{i}{\epsilon}\psi (x)\right)+{\rm c.c.} 
\end{equation}
and
\begin{equation}
\tilde{F}(x)=i({\bf k}\tilde{A} -\tilde{A} {\bf k})
          +\epsilon (\delr\tilde{A} -\tilde{A} \dell )
\label{pbe}
\end{equation}
(the ${\cal O}(\epsilon)$ term in $\tilde{F}$ is usually neglected in the
eikonal approximation).  Note that $F_0$ and $F_1$ are slowly varying
background fields, while $F_w$ is the rapidly varying wave field.
We must now construct the system action for a plasma of guiding/oscillation
centers immersed in this field.  The presence of the wave field has
two effects on the system action:  It means that the Hamiltonian must now
include the ponderomotive contribution, $K_2,$ and it means that the Maxwell
action must now include the wave field.

We first consider the effect on the Maxwell action.  We form
$-F:F/16\pi,$ and note that it contains the product of the slowly varying
terms, the product of the rapidly varying terms, and cross terms.  The
cross terms are oscillatory and vanish upon integration over $x.$
The remaining Maxwell action is then
\begin{equation}
S_m=(S_m)_0+\lambda^2{\tilde{S}}_m,
\end{equation}
where $(S_m)_0$ is the functional form of the Maxwell action with no wave
present (given by Eqs.~(\ref{pbb}) and (\ref{pbc})), and
\begin{equation}
{\tilde{S}}_m=-\frac{1}{8\pi}\int d^4x {\tilde{F}}^*:\tilde{F}
\label{pcn}
\end{equation}
is the contribution due to the wave.  Thus the {\it effective} (averaged)
Lagrangian density is
\begin{equation}
{\cal L}_m=({\cal L}_m)_0+\lambda^2{\tilde{\cal L}}_m,
\end{equation}
where $({\cal L}_m)_0$ is the functional form of the Lagrangian density
with no wave present (given by Eq.~(\ref{pbc})), and
\begin{equation}
{\tilde{\cal L}}_m=-\frac{1}{8\pi} {\tilde{F}}^*:\tilde{F}
\label{pco}
\end{equation}
is the contribution due to the wave.  Note that ${\tilde{\cal L}}_m$ is
quadratic in the field amplitude.

We now consider the modification of the action due to the presence of the
ponderomotive Hamiltonian.  Replacing $H$ by $H+\lambda^2K_2$ in
Eq.~(\ref{pai}), we see that
\begin{equation}
S_{gc}=(S_{gc})_0+\lambda^2 {\tilde{S}}_{gc},
\end{equation}
where $(S_{gc})_0$ is the functional form of the guiding-center action
with no wave present, and
\begin{equation}
{\tilde{S}}_{gc}=-\int d^4x \int dN(\eta)\int d\tau
   \delta^4 (x-R(\eta,\tau)) K_2
\end{equation}
is the contribution due to the wave.  Also note that the Lagrange multipliers
are altered by the introduction of $K_2$ (recall
that the Lagrange multipliers depend on the Hamiltonian).  Thus
$\lambda_\nu=(\lambda_\nu)_0+{\tilde{\lambda}}_\nu,$ where
\begin{equation}
{\tilde{\lambda}}_\nu=\xi_\nu\cdot\frac{\partial K_2}{\partial Z},
\end{equation}
and where the vectors $\xi_\nu$ were given in Eq.~(\ref{gjn})
at the end of Chapter~\ref{yac}.

Now $K_2$ can be expressed as  a real
function of the wave field amplitude, $\tilde{F},$ thanks to its manifest
gauge invariance.  Specifically, examination of Eq.~(\ref{oav}) shows
that it is a real quadratic form in the wave field amplitude.  Thus it can
be written
\begin{equation}
K_2(Z;F_i,\tilde{F},{\bf k})=\frac{1}{2}{\tilde{F}}_{\alpha\beta}^*
  {\cal K}^{\alpha\beta\xi\eta}(Z;F_0,{\bf k}){\tilde{F}}_{\xi\eta},
\label{pcp}
\end{equation}
where the antisymmetry of the field tensor imparts the following symmetry
properties to ${\cal K}$:
\begin{equation}
{\cal K}^{\alpha\beta\xi\eta}=-{\cal K}^{\beta\alpha\xi\eta}
   ={\cal K}^{\beta\alpha\eta\xi}=-{\cal K}^{\alpha\beta\eta\xi},
\end{equation}
and the reality of $K_2$ implies
\begin{equation}
{\cal K}^{\alpha\beta\xi\eta}=({\cal K}^{\xi\eta\alpha\beta})^*.
\end{equation}
It is clear that a kernel, ${\cal K},$ with the above properties is defined
implicitly by Eq.~(\ref{oav}).  Thus we can write
\begin{equation}
{\tilde{S}}_{gc}=-\frac{1}{2}\int d^4x \int dN(\eta)\int d\tau
   \delta^4 (x-R(\eta,\tau)) {\tilde{F}}_{\alpha\beta}^*
  {\cal K}^{\alpha\beta\xi\eta}(Z;F_i,{\bf k}){\tilde{F}}_{\xi\eta}.
\end{equation}
If we now define the fourth rank {\it generalized susceptibility tensor}
\begin{equation}
\chi^{\alpha\beta\xi\eta}(x,[Z,F_i,{\bf k}])
  \equiv \int dN(\eta)\int d\tau
   \delta^4 (x-R(\eta,\tau)) {\cal K}^{\alpha\beta\xi\eta}(Z;F_i,{\bf k}),
\label{pcq}
\end{equation}
(note that this differs from the more conventional definition of susceptibility
by a minus sign) then we can put this in still more compact form,
\begin{equation}
{\tilde{S}}_{gc}=-\frac{1}{2}\int d^4x
  {\tilde{F}}^*:\chi(x,[Z,F_i,{\bf k}]):{\tilde{F}}.
\end{equation}

Alternatively, we could write $K_2$ as a quadratic form
in the wave potential amplitude.  Using
$\tilde{F}=i({\bf k} \tilde{A}-\tilde{A} {\bf k}),$ we find
\begin{equation}
K_2(Z;F_i,\tilde{A},{\bf k})
   =2\tilde{A}^*_\alpha {\cal K}^{\alpha\xi} \tilde{A}_\xi,
\end{equation}
where the kernel
\begin{equation}
{\cal K}^{\alpha\xi}
   \equiv {\rm k}_\beta {\rm k}_\eta {\cal K}^{\alpha\beta\xi\eta}
\end{equation}
is a second rank tensor.  Note that we denote it by the same symbol
(${\cal K}$) that we use for the fourth rank kernel; which is meant
should be clear from either the context or the number of indices
adorning it.  The guiding-center action is then
\begin{equation}
{\tilde{S}}_{gc}=-2\int d^4x \int dN(\eta)\int d\tau
   \delta^4 (x-R(\eta,\tau)) {\tilde{A}}_\alpha^*
  {\cal K}^{\alpha\xi}(Z;F_i,{\bf k}){\tilde{A}}_\xi .
\end{equation}
We can then define the second rank {\it susceptibility tensor}
\begin{eqnarray}
\chi^{\alpha\xi}(x,[Z,F_i,{\bf k}])
  &\equiv& 2\int dN(\eta)\int d\tau
   \delta^4 (x-R(\eta,\tau)) {\cal K}^{\alpha\xi}(Z;F_i,{\bf k})\nonumber\\
  &=&2{\rm k}_\beta {\rm k}_\eta \chi^{\alpha\beta\xi\eta},
  \label{pcr}\nonumber\\
\end{eqnarray}
so that we may write
\begin{equation}
{\tilde{S}}_{gc}=-\int d^4x {\tilde{A}}^*\cdot\chi\cdot\tilde{A}.
\end{equation}
Once again note that we have used the same symbol to denote the fourth
order and second order versions of the susceptibility.

The guiding-center Lagrangian density is then clearly
\begin{equation}
{\cal L}_{gc}(x)=({\cal L}_{gc})_0(x)+\lambda^2 {\tilde{\cal L}}_{gc}(x),
\end{equation}
where $({\cal L}_{gc})_0(x)$ is the functional form of the Lagrangian density
when no wave is present, and
\begin{equation}
{\tilde{\cal L}}_{gc}(x)=-\frac{1}{2}
  {\tilde{F}}^*:\chi (x,[Z,F_i,{\bf k}]):{\tilde{F}}
\end{equation}
is the contribution due to the wave.

The {\it total} action is thus
\begin{equation}
S=(S)_0+\lambda^2\tilde{S},
\end{equation}
where
\begin{equation}
(S)_0\equiv (S_m)_0+(S_{gc})_0
\end{equation}
and
\begin{equation}
\tilde{S}={\tilde{S}}_m+{\tilde{S}}_{gc}=-\frac{1}{8\pi}\int d^4x
  {\tilde{F}}^*:\varepsilon (x,[Z,F_i,{\bf k}]):{\tilde{F}},
\end{equation}
and where in turn we have defined the fourth rank
{\it generalized dielectric tensor}
\begin{equation}
\varepsilon^{\alpha\beta}_{\phantom{\alpha\beta}\gamma\xi}\equiv
  \delta^\alpha_{\phantom{\alpha}\gamma}\delta^\beta_{\phantom{\beta}\xi}
  +4\pi\chi^{\alpha\beta}_{\phantom{\alpha\beta}\gamma\xi}.
\label{pbo}
\end{equation}
Alternatively, in terms of the wave potential amplitude, we have
\begin{equation}
\tilde{S}=-\frac{1}{4\pi}\int d^4x {\tilde{A}}^*\cdot
   {\cal D}(x,[Z,F_i,{\bf k}])\cdot\tilde{A},
\end{equation}
where we have defined the second rank {\it dispersion tensor}
\begin{equation}
{\cal D}^\alpha_{\phantom{\alpha}\xi}
  \equiv {\rm k}^2\delta^\alpha_{\phantom{\alpha}\xi}
  -{\rm k}^\alpha {\rm k}_\xi+4\pi\chi^\alpha_{\phantom{\alpha}\xi}.
\label{pct}
\end{equation}

Similarly, the {\it total} Lagrangian density is thus
\begin{equation}
{\cal L}=({\cal L})_0+\lambda^2\tilde{\cal L},
\end{equation}
where
\begin{equation}
({\cal L})_0\equiv ({\cal L}_m)_0+({\cal L}_{gc})_0
\end{equation}
and
\begin{eqnarray}
\tilde{\cal L}
  &=&{\tilde{\cal L}}_m+{\tilde{\cal L}}_{gc}\nonumber\\
  &=&-\frac{1}{8\pi}
  {\tilde{F}}^*:\varepsilon (x,[Z,F_i,{\bf k}]):{\tilde{F}}\nonumber\\
  &=&-\frac{1}{4\pi}
  {\tilde{A}}^*: {\cal D}   (x,[Z,F_i,{\bf k}]):{\tilde{A}}
  \label{pcs}\nonumber\\
\end{eqnarray}

The above action must be varied with respect to the particle coordinates and
the fields as before, but now we must also vary it with respect to the wave
fields, $\tilde{A}(x)$ and $\psi (x).$  Note that the action
depends on $\tilde{A}$ only through its dependence on $\tilde{F},$ thanks
to the manifest gauge invariance of $K_2$; variation with respect to
$\tilde{A}$ will yield the dispersion relation for linear plasma waves.
Note also that the action depends on $\psi$ only through its dependence
on ${\bf k}=\delr\psi,$ thanks to the averaging out of oscillating
terms; thus $\psi$ is an ignorable field coordinate, and variation with
respect to it will yield the conservation law for wave action.

Just as we found it useful to denote the dependence of a functional on $A_i$
and $F_i$ separately, we shall also find it useful to denote dependence on
$\tilde{A}$ and $\tilde{F}$ separately.  Using Eq.~(\ref{pbe}), the analog of
Eq.~(\ref{pan}) is easily found to be
\begin{equation}
{\left.\frac{\delta S}{\delta\tilde{A}}\right|}_{\rm total}
  =\frac{\delta S}{\delta\tilde{A}}
  -\frac{2i}{\epsilon} {\bf k}
  \cdot\left(\frac{\delta S}{\delta\tilde{F}}\right)
  -2\delr\cdot\left(\frac{\delta S}{\delta\tilde{F}}\right)
\label{pbd}
\end{equation}
(in the eikonal approximation, the third term on the right hand side is
usually neglected).  Similarly, we shall also find it useful to denote
dependence on $\psi$ and $\tilde{F}$ separately (note that $\tilde{F}$
contains ${\bf k}$ which is the gradient of $\psi$).  Once again, we use
Eq.~(\ref{pbe}) to write
\begin{equation}
{\left.\frac{\delta S}{\delta\psi}\right|}_{\rm total}
  =\frac{\delta S}{\delta\psi}
  -\frac{2i}{\epsilon}\delr\cdot\left({\tilde{A}}^*
  \cdot\frac{\delta S}{\delta\tilde{F}}\right).
\label{pbi}
\end{equation}
These results are very helpful in deriving what follows.

\subsection{The Vlasov Equation for Guiding/Oscillation Centers}
It is straightforward to see that
\begin{eqnarray}
\frac{\delta S}{\delta Z(\eta,\tau)}
  &=&\left(\frac{\delta S}{\delta Z(\eta,\tau)}\right)_0\nonumber\\
  &&\quad -\lambda^2\frac{\partial ({\tilde{\lambda}}_\nu C_\nu+K_2)}{\partial Z}
  (Z(\eta,\tau);F_i(R(\eta,\tau)),\nonumber\\
  &&\qquad\tilde{F}(R(\eta,\tau)),{\bf k}(R(\eta,\tau))),
  \label{pca}\nonumber\\
\end{eqnarray}
where, as usual, we have used a subscripted $0$ to denote
the functional form of a quantity when no wave is present.  The above result
yields the correction in the equations of motion due to the presence of
the ponderomotive Hamiltonian.  Thus, the only modification to the kinetic
equation due to the wave field is the inclusion of the ponderomotive effects
of the wave field on the guiding/oscillation centers of the plasma.

\subsection{The Field Equations}
Next, we use Eq.~(\ref{pan}) to take the functional derivative of $S$ with
respect to the $A_i$ to get
\begin{eqnarray}
\frac{\delta S}{\delta A_i}
  &=&\left(\frac{\delta S}{\delta A_i}\right)_0+2\lambda^2\delr\cdot
   \Bigl[ \int dN(\eta)\int d\tau \delta^4 (x-R(\eta,\tau))\nonumber\\
  &&\qquad \frac{1}{2} {\tilde{F}}^*
   \frac{:\partial {\cal K}(Z(\eta,\tau),F_i,{\bf k}):}{\partial F_i}
   {\tilde{F}}\Bigr].
  \label{pbq}\nonumber\\
\end{eqnarray}
Thus our field equation still follows from
\begin{equation}
\int d^4x ({\cal J}_0\cdot\delta A_0+{\cal J}_1\cdot\delta A_1)=0,
\end{equation}
but now:
\begin{eqnarray}
{\cal J}_i&=&({\cal J}_i)_0+2\lambda^2\delr\cdot
   \Bigl[ \int dN(\eta)\int d\tau \delta^4 (x-R(\eta,\tau))\nonumber\\
  &&\qquad \frac{1}{2} {\tilde{F}}^*
   \frac{:\partial {\cal K}(Z(\eta,\tau),F_i,{\bf k}):}{\partial F_i}
   {\tilde{F}}\Bigr].
  \label{pbr} \nonumber\\
\end{eqnarray}
Note that ${\cal K}$ has no explicit dependence on $F_1$ (the only effect of
$F_1$ is to alter the Poisson brackets), so only ${\cal J}_0$ is modified.
This may be interpreted as a modification to the guiding-center magnetization
density due to the presence of the wave field.  That is, our field equations
are still given by Eqs.~(\ref{pbf}) and (\ref{pbg}), but now
\begin{eqnarray}
M&=&(M)_0
  -2\lambda^2\int dN(\eta)\int d\tau \delta^4 (x-R(\eta,\tau))\nonumber\\
  &&\qquad \frac{1}{2} {\tilde{F}}^*
   \frac{:\partial {\cal K}(Z(\eta,\tau),F_i,{\bf k}):}{\partial F_i}
   {\tilde{F}}.
  \label{pbs}\nonumber\\
\end{eqnarray}
Note that the guiding-center current density is unaffected by the presence of
the wave; this is due to our neglect of resonant effects.

\subsection{The Linear Susceptibility}
We now have two additional equations of motion due to the variations with
respect to $\tilde{A}$ and $\psi.$  First we consider the variation with
respect to $\tilde{A}.$  We use Eq.~(\ref{pbd}), and in keeping with the
eikonal approximation, we neglect the third term on the right.  We
immediately get
\begin{equation}
0=\frac{\delta S}{\delta\tilde{A}}={\cal P}\left(
   \frac{i\lambda^2}{4\pi\epsilon}
   {\bf k}\cdot\varepsilon (x,[Z,F_i,{\bf k}]):{\tilde{F}}\right),
\end{equation}
so
\begin{equation}
{\cal P}\left({\bf k}\cdot
   \varepsilon (x,[Z,F_i,\delr\psi (x)]):{\tilde{F}}(x)\right)=0.
\label{pbk}
\end{equation}
This is the eikonal equation for linear plasma waves.  To see it in a
somewhat more familiar form, write
$\tilde{F}=i({\bf k}\tilde{A}-\tilde{A} {\bf k}),$ so after some
straightforward manipulation we arrive at
\begin{equation}
{\cal P}({\cal D}\cdot\tilde{A})=0,
\end{equation}
where we have used the {\it dispersion tensor} defined back in
Eq.~(\ref{pct}),
\begin{eqnarray}
{\cal D}^{\beta}_{\phantom{\beta} \xi}
 &\equiv& {\rm k}_\alpha {\rm k}^\gamma 
 (\varepsilon^{\alpha\beta}_{\phantom{\alpha\beta} \gamma\xi}
 -\varepsilon^{\alpha\beta}_{\phantom{\alpha\beta} \xi\gamma})\nonumber\\
 &=&{\rm k}^2\delta^\beta_{\phantom{\beta} \xi}-{\rm k}^\beta {\rm k}_\xi
 +8\pi {\rm k}_\alpha {\rm k}^\gamma
 \chi^{\alpha\beta}_{\phantom{\alpha\beta} \gamma\xi}\nonumber\\
 &=&{\rm k}^2\delta^\beta_{\phantom{\beta} \xi}-{\rm k}^\beta {\rm k}_\xi
 +4\pi \chi^{\beta}_{\phantom{\beta} \xi}
 \label{pbh}\nonumber\\
\end{eqnarray}

The dispersion relation for linear plasma waves is found by setting the
eigenvalues of the dispersion tensor equal to zero.  In ``three-plus-one''
notation, the dispersion tensor is three by three and so it has only three
eigenvalues that can be set to zero.  It seems that we are finding an
extra branch to the dispersion relation, and one might wonder why this
should be so.  By multiplying Eq.~(\ref{pbh}) by ${\rm k}_\beta,$ however,
it is easy to see that ${\bf k}$ is a null eigenvector of ${\cal D}.$
Thus, the extra eigenvalue is null, so setting it equal to zero does not
yield any new information.  The other three roots yield the more interesting
information about plasma waves.

\subsection{Conservation of Wave Action}
We next consider the equation of motion obtained by varying $\psi.$  Using
Eq.~(\ref{pbi}), we immediately find
\begin{equation}
0=\frac{\delta S}{\delta\psi}=\delr\cdot {\cal J},
\end{equation}
where we have defined the {\it wave action four flux}
\begin{equation}
{\cal J}\equiv {\cal P}\left(\frac{i\lambda^2}{2\pi\epsilon} {\tilde{A}}^*
  \cdot\varepsilon :\tilde{F}+\frac{\lambda^2}{8\pi} {\tilde{F}}^*
  \frac{:\partial\varepsilon :}{\partial {\bf k}}\tilde{F}\right).
\label{pbj}
\end{equation}
Our equation of motion thus expresses the conservation of this wave action.

The wave action takes on a much simpler form when written in terms of the
{\it dispersion tensor}, defined in Eq.~(\ref{pbh}).  We find
\begin{equation}
{\cal J}\equiv {\cal P}\left(\frac{\lambda^2}{8\pi} {\tilde{A}}^*
  \frac{\cdot\partial {\cal D}\cdot}{\partial {\rm k}} \tilde{A}\right).
\end{equation}

Finally note that the wave action is gauge invariant,
although this is not manifest in either
of the two forms presented above.  To prove this, we replace
${\tilde{A}}^*$ by ${\tilde{A}}^*-i{\bf k}\Lambda^*$ in Eq.~(\ref{pbj}).
Using the dispersion relation, Eq.~(\ref{pbk}), it is easy to see that the
term involving $\Lambda$ vanishes, leaving ${\cal J}$ unchanged.

\subsection{Applying the Noether Method}
We now consider what happens to the conservation laws obtained by the
Noether method when we include the effects of the wave field.  In this case,
Eq.~(\ref{par}) is altered in the following way:
\begin{eqnarray}
\delta {\cal L}
 &=&(\delta {\cal L})_0-\delr\cdot (\delta\psi {\cal J})
  -\frac{\lambda^2}{2\pi}\delr\cdot
  [(\varepsilon :\tilde{F})\cdot{\tilde{A}}^*]+\lambda^2\delr\cdot
  (\tilde{M}\cdot\delta A_0)\nonumber\\
 &&\qquad +\delr\cdot\Bigl\{\int dN(\eta)\int d\tau\delta R(\eta,\tau)
  \delta^4 (x-R(\eta,\tau))\nonumber\\
 &&\qquad \frac{\lambda^2}{2}{\tilde{F}}^*
  :{\cal K}(Z(\eta,\tau);F_i,{\rm k}):\tilde{F}\Bigr\}.
 \label{pbl}\nonumber\\
\end{eqnarray}
To derive this equation, we applied the variation to the full Lagrangian
density for the guiding/oscillation-center plasma in the presence of the
wave field.  We noted that
\begin{equation}
\tilde{F}=\frac{i}{\epsilon} {\bf k}\tilde{A}+\delr\tilde{A}-({\rm transpose}),
\label{pbm}
\end{equation}
so
\begin{equation}
\delta\tilde{F}=\frac{i}{\epsilon} {\bf k}\delta\tilde{A}
  +\frac{i}{\epsilon}(\delr\delta\psi)\tilde{A}+(\delr\delta\tilde{A})
  -({\rm transpose}).
\label{pbn}
\end{equation}
Finally, we used the equations of motion to simplify the result, just as we
did for the case in which there was no wave field present.

Note that the second term on the right hand side of Eq.~(\ref{pbm}) and the
third term on the right of Eq.~(\ref{pbn}) are usually neglected in the
eikonal approximation.  They are similar in this respect to the third term on
the right of Eq.~(\ref{pbd}), and the ${\cal O}(\epsilon)$ terms of
Eq.~(\ref{pbe}) (which also must be included in the analysis leading to
Eq.~(\ref{pbl})).  Up until now, we have consistently neglected these terms
in our analysis.  It will turn out that they are also unneccessary in deriving
the conservation law for energy-momentum, but they {\it are}
necessary in the derivation of the conservation law for angular momentum in
order to obtain the correct expression for the modification of the
guiding-center spin due to the presence of the wave.

\subsection{Conservation of Energy-Momentum}
We now use the same translational variation of the system that we did in
the case for which no wave was present, but now we add the variations of the
wave quantities,
\begin{equation}
\delta\psi=-\xi\cdot\delr\psi=-\xi\cdot {\bf k}
\end{equation}
and
\begin{equation}
\delta\tilde{A}=-\xi\cdot\delr\tilde{A}.
\end{equation}

There are five terms on the right hand side of Eq.~(\ref{pbl}).  The fifth
term cancels the portion of
$\delta {\cal L}_{gc}=-\xi\cdot\delr {\cal L}_{gc}$ (on the left hand side)
that is due to $K_2.$  The fourth term is the correction to the magnetization
density due to the wave, as defined in Eq.~(\ref{pap}).  It will simply cause
the magnetization density that appears in the conservation laws to be
corrected for the presence of the wave.  The third term is of the sort
discussed above that may be neglected in the usual eikonal approximation.
The new stuff comes from the second term,
$\delr\cdot ({\cal J}{\rm k}\cdot\xi),$
and from the portion of
$\delta {\cal L}_{m}=-\xi\cdot\delr {\cal L}_{m}$ (on the left hand side)
that is due to the wave.

The new stress-energy tensor is then
\begin{equation}
T=(T)_0+\lambda^2\tilde{T},
\end{equation}
where $(T)_0$ is the result with no wave field present (see Eq.~(\ref{pat})),
and $\tilde{T}$ is the modification due to the wave,
\begin{equation}
\tilde{T}=\tilde{M}\cdot F_0+{\tilde{\cal L}}_m {\bf 1}+{\cal J}{\bf k}.
\label{pcl}
\end{equation}
To recap, the first term on the right hand side above simply insures that
the magnetization that appears in the stress-energy tensor is that corrected
for the presence of the wave.  The second term on the right hand side above
similarly insures that the term ${\cal L}_m{\bf 1}$ that appears in the
stress-energy tensor is also corrected for the presence of the wave.  The
third term is the stress-energy due to the wave itself.  Note that it is the
tensor product of the wave action with the four wavevector.  This is sensible
since the wave action may be interpreted as the number flux of wave quanta
times some unit of action, and the unit of action times the four wavevector is
the energy-momentum per quantum.

\subsection{Conservation of Angular Momentum}
Finally, we examine the law of conservation of angular momentum.
We use the same rotational variation of the system that we did in
the case for which no wave was present, but now we add the variations of the
wave quantities,
\begin{equation}
\delta\psi=-(\Omega\cdot x)\cdot\delr\psi=-(\Omega\cdot x)\cdot {\bf k}
\end{equation}
and
\begin{equation}
\delta\tilde{A}=-(\Omega\cdot x)\cdot\delr\tilde{A}+\Omega\cdot\tilde{A}.
\end{equation}

Once again, we examine the five terms on the right hand side of
Eq.~(\ref{pbl}).  Now
$\delta {\cal L}=-(\Omega\cdot x)\cdot\delr {\cal L}
=-\delr\cdot [(\Omega\cdot x){\cal L}],$ so once again the fifth term will
cancel with the portion of $\delta {\cal L}_{gc}$ (on the left hand side)
that is due to $K_2.$  Similarly, it is straightforwardly shown that the
fourth term causes the magnetization density that appears in the angular
momentum tensor to be corrected for the presence of the wave, just as it did
in the stress-energy tensor.  The second term is
$\delr\cdot [{\cal J}{\bf k}\cdot\Omega\cdot x],$ and this contributes a new
term in the orbital angular momentum tensor; so
\begin{equation}
L=(L)_0+\lambda^2\tilde{L},
\end{equation}
where
\begin{equation}
{\tilde{L}}^{\alpha\beta\gamma}
 ={\tilde{T}}^{\alpha\beta}x^\gamma
 -{\tilde{T}}^{\alpha\gamma}x^\beta.
\label{pcm}
\end{equation}
Clearly, this is the orbital angular momentum due to the wave.

This time we retain the third term on the right hand side of Eq.~(\ref{pbl}).
It is
\begin{equation}
\frac{\lambda^2}{2\pi}\delr\cdot\left\{
  [-(\Omega\cdot x)\cdot\delr {\tilde{A}}^*+\Omega\cdot {\tilde{A}}^*]
  \cdot\varepsilon :\tilde{F}\right\}.
\end{equation}
We shall still ignore the first term in square brackets, as it contains a
gradient of the wave field amplitude, but we retain the second term.  After
some manipulation, it becomes
\begin{equation}
\frac{\lambda^2}{4\pi} [\Omega_{\beta\gamma} ({\tilde{A}}^{\gamma *}
  \varepsilon^{\beta\alpha}_{\phantom{\beta\alpha} \mu\nu}{\tilde{F}}^{\mu\nu}
  -{\tilde{A}}^{\beta *}
  \varepsilon^{\gamma\alpha}_{\phantom{\beta\alpha} \mu\nu}
  {\tilde{F}}^{\mu\nu})]_{,\alpha}.
\end{equation}
From this we can identify a correction to the spin angular momentum tensor.
We write
\begin{equation}
S=(S)_0+\lambda^2\tilde{S},
\end{equation}
where
\begin{equation}
{\tilde{S}}^{\alpha\beta\gamma}
 ={\tilde{A}}^{\gamma *}
  \varepsilon^{\beta\alpha}_{\phantom{\beta\alpha} \mu\nu}{\tilde{F}}^{\mu\nu}
  -{\tilde{A}}^{\beta *}
  \varepsilon^{\gamma\alpha}_{\phantom{\beta\alpha} \mu\nu}
  {\tilde{F}}^{\mu\nu}.
\label{pbp}
\end{equation}
This is the correction to the spin angular momentum tensor of a
guiding/oscillation-center plasma due to the presence of an eikonal wave
field.  This quantity is given by Soper~\cite{zcb} for oscillations in an
electromagnetic field in a vacuum.  He writes
\begin{equation}
{\tilde{S}}^{\alpha\beta\gamma}_{\rm vac}
 ={\tilde{A}}^{\gamma *} {\tilde{F}}^{\beta\alpha}
 -{\tilde{A}}^{\beta  *} {\tilde{F}}^{\gamma\alpha}.
\end{equation}
(see his Equation ~(9.3.14)).  If we set the susceptibility in Eq.~(\ref{pbo})
equal to zero, and plug the resulting vacuum dielectric into Eq.~(\ref{pbp}),
it is clear that our result will reduce to Soper's.  Thus, our result may
be considered to be an extension of his result to the case of dielectric
media.

The lack of gauge invariance of our result for $\tilde{S}$ is disturbing and
will be discussed further in Chapter~\ref{ybf}.

%% file: questions.tex
\chapter{Questions for Future Study}
\textheight=8.2truein
\pagestyle{myheadings}
\markboth{Questions for Future Study}{Questions for Future Study}
\label{ybf}
In this chapter, we discuss some questions raised by this study that could be
topics for future research.  These are in no particular order.

\begin{itemize}
\item The neglect of resonant effects is probably the most glaring omission of
this thesis, and probably that most likely to limit its utility.  There are
several schools of thought on how to deal with resonant effects, but they
break down into two major categories:

First, there are attempts to simply ``patch up'' the nonresonant treatment:
For example, since our nonresonant treatment has successfully
given us the hermitian part of the susceptibility
tensor, we could use the Kramers-Kronig relations to get the antihermitian
part.  Alternatively, we could simply dictate that all resonant denominators
are to be treated according to the Landau prescription.  These methods,
while successful in describing resonant particle effects on plasma waves, fall
far short of a unified description of the effects of resonant particles.
Furthermore, there is something aesthetically displeasing about tricks of this
sort.

Second, there are attempts to go back and redo the single particle analyses
to include resonant effects.  The general idea is that we first went astray
when we said that we could transform away the first order part of the
action due to the eikonal wave field.  While we can certainly do this far
away from the resonant regions of phase space, we certainly cannot do this
at (or even near) the resonance itself.  So we should go back and retain
the first order part of the action in the region of phase space near the
resonance.  Like the first technique, this approach explains certain things
nicely, but falls short of a unified description of resonant particles.
For example, the first order action that we retain will depend on the four
potential of the wave, and this will yield a modification to the current
density of a guiding-center plasma that is immersed in a wave field;
this is the {\it current drive} due to a wave field that tokamak researchers
study.  On the other hand, a good description of how
this residual piece of the first order action gives rise to Landau damping
does not seem to exist.  Furthermore, there is a great deal of arbitrariness
connected with how to decide just how much of this first order action to keep.
One approach uses ``window functions'' of some characteristic width, but there
is a great deal of freedom in just how these window functions should look
(square windows, gaussian windows, etc.); Dewar~\cite{zbz} gives a variational
principle for determining optimal window shape, but then we have to worry
about just what we mean by ``optimal.''  There is also a great deal of freedom
in choosing the width of such windows.  If we try to transform away the first
order action too close to the resonance, problems develop due to the presence
of the trapped particles, and the transformation ceases to be a near-identity
diffeomorphism.  Unfortunately, it is hard to quantify what we mean by
``too close'' in this regard.  Perhaps the window width should itself be
treated as a dynamical variable whose dynamics are given by some variational
principle (like that of Dewar); this might be a useful tool for the study of
``resonance broadening'' effects, where the width of the resonant region varies
in time.

\item Pursuing the oscillation-center Lie transforms to higher order
is a natural and obvious extension of this thesis.  In this way, one could
study induced scattering and three-wave phenomena.  Past attempts to study
these have either not used systematic perturbation theory (e.g. Lie
transforms), or have used Hamiltonian methods without manifest gauge
invariance.  This thesis should provide the tools needed to combine the
desiderata of systematic perturbation theory and manifest gauge invariance.
Central to this effort has been the use of the homotopy formula, and the
introduction of the pair of special functions, $Q_\ell$ and $R_\ell.$

It is interesting to note that this same program could have been carried out
for the nonrelativistic problem.  One must simply take the perturbation to
the action due to the wave (for which there now would be both a vector and a
scalar potential), and apply to it the guiding-center Lie transform, using
the homotopy formula in the same way that we did here.

\item The inclusion of dissipative effects (collisions, correlations, etc.)
would be an important generalization of the work presented here.  This is
undoubtedly related to the problems associated with the inclusion of
resonant effects.  A unified treatment of correlations would yield the
appropriate collision operator in the kinetic equation, and modify the
energy-momentum conservation law to describe the flow of energy into heat.

One way to approach this subject might be through the extended use of
projection operators.  We employed this technique in Chapter~\ref{yaq} to
show that it was possible for energy-momentum and angular momentum to flow
from one relevent region of Fourier space to another irrelevent one, and
thereby to effectively appear as a source term in the conservation laws.
We did not pursue this idea of partitioning Fourier space into one zone
for background fields, one zone for wave fields, and one zone for effects
of collisions (for example, we never introduced a second projection operator
for the wave fields, or a third one for fields arising in collisions).  This
approach may prove useful, but it quickly leads to great complication in the
procedure, and it is not clear how it might give rise to collision operators,
etc.

\item When we applied the Noether method to the action to obtain the
guiding-center spin angular momentum, we used the version of the action
that was both boostgauge and gyrogauge invariant.  There is a good reason
why we did this.  Other versions contain the quantity ${\cal R}$ that was
introduced back in Chapter~\ref{yac}.  If we had tried to apply Noether's
theorem to an action containing ${\cal R},$ we would at some point have been
faced with the question of how to vary ${\cal R}$ with respect to the four
potential.  It seems that ${\cal R}$ is not
independent of the four potential since it
was defined in terms of the unit vectors, ${\hat{e}}_\alpha,$ and
these, in turn, depend upon the background field.

We dodged the issue by going to the boostgauge and gyrogauge invariant
coordinates for which ${\cal R}$ does not appear in the action, but it is
interesting to contemplate the alternatives.  If we were to simply ignore this
term, we would not get guiding-center spin, and that would be unacceptable.
Though we had to go to higher order to find this term in our first derivation
of the guiding-center action, it has the same order as the $\mu d\theta$ term
which is obviously critically important.  Indeed, now that we have the
benefit of hindsight, we see that we could have avoided the higher order
guiding-center Lie transform altogether by examining the action at classical
order and asking what we would have to add to it to make the $\mu d\theta$
term gyrogauge invariant.  The answer would have been $-\mu {\cal R}\cdot dR,$
and this was really the only important term we found at higher order.  Thus,
the clever application of a gauge invariance requirement can save one from
going to higher order in a perturbation calculation!

So, since we can't ignore this term, how else could we have dealt with it?
There are a couple of possible avenues of approach.  First,
recall the well known result that the stress-energy tensor is given by
the derivative of the Lagrangian density with respect to the metric tensor
(this is true at least for spinless systems).
There seems to be an analogous theorem (or, at least, a conjecture)
enunciated by Hehl~\cite{zca}, that the spin
angular momentum tensor is the derivative
of the Lagrangian density with respect to torsion.  Torsion is the result
of an asymmetric affine connection, and the affine connection that we had
to introduce in Section~\ref{yal} to explain the ${\cal R}\cdot\dot{R}$ term
in $\dot{\Theta}$ is indeed asymmetric.  Now it is not clear to me that
${\cal R}$ is a torsion, but these remarks do make it clear that ${\cal R}$ has
at least something to do with torsion.  In any event, ${\cal R}$ appears in
our guiding-center action with a $\mu$ in front of it, so it is possible that
we could apply the above theorem (conjecture?) and derive guiding-center spin
directly (without recourse to Noether's theorem).  I suspect that, if this
were possible, it would be of more interest to researchers in quantum gravity
(which is the community to whom reference \cite{zca} was aimed) than it would
be to researchers in plasma physics.  It may be that guiding-center motion
provides a unique classical forum within which this topic of current research
in the field of quantum gravity may be applied, tested, and better understood.

Another possible approach to the spin problem is yet more speculative.
It is suggested by the {\it minimal coupling} idea of gauge field theory.
Recall that ${\cal R}$ is the gauge potential associated with the gyrogauge
group.  In Section~\ref{yal}, we even went one step further and derived the
corresponding gauge field, ${\cal N}.$  Using the techniques of gauge
field theory, it might be possible to use ${\cal R}$ to define a gauge
covariant derivative.  We could then add something like
${\cal N}:{\cal N}$ to the Lagrangian density, and treat $A$ and ${\cal R}$
as {\it independent} gauge fields.  Though these ideas are suggested by the
analogy with gauge field theories, they would all have to be rigorously
justified.  Furthermore, it is not obvious how guiding-center spin would
arise from these considerations.

\item Another mystery that should be mentioned is the apparant lack of gauge
invariance of the wave modification to guiding-center spin.  Our result is
clearly the extension to dielectric media of Soper's result for the
vacuum~\cite{zcb}.  The lack of gauge invariance did not seem to bother him,
except for a cryptic footnote that indicates that the result {\it is}
invariant with respect to a certain subgroup of the full gauge group.
One possible explanation might be that the {\it division} of angular momentum
into orbital and spin contributions is not a gauge-invariant division.  If
this were the case, however, one would expect that {\it neither} the orbital
nor the spin angular momentum should be gauge invariant by itself, but that
their sum should be gauge invariant.  Alas, the orbital angular momentum seems
to be gauge invariant all by itself, so the issue remains a mystery.

\item It would be nice to find a Hamiltonian field theoretical formulation of
the kinetic and field equations for the guiding-center and the
guiding/oscillation-center plasma.  Manifestly covariant Hamiltonian field
theories are, however, tricky to formulate.  We cannot give preference to the
time variable, and the proper time is not uniquely defined (every particle in
the system has its own proper time).  There may be ways of getting around this
difficulty by generalizing the form of Hamiltonian equations of motion for
such systems.  If this could be done, it might be possible to use the
energy-casimir method to study plasma stability to nonlinear perturbations.

\item We have developed conservation laws for energy-momentum and angular
momentum for the guiding/oscillation-center plasma.  In most studies of
plasma dynamics, use is made of energy conservation, but not of
momentum or angular momentum conservation (of course, in a covariant
relativistic treatment energy and momentum are inseparable).  It is
possible that these conserved quantities could play a far greater role
in the study of, say, plasma stability theory than they have until now.
For example, the Lyapunov method for assessing stability rests heavily on the
discovery of conserved quantities.  Just how to go about doing this is not
immediately clear.
\end{itemize}

%% file: glossary.tex
\chapter{Glossary of Notation}
\textheight=8.64truein
\setcounter{equation}{924}
\pagestyle{myheadings}
\markboth{Glossary of Notation}{Glossary of Notation}
\label{yuk}

In this appendix, we list all the important symbols used in this thesis,
giving the number of the equation where they were first used (if appropriate)
and a brief description (if appropriate).
\newcommand{\glossitem}[3]{$#1$ \> #2 \> \parbox[t]{3.5truein}{#3} \\}
\begin{tabbing}
\hspace{1.0truein} \= \hspace{1.2truein} \= \hspace{3.5truein} \kill
{\bf SYMBOL} \> {\bf EQUATION} \> {\bf DESCRIPTION} \\
\vspace{0.5truein}
\glossitem{\alpha (x)}{(\ref{pce})}{Clebsch potential for field}
\glossitem{\hat{\bf \alpha}}{(\ref{gje})}{Gyrogauge-invariant coordinatization of
   gyroangle}
\glossitem{\beta}{(\ref{gii})}{Angular hyperbolic polar
   coordinate for parallel part of particle four velocity}
\glossitem{\beta (x)}{(\ref{pce})}{Clebsch potential for field}
\glossitem{\bfbet_E}{(\ref{gij})}{${\bf E}\times {\bf B}/B^2$}
\glossitem{\bfbet_v}{(\ref{gil})}{${\bf v}/c$}
\glossitem{\Gamma_{gc}}{(\ref{gfy})}{Guiding-center action one form}
\glossitem{\gamma}{}{Action one-form}
\glossitem{\gamma_v}{}{Relativistic gamma factor: $\gamma_v=1/\sqrt{1-\beta_v^2}.$}
\glossitem{\delta^\alpha_{\phantom{\alpha}\beta}}{}{Kronecker delta}
\glossitem{\epsilon}{}{Guiding-center expansion parameter}
\glossitem{\epsilon^{\nu_1\cdots\nu_n}}{}{Levi-Civita tensor in $n$ dimensions}
\glossitem{\varepsilon^{\alpha\beta}_{\phantom{\alpha\beta}\gamma\xi}}{(\ref{pbo})}
   {Generalized dielectric tensor}
\glossitem{\eta}{(\ref{pbu})}{Continuous particle label}
\glossitem{\Theta}{(\ref{gfw})}{Angular polar coordinate for perpendicular
   part of guiding-center four velocity}
\glossitem{\theta}{(\ref{gii})}{Angular polar coordinate for perpendicular
   part of particle four velocity}
\glossitem{\kappa (x)}{(\ref{pcf})}{Clebsch potential for field}
\glossitem{\lambda}{(\ref{obx})}{Oscillation-center expansion parameter}
\glossitem{\lambda_1}{(\ref{gac})}{Lorentz scalar for electromagnetic field}
\glossitem{\lambda_2}{(\ref{gad})}{Lorentz pseudoscalar for electromagnetic field}
\glossitem{\lambda_\nu}{(\ref{gjj})}{Lagrange multiplier}
\glossitem{\lambda_B}{(\ref{gah})}{Related to eigenvalues of $F$}
\glossitem{\lambda_E}{(\ref{gag})}{Related to eigenvalues of $F$}
\glossitem{\mu}{(\ref{gfz})}{Gyromomentum}
\glossitem{\nu}{}{Constraint label}
\glossitem{\Xi}{(\ref{gkh})}{}
\glossitem{\Xi'}{(\ref{gki})}{}
\glossitem{\xi}{(\ref{pcj})}{Generator of infinitesimal translation in spacetime}
\glossitem{\xi^\alpha_\nu}{(\ref{gjn})}{}
\glossitem{\sigma (x)}{(\ref{pcf})}{Clebsch potential for field}
\glossitem{\tau}{}{Proper time}
\glossitem{\Upsilon}{(\ref{gkg})}{}
\glossitem{\Upsilon'}{(\ref{gkd})}{}
\glossitem{\chi^{\alpha\beta}_{\phantom{\alpha\beta}\gamma\xi}}{(\ref{pcq})}
   {Generalized susceptibility tensor}
\glossitem{\Psi_\ell}{(\ref{oca})}{}
\glossitem{\psi}{(\ref{oay})}{Phase of eikonal wave}
\glossitem{\Omega}{(\ref{pci})}{Generator of infinitesimal rotation in spacetime}
\glossitem{\Omega_B}{(\ref{gkf})}{Gyrofrequency with respect to proper time}
\glossitem{\Omega_{gc}}{(\ref{gga})}{Guiding-center Lagrangian two form}
\glossitem{\omega}{}{Lagrangian two-form}
\glossitem{A}{(\ref{gaa})}{Four-vector potential}
\glossitem{A_0}{}{Zero-order four-vector potential}
\glossitem{A_1}{}{First-order four-vector potential}
\glossitem{A_w}{(\ref{oay})}{Eikonal wave potential}
\glossitem{\tilde{A}}{(\ref{oay})}{Amplitude of eikonal wave potential}
\glossitem{\bf A}{(\ref{gaa})}{Three-vector potential}
\glossitem{\hat{\bf a}}{(\ref{gat})}{Member of orthonormal basis tetrad}
\glossitem{\bf B}{(\ref{gaa})}{Magnetic field pseudovector}
\glossitem{\cal B}{(\ref{gfw})}{Angular hyperbolic polar
   coordinate for parallel part of guiding-center four velocity}
\glossitem{\bf b}{}{Unit three-vector in direction of magnetic field}
\glossitem{\hat{\bf b}}{(\ref{gas})}{Member of orthonormal basis tetrad}
\glossitem{C_\nu}{(\ref{gjf})}{Constraints}
\glossitem{c}{}{Speed of light}
\glossitem{\hat{\bf c}}{(\ref{gar})}{Member of orthonormal basis tetrad}
\glossitem{D}{(\ref{pad})}{Jacobian}
\glossitem{{\cal D}^\alpha_{\phantom{\alpha}\beta}}{(\ref{pbh})}{Dispersion Tensor}
\glossitem{D_\ell}{(\ref{ord})}{Resonant denominator}
\glossitem{e}{}{Charge}
\glossitem{\bf E}{(\ref{gaa})}{Electric field vector}
\glossitem{F}{(\ref{gaa})}{Field tensor}
\glossitem{F'}{(\ref{gjs})}{}
\glossitem{F''}{(\ref{gkj})}{}
\glossitem{F'''}{(\ref{gkk})}{}
\glossitem{F_0}{}{Zero-order field tensor}
\glossitem{F_1}{}{First-order field tensor}
\glossitem{F_w}{(\ref{pbe})}{Eikonal wave field}
\glossitem{\tilde{F}}{(\ref{pbe})}{Amplitude of eikonal wave field}
\glossitem{\cal F}{(\ref{gab})}{Dual field tensor}
\glossitem{f_n}{(\ref{pbu})}{Pseudoscalar Eulerian particle distribution function}
\glossitem{{\sf f}_n}{(\ref{pac})}{Scalar Eulerian particle distribution function}
\glossitem{\overline{f}_n}{(\ref{pbu})}{Pseudoscalar Eulerian guiding-center
   distribution function}
\glossitem{\overline{\sf f}_n}{(\ref{pac})}{Scalar Eulerian guiding-center
   distribution function}
\glossitem{g_{\mu\nu}}{}{Metric tensor}
\glossitem{G_0}{(\ref{pcb})}{Macroscopic field tensor for perpendicular current}
\glossitem{G_1}{(\ref{pcc})}{Macroscopic field tensor for parallel current}
\glossitem{H}{}{Hamiltonian}
\glossitem{H_{gc}}{(\ref{gfz})}{Guiding-center Hamiltonian}
\glossitem{i}{}{$\sqrt{-1}$}
\glossitem{i_g}{(\ref{tac})}{Interior product with respect to vector field $g$}
\glossitem{J}{(\ref{pao})}{Four-current density}
\glossitem{J_{gc}}{}{Guiding-center poisson tensor}
\glossitem{{\cal J}_0}{(\ref{pcg})}{}
\glossitem{{\cal J}_1}{(\ref{pch})}{}
\glossitem{{\cal J}_\ell^\pm}{(\ref{ojpm})}{}
\glossitem{\bf k}{}{Wave four vector}
\glossitem{K}{(\ref{gii})}{Radial hyperbolic polar
   coordinate for parallel part of guiding-center four velocity}
\glossitem{K_2}{(\ref{obd})}{Ponderomotive Hamiltonian}
\glossitem{\cal K}{(\ref{pcp})}{Kernel of ponderomotive Hamiltonian}
\glossitem{k}{(\ref{gfw})}{Radial hyperbolic polar
   coordinate for parallel part of particle four velocity}
\glossitem{L}{(\ref{pck})}{Guiding-center orbital angular momentum tensor}
\glossitem{\tilde{L}}{(\ref{pcm})}{Wave contribution to guiding-center
   orbital angular momentum tensor}
\glossitem{L_{gc}}{(\ref{gjm})}{Guiding-center Lagrangian}
\glossitem{{\cal L}_g}{(\ref{mcn})}{Lie derivative with respect to vector field $g$}
\glossitem{{\cal L}_m}{(\ref{pbc})}{Lagrangian density of Maxwell field}
\glossitem{{\tilde{\cal L}}_m}{(\ref{pco})}{Lagrangian density of eikonal wave field}
\glossitem{\ell}{(\ref{oaz})}{Index for Fourier expansion in gyroangle}
\glossitem{M_0}{(\ref{pap})}{Magnetization density tensor for perpendicular current}
\glossitem{M_1}{(\ref{pbw})}{Magnetization density tensor for parallel current}
\glossitem{\cal M}{(\ref{taa})}{Boostgauge field}
\glossitem{m}{}{Mass}
\glossitem{dN(\eta)}{pbu}{Measure of particles with labels between $\eta$
   and $\eta+d\eta$}
\glossitem{\cal N}{(\ref{tab})}{Gyrogauge field}
\glossitem{P_\parallel}{(\ref{gal})}{Parallel projection operator}
\glossitem{P_\perp}{(\ref{gam})}{Perpendicular projection operator}
\glossitem{\cal P}{(\ref{pcd})}{Smoothing projection operator}
\glossitem{Q_\ell}{(\ref{oqdf})}{Special Function}
\glossitem{\cal Q}{(\ref{gzq})}{}
\glossitem{r}{}{Particle spacetime position}
\glossitem{R_\ell}{(\ref{ordf})}{Special function}
\glossitem{\cal R}{(\ref{gzr})}{}
\glossitem{\Re}{}{The set of real numbers}
\glossitem{S}{(\ref{pax})}{Guiding-center spin angular momentum tensor}
\glossitem{\tilde{S}}{(\ref{pbp})}{Wave contribution to guiding-center
   spin angular momentum tensor}
\glossitem{S_{gc}}{(\ref{pai})}{Guiding-center action}
\glossitem{S_m}{(\ref{pbb})}{Maxwell action}
\glossitem{\tilde{S}_m}{(\ref{pcn})}{Maxwell action due to eikonal wave}
\glossitem{T}{(\ref{pat})}{Guiding-center stress-energy tensor}
\glossitem{\tilde{T}}{(\ref{pcl})}{Wave contribution to guiding-center
   stress-energy tensor}
\glossitem{\hat{\bf t}}{(\ref{gaq})}{Member of orthonormal basis tetrad}
\glossitem{U}{(\ref{gjc})}{Boostgauge-invariant coordinatization of guiding-center
   parallel velocity}
\glossitem{u}{(\ref{gae})}{Particle four-velocity}
\glossitem{\bf v}{}{Three-velocity}
\glossitem{W}{(\ref{gfw})}{Radial polar coordinate for perpendicular
   part of guiding-center four velocity}
\glossitem{w}{(\ref{gii})}{Radial polar coordinate for perpendicular
   part of particle four velocity}
\glossitem{x}{}{Spacetime coordinates}
\glossitem{Z}{}{Generic coordinates}
\glossitem{:}{}{Double index contraction: $A:B\equiv A_{\mu\nu}B^{\mu\nu}.$}
\end{tabbing}

%% file: vecspaces.tex
\chapter[Vector Spaces, etc.]{Vector Spaces, Dual Spaces,
Algebras, and Modules}
\setcounter{equation}{924}
\pagestyle{myheadings}
\markboth{Vector Spaces, etc.}{Vector Spaces, etc.}
\label{yav}

This appendix is included to establish the set-theoretical foundations
of tensor calculus and exterior
algebra, as these ideas are used extensively in this thesis.
It is intended to provide a review 
for people already familiar with these topics, and to establish notation.
The reader is expected to be familiar with linear algebra and with the
topology of the real numbers.
If anything herein is unfamiliar, the reader is urged to consult
one of the above-mentioned introductory references.

We begin with some set-theoretical notation:  Given two sets, $A$ and $B,$ 
we define the {\it Cartesian product}, $A\times B,$ to be the set of all 
ordered pairs, $(a,b),$ such that $a\in A$ and $b\in B.$  
The symbol $\forall$ is read ``for all,'' and the symbol
$\exists$ is read ``there exists.''
A set is said to be {\it partitioned} if there exist subsets
such that each and every element of the set is a member of one and only one
subset.  A map that associates an element of a set, $B,$ to each element of 
a set, $A,$ is denoted by $A\mapsto B.$

A {\it relation}, $R,$ among the elements of a set, $A,$ is defined
to be a subset of $A\times A$; we write $R\subset A\times A.$
Two elements of $A,$ say $a_1$ and $a_2,$ are then said to be {\it related} if 
$(a_1,a_2)\in R.$  In this case, we may write $a_1\sim a_2.$  A relation is
{\it reflexive} if $a\sim a$ for all $a\in A.$  A relation is {\it symmetric}
if $a\sim b$ implies $b\sim a$ for all $a,b\in A.$  A relation is 
{\it transitive} if $a\sim b$ and $b\sim c$ implies $a\sim c$ for all
$a,b,c\in A.$  A relation that is reflexive, symmetric and transitive is called
an {\it equivalence relation}.  An equivalence relation naturally partitions a
set into subsets called {\it equivalence classes}. Any two members of
the same equivalence class are related to each other by the equivalence 
relation, and members of different equivalence classes are not related by the
equivalence relation.  For example, the equivalence relation of ``similarity''
partitions the set of all triangles into an infinity of equivalence classes, 
and the equivalence relation of ``equality modulo three'' partitions the set 
of integers into three classes.  The relation ``is the same height or taller 
than'' is {\it not} an equivalence relation on the set of all trees, because,
although it is reflexive and transitive, it is not symmetric, etc.

The set of all real numbers will be denoted by $\Re.$  The set of all
$n$-tuples of real numbers will be denoted by $\Re^n,$ and the reader is
assumed to have some familiarity with its usual topology.  In particular, by 
using, say, the Euclidean norm, it is possible to define open sets as 
neighborhoods, and thus to have a concept of nearness, continuity, convergence,
etc.

Let ${\cal V}$ be a set with $U,V,W,\ldots\in {\cal V},$ and let
$a,b,c,\ldots\in\Re.$  Let $+$ denote an operation that takes two elements of
${\cal V}$ and returns a third one; that is,
$+$ is a map ${\cal V}\times {\cal V}\mapsto {\cal V}.$  Let $\cdot$
denote an operation that takes an element of $\Re$ and an element of ${\cal V}$
and returns an element of ${\cal V};$ that is, $\cdot$ is a map
$\Re\times {\cal V}\mapsto {\cal V}.$  Then ${\cal V}$ is a {\it vector space}
over the field of real numbers if and only if the following conditions hold:
\newtheorem{cond}{Condition}[subsection]
\begin{cond}
$\forall U,V,W\in {\cal V}:  U+(V+W)=(U+V)+W.$
\end{cond}
\begin{cond}
$\forall U,V\in {\cal V}:  U+V=V+U.$
\end{cond}
\begin{cond}
$\exists 0\in {\cal V}:  \forall V\in {\cal V}:  V+0=V.$
\end{cond}
\begin{cond}
$\forall U\in {\cal V}:  \exists V\in {\cal V}:  U+V=0.$
\end{cond}
\begin{cond}
$\forall a,b\in\Re ,U\in {\cal V}:  (ab)\cdot U=a\cdot (b\cdot U).$
\label{mci}
\end{cond}
\begin{cond}
$\forall a,b\in\Re ,U\in {\cal V}:  (a+b)\cdot U=a\cdot U+b\cdot U.$
\end{cond}
\begin{cond}
$\forall a\in\Re ,U,V\in {\cal V}:  a\cdot (U+V)=a\cdot U+a\cdot V.$
\end{cond}
\begin{cond}
$\forall U\in {\cal V}: 1\cdot U=U.$
\label{mcj}
\end{cond}

A set of vectors, $U_1,\ldots,U_n,$ is said to be {\it linearly independent}
if and only if the only real numbers, $c_1,\ldots,c_n,$ satisfying
\begin{equation}
c_1\cdot U_1+\cdots c_n\cdot U_n=0
\end{equation}
are $c_1=\cdots =c_n=0.$  Otherwise, the vectors are said to be {\it linearly
dependent}.  The number of elements in the largest possible set of linearly
independent vectors is called the {\it dimension} of the vector space.
If a vector space has dimension $n,$ then any set of $n$ linearly independent
vectors constitutes a {\it basis} for that vector space.  If $V_1,\ldots,V_n$
is a basis for ${\cal V},$ then any vector, $U,$ in ${\cal V}$ can be expressed
\begin{equation}
U=a_1\cdot V_1+\cdots a_n\cdot V_n,
\end{equation}
where the real constants, $a,$ are uniquely determined by $U,$ and can be
computed by standard techniques of linear algebra.  In this case, we say
that the basis {\it spans} the vector space.
A {\it vector subspace} of a vector space, ${\cal V},$ is a subset of 
${\cal V}$ that is itself a vector space closed under $\cdot$ and $+.$  The 
dimension of the vector subspace is the minimal number of basis vectors needed 
to span it.

Vector spaces can be finite or infinite dimensional.  An example of an infinite
dimensional vector space is the space of all infinitely differentiable 
($C^\infty$) real-valued functions on $\Re.$  The addition and multiplication
operations are then
\begin{equation}
(f+g)(x)=f(x)+g(x)
\end{equation}
and
\begin{equation}
(a\cdot f)(x)=a\cdot f(x).
\end{equation}
This very important space will be called $\Lambda (\Re ).$
A basis for this vector space would have to contain an infinite number of
elements; the theory of Fourier series provides an example of how to go
about constructing and using such bases.  The set of all polynomial functions
of a real argument is a vector subspace of $\Lambda (\Re ).$

A {\it functional}, $U^*,$ operating on a vector space, ${\cal V},$ is a map 
${\cal V}\mapsto\Re.$  Equivalently, we can think of functionals as objects
which pair with vectors to yield real numbers.  The notation for this pairing
is $\langle U^*,V \rangle\in\Re.$  Note that we frequently denote functionals
with superscripted stars.  It is possible to define operations of addition and
real number multiplication on the space of functionals as follows:
\begin{equation}
\langle U^*+V^*,W \rangle=\langle U^*,W \rangle +\langle V^*,W \rangle
\end{equation}
and
\begin{equation}
\langle a\cdot U^*,W \rangle=a\langle U^*,W \rangle.
\end{equation}
It is readily verified that these operations make the space of all functionals
operating on ${\cal V}$ into a vector space which we shall denote by 
${\cal V}^*,$ and which we shall call the {\it dual space} to the vector
space, ${\cal V}.$  Furthermore, it is also readily verified that the 
dimensions of ${\cal V}$ and ${\cal V}^*$ are equal.  An example of this from
linear algebra may be instructive:  The dual space to the vector space of
column vectors may be identified with the vector space of row vectors, since
a row vector and a column vector pair to yield a real number under matrix
multiplication.

If a vector space, ${\cal V}$ is endowed with a further bilinear operation 
that maps ${\cal V}\times {\cal V}\mapsto {\cal V},$ then it is called an
{\it algebra}.  Since this operation pairs vectors with other vectors, it
can be written in the form $(U,V)\in {\cal V}.$  By ``bilinear,'' we mean
\begin{equation}
(a\cdot U+b\cdot V,W)=a\cdot (U,W)+b\cdot (V,W)
\end{equation}
and
\begin{equation}
(U,a\cdot V+b\cdot W)=a\cdot (U,V)+b\cdot (U,W).
\end{equation}
An algebra is {\it commutative} if $\forall U,V\in {\cal V}:  (U,V)=(V,U).$
An algebra is associative if $\forall U,V,W\in {\cal V}:  (U,(V,W))=((U,V),W).$
The set of real numbers, $\Re,$ becomes a commutative, associative algebra
when equipped with the operation of multiplication of real numbers.
The space $\Lambda (\Re )$ described above is also a commutative,
associative algebra if we equip it with the multiplication
\begin{equation}
(fg)(x)=f(x)g(x).
\end{equation}
In linear algebra, the set of all $n$ by $n$ square matrices is a vector
space of dimension $n^2$ with the usual definitions of matrix addition and 
multiplication by real numbers; it becomes an associative (but not commutative)
algebra when equipped with matrix multiplication.

An algebra, ${\cal V},$ is called a {\it Lie algebra} if and only if it is 
anticommutative
\begin{equation}
\forall U,V\in {\cal V}:  (U,V)=-(V,U),
\end{equation}
and satisfies the {\it Jacobi identity}
\begin{equation}
\forall U,V,W\in {\cal V}:  (U,(V,W))+(V,(W,U))+(W,(U,V))=0.
\end{equation}
The space of vectors in $\Re^3$ becomes a Lie algebra when equipped with the
usual cross product.

A vector subspace of an algebra is called a {\it subalgebra} if it is closed
under the algebra's multiplication rule.  For example, the space of all
polynomial functions of a real argument is a subalgebra of $\Lambda (\Re ).$
A subalgebra of a Lie algebra is called a {\it Lie subalgebra}.

We can generalize the concept of a vector field somewhat by relaxing the
requirement that $a$ and $b$ in Conditions~\ref{mci} through \ref{mcj} above
are real numbers.  Suppose instead that they are members of any associative
algebra, $A.$  Then Conditions~\ref{mci} through \ref{mcj} still make sense,
though the number $1$ that appears in Condition~\ref{mcj} must be reinterpreted
to refer to the identity element of the algebra, $A.$  In this case, ${\cal V}$
is said to be a {\it module} over the algebra, $A.$  For example, in linear
algebra, the space of column vectors is a module over the above-described 
algebra of square matrices.

Given an algebra, ${\cal V},$ with subspace, ${\cal U},$ we say that ${\cal U}$
is an {\it ideal} of ${\cal V}$ if and only if $(U,V)\in {\cal U}$ and
$(V,U)\in {\cal U}$ for all $U\in {\cal U},$ and $V\in {\cal V}.$  For example,
let ${\cal V}$ be the vector space of all polynomial functions of a real 
argument, $x;$  Recall that this is a subalgebra of $\Lambda (\Re ).$
Then, the subspace, ${\cal U}\subset {\cal V},$ of all polynomials with zeros
at some particular location(s) is an ideal of ${\cal V}.$

Throughout this thesis, when a scalar multiplies a vector, the dot is
suppressed; that is, $a\cdot V$ is written simply $aV.$
The dot notation is used for other things.  Also, boldface type is used to
denote a vector, though its components in a given coordinate system are
denoted by the same letter in ordinary typeface (with a superscripted index
to label components).

%% file: gyrofreq.tex
\chapter[Gyrofrequency Shift, etc.]{Gyrofrequency Shift for Two-Dimensional
Nonrelativistic Guiding-Center Motion}
\textheight=8.64truein
\setcounter{equation}{935}
\pagestyle{myheadings}
\markboth{Gyrofrequency Shift, etc.}{Gyrofrequency Shift, etc.}
\label{ybd}

As a straightforward but nontrivial example of the vector Lie transform
technique, we consider two-dimensional nonrelativistic guiding-center
motion in a magnetic field of the form
\begin{equation}
{\bf B}=B(x,y)\hat{\bf z},
\end{equation}
and a perpendicular electric field of the form
\begin{equation}
{\bf E}=E_x(x,y)\hat{\bf x}+E_y(x,y)\hat{\bf y}.
\end{equation}
To lowest order, the gyrofrequency is given by $\Omega=eB/mc.$  We shall
address the problem of computing the correction to this quantity due to the
spatial dependence of ${\bf B}$ and ${\bf E}.$

The single-particle equations of motion are
\begin{eqnarray}
\dot{x} &=&u\nonumber\\
\dot{y} &=&v\nonumber\\
\dot{u} &=&\frac{e}{m}E_x+\Omega v\nonumber\\
\dot{v} &=&\frac{e}{m}E_y-\Omega u. \label{apa}\\
\noalign{\hbox{Introduce the perpendicular velocity and the gyroangle,}}
w &=&\sqrt{u^2+v^2}\nonumber\\
\theta &=&{\rm arg}(-v-iu), \label{apb}\\
\noalign{\hbox{so that}}
u &=&-w\sin\theta\nonumber\\
v &=&-w\cos\theta .\label{apc}\\
\noalign{\hbox{In terms of $w$ and $\theta$ the equations of motion are found to be}}
\dot{x}&=&-w\sin\theta\nonumber\\
\dot{y}&=&-w\cos\theta\nonumber\\
\dot{w}&=&-\frac{e}{m}\left( E_x\sin\theta+E_y\cos\theta \right)\nonumber\\
\dot{\theta}&=&\frac{1}{\epsilon}\Omega-\frac{e}{mw}
                     \left( E_x\cos\theta-E_y\sin\theta \right)
                     \label{apd}
\end{eqnarray}
Here we have introduced the formal ordering parameter $\epsilon,$ and have
ordered the equations of motion by the prescription $e\mapsto e/\epsilon$ and
${\bf E}\mapsto\epsilon {\bf E}.$

Though it is most useful and quite elegant to treat this problem with Hamiltonian
perturbation theory, we shall instead use Lie transforms directly on the dynamical
vector field.  We do this for the purposes of illustration.  In Chapter~\ref{yac}
of this thesis, we treat the much more general problem of relativistic guiding-center
motion in arbitrary electromagnetic field geometry in space-time (including
perpendicular electric fields that may be order unity in the guiding-center
expansion parameter, $\epsilon$), and there we make full use of the Hamiltonian nature
of the equations of motion and we spend a great deal of time studying the associated
Poisson structure.  It is useful to compare the two approaches.

We denote the phase-space coordinates by ${\bf z}=(x,y,w,\theta ),$ and the
equations of motion by
\begin{equation}
\dot{\bf z}=\frac{1}{\epsilon}{\bf V}_0+{\bf V}_1,
\end{equation}
where the dynamical vector field is described by
\begin{eqnarray}
V_0^x&=&0\nonumber\\
V_0^y&=&0\nonumber\\
V_0^w&=&0\nonumber\\
V_0^\theta&=&\Omega \label{apj}\\
\noalign{\hbox{and}}
V_1^x&=&-w\sin\theta\nonumber\\
V_1^y&=&-w\cos\theta\nonumber\\
V_1^w&=&-\frac{e}{m}\left( E_x\sin\theta+E_y\cos\theta \right)\nonumber\\
V_1^\theta&=&-\frac{e}{mw}
   \left( E_x\cos\theta-E_y\sin\theta \right). \label{apf}\\
\noalign{\hbox{The unperturbed problem, $\dot{\bf z}={\bf V}_0/\epsilon,$ thus
   has the solution}}
x&=&x_0\nonumber\\
y&=&y_0\nonumber\\
w&=&w_0\nonumber\\
\theta &=&\theta_0+\Omega t/\epsilon, \label{apg}
\end{eqnarray}
so that averages over the unperturbed motion are equivalent to averages over
$\theta.$

At first order, Eq.~(\ref{mdh}) tells us that
\begin{equation}
{\cal V}_1={\bf V}_1-{\cal L}_1 {\bf V}_0,
\end{equation}
where ${\cal V}$ denotes the Lie transformed dynamical vector field at first order.
The separate components of the above equation are then
\begin{eqnarray}
\Omega\frac{\partial}{\partial\theta}g_1^x&=&{\cal V}_1^x+w\sin\theta\nonumber\\
\Omega\frac{\partial}{\partial\theta}g_1^y&=&{\cal V}_1^y+w\cos\theta\nonumber\\
\Omega\frac{\partial}{\partial\theta}g_1^w
   &=&{\cal V}_1^w+\frac{e}{m}\left( E_x\sin\theta+E_y\cos\theta \right)\nonumber\\
\Omega\frac{\partial}{\partial\theta}g_1^\theta
   &=&{\cal V}_1^\theta+\frac{e}{mw}
   \left( E_x\cos\theta-E_y\sin\theta \right)
   +g_1^x\Omega_{,x}+g_1^y\Omega_{,y}. \label{aph}
\end{eqnarray}
We demand that the generator vector ${\bf g}_1$ be purely oscillatory (single-valued
in $\theta$).  Thus, averaging the above equations immediately yields
\begin{equation}
{\cal V}_1=0.
\end{equation}
Then, we can solve Eqs.~(\ref{aph}) for the components of ${\bf g}_1.$  We get
\begin{eqnarray}
g_1^x&=&-\frac{w}{\Omega}\cos\theta\nonumber\\
g_1^y&=& \frac{w}{\Omega}\sin\theta\nonumber\\
g_1^w&=&\frac{e}{m\Omega}\left( -E_x\cos\theta+E_y\sin\theta \right)\nonumber\\
g_1^\theta &=&\frac{e}{mw\Omega}
   \left( E_x\sin\theta+E_y\cos\theta \right)
   -\frac{w\Omega_{,x}}{\Omega^2}\sin\theta
   -\frac{w\Omega_{,y}}{\Omega^2}\cos\theta. \label{api}
\end{eqnarray}
Thus we have completely removed the perturbation in the dynamical vector
field at first order.  The guiding-center equations of motion will appear at
the next order, as will the desired correction to the gyrofrequency.

At second order, Eq.~(\ref{mbt}) tells us that
\begin{equation}
{\cal V}_2=-{\cal L}_2 {\bf V}_0-{\cal L}_1 {\bf V}_1+\frac{1}{2}{\cal L}_1^2 {\bf V}_0
   =-{\cal L}_2 {\bf V}_0-\frac{1}{2}{\cal L}_1 {\bf V}_1.
\end{equation}
The generator ${\bf g}_2$ must be chosen so that ${\cal V}_2$ is purely averaged.
Thus, without having to actually compute ${\bf g}_2,$ we can deduce
\begin{equation}
{\cal V}_2=\left\langle -\frac{1}{2}{\cal L}_1 {\bf V}_1 \right\rangle.
\end{equation}
To get the shift in gyrofrequency, we need only ${\cal V}_2^\theta.$  Because
both ${\bf V}_1$ and ${\bf g}_1$ contain oscillatory terms, the Lie derivative
of one with respect to the other will contain products of oscillatory terms,
and some of these will not average to zero.  After some tedious algebra, we find
\begin{equation}
{\cal V}_2^\theta=-\frac{e\Omega}{2m}\nabla\cdot\left(\frac{{\bf E}}{\Omega^2}\right)
   +\frac{w^2}{4}\nabla\cdot\left(\frac{\nabla\Omega}{\Omega^2}\right).
\end{equation}
This is the gyrofrequency shift.  The first term is the shift due to the spatial
dependence of the perpendicular electric field, and the second term is the shift
due to the spatial dependence of the magnetic field.  The first of these terms
was discovered by Kaufman~\cite{zcc} in 1960, who also showed that it gives rise
to the phenomenon of {\it gyroviscosity}.

It is interesting to note that, when the results of Chapter~\ref{yac} are cast
into ``$1+3$'' notation and the nonrelativistic limit is taken, the first of the
above pair of terms is present but the second is not.  This is because the
ordering scheme used is quite different.  In this appendix, we treated the
perpendicular electric field as an order $\epsilon$ quantity, whereas in
Chapter~\ref{yac} we took it to be order unity.  Thus both terms appear at the
same order above (the first term has a spatial gradient and an electric field,
and the second term has two spatial gradients), whereas in Chapter~\ref{yac} the
second term would appear at one higher order than the first term (and we did not
calculate to high enough order there to see it).  It is also interesting to note
that the term involving ${\cal R}$ in Eq.~(\ref{ggs}) of Chapter~\ref{yac} is
a three (or higher) dimensional effect, and has no analog in two-dimensional
guiding-center motion.

%% file: specfuns.tex
\chapter{Properties of the Special Functions}
\setcounter{equation}{952}
\pagestyle{myheadings}
\markboth{Properties of the Special Functions}{Properties of the Special Functions}
\label{yaw}

The following is a list of properties of the $Q_\ell$ and $R_\ell$
functions that follow directly from their definitions given in
Section~\ref{yax}.
\newtheorem{prop}{Property}[subsection]
\subsection{The Q Functions}
\begin{figure}[p]
\center{
\vspace{1.21truein}
\mbox{\includegraphics[bbllx=0,bblly=0,bburx=287,bbury=251,width=5.82truein]{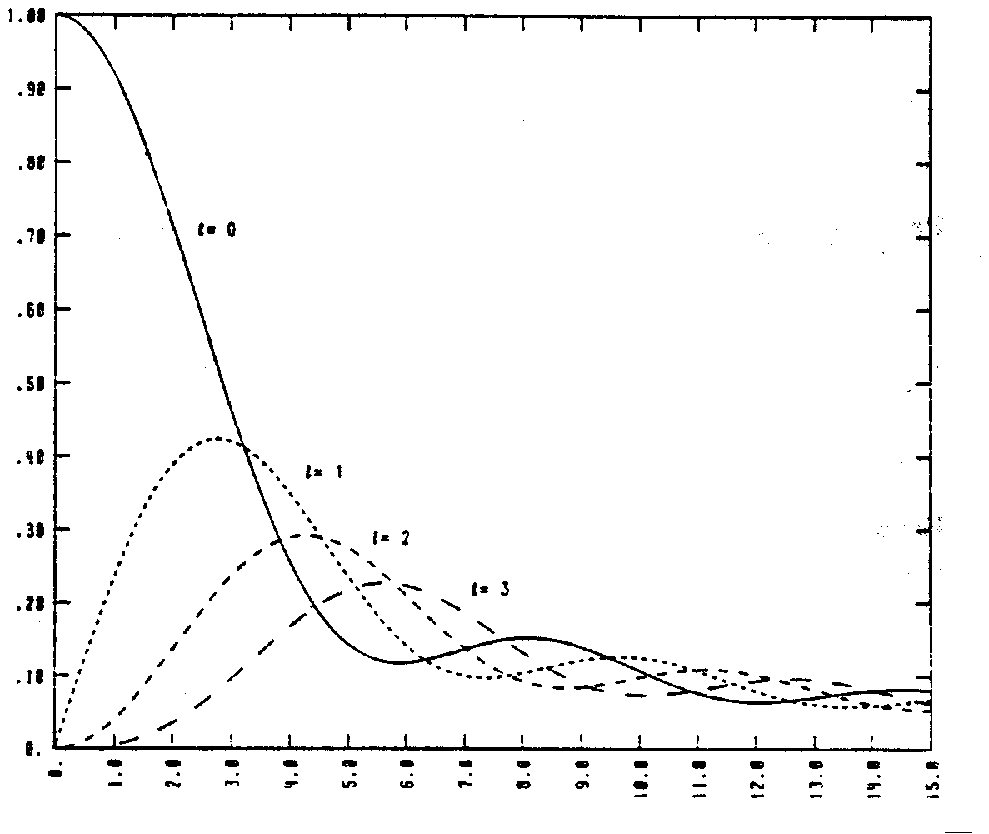}}
\vspace{1.21truein}
}
\caption{The Q Functions}
\label{oaj}
\end{figure}

\begin{prop}[Defining Integral]
\begin{displaymath}
Q_\ell(x)\equiv\frac{1}{2\pi}\int_0^{2\pi}d\xi\left(\frac{e^{ix\sin\xi}-1}{ix
   \sin\xi}\right)e^{-i\ell\xi}
\end{displaymath}
\end{prop}
\begin{prop}[Relationship with Bessel Functions]
\begin{displaymath}
\frac{d}{dx}\left[xQ_\ell(x)\right]=J_\ell(x)
\end{displaymath}
\end{prop}
\begin{prop}[Power Series]
\begin{displaymath}
Q_\ell(x)=\sum_{j=0}^{\infty}\frac{(-1)^j(x/2)^{2j+\ell}}{(2j+\ell+1)j!(\ell
   +j)!}=\frac{(x/2)^\ell}{(\ell+1)!}+\cdots
\end{displaymath}
\end{prop}
\begin{prop}[Asymptotic Behavior for Large Argument]
\begin{displaymath}
Q_\ell(x)\sim\frac{1}{x}+\sqrt{\frac{2}{\pi x^3}}\sin
\left( x-\frac{\pi}{2}\ell-\frac{\pi}{4} \right)+\cdots
\end{displaymath}
\end{prop}
\begin{prop}[Recursion Relations]
\begin{displaymath}
Q_{\ell-1}(x)+Q_{\ell+1}(x)=\frac{2\ell}{x}\int_0^x dy\frac{J_\ell(y)}{y}
\end{displaymath}
\begin{displaymath}
Q_{\ell-1}(x)-Q_{\ell+1}(x)=\frac{2}{x}J_\ell(x)
\end{displaymath}
\end{prop}
\begin{prop}[Formula for Derivative]
\begin{displaymath}
Q_\ell^\prime(x)=\frac{1}{x}\left[J_\ell(x)-Q_\ell(x)\right]
\end{displaymath}
\end{prop}
Graphs of the Q functions are presented in Fig.~\ref{oaj}.

\subsection{The R Functions}
\begin{figure}[p]
\center{
\vspace{1.16truein}
\mbox{\includegraphics[bbllx=0,bblly=0,bburx=282,bbury=251,width=5.82truein]{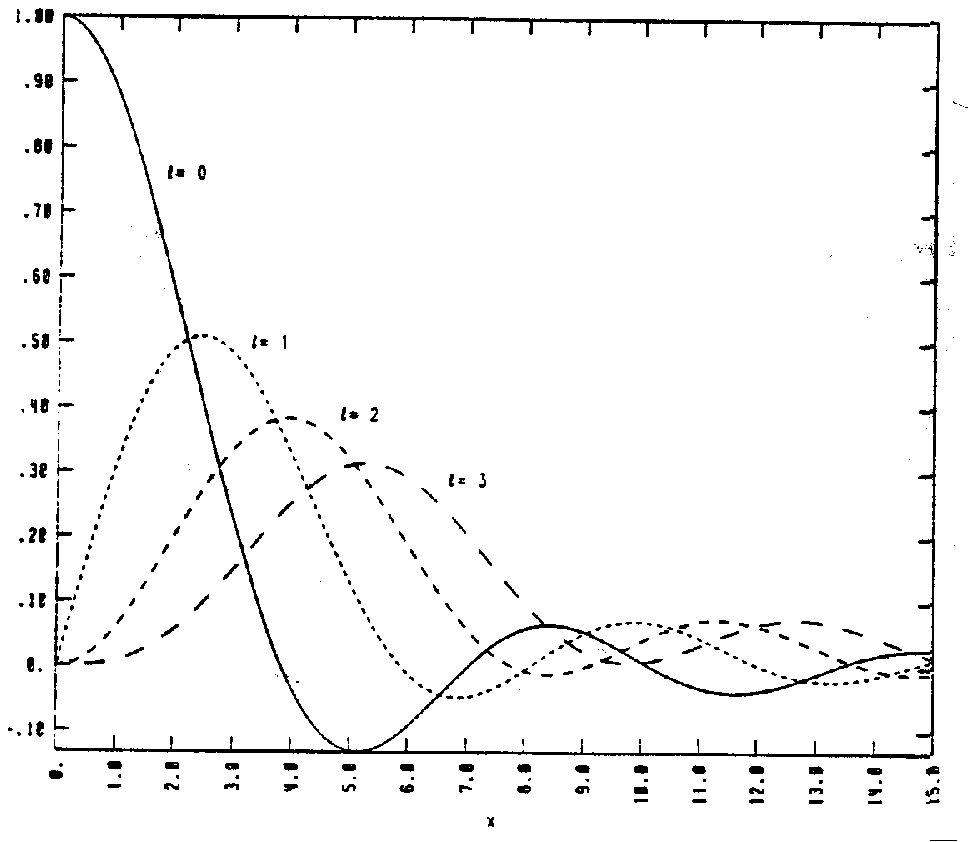}}
\vspace{1.16truein}
}
\caption{The R Functions}
\label{oak}
\end{figure}

\begin{prop}[Defining Integral]
\begin{displaymath}
R_\ell(x)\equiv\frac{1}{\pi}\int_0^{2\pi}d\xi\left(\frac{(1-ix\sin\xi)
   e^{ix\sin\xi}-1}{x^2\sin^2\xi}\right)e^{-i\ell\xi}
\end{displaymath}
\end{prop}
\begin{prop}[Relationship with Bessel Functions]
\begin{displaymath}
\frac{d}{dx}\left[x^2R_\ell(x)\right]=2xJ_\ell(x)
\end{displaymath}
\end{prop}
\begin{prop}[Power Series]
\begin{displaymath}
R_\ell(x)=2\sum_{j=0}^{\infty}\frac{(-1)^j(x/2)^{2j+\ell}}{(2j+\ell+2)j!(\ell
   +j)!}=\frac{2(x/2)^\ell}{(\ell+2)\ell !}+\cdots
\end{displaymath}
\end{prop}
\begin{prop}[Asymptotic Behavior for Large Argument]
\begin{displaymath}
R_\ell(x)\sim \sqrt{\frac{2}{\pi x^3}}\sin
   \left( x-\frac{\pi}{2}\ell-\frac{\pi}{4} \right)+\cdots
\end{displaymath}
\end{prop}
\begin{prop}[Recursion Relations]
\begin{displaymath}
R_{\ell-1}(x)+R_{\ell+1}(x)=\frac{4\ell}{x}Q_\ell(x)
\end{displaymath}
\begin{displaymath}
R_{\ell-1}(x)-R_{\ell+1}(x)=\frac{4}{x}\left[J_\ell(x)-Q_\ell(x)\right]
\end{displaymath}
\end{prop}
\begin{prop}[Formula for Derivative]
\begin{displaymath}
R_\ell^\prime(x)=\frac{2}{x}\left[J_\ell(x)-R_\ell(x)\right]
\end{displaymath}
\end{prop}
Graphs of the R functions are presented in Fig.~\ref{oak}.

%% file: besssums.tex
\chapter{Useful Bessel Function Sums}
\setcounter{equation}{952}
\pagestyle{myheadings}
\markboth{Useful Bessel Function Sums}{Useful Bessel Function Sums}
\label{ybe}
All of the Bessel function summation formulas used in Chapter~\ref{yap}
can be derived from the following theorems:
\begin{equation}
\sum_\ell J_{\ell+k}(z)J_{\ell-k}(z)=\delta_{k0}
\label{aqa}
\end{equation}
and
\begin{equation}
\sum_\ell J_{\ell+k+1}(z)J_{\ell-k}(z)=0,
\label{aqf}
\end{equation}
the usual Bessel function recursion relations
\begin{equation}
J_{\ell-1}(z)+J_{\ell+1}(z)=\frac{2\ell}{z} J_\ell (z)
\label{aqg}
\end{equation}
and
\begin{equation}
J_{\ell-1}(z)-J_{\ell+1}(z)=2J'_\ell (z),
\label{aqc}
\end{equation}
and the parity rule
\begin{equation}
J_{-\ell}(z)=(-1)^\ell J_\ell(z).
\label{aqd}
\end{equation}

To verify Eq.~(\ref{aqa}), let us define
\begin{equation}
f_k(z)\equiv\sum_\ell J_{\ell+k}(z)J_{\ell-k}(z),
\end{equation}
and differentiate with respect to $z$ to get
\begin{eqnarray}
f'_k(z)&=&\sum_\ell\left(J'_{\ell+k}J_{\ell-k}+J_{\ell+k}J'_{\ell-k}\right)\nonumber\\
  &=&\frac{1}{2}\sum_\ell\left[\left(J_{\ell+k-1}-J_{\ell+k+1}\right)J_{\ell-k}
                   +J_{\ell+k}\left(J_{\ell-k-1}-J_{\ell-k+1}\right)\right]\nonumber\\
  &=&\frac{1}{2}\sum_\ell\left(J_{\ell+k-1}J_{\ell-k}-J_{\ell+k+1}J_{\ell-k}
                             +J_{\ell+k+1}J_{\ell-k}-J_{\ell+k-1}J_{\ell-k}
                        \right)\nonumber\\
  &=&0, \label{aqb}
\end{eqnarray}
where we have used Eq.~(\ref{aqc}) in the second line and have redefined
the summation variable in the third line (we have also omitted explicit indication
of the functional dependence of $J_\ell$ on $z$ since no ambiguity can result from
doing so).  This means that $f_k(z)$ cannot depend
on $z,$ so it is a constant for each value of $k.$  To find the value of this
constant, set $z$ equal to zero in Eq.~(\ref{aqa}).  Recalling that
$J_\ell (0)=\delta_{\ell 0},$ we see that $f_k(z)=\delta_{k0},$ and the theorem
is proved.

To verify Eq.~(\ref{aqf}), use the parity rule, Eq.~(\ref{aqd}).  We have
\begin{eqnarray}
\sum_\ell J_{\ell+k+1}J_{\ell-k}
  &=&\frac{1}{2}\sum_\ell \left(J_{ \ell+k+1} J_{ \ell-k}
                             + J_{-\ell-k-1} J_{-\ell+k}\right)\nonumber\\
  &=&\frac{1}{2}\sum_\ell \left(J_{ \ell+k+1} J_{ \ell-k}
               +(-1)^{2\ell+1} J_{ \ell+k+1} J_{ \ell-k}\right)\nonumber\\
  &=&\frac{1}{2}\sum_\ell \left(J_{ \ell+k+1} J_{ \ell-k}
                             - J_{ \ell+k+1} J_{ \ell-k}\right)\nonumber\\
  &=&0, \label{aqe}
\end{eqnarray}
where we have redefined the summation variable in the first line
($\ell\mapsto -\ell$ in the second term), and used the parity rule
in the second line.

These theorems can be used to derive sum rules with summands that are quadratic
in the Bessel functions.  To do this, note first that setting
$k=0$ in Eqs.~(\ref{aqa}) and (\ref{aqf}) immediately yields
\begin{equation}
\sum_\ell J_\ell^2=1
\end{equation}
and
\begin{equation}
\sum_\ell J_{\ell+1} J_\ell=\sum_\ell J_\ell J_{\ell-1}=0.
\end{equation}
To derive a sum rule that includes $\ell$ raised to some power, first use
Eq.~(\ref{aqg}) to get rid of the power of $\ell.$  To derive a sum rule that
includes a derivative of a Bessel function, first use Eq.~(\ref{aqc}) to
express the Bessel function derivative in terms of undifferentiated Bessel
functions; alternatively, if a sum rule that includes a Bessel function
derivative can be expressed as the derivative of another sum rule with
undifferentiated Bessel functions, then this is usually a better way to
proceed.

As an example of some generality, consider the sum over $\ell$ of
$\ell^4 J_\ell J'_\ell.$  This can be expressed as follows:
\begin{equation}
\sum_\ell \ell^4 J_\ell J'_\ell=\frac{1}{2}\frac{d}{dz}\sum_\ell \ell^4 J^2_\ell.
\end{equation}
Now note
\begin{eqnarray}
\ell^4 J^2_\ell
  &=&\ell^2 \left(\ell J_\ell\right)^2\nonumber\\
  &=&\frac{\ell^2 z^2}{4}
    \left(J^2_{\ell-1}+2J_{\ell-1}J_{\ell+1}+J^2_{\ell+1}\right)\nonumber\\
  &=&\frac{z^2}{4}
    \Bigl\{
    \left[ (\ell-1)^2+2(\ell-1)+1 \right] J^2_{\ell-1}\nonumber\\
  &&\qquad +2\left[ (\ell-1)(\ell+1)+1     \right] J_{\ell-1}J_{\ell+1}\nonumber\\
  &&\qquad +\left[ (\ell+1)^2-2(\ell+1)+1 \right] J^2_{\ell+1}
    \Bigr\}\nonumber\\
  &=&\frac{z^2}{4}
    \Bigl\{
      \left[ \frac{z^2}{4}\left( J_{\ell-2}+J_\ell \right)^2
             +2\frac{z}{2}\left( J_{\ell-2}+J_\ell \right)J_{\ell-1}
             +J^2_{\ell-1}
      \right]\nonumber\\
  &&\qquad +2
      \left[ \frac{z^2}{4}\left( J_{\ell-2}+J_\ell \right)
                          \left( J_\ell+J_{\ell+2} \right)
             +J_{\ell-1}J_{\ell+1}
      \right]\nonumber\\
  &&\qquad +
      \left[ \frac{z^2}{4}\left( J_\ell+J_{\ell+2} \right)^2
             -2\frac{z}{2}\left( J_\ell+J_{\ell+2} \right)J_{\ell+1}
             +J^2_{\ell+1}
      \right]
    \Bigr\}, \label{aqh}
\end{eqnarray}
so that application of our theorems to this last equation yields
\begin{eqnarray}
\sum_\ell \ell^4 J^2_\ell
  &=&\frac{z^2}{4}
    \Bigl\{
      \left[ \frac{z^2}{4}\left( 1+2\cdot 0+1 \right)
             +z\left( 0+0 \right)
             +1
      \right]\nonumber\\
  &&\qquad +2
      \left[ \frac{z^2}{4}\left( 0+0+1+0 \right)
             +0
      \right]\nonumber\\
  &&\qquad +
      \left[ \frac{z^2}{4}\left( 1+2\cdot 0+1 \right)
             -z\left( 0+0 \right)
             +1
      \right]
    \Bigr\}\nonumber\\
  &=&\frac{z^2}{2}+\frac{3z^4}{8}.
  \label{aqn}
\end{eqnarray}
Thus, we finally get
\begin{equation}
\sum_\ell \ell^4 J_\ell J'_\ell=\frac{z}{2}+\frac{3z^3}{4}.
\end{equation}

The following is a list of useful results that can be established in
the above manner:
\begin{eqnarray}
\sum_\ell && J_\ell {\cal J}^{+}_\ell 
  =0
  \label{aqi}\\
\sum_\ell && \ell J_\ell {\cal J}^{+}_\ell
  =-\frac{\rho}{\sqrt{2}\lambda_B} F_0 \cdot {\bf k}
  \label{aqj}\\
\sum_\ell && {\cal J}^{- *}_\ell {\cal J}^{+}_\ell 
  =\frac{i}{\lambda_B} F_0
  \label{aqk}\\
\sum_\ell && \ell {\cal J}^{- *}_\ell {\cal J}^{+}_\ell
  =P_\perp
  \label{aql}\\
\sum_\ell && \left( J_{\ell-1}-J_{\ell+1} \right) {\cal J}^{- *}_\ell
  =-\frac{\sqrt{2}}{k_\perp\lambda_B} F_0\cdot {\bf k}.
  \label{aqm}
\end{eqnarray}
These sum rules are needed in the proof that the results for $K_2$ in
Eqs.~(\ref{obz}), (\ref{oav}) and (\ref{oby}) are indeed the same.

%% file: thesis.bbl
\begin{thebibliography}{99}
\textheight=8.4truein
\thispagestyle{myheadings}
\addcontentsline{toc}{chapter}{Bibliography}
\parindent=.6em
\bibitem{zbw} Dewar, R. L., {\it J. Phys. A:  Math. Gen.}, {\bf 9}:2043 (1976).
\bibitem{zbj} Johnston, S., Kaufman, A. N., {\it J. Plasma Physics},
{\bf 22}:105 (1979).
\bibitem{zbi} Johnston, S., {\it Phys. Fluids}, {\bf 19} (1976).
\bibitem{zbk} Cary, J. R., ``Nonlinear Wave Evolution in Vlasov Plasma:
A Lie-Transform Analysis,'' PhD Dissertation, University of California at
Berkeley, University Microfilms International number 80-14628,
or Lawrence Berkeley Laboratory Report LBL-8185 (August, 1979).
\bibitem{zbl} Cary, J. R., Kaufman, A. N., {\it Phys. Fluids}, {\bf 24} (1981).
\bibitem{zbm} Littlejohn, R. G., ``Hamiltonian Theory
of Guiding-Center Motion,'' PhD Dissertation, University of California at
Berkeley, University Microfilms International number 80-29478,
or Lawrence Berkeley Laboratory Report LBL-12942 (May, 1980).
\bibitem{zaq} Littlejohn, R. G., {\it J. Plasma Physics}, {\bf 29}:111-125 (1983).
\bibitem{zbn}Grebogi, C., Littlejohn, R. G., {\it Phys. Fluids},
{\bf 27}:1996 (1984).
\bibitem{zbo} Cary, J. R., Newberger, B. S., {\it Phys. Fluids},
{\bf 28}:423 (1985).
\bibitem{zbp} Dubin, D. H. E., Krommes, J. A., Oberman, C., Lee, W. W.,
{\it Phys. Fluids}, {\bf 26}:3524 (1983).
\bibitem{zbq} Kaufman, A. N., Boghosian, B. M., in {\it Contemporary
Mathematics,} American Mathematical Society, Providence, R.I., {\bf 28}, {\it
Fluids and Plasmas:  Geometry and Dynamics,} edited by J. E. Marsden (1984).
\bibitem{zbr} Similon, P. L., {\it Phys. Lett.}, {\bf 112}A:33 (1985).
\bibitem{zae} Fradkin, D. M., {\it J. Phys. A: Math. Gen.}, 
{\bf 11}:1069-1086 (1978).
\bibitem{zbs} Dumais, J.-F., {\it Am. J. Phys.}, {\bf 53}:264 (1985).
\bibitem{zaa} Schutz, B. F., {\it Geometrical Methods of
Mathematical Physics}, Cambridge University Press (1980).
\bibitem{zab} Edelen, D. G. B., {\it Applied Exterior Calculus},
John Wiley and Sons (1985).
\bibitem{zac} Singer, I. M. and Thorpe J. A., {\it Lecture Notes on
Elementary Topology and Geometry}, Springer-Verlag (1967).
\bibitem{zad} Burke, W. L., {\it Applied Differential Geometry},
Cambridge University Press (1985).
\bibitem{zbc} Flanders, H., {\it Differential Forms}, Academic Press,
New York (1963).
\bibitem{zbd} Misner, C. W., Thorne, K. S., Wheeler, J. A., {\it Gravitation},
W. H. Freeman and Company (1973).
\bibitem{zbe} Chandrasekhar, S., {\it The Mathematical Theory of Black Holes},
Oxford University Press, New York (1983).
\bibitem{zas} Littlejohn, R. G., {\it J. Math. Phys.} {\bf 20}:2445 (1979).
\bibitem{zaz} Marsden, J. E., Weinstein, A., Ratiu, T., Schmid R. and
Spencer, R. G., in the {\it Proceedings of the IUTAM Symposium on Modern 
Developments in Analytical Mechanics,} Torino, Italy (June 7-11, 1982).
\bibitem{zbb} Goldstein, H., {\it Classical Mechanics}, second edition,
Addison-Wesley, Chapters 4 and 5 (1980).
\bibitem{zbf} Arnold, V. I., {\it Ann. Inst. Fourier, Grenoble},
{\bf 16}:319 (1966).
\bibitem{zbg} Low, F. E., {\it Proc. Roy. Soc. A}, {\bf 248}:282 (1958).
\bibitem{zbx} Iwinski, Z. R. and Turski, L. A., {\it Lett. in Appl. and Eng. Sci.},
{\bf 4}:179 (1976).
\bibitem{zbh} Morrison P. J., {\it Phys. Lett.}, {\bf 80A}:383 (1980).
Footnote 1 credits Kaufman.
\bibitem{zgb} Gibbons, J., {\it Physica}, {\bf 3D}:503 (1981).
\bibitem{zgc} Kaufman, A. N. and Dewar, R. L., in {\it Contemporary Mathematics,}
American Mathematical Society, Providence, R.I., {\bf 28}, {\it Fluids
and Plasmas:  Geometry and Dynamics,} edited by J. E. Marsden (1984).
\bibitem{zat} Littlejohn, R. G., in {\it Mathematical Methods in
Hydrodynamics and Integrability in Dynamical Systems,} edited by M. Tabor
and Y. M. Treve, American Institute of Physics Conference Proceedings
Number 88 (New York, 1982).
\bibitem{zba} Weinstein, A., in {\it Contemporary Mathematics,}
American Mathematical Society, Providence, R.I., {\bf 28}, {\it Fluids
and Plasmas:  Geometry and Dynamics,} edited by J. E. Marsden (1984).
\bibitem{zar} Littlejohn, R. G., {\it J. Math. Phys.} {\bf 23}:742 (1982).
\bibitem{zaf} Littlejohn, R. G. and Cary, J. R., {\it Annals
of Physics}, {\bf 151}:1-34 (1983).
\bibitem{zbt} Dirac, P. A. M., {\it Lectures on Quantum Mechanics}, Belfer
Graduate School of Science, Yeshiva University, New York (1964).
\bibitem{zai} Deprit, A., {\it Celestial Mechanics},
{\bf 1}:12-30 (1969).
\bibitem{zaj} Dragt, A. J. and Finn, J. M., {\it J. Math. Phys.},
{\bf 17}:2215-2227 (1976).
\bibitem{zau} Cary, J. R.,{\it Phys. Rep.}, {\bf 79}:131 (1981). 
\bibitem{zbu} Kolmogorov, A. N., {\it Dokl. Akad. Nauk. SSSR},
{\bf 98}:527 (1954).
\bibitem{zao} Hori, G., {\it Pub. Astron. Soc. Japan},
{\bf 18}:287 (1966).
\bibitem{zak} Littlejohn, R. G., in {\it Contemporary 
Mathematics}, American Mathematical Society, Providence, R. I.,
{\bf 28}:151-167, {\it Fluids and Plasmas:  Geometry and Dynamics},
edited by J. E. Marsden (1984).
\bibitem{zan} Hagan, W. K. and Frieman E. A., {\it Phys. Fluids},
{\bf 28}:2641-2643 (1985).
\bibitem{zap} Northrop, T. G., {\it The Adiabatic Motion of Charged
Particles,} Interscience, New York (1963).
\bibitem{zag} Jackson, J. D., {\it Classical Electrodynamics},
second edition, John Wiley and Sons, pp. 503-578 (1975).
\bibitem{zal} Kruskal, M., in {\it Plasma Physics}, I.A.E.A.,
Vienna, pp. 67-102 (1965).
\bibitem{zam} Kruskal, M., in {\it Mathematical Models in the
Physical Sciences}, Eaglewood Cliffs, N.J., edited by S. Drobot,
pp. 17-48 (1963).
\bibitem{zcc} Kaufman, A. N., {\it Phys. Fluids}, {\bf 3}:610 (1960).
\bibitem{zby} Vandervoort, P., {\it Ann. Phys.}, {\bf 10}:401 (1960).
\bibitem{zcd} Achterberg, A., {\it J. Plasma Physics}, {\bf 35}:257 (1986).
\bibitem{zcb} Soper, D. E., {\it Classical Field Theory},
John Wiley and Sons (1978).
\bibitem{zbz} Dewar, R. L., {\it Physica}, {\bf 17}D:37 (1985).
\bibitem{zca} Hehl, F. W., {\it Spin and Torsion in General Relativity:
I. Foundations}, {\it Gen. Rel. and Grav.}, {\bf 4}:333 (1973).
\end{thebibliography}
